\documentclass{aa}
\pdfoutput=1

\usepackage{graphicx}
\usepackage{txfonts}
\usepackage{pictex}
\usepackage{latexsym}
\usepackage{graphics}
\usepackage{afterpage}
\usepackage{units}

\DeclareUnicodeCharacter{2212}{-}

\begin{document}

   \title{The galaxy population within the virial radius of the \\ 
          Perseus cluster\thanks{The full galaxy catalogue is only available in electronic form
          at the CDS via anonymous ftp to cdsarc.u-strasbg.fr (130.79.128.5) or via http://cdsweb.u-strasbg.fr/cgi-bin/qcat?J/A+A/}
          }

   \author{H. Meusinger\inst{1,2}
                    \and
          C. Rudolf\inst{1}
                    \and
          B. Stecklum\inst{1}   
                    \and 
          M. Hoeft\inst{1}   
                    \and           
          R. Mauersberger\inst{3}
                    \and
          D. Apai\inst{4}          
          }

   \institute{Th\"uringer Landessternwarte, Sternwarte 5, 07778 Tautenburg, Germany, 
              \email{meus@tls-tautenburg.de}
         \and
              Universit\"at Leipzig, Fakult\"at f\"ur Physik und Geowissenschaften, Linnestra{\ss}e 5, 04103 Leipzig, Germany 
         \and 
              Max Planck Institute for Radio Astronomy, Auf dem H\"ugel 69, 53121 Bonn (Endenich),
              Germany
         \and
              Steward Observatory and the Lunar and Planetary Laboratory, The University of Arizona, Tucson, AZ 85721, USA
             }

   \date{Received XXXXXX; accepted XXXXXX}

 
  \abstract
   {
The Perseus cluster is one of the most massive nearby galaxy clusters and is fascinating in various respects. 
Though the galaxies in the central cluster region have been intensively investigated, an analysis of the galaxy population in a larger field is still outstanding.  
  }
  {
This paper investigates the galaxies that are brighter than $B \approx 20$ within a field corresponding to the Abell radius of the Perseus cluster. Our first aim is to compile a new catalogue in a wide field around the centre of the Perseus cluster. The second aim of this study is to employ this catalogue for a systematic study of the cluster galaxy population with an emphasis on morphology and activity.
  }
  {
We selected the galaxies in a 10 square degrees field of the Perseus cluster on Schmidt CCD images in B and H$\alpha$ in combination with SDSS images. 
Morphological information was obtained both from the `eyeball' inspection and the surface brightness profile analysis.
We obtained low-resolution spectra for 82 galaxies and exploited the spectra archive of SDSS and redshift data from the literature. 
}
  {
We present a catalogue of 1294 galaxies with morphological information for $90$\% of the galaxies and spectroscopic redshifts for $24$\% of them.   
We selected a heterogeneous sample of 313 spectroscopically confirmed cluster members and
two different magnitude-limited samples with incomplete redshift data. These galaxy samples were used to derive such properties as the projected radial velocity dispersion profile, projected radial density profile, galaxy luminosity function, supermassive black hole mass function, total stellar mass, virial mass, and virial radius, 
to search for indications of substructure, to select active galaxies, and to study the relation between morphology, activity, density, and position. 
In addition, we present brief individual descriptions of 18 cluster galaxies with conspicuous morphological peculiarities.  
   }
   {}

  \titlerunning{Galaxies in the Perseus cluster field}
  \authorrunning{H. Meusinger et al.}

   \maketitle
%
%

%

%
\section{Introduction}\label{sect:Introduction}
%

Galaxy clusters are the largest gravitationally bound structures in the Universe and they are powerful tracers of the structures of matter on the largest scales. The mass of galaxy clusters is thought to be dominated by an unseen and presumably collisionless component of dark matter ($\sim 85$\%) and a  diffuse, hot intracluster gas (intracluster medium, or ICM). The baryonic matter in the cluster galaxies contributes only a few per cent to the cluster mass. However, rich clusters host enough galaxies to provide a fascinating laboratory for studying not only cluster properties, but also key processes of galaxy evolution under diverse external influences.

Understanding the role of environmental effects on galaxy properties is crucial to understand galaxy evolution. In the local Universe, the galaxies populate two distinct areas in diagrams showing colour versus luminosity or star formation rate (SFR) versus stellar mass: a red sequence of passive, mostly early-type galaxies and a blue cloud of star-forming galaxies, which are mostly of late morphological types
\citep{Strateva_2001, Blanton_2003, Baldry_2004, Eales_2018}. 
Many studies have found that galaxy properties depend on local density, with red-sequence galaxies preferably found in dense regions of galaxy clusters and  blue-cloud  galaxies  in  less  dense environments 
\citep[e.g.][]{Dressler_1980, Postman_1984, Balogh_2004, Rawle_2013, Odekon_2018, Liu_2019, Mishra_2019}, 
though other studies concluded that environment only has a little effect once the morphology and luminosity of a galaxy is fixed \citep[e.g.][]{Park_2008, Davies_2019, Man_2019}.
It has also been suggested that large-scale structure has an important affect on galaxy evolution beyond the known trends with local density \citep[e.g.][]{Rojas_2004, Hoyle_2005, Fadda_2008, Biviano_2011, Koyama_2011, Chen_2017, Vulcani_2019}. 

In a cluster environment, galaxies are subject to interactions with other galaxies, with the cluster gravitational potential well, and with the ICM \citep[for a review see][]{Boselli_2006}. These processes may influence fundamental properties like SFR, stellar mass assembly, colour, luminosity, and morphology, and they may cause galaxies to migrate from the blue cloud to the red sequence, and probably back.
Tidal interactions and mergers of galaxies, a key mechanism of structure evolution in hierarchical models, can act as triggers for star formation \citep[SF;][]{Toomre_1972, Larson_1978, Barton_2000, Hopkins_2008, Holincheck_2016} and can perturb the structure of the involved galaxies on time-scales of the order of a gigayear
\citep[Gyr; e.g.][]{Mihos_1996, DiMatteo_2007, Duc_2013}. 
Slow encounters and major mergers are unlikely to occur in the dense cluster core regions where the velocity dispersion is high ($\sim 1000$ km\,s$^{-1}$), 
but may play a role in the cluster outskirts where galaxy groups are falling into the cluster potential. 
As a galaxy passes through the ICM, ram pressure can strip out its interstellar medium 
\citep{Gunn_1972} and gas-rich galaxies can lose more and more of their cold gas reservoir for SF.  
Ram pressure stripping seems evident both from dynamical simulations \citep[e.g.][]{Abadi_1999, Roediger_2006} and
from observations of cluster galaxies \citep[e.g.][]{Oosterloo_2005, Lee_2017} that show tails of neutral hydrogen and also of molecular gas clouds. 
Interactions may also stall the supply of external gas. 
Such `strangulation' or `starvation' has been identified as one of the main mechanisms for quenching star formation in cluster galaxies \citep[e.g.][]{Larson_1980, Peng_2015}, 
although the exact physical mechanism is still unknown.
The combined effect of tidal interactions of an infalling galaxy with the cluster potential and of subsequent high-speed encounters with other cluster members (`galaxy harassment') 
is also thought to be suitable for quenching the SF activity and causing morphological transformations from late-type to early types 
\citep[e.g.][]{Moore_1996, Bialas_2015}. 
Other relevant processes include  minor mergers
\citep[e.g.][]{Kaviraj_2014, Martin_2018} and gas outflow caused by the feedback of an active galactic nucleus \citep[AGN; e.g.][]{Fabian_2012, Heckman_2014, Harrison_2018}. 
Although the relative importance of the various processes remains a matter of debate, 
it is very likely that at least some of them play a role in the evolution of cluster galaxies.
The environmental dependence of galaxy properties is discernible from the comparison of the galaxy populations in the dense core and the outer regions where infalling galaxies begin to experience the influence of the cluster environment 
\citep[e.g.][]{Diaferio_1997, Meusinger_2000, Verdugo_2008, Behroozi_2014, Zinger_2018}. 

The present paper deals with the Perseus cluster (\object{Abell\,426}; hereafter A\,426), a nearby ($z \approx 0.0179$), massive, rich galaxy cluster of Bautz-Morgan class II-III and Rood-Sastry type L \citep{Struble_1999}. 
It has long been known that the galaxy population in the core region of the Perseus cluster is dominated by passive, early-type galaxies and exhibits an exceptionally strong deficiency of spirals \citep[e.g.][]{Kent_1983}.  
The Perseus cluster marks the east end of the about 100 Mpc large Perseus-Pisces supercluster and lies
close to the border of the Taurus void \citep{Batuski_1985}. There is observational evidence of the galaxy population in the surroundings of cosmic voids to have a higher specific star formation rates (sSFR) and a higher percentage of later 
morphological types compared to regions of higher density, 
which is probably not fully explained by the increased proportion of low-mass galaxies in low-density regions \citep[e.g.][]{Moorman_2016, Ricciardelli_2017}. 

The Perseus cluster is the brightest cluster of galaxies in the X-ray sky \citep{Edge_1990}. 
The X-ray emission from the hot intracluster medium (ICM), which dominates the baryonic mass in the cluster,
is sharply peaked on NGC\,1275, but the X-ray centroid is not identical with the centre of the optical galaxy \citep{Branduardi_1981}. 
Because the radiative cooling time of the ICM in the core is much less than the Hubble time, a cooling flow of more than 100 $M_\odot$\,yr$^{-1}$ has been predicted if there would be no re-heating source \citep{Fabian_1994}.
The Perseus cluster was one of the first galaxy clusters where an example of energetic feedback from the central active galactic nucleus (AGN) 
was observed: Cavities in the ICM are filled with radio emission from the AGN, which is pumping out relativistic electrons into the surrounding X-ray gas \citep{Branduardi_1981, Boehringer_1993, Fabian_2011a}.
Deep Chandra images revealed fine structures in the ICM, such as bubbles, ripples and weak shock fronts within about $100$\,kpc from the core that are thought to be related to the activity of the AGN in NGC\,1275 and to the re-heating of the cooling gas \citep{Mathews_2006, Fabian_2011a}. 

Although appearing relatively relaxed, e.g. compared to other nearby clusters, 
there is some indication that the Perseus cluster is not yet virialised: 
the deviation from spherical symmetry indicated by the chain of bright galaxies,
the non-uniform distribution of morphological types \citep{Andreon_1994}, 
the displacement of the X-ray centroid from the centre of the optical galaxy positions \citep{Ulmer_1992},
and the large-scale substructures in the X-ray emission \citep{Churazov_2003}.  
The global east-west asymmetry, along with other structures in the distributions 
of the temperature and surface brightness of the hot gas can be modelled as due to an ongoing merger close to the direction of the chain of bright galaxies \citep{Churazov_2003, Simionescu_2012}.  
A spiral-like feature extending from the core outwards is interpreted as a sloshing cold front  \citep[][and references therein]{Ichinohe_2019}.
Gas sloshing observed on larger scale might be related to a disturbance of the cluster potential caused by a recent or ongoing merger \citep{Markevitch_2007}.
Cluster mergers produce large-scale shock waves in the intracluster medium (ICM) where frequently Mach numbers two to four are measured. These shock fronts may affect the interstellar medium and thus the SF properties of cluster galaxies \citep{Mulroy_2017}.

It has long been known that the Perseus cluster harbours a radio mini-halo \citep[e.g.][]{Ryle_1968, Soboleva_1983}. 
The rare phenomenon of a radio mini-halo typically occurs in a cool-core cluster and is thought to characterise a relaxed system.
\citet{Fabian_2011a} noted interesting correlations between structures in the radio and X-ray maps. 
More recently, \citet{Gendron_Marsolais_2017} presented low-frequency VLA observations at 230-470 MHz 
that reveal a multitude of structures related to the mini-halo in the Perseus cluster.  
These structures seem to indicate the influence of both the AGN activity and 
a sloshing motion of the cool core cluster's gas. 
In addition, these authors argue that there is a filamentary structure similar to that seen in radio relics, which are found in merging clusters only where they are thought to be related to shock fronts. There is, however, no indication of a shock front at this position in Perseus. 
A several Mpc long diffuse polarisation structure has been detected at 350 MHz and tentatively attributed to a 
shock front caused by gas infall into the Perseus cluster along the Perseus-Pisces filamentary structure \citep{deBruyn_2005}. 
However, though confirmed in a re-investigation, this structure was found to be more likely related to the Galactic foreground than to the Perseus cluster \citep{Brentjens_2011}. 

The brightest cluster galaxy (BCG) NGC\,1275, close to the centre of the Perseus cluster, exhibits a wide range of peculiarities and has been the subject of a huge number of studies. 
(The NASA/IPAC Extragalactic Database\footnote{https://ned.ipac.caltech.edu/}, NED, lists 1892 references.) 
Its morphological classification varies between the types pec (peculiar), E pec, cD in NED, and S0 in HyperLeda. NGC\,1275 is the host of the strong radio source 3C84 and of an optical AGN of type S1.5 powered by a supermassive black hole (SMBH) of approximately $8\,10^8\ M_\odot$ \citep{Scharwachter_2013}. 
The galaxy is surrounded by a huge, complex system of emission-line filaments \citep{Minkowski_1957, Lynds_1970, Conselice_2001, Gendron_Marsolais_2018}. 
A high-velocity filament system has been related to the infall of a giant low-surface brightness galaxy towards the cluster centre \citep{Boroson_1990, Yu_2015}.
The low-velocity system is apparently dragged out by the rising bubbles of relativistic plasma from the AGN and supported by magnetic fields \citep{Hatch_2006, Fabian_2008, Fabian_2011b}. 
BCGs are often found to possess extended emission line nebulae and filaments that appear to be related to the ICM \citep{Hamer_2016}.
The morphological diversity of the filamentary and clumpy emission, 
as revealed by HST far-ultraviolet imaging, is reproduced by hydrodynamical simulations in which the AGN powers a self-regulating process of SF in gas clouds that precipitate from the hot atmosphere \citep{Tremblay_2015}. 

The galaxy population in the Perseus cluster core region has been investigated in several studies. 
\citet{dePropris_1998} derived the I-band luminosity function down to $I = 20$ in a 147\,arcmin$^2$ field centred on NGC\,1275. 
\citet{Conselice_2002} analysed the galaxy population in a field of about $170$\,arcmin$^{2}$ that is centred on the second brightest cluster galaxy, NGC\,1272, observed with the WIYN 3.5 m telescope in U, B and R. They found that the galaxies in the cluster centre are mostly relaxed and composed of old stellar populations, whereas only a few galaxies were found that are unusual and undergoing rapid evolution. Based on the same observations, \citet{Conselice_2003} demonstrated that the low-mass cluster galaxies  have multiple origins where roughly half of them have stellar populations and kinematic properties consistent with being the remnants of stripped galaxies accreted into clusters several Gyr ago.
\citet{Penny_2014} provided evidence that SA 0426-002, one of the peculiar galaxies from \citet{Conselice_2002}, has morphologically transformed from a low-mass disc to dE via harassment. 
Based on HST imaging, \citet{Rijcke_2009} investigated photometric scaling relations for early-type galaxies, ranging from dwarf spheroidals, over dwarf elliptical galaxies, up to giant ellipticals, in different environments including the Perseus cluster core. \citet{Wittmann_2017} identified ultra-diffuse low-surface brightness galaxies on deep V band images of a 0.3\,deg$^2$ field observed with the 4\,m William Herschel Telescope and concluded that these galaxies cannot survive in the central region of the cluster because of the strong tidal forces there. Primarily based on the same observational material, \citet{Wittmann_2019} presented a large and deep catalogue of 5437 morphologically classified sources.

Because of its size and proximity, the Perseus cluster covers a large sky area, 
the angular size of the Abell diameter 
\footnote{The Abell radius is $R_{\rm A} = 1.5 h^{-1}$\,Mpc, i.e  2.14\,Mpc for $h = 0.7$.} 
amounts to approximately $3.3$ degrees. 
The wide field, in combination with the uncomfortable low Galactic latitude ($b \approx -15$\degr), is probably
the reason why the galaxy population in a larger area of the cluster has been only marginally researched. 
Studies of the galaxy population that include the outer cluster regions were based on relatively shallow surveys with Schmidt telescopes \citep{Melnick_1977, Bucknell_1979, Kent_1983, Poulain_1992, Andreon_1994}. 
\citet{Kent_1983} analysed the structure and dynamics of the Perseus cluster based on a combined sample of 201 galaxies with spectroscopic redshifts including a complete sample of 119 galaxies brighter than $m_p \approx 16$ within three degrees from the cluster core selected on photographic plates from the Palomar 122\,cm Schmidt telescope.
\citet{Brunzendorf_1999} used co-added digitised photographic plates from the Tautenburg 134\,cm Schmidt camera to
compile a morphological catalogue of 660 galaxies brighter than $B \approx 19.5$ in a field of about 10 square degrees centred on a position about 13 arcmin west of NGC\,1275. The catalogue was used in a subsequent study on star-forming galaxies identified by their far-infrared (FIR) emission \citep{Meusinger_2000}. 

The aim of the present study is twofold: first, to create a revised compilation of galaxies within about one Abell radius of the Perseus cluster and secondly to provide a statistical analysis of this new database. 
The new catalogue is based on a combination of CCD imaging and spectroscopic observations from the 
Tautenburg 2\,m telescope, the Sloan Digital Sky Survey \citep[SDSS,][]{York_2000, Abolfathi_2018}, and other telescopes. The major improvements over the previous database are a fainter magnitude limit, which leads to nearly twice as many galaxies, an improved morphological classification, a significantly increased number of galaxies with redshift information, and the combination with photometric data in the near infrared (NIR) and mid infrared (MIR) from 2MASS \citep{Skrutzkie_2006} and WISE \citep{Wright_2010}. 

The paper is structured as follows. The observational database is described in 
Sect.\,\ref{Observations}, followed by the description of the catalogue in Sect.\,\ref{Catalogue} and the selection of suitable cluster galaxy samples in Sect.\,\ref{sect:galaxy_samples}. 
These galaxy samples are then used to analyse the cluster profiles and substructures
(Sect.\,\ref{sect:Cluster_profiles}), the morphology-density-position relation, the galaxy luminosity function, and the mass function of supermassive black holes  
(Sect.\,\ref{sect:Cluster_galaxies}). Some properties of the sub-samples of star-forming galaxies and AGN galaxies are discussed separately in Sect.\,\ref{sect:Active}, 
Morphologically peculiar galaxies are the subject of Sect.\,\ref{sect:peculiar}. 
particularly interesting systems are discussed individually in Appendix\,\ref{sect:individual}. 
Finally, the results are summarised in Sect.\,\ref{sect:summary}.

%
\section{Observational data}\label{Observations}
%

\subsection{Imaging}\label{Imaging}

\subsubsection{Tautenburg Schmidt CCD observations}\label{sec:TLS_imaging}

We observed the Perseus cluster field repeatedly with the CCD camera in the Schmidt focus of the TLS Tautenburg 
2\,m Alfred Jensch telescope in the framework of a long-term monitoring programme. In its Schmidt version, the telescope has a free aperture of 134\,cm, an unvignetted field of $3\degr3$, and a scale of 51.4 arcsec/mm. A 2k $\times$ 2k SITe\#T5b CCD chip was used with 24\,$\mu$m pixels and a field size of $42\arcmin \times 42\arcmin$. 
Because this programme continued a preceding monitoring with photographic plates in the B band, the same band was used also for the CCD imaging. 
The survey field was centred on NGC\,1275 and adapted to the $10$ square degrees field of the photographic plates. The field was covered by $5 \times 5$ overlapping CCD frames.
The CCD monitoring consisted of seven campaigns between 2007 and 2013. On average, each sub-field was observed 17 times with a higher cadence for the central fields. The exposure time was 300\,sec per single exposure.  Dome flats were taken for the flat-field correction.
The usual technique for debiasing, flat-field correction and cosmic filtering was applied. 
Finally, to improve the signal-to-noise ratio, the corrected images of each sub-field were co-added using weighting factors that take account of the different quality  \citep{Froebrich_2000}.
Although about one third of the observations were taken under only moderate sky conditions, the quality-weighted co-addition results in images that are about one mag deeper than the best single exposures. These co-adds are useful for the galaxy selection and the evaluation of faint extended morphological features, but suffer from an only moderate spatial resolution. The astrometric calibration was based on the USNO-B1.0 catalogue \citep{Monet_2003} and was performed with the Graphical Astronomy and Image Analysis Tool (GAIA)\footnote{http://star-www.dur.ac.uk/~pdraper/gaia/gaia.html}.

\begin{figure}[htbp]
\begin{center}
\fbox{\includegraphics[viewport= 3 5 466 435,width=7.0cm,angle=0]{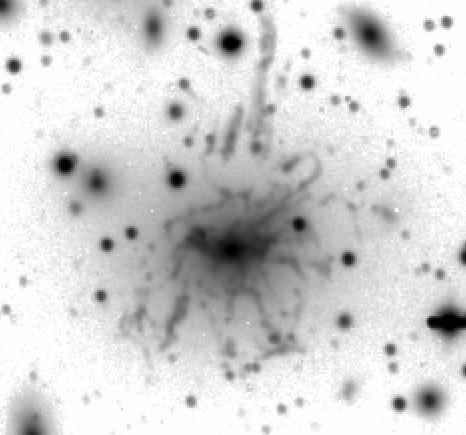}}
\end{center}
\vspace{0.3cm}
\caption[NGC1275]{H$\alpha$ image of the $4\arcmin5 \times 4\arcmin5$ region of the Perseus cluster around NGC\,1275 taken with the Tautenburg Schmidt camera.}
\label{fig:NGC1275_Halpha}
\end{figure}

In addition to the B band imaging, the Perseus cluster field was observed through a narrow-band filter centred on redshifted H$\alpha$ at 6670\,\AA\ with a FWHM of 100\,\AA.
At least two (but up to five) 600\,sec exposures were taken per sub-field. 
For a rough continuum subtraction, additional R band images were obtained with exposure times of 150\,sec per frame. The most eye-catching phenomenon in the H$\alpha$ image is of course the well-known giant system of emission-line filaments around NGC\,1275 (Fig.\,\ref{fig:NGC1275_Halpha}).

\subsubsection{Sloan Digital Sky Survey}\label{sec:SDSS_imaging}

The Sloan Digital Sky Survey (SDSS) has mapped roughly a quarter of the high Galactic latitude sky in the five broad bands u,g,r,i, and z with the 2.5 m telescope at Apache Point Observatory \citep{York_2000}. The imaging scans were mostly taken under good seeing conditions in moonless photometric nights.  Although not part of the main surveys of the SDSS, the Perseus cluster region has been observed as one of the supplementary fields. The data has been made available with the Data Release 5 \citep[DR\,5][]{Adelman_2007} and DR\,6 \citep{Adelman_2008}. The imaging scans were centred roughly on the cluster core.

\subsubsection{CAFOS at Calar Alto}\label{sec:CA}

The focal reducer camera CAFOS at the 2.2\,m telescope of the Centro Astron\'omico Hispano-Aleman (CAHA)
at the Calar Alto (CA) observatory,
Spain, was used for the imaging of a relatively small number of galaxies under good seeing conditions. The galaxies were selected because of indications of uncommon morphology or activity. 19 galaxies were imaged for a study of IRAS galaxies in the Perseus cluster \citep{Meusinger_2000} and
about the same number of other galaxies were observed in subsequent observation runs at a typical seeing of  about $1\farcs0$ to $1\farcs5$.
CAFOS was equipped with a SITe detector with a pixel scale of $0\farcs5$/pixel. The B, V, and R filters were used. 
The  total exposure times were between 600\,s and 2000\,s.

\subsection{Spectroscopy}\label{Spectra}

\subsubsection{Sloan Digital Sky Survey}\label{sec:SDSS_spectra}

\begin{figure}[htbp]
\begin{center}
\includegraphics[viewport= 0 10 570 570,width=8.0cm,angle=0]{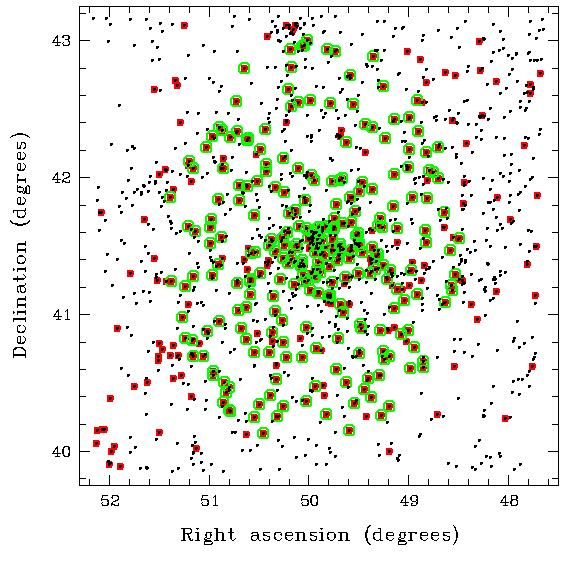}
\end{center}
\caption[Spectra field]{
Survey field with all galaxies in the final catalogue (black dots). The galaxies with redshift information from all available sources are labeled red. Larger green circles mark the galaxies with SDSS spectroscopy. 
}
\label{fig:Spectra_field}
\end{figure}

In addition to the imaging survey, SDSS obtained spectroscopy for huge number of galaxies. 
The SDSS DR\,14 \citep{Abolfathi_2018} contains useful spectra for roughly three million galaxies and quasars.
The imaging scans of the  Perseus cluster region were used by SDSS to target approximately 400 galaxies and 300 foreground F stars with the two special spectroscopic plates 1665 and 1666. The majority of the redshifts used in the present study comes from SDSS (Sect.\,\ref{sec:redshifts}). However, the two spectroscopic plates cover only about half the survey field (Fig.\,\ref{fig:Spectra_field}).

The spectra from the SDSS double fibre-fed spectrographs cover a wavelength range from 3\,800 \AA\ to 9\,200 \AA\ with a resolution of $2\,000$ and sampling of $2.4$ pixels per resolution element. 
For a galaxy near the main sample limit, the typical signal-to-noise ratio (S/N) is approximately $10$ per pixel. 
For the regular plates, the SDSS spectroscopic pipeline provides flux and wavelength calibrated spectra, redshifts, and line data, which are available both from the SDSS Explorer pages and from tables in the Catalogue Archive Server (CAS). 
For  special plates, however, not all measurements are available from the SDSS table {\tt specPhotoAll}\footnote{See 
{\tt http://classic.sdss.org/dr6/products/spectra/ index.html\#specialquery}}. 
We downloaded redshifts and equivalent widths of the lines H$\beta$, H$\alpha$, [\ion{O}{iii}]\,5007, and [\ion{N}{ii}]\,6584 via SQL queries from the CAS tables {\tt specObjAll} and {\tt galSpecLine}.

\subsubsection{Tautenburg and Calar Alto spectroscopy}\label{sec:TLS_CA_spectra}

Optical low-resolution spectroscopy data were collected also with the 2.2\,m CAHA telescope on Calar Alto and with the TLS Tautenburg 2\,m  telescope in 
several observation runs. All spectra were obtained in long-slit mode using a slit width of $1\arcsec$ to $2\arcsec$, depending on the seeing conditions. The slit centre was always positioned at the core of the galaxy.
 
For 17 galaxies spectra were obtained with CAFOS at the Calar Alto observatory in the framework of a backup programme. 
The grism B-400 was used with a wavelength coverage from $3600$\,\AA\ to 8000\,\AA\ and a dispersion of 10\,\AA/px on the SITe1d CCD. 
For six galaxies additional spectra were taken with the grisms R-200 (6000\,\AA\ to 11000\,\AA, 4\,\AA/px) 
and/or G-100 (4900\,\AA\ to 7800\,\AA, 2\,\AA/px) for the analysis of the H$\alpha$+[\ion{N}{ii}] region (Sect.\,\ref{sect:SF_WHAN_diagram}). 
With the Nasmyth Focal Reducer Spectrograph (NASPEC) at the TLS 2\,m telescope we observed
65 galaxies. The spectrograph was equipped with a  $2800 \times 800$ pixel SITe CCD. 
The V-200 grism was used in most cases, which gives a wavelength coverage from 4000\,\AA\ to 8500\,\AA\ and a 
spectral resolution of $\approx 600$ and 300 for the selected slit width of $2\arcsec$ and 
$3\arcsec$. 

The targets for the TLS and CA observations were primarily selected among the galaxies that were found to show some kind of distorted morphology. There is no explicit selection bias with regard to the distribution across the survey field. All spectra were reduced with standard routines from the ESO MIDAS data reduction package with standard procedures including bias subtraction, flat-fielding, cosmic ray removal, wavelength calibration, and a rough flux calibration. The optimal extraction algorithm of \citet{Horne_1986} was applied to extract one-dimensional spectra from the two-dimensional ones. The wavelength calibration of the CAFOS spectra was done with calibration lamps. For the TLS spectra, the wavelength calibration is based on night-sky lines in the target spectra following \citet{Osterbrock_1992}.

%
\section{The galaxy catalogue}\label{Catalogue}
%

\subsection{Galaxy selection}\label{sec:gal_selection}

Galaxies were selected, in a first step, by a combination of automated analysis and visual inspection of the co-added B band images from the 
Tautenburg Schmidt CCD camera. 
The automated source detection was performed running SExtractor \citep{Bertin_1996}. The 
SExtractor parameters were determined for each field separately by trial and error in an iterative process.  
The vast majority of the approximately $125\,000$ extracted sources are stars or faint background galaxies. Sources 
with a SExtractor parameter {\tt class\_star} $ < 0.1$ and larger than about $20\arcsec$ were selected as galaxies that possibly belong to the Perseus cluster. 

A high error rate was to be expected for the automated selection of galaxies at the low Galactic latitude of our  field. 
Therefore, the resulting galaxy catalogue was thoroughly checked in two separate visual inspections of the images. 
The aim of the first inspection was to find out and reject sources that were wrongly classified as galaxies, 
such as not de-blended merged star-like objects, faint stars projected on an outer part of a faint and small galaxy, or artefacts due to an imperfect flat-field correction. 
A second inspection was performed to search for galaxies that were not selected by the SExtractor. 
This was the case, for instance, when a foreground star is projected onto the core of a galaxy, 
which leads to a {\tt class\_star} parameter $ > 0.1$ or when a faint galaxy is close to bright foreground star. 
Finally, we checked the SDSS Object Explorer page for every single source.
The final catalogue contains 1294 galaxies. The catalogue will be published in electronic form in the VizieR Service of the CDS Strasbourg. A column-by-column description is given in Appendix\,\ref{sect:catalogue}.

\begin{figure}[htbp]
\includegraphics[viewport= 25 30 570 580,width=6.5cm,angle=270]{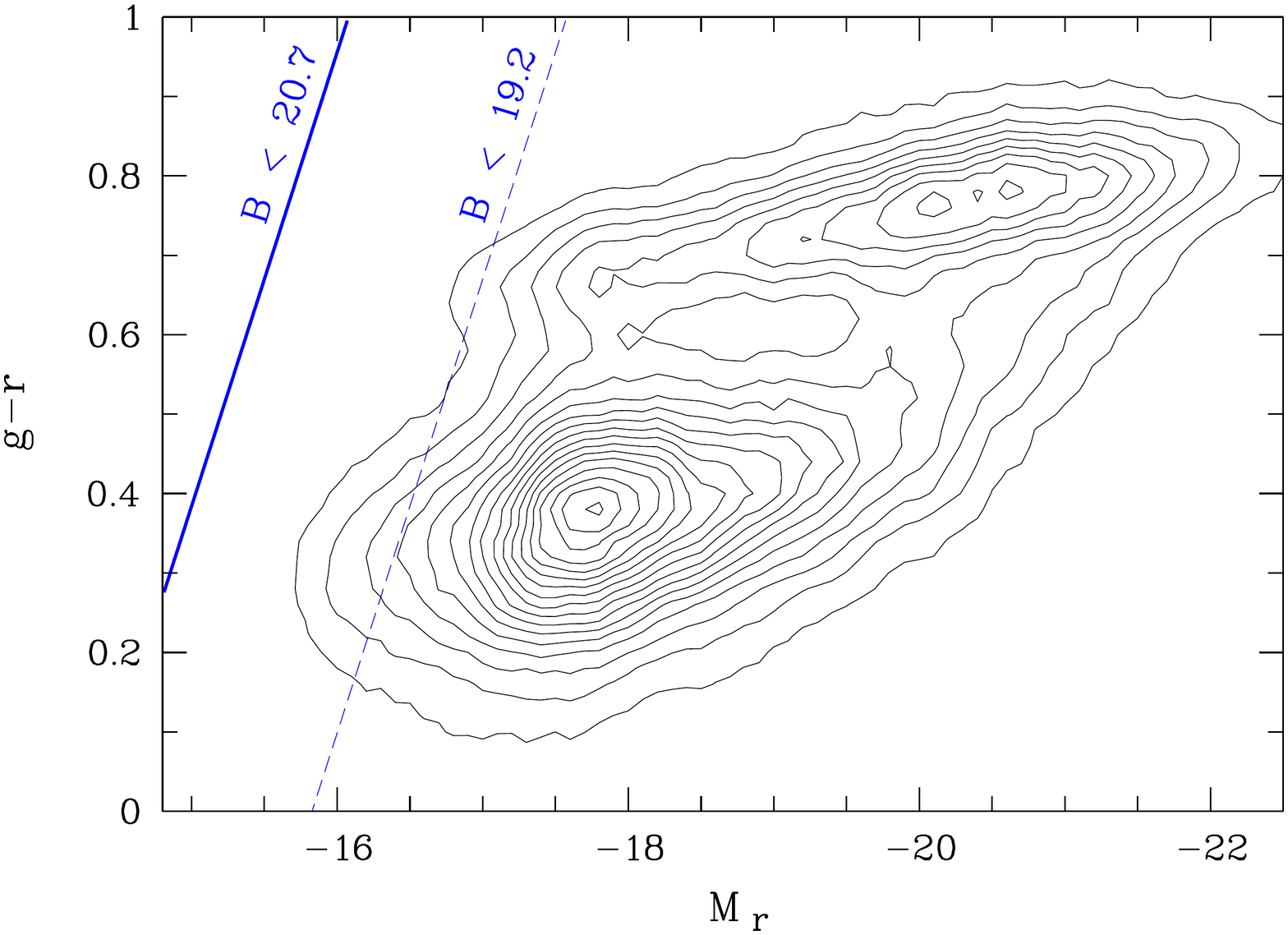}
\caption[CMD_gr_selection]{
Colour magnitude diagram for 24\,155 SDSS galaxies (contours) in the redshift range $z = 0.010 \ldots 0.26$. 
The blue diagonal lines mark the completeness limit (dashed) and the detection threshold (solid) of the galaxy selection.
}
\label{fig:CMD_gr_selection}
\end{figure}
 
The galaxy selection in the B band guarantees a high completeness of the population of blue, 
star forming galaxies without producing a substantial selection bias against red galaxies. Figure\,\ref{fig:CMD_gr_selection} 
shows the distribution of 24\,155 nearby SDSS galaxies in the $(g-r)$ versus $M_r$
colour-magnitude diagram (CMD) represented by equally spaced local point density contours computed for a grid size of 
$\Delta M_r = 0.1$\,mag and $\Delta (u-r) = 0.05$\,mag.
The bimodality of the SDSS galaxy population is clearly indicated where the galaxies from the red sequence are typically an order of magnitude more luminous than those from the blue cloud. 
Using the B magnitudes from \citet{Brunzendorf_1999} we found the transformation relation $B = g + 0.75\,(g-r) + 0.07$ for the extinction corrected magnitudes. 
A magnitude cut in the B band produces therewith a colour-dependent limiting absolute r magnitude 
$M_{\rm r,lim} = B_{\rm lim, obs} -1.75\,(g-r) - 35.03$ for the cluster galaxies, 
where the distance modulus $DM = 34.34$\,mag (Sect.\,\ref{sect:redshift}) and a mean reddening $E(B-V) = 0.15$\,mag were adopted,
$B_{\rm lim, obs}$ is the observed (not corrected for Galactic foreground extinction) B-band magnitude limit.
This relation is plotted in Fig.\,\ref{fig:CMD_gr_selection} for the estimated completeness limit at $B_{\rm lim, obs} = 19.2$ and the detection limit at $20.7$. Both limits are not sharp for several reasons.
The completeness limit corresponds to the maximum in the distribution of the W1 magnitudes and the detection limit is defined by the strong decline at $W1 \approx 16.5$ (Fig.\,\ref{fig:hist_W1}).

\subsection{Photometric data}\label{sect:obs_photo}

We used photometric data from the SDSS Photometric Catalogue \citep{Alam_2015}, 
the Extended Source Catalogue \citep[2MASS XSC;][]{Jarrett_2000} from the Two-Micron All-Sky Survey \citep[2MASS;][]{Skrutzkie_2006}, 
and the All-Sky Source Catalogue from the Wide-field Infrared Survey Explorer \citep[WISE;][]{Wright_2010}.  

The SDSS cautions that supplementary fields of low Galactic latitude, such as A\,426, are highly crowded and of high extinction. These data were processed with the standard SDSS photo pipelines, which were not designed however to work in such crowded regions. The quality of the photometry in these areas is thus not guaranteed to be as accurate as in the SDSS main part. In addition, the frames are not fully de-blended and thus not completely catalogued. SDSS magnitudes are available for 1202 galaxies, that is, 93\% of the whole sample. The database becomes more incomplete towards the cluster centre.  Our catalogue lists 140 galaxies within $20\arcmin$ from the centre where only 116 galaxies (83\%) have SDSS magnitudes. The completeness drops to 65\% within $10\arcmin$ and only 29\% within $5\arcmin$. In  particular, SDSS magnitudes are not available for the bright NGC galaxies 1272, 1273, 1274, 1275, 1277, and 1278. 
Photometric data for NGC\,1275 were taken from \citet{Brown_2014}.

2MASS magnitudes in the NIR bands J, H and K were downloaded via SQL query from the SDSS table {\tt TwoMassXSC}. This table contains one entry for each match between the SDSS photometric catalogue {\tt photoObjAll} and the 2MASS XSC\footnote{http://tdc-www.harvard.edu/catalogs/tmx.format.html}.
The 2MASS\,XSC reaches a completeness of $>90$\% at an S/N = 10 flux limit fainter than 15.0, 14.3 and 13.5 at 
$J, H$ and $K_{\rm s}$ for sources more than 30$\degr$ from the Galactic plane.
From now on we refer to the 2MASS K$_{\rm s}$ band as K the band.
For $|b| > 20\degr$, the star-galaxy separation is expected to be reliable to 
$K \approx 12.8$ and only falling to 97 per cent by 
$K = 13.5$.\footnote{http://www.ipac.caltech.edu/2mass/releases/allsky/doc/sec6\_5b2.html}
We used the standard (recommended) 2MASS photometry, that is the isophotal fiducial elliptical aperture magnitudes 
with apertures set by the 20 mag\,arcsec$^{-2}$ isophote in the K band. 
Matches were found for 49\% of our catalogue galaxies with a mean positional distance of $0\farcs7$, 96\% of them have positional distances $<1\farcs5$. There is no trend towards an increased incompleteness in the cluster core. 
However, only 19\% of the galaxies with r band magnitudes fainter than the catalogue mean value have matches in the 2MASS\,XSC. 

WISE performed an all-sky astronomical survey in the four wavelength bands W1 at 3.4\,$\mu$m, W2 at 4.6\,$\mu$m, W3 at 
12\,$\mu$m, and W4 at 22\,$\mu$m. 
We cross-matched our galaxy catalogue with the WISE All-Sky Source Catalogue using the NASA/IPAC Infrared Science Archive IRSA. 
With a cone search radius of 3\arcsec, WISE counterparts were found for 91\% of the catalogued galaxies. The mean positional distance is  
$0\farcs7$, 92\% have positional distances less than $1\farcs5$. 
The completeness is comparable to that of the SDSS photometry. However, the WISE data are more complete for the brighter galaxies: 99.6\% for $K_{\rm s} < 13.5$ compared to 96\% for SDSS magnitudes. Of particular importance is the high completeness in the cluster core with 100\% within $5\arcmin$ and 96\% within 20\arcmin.

\begin{figure}[htbp]
\includegraphics[viewport= -10 0 770 600,width=8.7cm,angle=0]{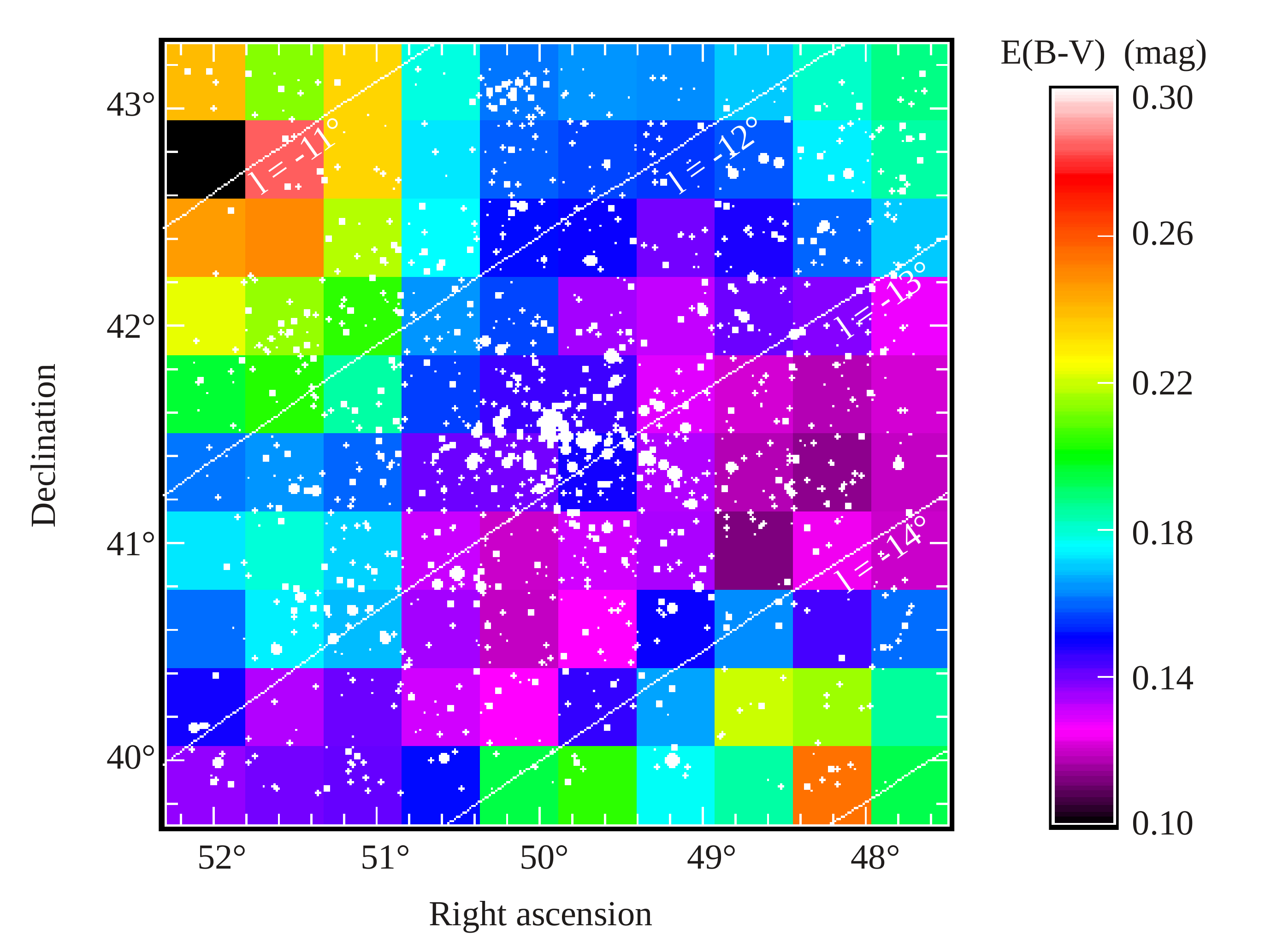}
\caption[Reddening]{Reddening map of the Perseus cluster field. 
The colour indicates the reddening value $E(B-V)$ from 0.1 (magenta) to 0.3 mag (red; see colour bar).  
The positions of the catalogued galaxies are over-plotted by white symbols 
where the symbol size scales with the brightness in the WISE W1 band. 
The diagonal white lines are curves of equal Galactic latitude from $l = -11\degr$ (top left) to $-15\degr$ (bottom right) in steps of one degree.
}
\label{fig:reddening}
\end{figure}

\begin{figure}[htbp]
\includegraphics[viewport= 25 30 570 580,width=6.5cm,angle=270]{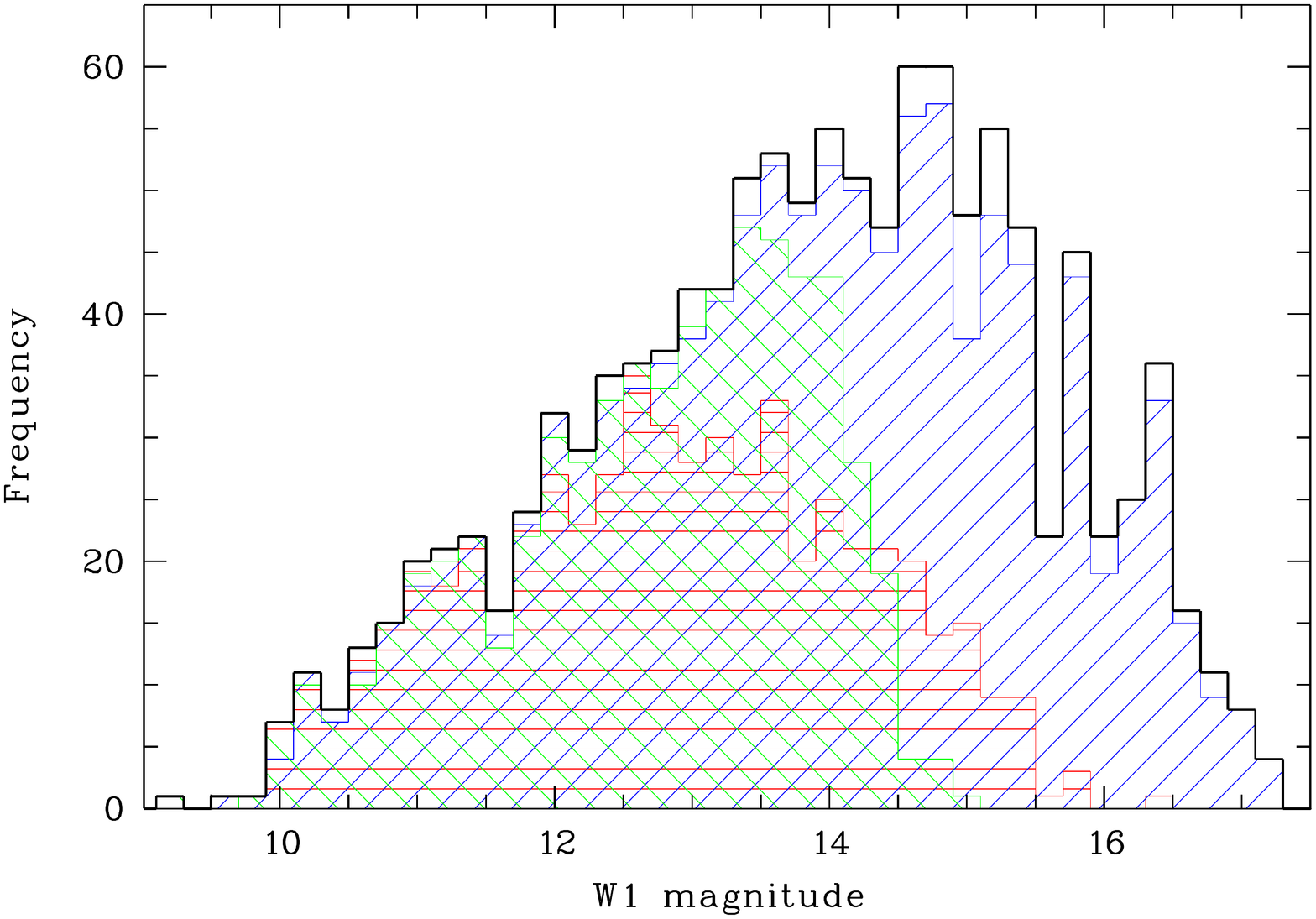}
\caption[hist_W1]{
Histogram of the W1 magnitudes for the selected galaxies with WISE photometric data (black). 
Overplotted are the histograms of the sub-samples with photometry from SDSS (blue) and 2MASS\,XSC (green).  
For comparison, the sample from \citep{Brunzendorf_1999} is shown (red).
}
\label{fig:hist_W1}
\end{figure}

The magnitudes in all bands were corrected for Galactic foreground extinction using the colour excess $E(B-V)$ 
from \citet{Schlafly_2011}\footnote{http://irsa.ipac.caltech.edu/applications/DUST}.
The Galactic foreground reddening map of the survey field is shown in Fig.\,\ref{fig:reddening}. 
$E(B-V)$ varies strongly over the field from approximately $0.1$ to $0.3$\,mag. As expected, reddening is strongest at lowest Galactic latitudes (top left corner). It is particularly notable, however,
that there are also considerable local fluctuations, significant reddening is seen also at the bottom right. 
Figure\,\ref{fig:hist_W1} shows the histogram of the $W1$ magnitudes for the selected galaxies (black) and for the sub-samples with photometry from SDSS (blue) and 2MASS\,XSC (green). For comparison, the old catalogue  of Perseus cluster galaxies \citep{Brunzendorf_1999} is overplotted (red).

\subsection{Redshifts}\label{sec:redshifts}

Heliocentric redshifts are available from the spectroscopic observations described above.
The SDSS spectra in the Perseus field have been made public only in the SDSS DR 7 \citep{Abazajian_2009}. 
There were several targets from our preceding spectroscopic observation runs at TLS and Calar Alto 
among the galaxies with SDSS spectra. 
For those galaxies the redshifts derived from our spectra are generally consistent with those from SDSS. 
The mean redshift difference (SDSS - here) is $\Delta z = 0.0003 \pm 0.0002$ (standard error). 
We also checked the NED, Version November 2017, for spectroscopic redshifts of additional catalogue galaxies. 
If a galaxy was found to have redshifts from more than one source, we always preferred SDSS redshifts, if available. 
Whenever different values were reported from different data sources, the available spectra were checked in detail. 
In particular, we found a few redshift data from \ion{H}{I}\, 21\,cm observations to be wrong.    
Altogether, our catalogue includes spectroscopic redshifts for 384 galaxies where 254 are from SDSS, 41 from TLS and CA,
3 from \citet{Sakai_2012}, and 86 from the NED. 
The spectroscopic redshift sample thus comprises 29\% of the catalogue. 
Additional redshift information can be derived from H$\alpha$ surveys, as is discussed in Sect.\,\ref{sect:eSCGS} below.

We also checked the usefulness of photometric redshifts for our project. Photometric redshifts estimated by robust fit to nearest neighbours in a reference set were made available by SDSS DR14 \citep{Abolfathi_2018}. 
SDSS provides seven categories of error flags. It is recommended by the SDSS to use {\tt photoErrorClass} = 1 only (average RMS 0.043)\footnote{\tt http://www.sdss.org/dr14/algorithms/photo-z}, which is the case for 1055 catalogue entries. 
For 350 galaxies we have both spectroscopic redshifts and photometric redshifts in {\tt photoErrorClass} = 1. 
With a median difference of 0.014 and a larger (due to the non-Gaussian distribution) RMS value of 0.043, the photometric redshifts are clearly too uncertain to be useful for discriminating cluster members from background galaxies. 

The percentage of galaxies with available redshifts is high for the bright galaxies, 98\% for $K < 11$, and decreases systematically towards fainter magnitudes: 94\% for $K < 12$, 76\% for $K < 13$, and still 61\% for $K < 14$.   
However, given its heterogeneous nature, the selection effects of the spectroscopic sample are difficult to grasp. 
The selection criteria for the galaxies tiled in SDSS are basically unknown. However, because of the incomplete field coverage by the two SDSS spectroscopic plates (Fig.\,\ref{fig:Spectra_field}) it is clear that the sample of galaxies with SDSS redshifts is biased towards the cluster centre region. The targets of our spectroscopic observing runs are predominantly galaxies with conspicuous morphological or photometric features.  Therefore, we must note that the spectroscopic sample does not present a well-defined sub-sample.

\subsection{Morphological classification}\label{sect:morph_class}

The morphological types of about $900\,000$ galaxies from the SDSS main survey were made available by the GalaxyZoo project \citep{Lintott_2011, Willett_2013}. 
Unfortunately, GalaxyZoo does not cover the Perseus field. 
We performed a target-oriented detailed visual analysis of all selected galaxies on the SDSS colour images and the Tautenburg B and H$\alpha$ images simultaneously.
The `eyeball' inspection is probably still the most robust method to classify galaxy samples that are not too extensive. 
In many cases, however, the visual inspection alone did not enable us to distinguish between the types E and S0. 
For 143 of these galaxies the final classification was based on an analysis of the radial surface brightness profile. 
The radial profile, measured in a clean area of the galaxy on Tautenburg images, was fitted both by an exponential and a de Vaucouleurs law. If the exponential profile provides the better fit in the outer part, the galaxy was classified as S0, otherwise as E.

The classification consists of three parts. 
First, each galaxy was categorised by one of the fundamental morphological classes: 
E, S0, S, Irr or an intermediate class in ambiguous cases with approximately the same probability for both classes, such as E/S0. 
The result is stored in the parameter ${\tt cl1}$ 
with the values $1 \ldots 8$ for the sequence E - E/S0 - S0 - S0/S - Sa - Sb/Sc - Sc/Irr - Irr and 
${\tt cl1} = 9$ for a galaxy merger where the involved components are not separated. 
The reliability of {\tt cl1} is  stored in ${\tt flag\_cl1}= 0 \ldots 3$, with 3 for highest reliability.
12\% of the catalogue galaxies could not be classified, that is ${\tt cl1}=0$ and ${\tt flag\_cl1}= 0$.
For about 76\% of them, the classification reliability flag is  ${\tt flag\_cl1} \ge 2$.
Secondly, more detailed information is stored in a second morphological parameter, {\tt cl2}.
Following \citet{vandenBergh_1976}, galaxies classified as spirals are subdivided 
into `anemic' spirals with weak and red spiral arms ({\tt cl2} = A) and normal spirals 
with pronounced, blue spiral arms  ({\tt cl2} = S).
Whenever possible, S and A galaxies and also the intermediate types A/S, A/S0 and S/S0, were assigned to sub-types a, b, or c for a decreasing bulge-to-disc ratio. 
E galaxies are simply subdivided according to their apparent flatness.
The presence of a ring (R), a bar (B), or other remarkable features is noted. 

Finally, morphological peculiarities are encoded into the 
peculiarity parameter ${\tt pec} = 1 \ldots 9$. 
For 244 catalogue galaxies (19\%), the following types of morphological peculiarities were registered: 
weakly or strongly lopsided (${\tt pec} = 1$ or 2), 
minor merger (3), 
early, intermediate, or late major merger (4-6), 
M51 type (7), 
collisional ring (8), 
and others (9). 
More information relevant either to the classification process (for instance, nearby foreground star) or to the morphology itself (for example, nearby galaxy, tidal tale, unusual dust absorption, etc.) is stored in the catalogue column {\tt remarkMorph} for 26\% of the galaxies.

The SDSS photometric pipeline takes the best exponential and de Vaucouleurs fits in each band and asks for the linear combination of the two that best fits the image. 
The coefficient of the de Vaucouleurs term, clipped between zero and one, is stored in the parameter {\tt fracDeV} 
that indicates whether the luminosity profile is closer to a galaxy disc (lower value) or an E galaxy (higher value). 
A number of previous studies  has demonstrated {\tt fracDeV}  to be a useful tool for separating galaxy samples into disc-dominated and bulge-dominated or ellipticals \citep[e.g.][]{Rodriguez_2013, Wilkinson_2017}. The SDSS table {\tt galaxy} provides {\tt fracDeV}  for 94\% of the catalogue galaxies in each of the five SDSS bands. We simply used the mean value of {\tt fracDeV}  from the g and r bands. For another 15 galaxies, we estimated {\tt fracDeV} from our measured surface brightness profiles.

\begin{figure}[htbp]
\includegraphics[viewport= 50 30 600 790,width=6.4cm,angle=270]{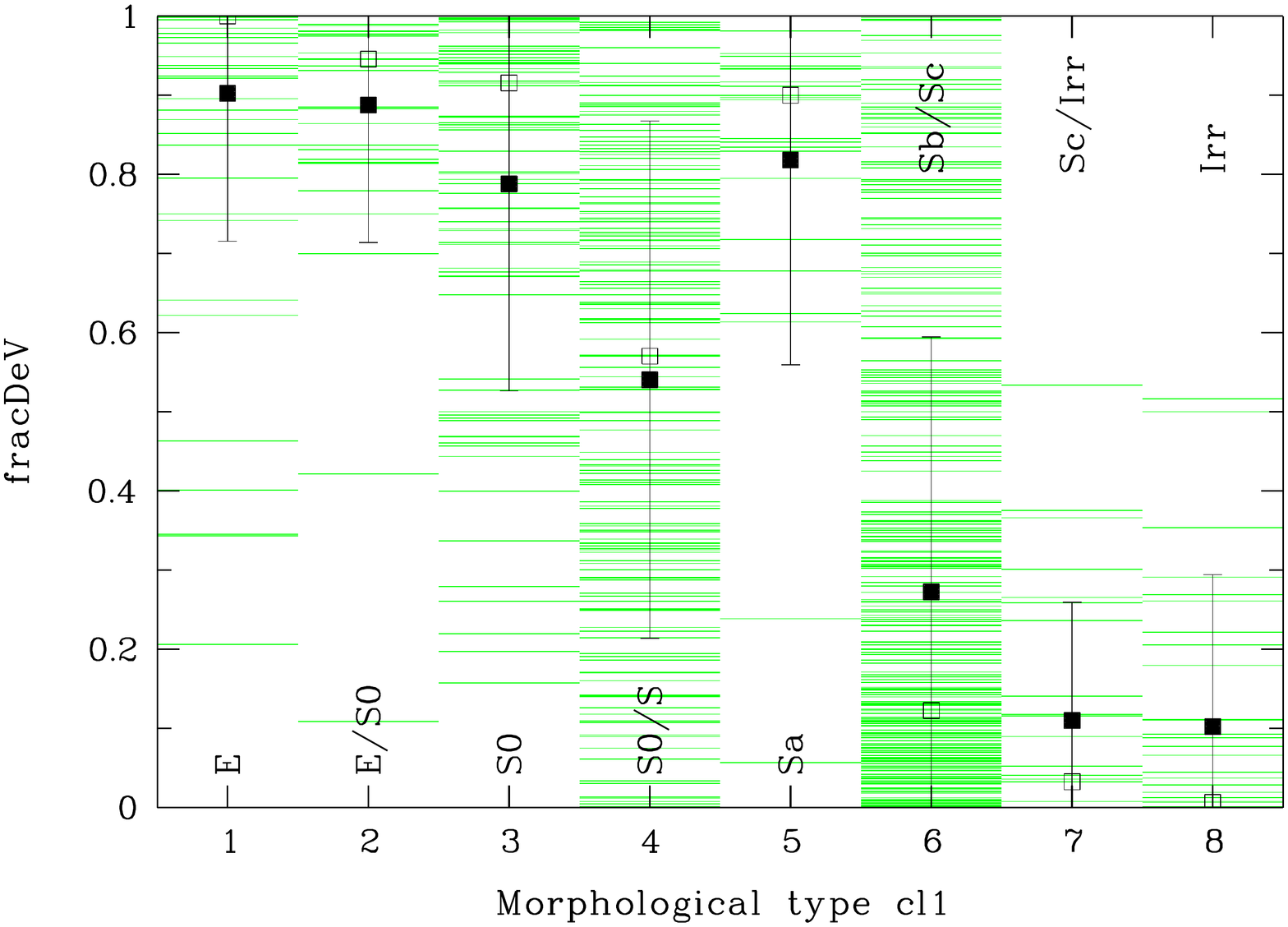}
\caption[Morphology]{
SDSS {\tt fracDeV} parameter versus morphological class {\tt cl1}. 
Each horizontal green line represents an individual galaxy, the black squares are the mean (filled) and median (open) {\tt fracDeV} in each class with standard deviation (error bar).    
}
\label{fig:fracDeV_cl1}
\end{figure}

In Fig.\,\ref{fig:fracDeV_cl1}, the morphology class {\tt cl1} from our classification is compared with 
{\tt fracDeV} for 847 catalogue galaxies with reliable classification (${\tt flag\_cl1} \ge 2$) 
and without strong peculiarities (${\tt pec} \le 1$). 
Each galaxy is represented in this diagram by a vertical line in the corresponding {\tt cl1} bin. 
The mean values (filled squares) and the medians (open squares) indicate a clear trend along the {\tt cl1} sequence.  
The types  ${\tt cl1} \le 5$ have on average a high ${\tt fracDeV} \ga 0.5$ with ${\tt fracDeV}  \ga 0.8$ for E and E/S0.  On the other side, the classes ${\tt cl1} = 6 - 8$ have on average low values with  ${\tt fracDeV} \la 0.1$ for Sc/Irr and Irr. 
If not stated otherwise, we consider ${\tt cl1} = 1 - 5$ as early types and ${\tt cl1} = 6 - 8$ as late types 
in this paper. Figure\,\ref{fig:fracDeV_cl1} indicates a remarkable scatter. 
Possible contamination by spirals of a sample of elliptical galaxies selected by {\tt fracDeV} , 
and vice versa, was discussed at length by \citet{Padilla_2008} and  \citet{Rodriguez_2013}.
In addition, another two effects play a role in our study. 
First, following \citet{vandenBergh_1976}, S0 galaxies are not restricted to high bulge-to-disc ratios
and a wide range of {\tt fracDeV} values is thus expected. 
This is the main reason for the relatively high portion of low  {\tt fracDeV}   values among the galaxies in the ${\tt cl1} = 4$ bin. The second source of the scatter is the above-mentioned relatively high probability of blending by foreground stars, which may have an effect on {\tt fracDeV} for a substantial number of the galaxies.

%
\section{Galaxy samples}\label{sect:galaxy_samples}
%

The construction of a reliable cluster galaxy sample is faced with several problems. A cluster membership criterion based on both position on the sky and in redshift space is desirable but suffers from incompleteness in the available spectroscopic data. Photometric criteria such as colours and limiting magnitude produce biased samples that are very likely contaminated by non-cluster members. Moreover, as a part of the Perseus-Pisces supercluster, A\,426 is not a closed, isolated structure with a clear demarcation between the cluster and its surroundings

\begin{figure}[bhtp]
\includegraphics[viewport= 30 0 765 585,width=9.0cm,angle=0]{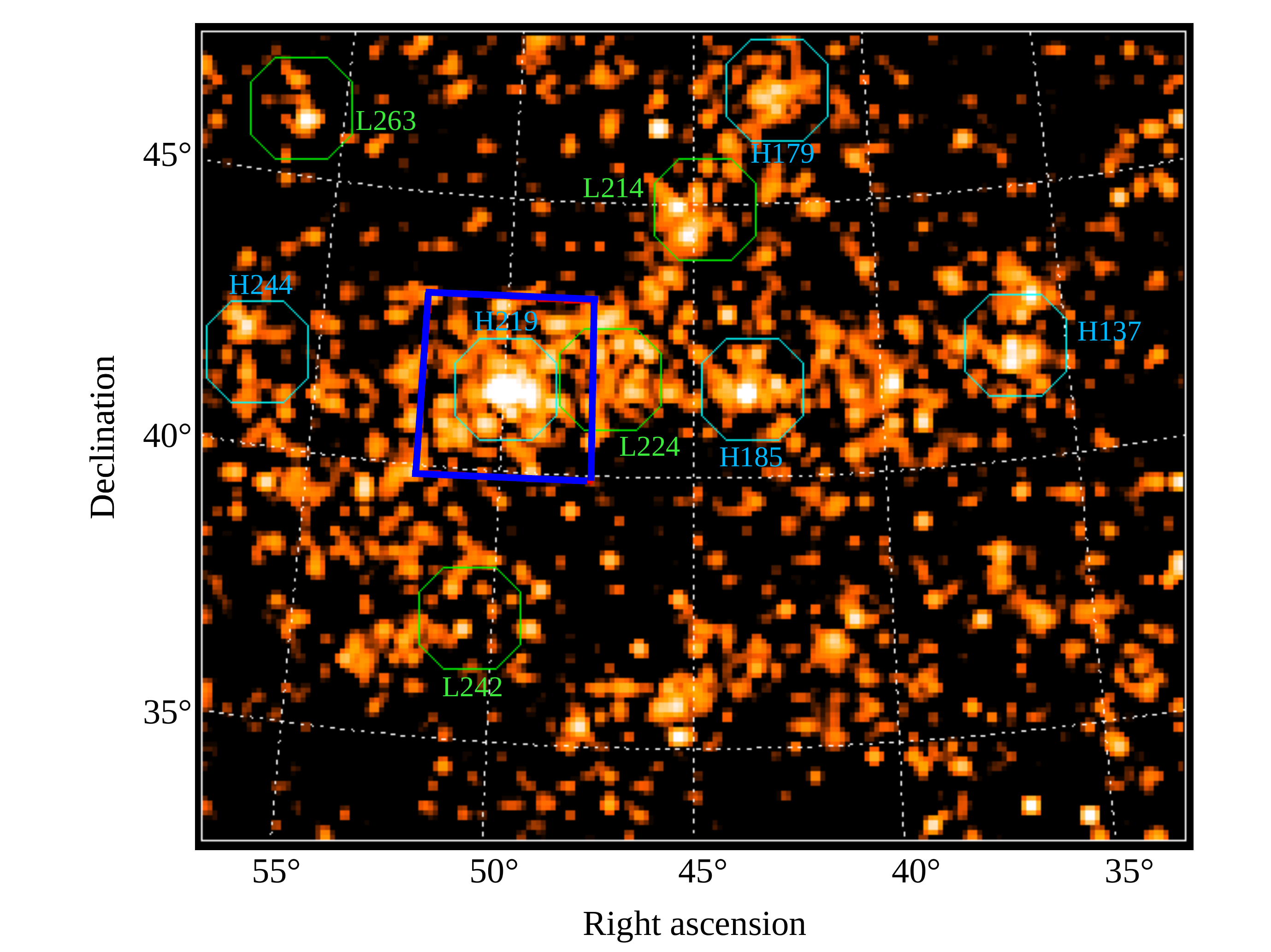}
\caption[Supercluster]{
K band surface brightness sky map (orthographic projection) of the eastern part of the Perseus-Pisces supercluster
from 2MASS\,XSC objects with $8 \le K \le 14$. The present survey field is indicated by the dark blue box, 
the hexagons indicate galaxy groups from \citet{Crook_2007} with $0.016 \la z \la 0.030$, each labelled with the number in the Low Density Contrast (LDC = L, green) or High Density Contrast (HDC = H; cyan) catalogue.
}
\label{fig:supercluster}
\end{figure}

\subsection{Survey field}\label{sect:survey_field}

Rich galaxy clusters are the dense nodes in the cosmic web of filaments and sheets of matter.
Figure\,\ref{fig:supercluster} shows the K band surface brightness map of the larger environment of the Perseus cluster that contains the eastern part of the Perseus-Pisces supercluster. The map contains around 11\,000 entries from the 2MASS\,XSC with $8 \le K \le 14$ and larger than ${\tt r\_ext} = 5$\,arcsec.  

\citet{Crook_2007} applied a variable linking-length percolation to identify groups of galaxies from the Two Micron All-Sky Redshift Survey based on both sky position and redshift. 
The groups within  the  redshift range $0.016 \la z \la 0.030$ are labeled in Fig.\,\ref{fig:supercluster} by hexagons and labelled by the catalogue number, where `H' stands for HDC = high density contrast and `L' for LGC = low density contrast. The two HDC groups H\,219 and H\,137 are identical with the Abell clusters A\,426 and A\,347. They were identified as two separate groups in the HDC catalogue and
make up the third largest group in the LDC catalogue. H\,179 and L\,214 are at $z = 0.0276$ and 0.031, respectively, all other groups are in the narrow redshift range $z = 0.0166$ to 0.0186. 
The field covered by our galaxy catalogue is indicated by the red box around the brightest structure in the map.  All groups are part of an extended structure without clear borders.

\subsection{Spectroscopic cluster galaxy sample}\label{sect:SCGS}

\vspace{1.3cm}

\begin{figure}[htbp]
\includegraphics[viewport= 10 0 400 300,width=6.9cm,angle=0]{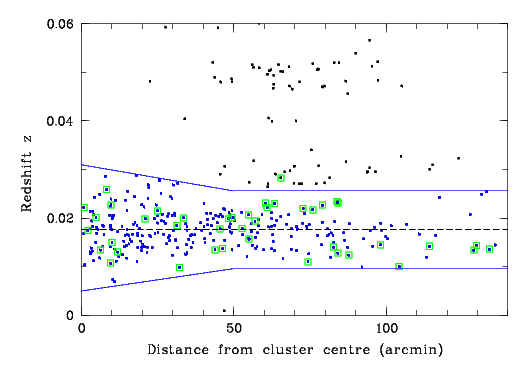}
\caption[Cluster membership]{
Redshift $z$ versus cluster-centric distance for the catalogue galaxies with $z < 0.06$ (filled squares).
The blue solid lines represent the cluster membership criterion adopted in the present study. The 
dashed horizontal line marks the median redshift of the selected cluster members (blue filled squares).
Green open squares indicate galaxies classified as H$\alpha$ emitters from H$\alpha$ imaging surveys (Sect.\,\ref{sect:eSCGS}).  
At the distance of the Perseus cluster, 140$\arcmin$ correspond to $\sim 3$\,Mpc. 
}
\label{fig:membership}
\end{figure}

The redshifts of our catalogue galaxies cover the range $0.009 < z < 0.155$. 
Figure\,\ref{fig:membership} shows redshift versus cluster-centric distance $R$ for the 358 galaxies with $z < 0.06$, where the centre position from \citet{Ulmer_1992} was adopted. 
The distribution of the galaxies in the $z - R$ plane indicates a clear separation of cluster galaxies from the background. 
We adopted the simple membership criterion indicated by the solid lines in Fig.\,\ref{fig:membership}, which corresponds roughly to the $R$-dependent 2.5$\sigma$ deviation from the median. This criterion defines our spectroscopic cluster galaxy sample (SCGS) of 286 certain cluster galaxies.  
Of course, it cannot be excluded that some of the galaxies with $z \ga 0.025$ are, in fact, background objects.
The mean SCGS redshift is  $\bar{z} = 0.0177 \pm 0.0039$, the median redshift is 0.0176 (dashed horizontal line). These values change only slightly when the spatial extension of the sample is restricted. For the 206 cluster members with $R \le 1$\degr, the mean value amounts to $\bar{z} = 0.0179 \pm 0.0040$. 

Compared to \citep{Brunzendorf_1999}, our galaxy catalogue (Appendix\,\ref{sect:catalogue}) lists new redshifts for 133 (47\%) of the SCGS galaxies corresponding to an increase by a factor of 1.9.
However, it must be noted that the SCGS is composed of data from various sources and is thus not homogeneous. The selection effects from the different sources are difficult to grasp.  The great impact of the SDSS spectroscopy clearly introduces a selection bias against the outermost parts of the field (see Fig.\,\ref{fig:Spectra_field}). The mean cluster-centric distance of the galaxies with redshifts from SDSS amounts to $\langle R \rangle = 36 \pm 2$ arcmin, clearly less than for the galaxy samples in Table\,\ref{tab:samples}. On the other side, the galaxies from the TLS-CA spectroscopic sample (Sect.\,\ref{sec:TLS_CA_spectra}) have a much larger mean distance of $\langle R \rangle = 60 \pm 7$ arcmin. The use of the spectroscopic sample for the study of trends with $R$ must thus be considered with caution.

\subsection{Extended spectroscopic cluster galaxy sample}\label{sect:eSCGS}

The H$\alpha$ survey  (Sect.\,\ref{sec:TLS_imaging}) provides an additional approach to select cluster members among the galaxies without available spectra. The narrow-band filter for the redshifted H$\alpha$ efficiently selects H$\alpha$ emitters within a redshift interval of $\pm 0.003$ around the mean cluster redshift. Such galaxies are most likely cluster members.  
57 catalogued galaxies  are flagged as H$\alpha$ sources in our H$\alpha$ survey. 
An additional nine H$\alpha$ sources were added from the H$\alpha$ survey performed by \citet{Sakai_2012} with the MOSAIC-1 Imager on the Kitt Peak National Observatory 0.9\,m telescope. These authors imaged A\,426 with three narrow-band filters for the redshifts $z = 0.0122 \pm 0.006, 0.0183\pm0.006$ and $0.0244\pm0.006$.
Another H$\alpha$ survey of galaxy clusters was undertaken by \citet{Moss_2000, Moss_2006} using objective prism observations with the 61/94-cm Burrell Schmidt Telescope on Kitt Peak. 
Four A\,426 galaxies are classified as H$\alpha$ emitters from the objective prism survey but not flagged as such in our catalogue. Their spectroscopic redshifts are consistent with cluster membership for three galaxies and with background in the other case.  We set the H$\alpha$ flag for the three cluster members only.  

Among the 69 galaxies flagged as H$\alpha$ sources, spectroscopic redshifts are available for 42 with a mean value $\bar{z} = 0.0178 \pm 0.0047$.  
As shown in Fig.\,\ref{fig:membership}, 41 H$\alpha$ galaxies (98\%) fall within the cluster region. 
Only for one H$\alpha$  galaxy from the \citet{Sakai_2012} survey the cluster membership criterion is closely missed. 
It may be assumed that the vast majority of the remaining 27 H$\alpha$ sources without spectroscopic redshifts belong to the cluster. It is thus reasonable to add all these H$\alpha$ sources to the sub-sample of galaxies classified as cluster members. It goes without saying that the cluster member selection technique based on H$\alpha$ imaging is heavily biased towards strong emission line galaxies. 

The total number of galaxies in this extended spectroscopic cluster galaxy sample (eSCGS) is thus $N = 313$. 
For comparison, \citet{Brunzendorf_1999} listed 660 galaxies in the Perseus cluster field with radial velocities for 171 entries. The eSCGS makes up 24\% of our present catalogue. This percentage increases to 39\% when we restrict the sample to galaxies with a Kron radius larger than $30\arcsec$, 
and to 59\% if we further restrict the sample to the galaxies with $R \le 100 \arcmin$, which means, within the Abell radius. It should be noted, on the other hand, that 26\% of the spectroscopically observed galaxies in the field are not cluster members (Sect.\,\ref{sect:background}).

\subsection{Magnitude-limited galaxy samples}\label{sect:MLGS}

The cluster membership criterion based on spectroscopic redshifts ensures that the samples SCGS and eSCGS are free from strong contamination by background galaxies. On the other hand, these samples suffer from different selection effects in the process of targeting galaxies for spectroscopic observations that are difficult to assess.
Given the heterogeneous nature of the spectroscopic target selection, it seems reasonable to consider, in addition, 
galaxy samples based on strict magnitude and field limitations where no other redshift constraints are used than the exclusion of the spectroscopically confirmed foreground and background galaxies. 
The field limitation is simply set by $R \le 100\arcmin$, which is the maximum distance from the centre covered in all directions and corresponds roughly to one Abell radius.  A small region around the background cluster at 
$\mbox{RA, DEC} =  50\fdg2, 43\fdg1$ (Sect.\,\ref{sect:background}) was excluded.

\subsubsection{W1 magnitude-limited sample MLS\_W1}

\begin{figure}[htbp]
\includegraphics[viewport= 25 30 570 580,width=6.5cm,angle=270]{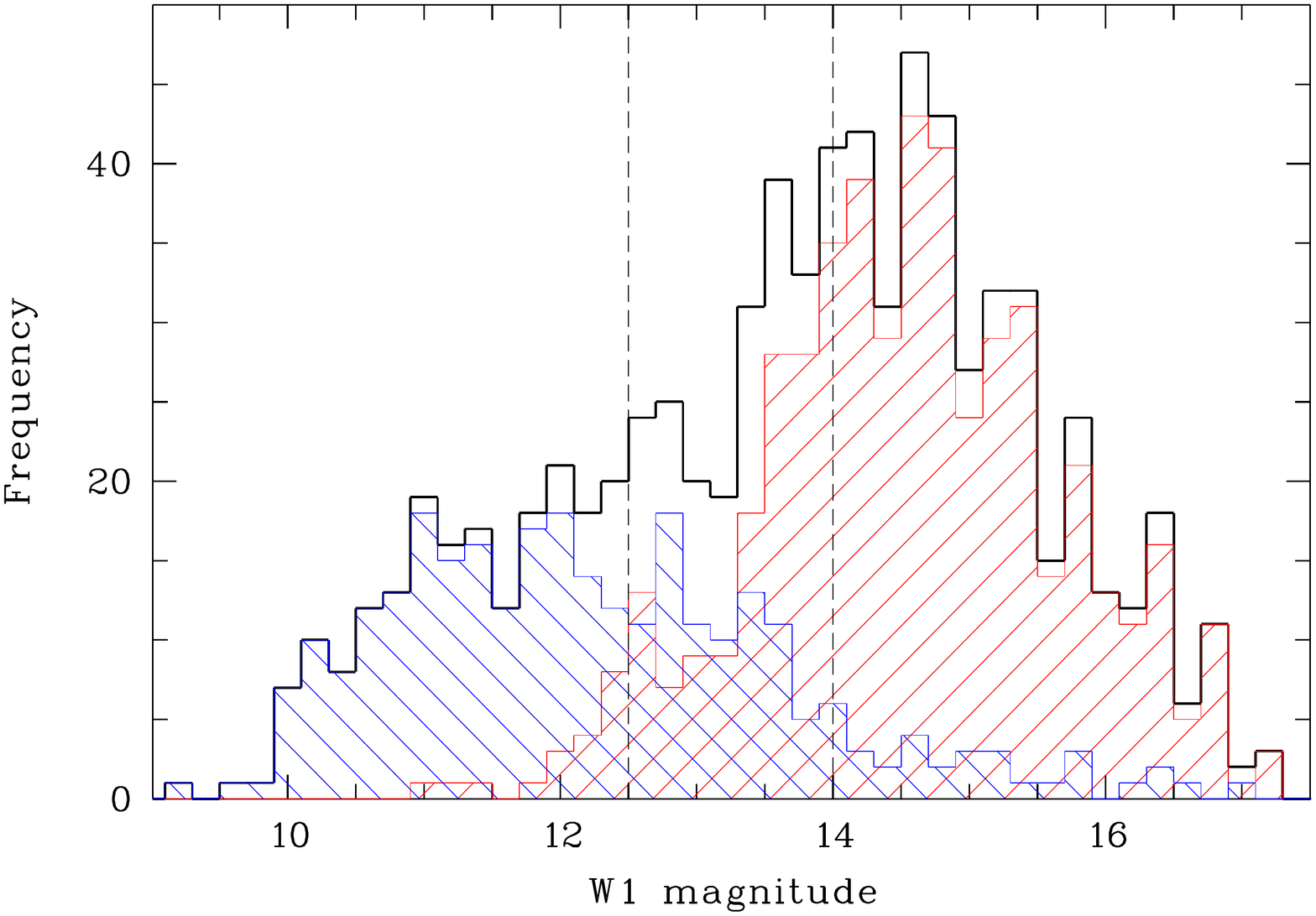}
\caption[hist_W1_lim_samples]{
Histograms of the W1 magnitudes for the morphologically classified possible cluster members 
within a cluster-centric distance $R = 100\arcmin$ (black), the sub-sample
of spectroscopically confirmed cluster members (blue), 
and the sub-sample of galaxies without spectroscopic redshifts (red).
The two dashed vertical lines indicate the magnitude limits for MLS\_W1b and MLS\_W1.
}
\label{fig:hist_W1_lim_samples}
\end{figure}

The WISE 2.4 $\mu$m magnitude $W1$ was selected to define a magnitude limit mainly for the sake of completeness 
(see Sect.\,\ref{sect:obs_photo}). In addition, the W1 band is a factor of 10 less affected by reddening than the SDSS r band. This is a clear advantage with regard to the irregular reddening over the survey field 
(Fig.\,\ref{fig:reddening}). 
Figure\,\ref{fig:hist_W1_lim_samples} shows the histogram of the WISE $W1$ magnitudes for the galaxies from this selection. Overplotted are the sub-samples of the spectroscopically confirmed cluster members (blue) and of the
galaxies without spectroscopic redshifts (red). At fainter magnitudes, $W1 \ga 14.5$, the catalogue becomes 
increasingly incomplete.

We adopted the magnitude limit $W1_{\rm lim} = 14$. 
With mean colour indexes $V-K \approx 3.3$ from 
NED\footnote{https://ned.ipac.caltech.edu/level5/Sept04/Aaronson/Aaronson4\_3.html}
and $K-W1 \approx -0.2$ from our catalogue data, $W1 \le 14$ corresponds to a $V$ limit close to that of the \citet{Kent_1983} sample, which is however highly incomplete. The magnitude-limited sample MLS\_W1 thus includes the Perseus cluster galaxies with absolute magnitudes $M_{\rm v} \la -17$, that is, between the SMC and the LMC.  
For the sake of comparison, we also constructed a brighter sub-sample, MLS\_W1b, of 194 galaxies with $W1 \le 12.5$.

\subsubsection{ur magnitude-limited sample MLS\_ur}

Despite the advantage of the $W1$ selection, a magnitude limit set in the NIR produces 
a bias towards the brighter galaxies in the red sequence (see Sect.\,\ref{sect:cmd_morph} below). 
We decided therefore to construct, in addition, an alternative magnitude-limited sample 
based on the constraint $r < 20.6 - 2.0 (u-r)$. The selection criterion is defined in such a way that 
the galaxies from the blue cloud have a similar chance to be collected as those from the red sequence
(see Figs.\,\ref{fig:CMD_gr_selection} and \ref{fig:CMD}). 
This sample is referred to as MLS\_ur in this paper. 
The drawback of such a selection is the incompleteness and uncertainties of the SDSS photometry 
(see Sect.\,\ref{sect:obs_photo} above and Sect.\,\ref{sect:cmd_morph} below). 
The sample size is comparable to that of the MLS\_W1.

\begin{table}[h]\centering
\caption{Summary of the samples of galaxies in the Perseus cluster field used in the present study.}
\begin{tabular}{lrclccc}
\hline\hline
Sample 
& $N$ 
& $f_z$ 
& $f_{\rm b}$  
& $\langle r \rangle$ 
& $\langle W1 \rangle$  
& $\langle R \rangle$ (\arcmin) \\
\hline
MLS\_W1b      &  194   & 0.89   & 0.01   & 13.9  & 11.4  & $41\pm2$ \\
SCGS          &  286   & 1.00   & 0.00   & 14.4  & 12.0  & $43\pm2$ \\
eSCGS         &  313   & 0.91   & 0.00   & 14.6  & 12.3  & $45\pm2$ \\
MLS\_ur       &  372   & 0.62   & 0.06   & 14.9  & 12.6  & $50\pm1$ \\
MLS\_W1       &  412   & 0.61   & 0.09   & 14.9  & 12.4  & $50\pm1$ \\
MaxS          & 1196   & 0.24   & 0.17$^\ast$   & 16.5  & 13.8  & $64\pm1$ \\
\hline
\end{tabular}
\tablefoot{$^\ast$ most likely a lower limit only (see Sect.\,\ref{sect:background_cont}).}
\label{tab:samples}
\end{table}

\subsubsection{Background contamination and incompleteness}\label{sect:background_cont}

Background contamination is an unavoidable effect in a magnitude-limited sample.  
For simplicity, we refer to both foreground and background galaxies as background galaxies in the following.
Here we limit the discussion to the estimation of the background fraction $f_{\rm b} = N_{\rm b}/N$ in the MLS\_W1, where $N_{\rm b}$ and $N$ are the (unknown) number of background galaxies and the number of all galaxies in the sample, respectively. 

The background fraction can be roughly estimated as follows. 
MLS\_W1 consists of $N = 412$ galaxies, of which 259 are established cluster members. 
The fraction of galaxies without spectroscopic redshifts $z$ is thus $f_{\rm nz} = (412-259)/412 = 0.37$. 
In the extreme case that none of the galaxies without $z$ is a cluster member, the background fraction 
would be $f_{\rm b} = f_{\rm nz} = 0.37$, which is as an upper limit if the sample is complete. A more realistic assumption is that a substantial number of the galaxies without known redshifts belong to the cluster. 
When we apply the same selection criteria as for the magnitude-limited sample but without excluding the spectroscopic non-members, we find 491 galaxies. Among them are 332 galaxies with spectroscopic redshifts, of which 81 are non-members. The non-member fraction of this spectroscopic sub-sample is thus $f_{\rm nm} = 81/332 = 0.24$. Assuming that the same fraction applies to the sub-sample with unknown redshifts, we end up with a background galaxy fraction 
$f_{\rm b} = f_{\rm nm}\cdot\,f_{\rm nz} = 0.09$. The same computation leads to $f_{\rm nm} = 0.17$ for the whole catalogue. The real background fractions are probably higher because the non-spectroscopic sub-samples are always fainter than the corresponding spectroscopic sub-samples (Fig.\,\ref{fig:hist_W1_lim_samples}, see also Sect.\,\ref{sect:surf_density} below). 
The background contamination is assumed to be significantly lower at brighter magnitudes. 
For MLS\_W1b we estimated a contamination fraction $f_{\rm b} \approx 0.01$.

\begin{figure}[htbp]
\centering
\includegraphics[viewport= 25 30 590 790,width=6.2cm,angle=270]{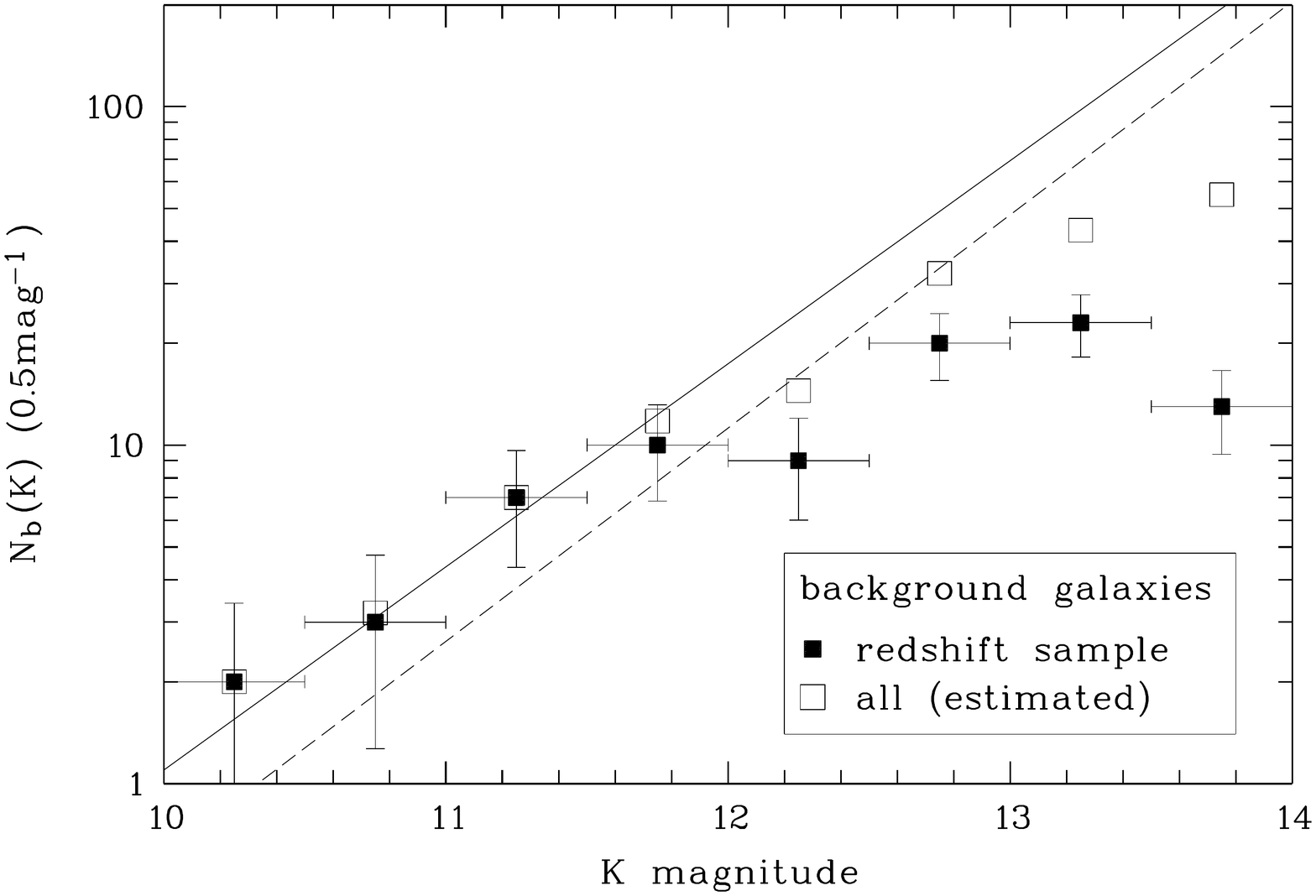}
\vspace{0.3cm}
\caption[background_corr]{Number counts of background galaxies per K magnitude interval of the width 0.5 mag. Filled squares with vertical error bars: spectroscopically established background galaxies, the horizontal bars indicate the bin widths. Open squares: Total number of expected background galaxies. Diagonal lines: galaxy number counts $N(K)$ from 2MASS  \citep[][solid]{Frith_2003} and GAMA \citep[][dashed]{Whitbourn_2014}. 
}
\label{fig:background_K}
\end{figure}

\begin{figure*}[bhpt]
\centering
\includegraphics[viewport= 0  0 570 570,width=6.0cm,angle=0]{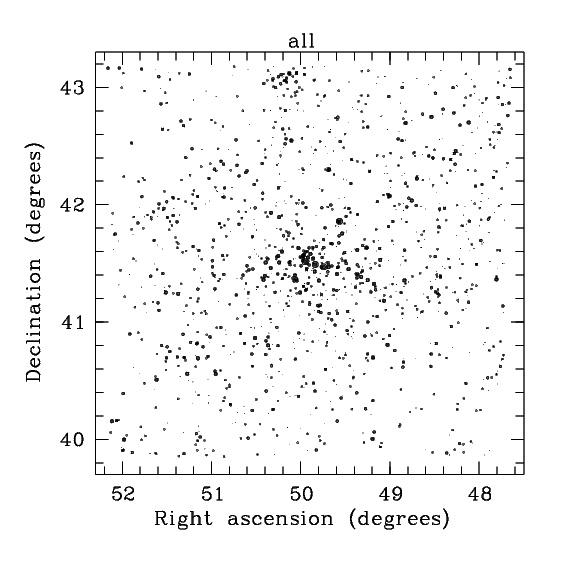}
\includegraphics[viewport= 0  0 570 570,width=6.0cm,angle=0]{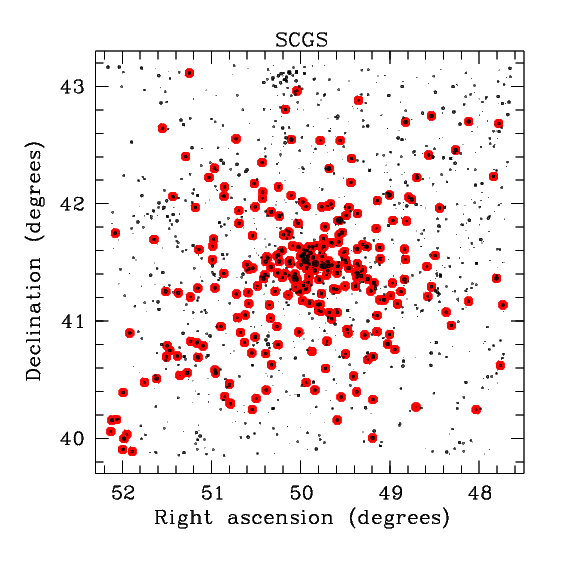}
\includegraphics[viewport= 0  0 570 570,width=6.0cm,angle=0]{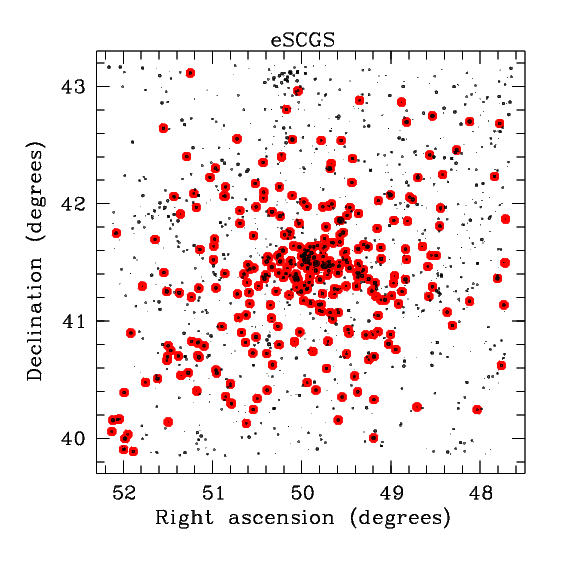}\\
\includegraphics[viewport= 0 16 570 570,width=6.0cm,angle=0]{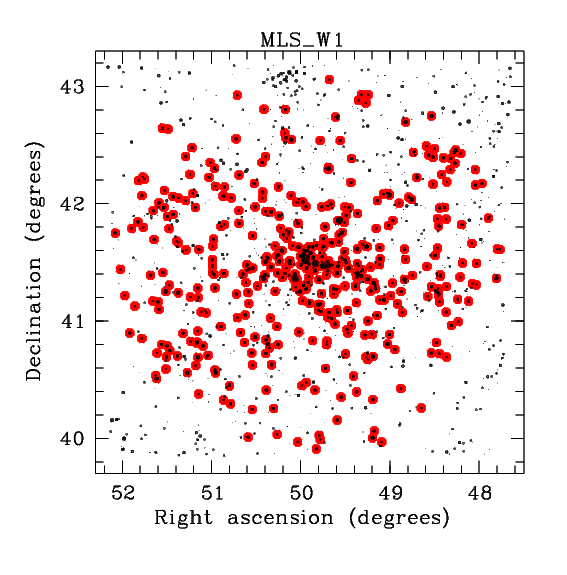}
\includegraphics[viewport= 0 17 570 570,width=6.0cm,angle=0]{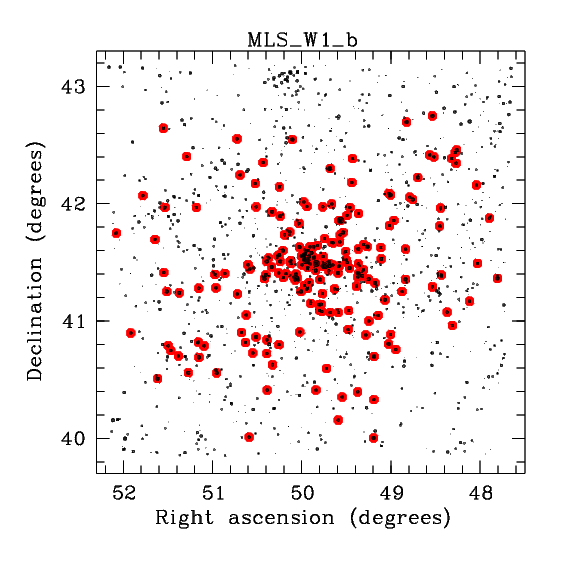}
\includegraphics[viewport= 0  0 570 570,width=6.0cm,angle=0]{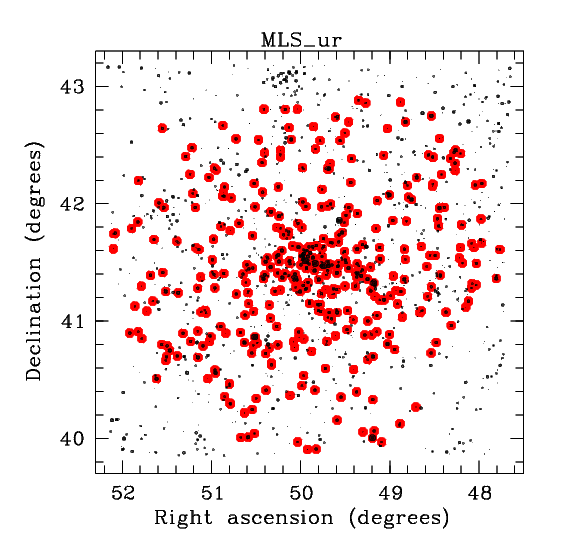} 
\caption[Spatial distribution 1]{Sky map of all catalogued galaxies (black). The red boxes indicate the galaxies in the different samples.}
\label{fig:spatial_distribution}
\end{figure*}

Alternatively, the number density of background galaxies can be estimated from the number counts in the 2MASS database. 
Within our survey field, the 2MASS\,XSC lists 657 sources with $K < 13.8$ and radius ${\tt r\_ext} > 10$ arcsec  
compared to 515 galaxies from our visual search.\footnote{We use $K<13.8$ because of $W1_{\rm lim} = 14$ and $K-W1 \approx -0.2$.} The difference seems to indicate that either MLS\_W1 is 22\% incomplete or that 22\% of the 2MASS\,XSC sources in the field are not galaxies. Although designed for completeness down to low Galactic latitudes, the 2MASS\,XSC is known to suffer from confusion near the Galactic plane.                                                   
From the low-density regions in Fig.\,\ref{fig:supercluster} we estimated a mean surface density of $21$ objects  
with $K < 13.8$ per square degree. 
After correcting for 22\% contamination, we would expect 144 background galaxies in the MLS\_W1 field. 
Because 81 galaxies were already confirmed as background (not included in MLS\_W1), the resulting background fraction is $f_{\rm b} = (144-81)/412 = 0.15$. However, this value may turn out higher (by a factor of two) when a possible incompleteness of the galaxy selection for the present catalogue is taken into account (Sect.\,\ref{sect:LF}, below).

Figure\,\ref{fig:background_K} displays the $N-K$ relation for the spectroscopic background galaxies within $R = 100\arcmin$. The filled squares show the numbers of galaxies per binning interval of the width $\Delta K = 0.5$\,mag. 
We applied the background correction as discussed above for each bin. 
The open squares indicate the sum of the spectroscopic background systems plus the expected number of background galaxies from the non-spectroscopic sub-sample. These total numbers of background galaxies, $N_{\rm b}$, can be compared with K-band number counts of field galaxies from  the literature. 
\citet{Frith_2003} presented galaxy number counts from 2MASS\,XSC in a number of fields. 
For the Northern Galactic Cap region their counts are roughly fitted by their homogeneous model with $\log N (\mbox{deg}^{-2}\,\mbox{0.5 mag}^{-1}) = 0.6 K - 6.9$. This relation, applied to the survey area $\pi R^2$, is shown as the solid diagonal line in Fig.\,\ref{fig:background_K}. The dashed line corresponds to the $N-K$ relation for the deeper K-band GAMA counts from \citet[][their Figure 6]{Whitbourn_2014}. 
At $K \la 12$, our background-corrected counts, $N_{\rm b}$, are nicely fitted by the solid line, but at fainter magnitudes an increasing difference is observed that might indicate an increasing incompleteness in our catalogue. 

One obvious reason for the assumption of incompleteness of the galaxy selection is again the low Galactic latitude of the survey field: a faint galaxy close to a brighter foreground star can be easily overlooked. 
Further, at faint magnitudes, the SExtractor parameter {\tt class\_star} does not provide an efficient star-galaxy separation (Sect.\,\ref{sec:gal_selection}). Whereas stars wrongly classified as galaxies were sorted out in the subsequent visual examination, faint compact galaxies misclassified as stars are more likely to get lost. Hence, we can only assume that there is an increasing incompleteness towards the sample limit. In principle, the $K$-dependent incompleteness correction can be derived from
the comparison of our $N_{\rm b}$ with the solid line in Fig.\,\ref{fig:background_K} (see Sect.\,\ref{sect:LF}). However, the background galaxy density differs from field to field 
\citep[see e.g.][their Figure 5]{Frith_2003} and also between different studies, and the resulting uncertainty is thus accordingly large.

\subsubsection{Overview of the galaxy samples}

Figure\,\ref{fig:spatial_distribution} shows the distributions of the galaxies over the survey field for all samples.  
The small black symbols mark the catalogued galaxies where the symbol size scales with the WISE W1 magnitude. The members
of the corresponding sample are highlighted by the larger red symbol. The top left panel shows the whole catalogue. A lenticular structure with the famous chain of bright cluster galaxies is clearly indicated in the centre.

\begin{figure}[htbp]
\begin{center}
\fbox{\includegraphics[viewport = 0 0 400 350,width=6.0cm,angle=0]{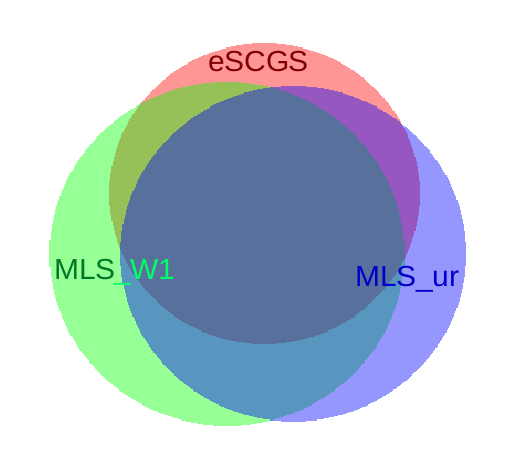}}
\vspace{0.5cm}
\caption[Venn diagram galaxy samples.]{Venn diagram of the three galaxy samples eSCGS (red), MLS\_W1 (green) and MLS\_ur (blue).}
\label{fig:Venn_samples} 
\end{center}
\end{figure}

\begin{figure*}[htbp]
\centering
\includegraphics[viewport= 0 0 570 570,width=6.0cm,angle=0]{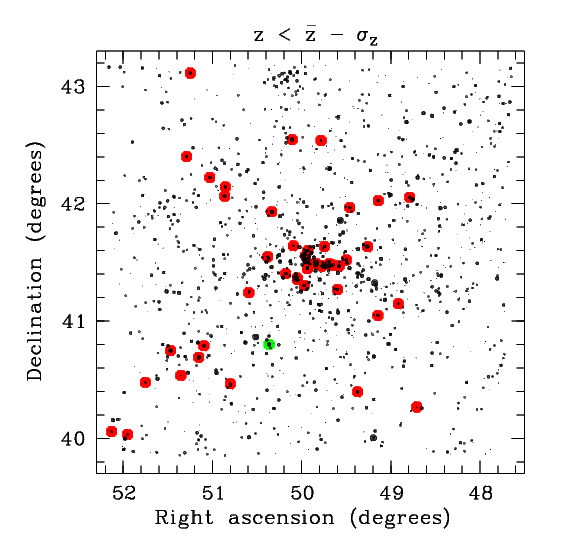}
\includegraphics[viewport= 0 0 570 570,width=6.0cm,angle=0]{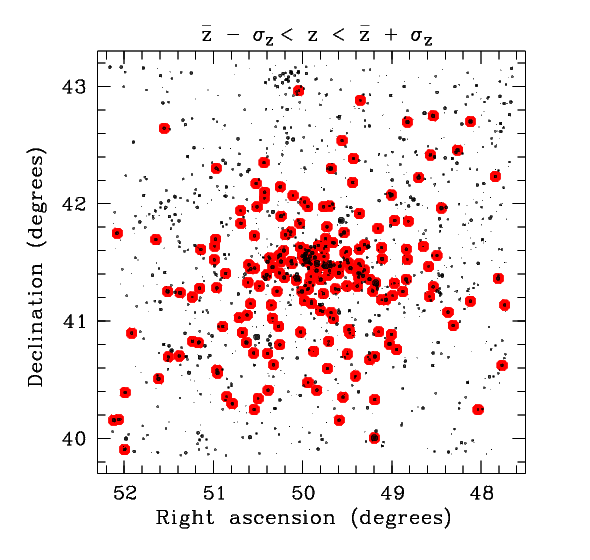}
\includegraphics[viewport= 0 0 570 570,width=6.0cm,angle=0]{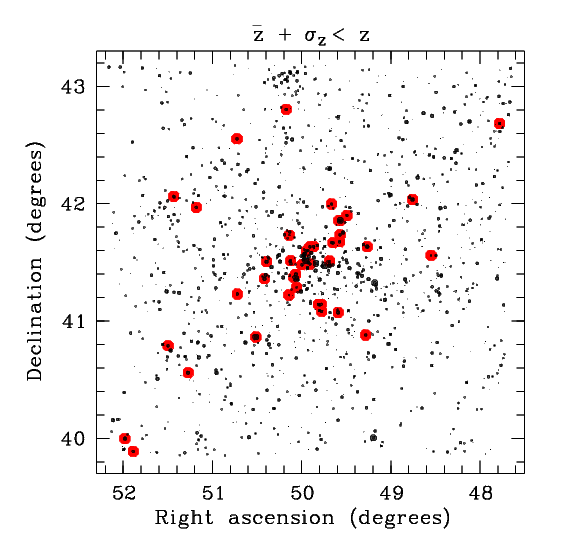} \\
\includegraphics[viewport= 0 0 570 570,width=6.0cm,angle=0]{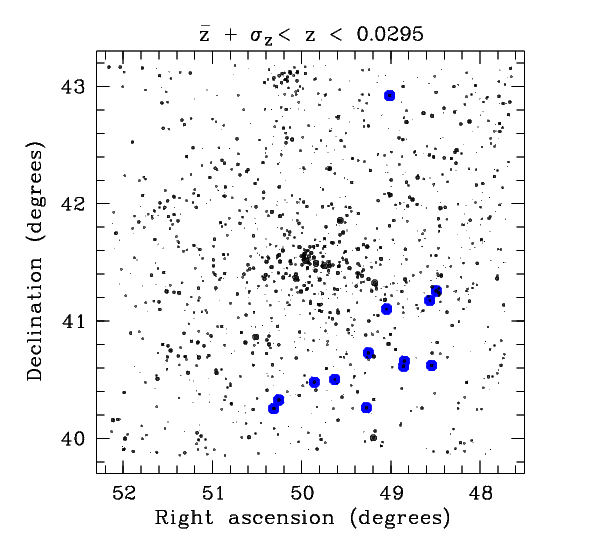} 
\includegraphics[viewport= 0 0 570 570,width=6.0cm,angle=0]{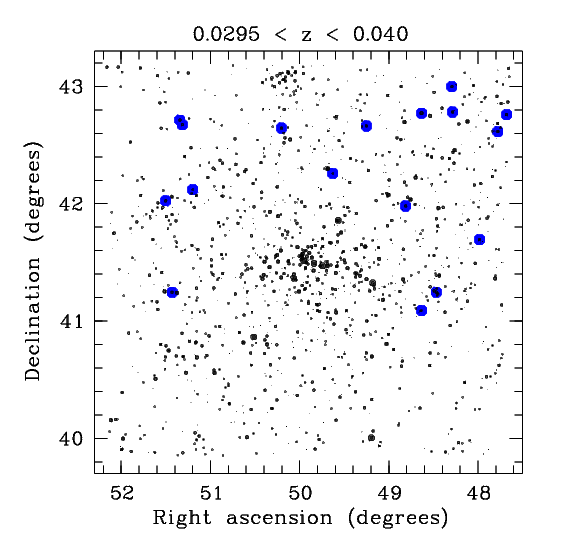} 
\includegraphics[viewport= 0 0 570 570,width=6.0cm,angle=0]{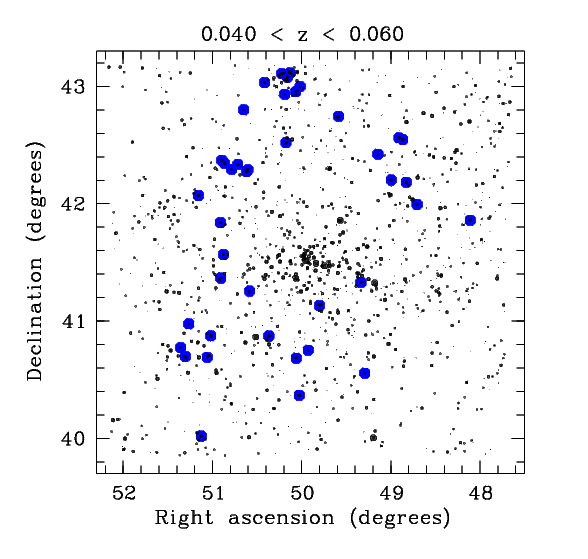} \\
\caption[Spatial distribution]{Sky map of all catalogued galaxies (black). 
The galaxies with known redshifts are indicated by larger coloured dots in six different $z$ intervals. 
The redshift intervals are indicated at the top of each panel, where $\bar{z}$ and $\sigma_z$ are the mean and the standard deviation, respectively, of the redshifts for the spectroscopic cluster members (Table\,\ref{tab:virial_radius}). In the top row, the colours highlight the cluster members (red) and the one foreground galaxy (green). 
In the bottom row, background galaxies are labeled blue.}
\label{fig:background}
\end{figure*}

The various galaxy samples used in the present paper are briefly summarised in Tab.\,\ref{tab:samples} sorted by the sample size $N$, which correlates with the sample averaged magnitudes in the r and W1 bands.
The first two columns give the name and the sample size $N$. The next two columns list the proportion $f_z$ of galaxies with spectroscopic redshift and the estimated background fraction $f_{\rm b}$. The last three columns are the mean r magnitude, the mean W1 magnitude and the mean cluster-centric distance. The bottom line contains the data for a `maximum sample' (MaxS) of all catalogued galaxies except of the spectroscopically established background systems. The relative sizes and the overlap of the three main samples eSCGS, MLS\_W1 and MLS\_ur are illustrated by the Venn diagram \citep{Hulsen_2008} in Fig.\,\ref{fig:Venn_samples}.
\footnote{created with BioVenn under http://www.biovenn.nl/}

\subsection{Established background galaxies}\label{sect:background}

Based on their redshifts, one galaxy is a foreground system and 97 galaxies are in the background of the Perseus cluster. 
Figure\,\ref{fig:membership} indicates two galaxy concentrations in the background, namely at $z \approx 0.03$ and 0.06.
The distribution of the galaxies over the field is shown in Fig.\,\ref{fig:background}  for six $z$ slices.
As in Fig.\,\ref{fig:spatial_distribution}, all catalogue galaxies are plotted as small black symbols in each panel. 
The first three panels (top row) show the cluster members (red) in three different parts of the 
redshift distribution, the core ($\bar{z} - \sigma_z \le z \le \bar{z} + \sigma_z$, middle), 
the blue wing ($z < \bar{z} - \sigma_z$, left), and the red wing  ($z > \bar{z} + \sigma_z$, right).
A remarkable structure, which is seen in all three panels, is the extension from the cluster core to the SE corner that coincides with the diagonal structure through A\,426 indicated in Fig.\,\ref{fig:supercluster}. 
It cannot be excluded that the wide redshift interval $z = 0.010 \ldots 0.025$ indicates an extended ($\sim 50$\,Mpc) sheet-like structure rather than a larger velocity scatter of cluster members. 
The one foreground galaxy (green symbol in the left panel) is also located in this area.

The lower panels of Fig.\,\ref{fig:spatial_distribution} show the positions of the background galaxies (magenta) with $z < 0.06$. The CfA survey \citep{Huchra_1999} has found a huge under-density between the Perseus-Pisces supercluster 
in the foreground and a dense structure at $z \approx 0.04$ as the distant boundary. 
In fact, the panel for $0.04 < z < 0.06$ indicates a coherent background structure of a comparatively high galaxy density. This redshift slice includes the background cluster at $\mbox{RA, DEC} = 50\fdg1, 43\fdg1$ that was first reported by \citet{Brunzendorf_1999}. Spectroscopic redshifts are available for eight galaxies in the field of this cluster. 
One of them is found to belong to the  Perseus cluster, the mean redshift of the remaining seven galaxies is 
$z = 0.051\pm0.004$. It is worth mentioning that there are two broad-line galaxies in this redshift interval, J032006.3+402159 and J032512.9+404153, both at $z=0.0472$.

%
\section{Cluster profiles and substructure}\label{sect:Cluster_profiles}
%

\subsection{Line-of-sight velocities}\label{sect:velocity_distribution}

\begin{figure}[htbp]
\includegraphics[viewport= 0 0 560 800,width=4.4cm,angle=0]{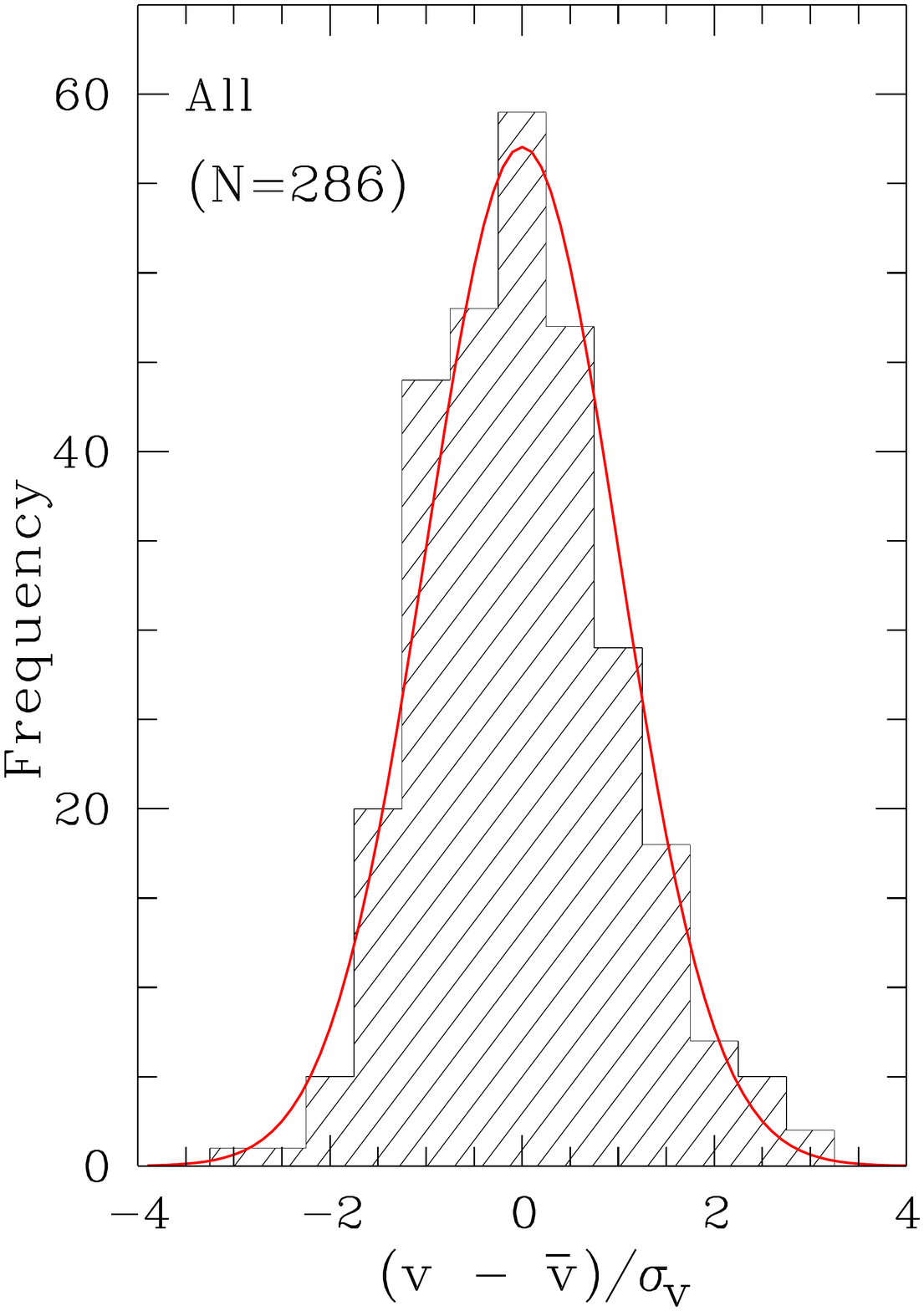}
\includegraphics[viewport= 0 0 560 800,width=4.4cm,angle=0]{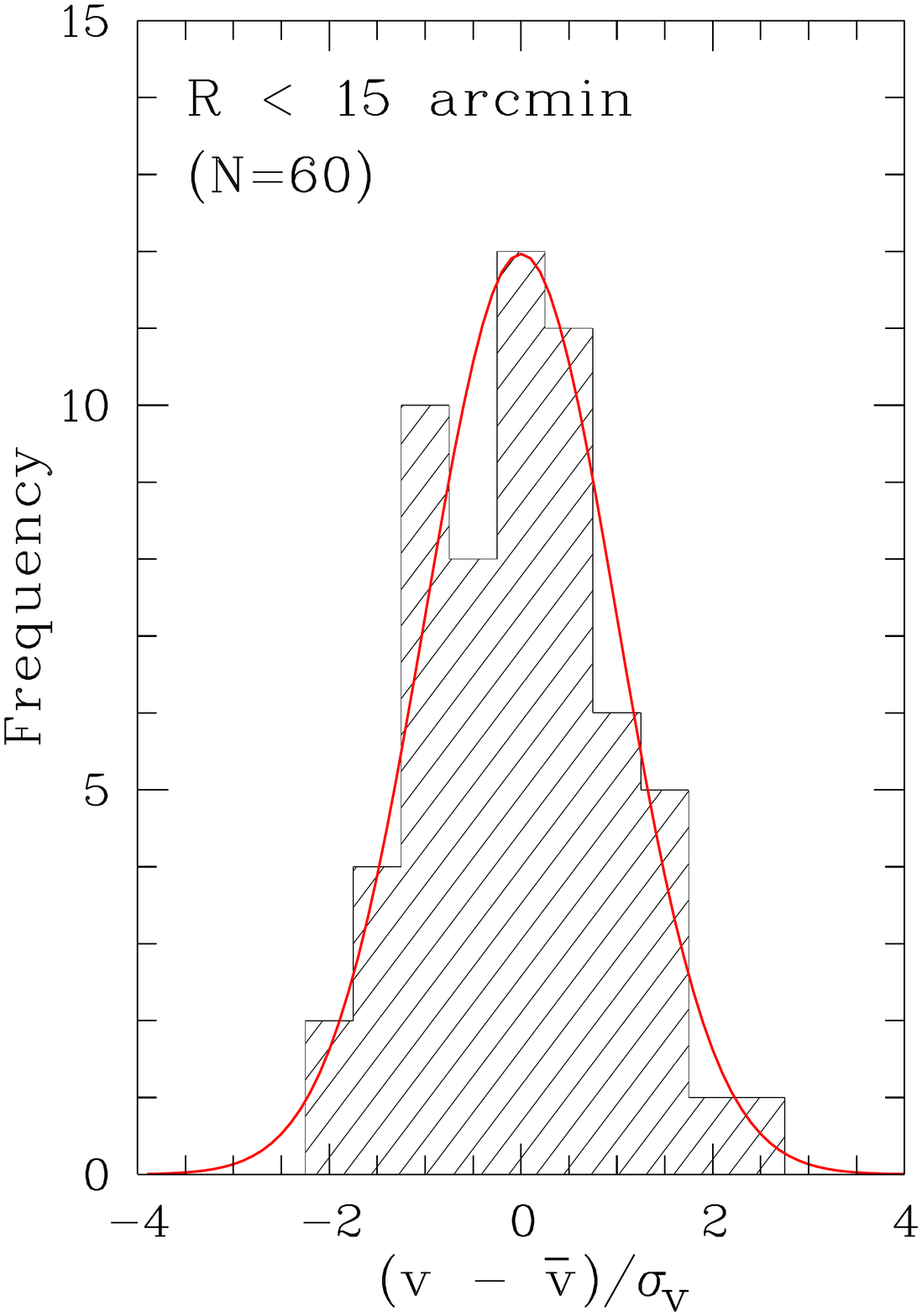}
\caption[Velocity distribution]{
Velocity distribution of the SCGS galaxies (black) and best-fitting Gaussian (red) for all sample members (left) and  the inner 15\,arcmin (right). 
}
\label{fig:v_distribution}
\end{figure}

Dynamically relaxed clusters are expected to have Gaussian line-of-sight velocity distributions, 
whereas departures from the Gaussian distribution indicate unrelaxed systems. 
Figure\,\ref{fig:v_distribution} shows the histograms of the relative velocity ${\rm v} - \bar{\rm v}$ in units of the standard deviation $\sigma_{\rm v}$ for all SCGS galaxies (left) 
and for the inner 15\,arcmin, which roughly corresponds to the core region (right), 
overplotted by the best-fitting Gaussian. 
We performed chi-square goodness of fit tests to check the null hypothesis $H^0$ that the
observed velocities are normal distributed. We computed the test statistic 
$ T = \sum_{i=1}^{n} {(N_i - N \cdot p_i)^2}/{p_i}$,
where $n$ is the number of non-empty bins, $N_i$ is the number of galaxies in bin $i$, 
$p_i$ is the Gaussian probability of falling into bin $i$ 
and $N = \sum_i N_i$.  
The null hypothesis is rejected if $T$ is larger than the critical value $\chi^2_c$ for a given significance level $\alpha$. For the two samples from Fig.\,\ref{fig:v_distribution},
we find $(T,\chi^2_c) = (11.9, 21.7)$ (left) and $(3.8, 18.5)$ (right).

Cluster substructure or an ongoing cluster merger may result in a bimodal or multimodal velocity distribution.  
A straightforward approach to probe distributions for unimodality versus bimodality is provided by Sarle's bimodality coefficient $b$ \citep{Pfister_2013} where $b$ larger than the benchmark value $b_{\rm crit} = 5/9 \approx 0.55$ points towards bimodality or multimodality. 
The two samples from Fig.\,\ref{fig:v_distribution} have $b=0.30$ and $0.38$.
There is obviously no reason to reject the assumption of a uniform Gaussian velocity distribution.

\begin{figure}[htbp]
\includegraphics[viewport= 40 30 600 800,width=6.8cm,angle=270]{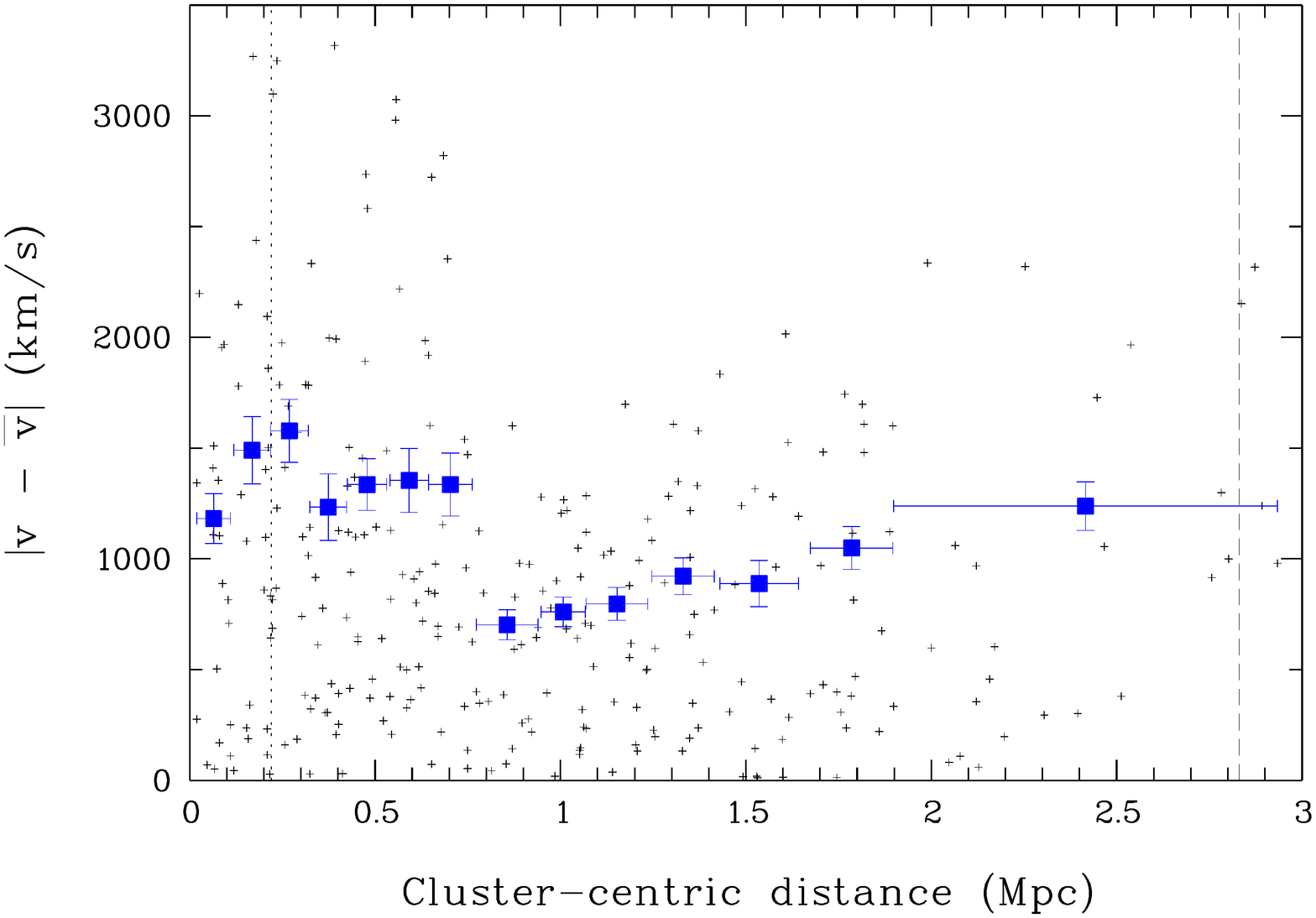}
\caption[Radial velocity dispersion distance]{
Absolute value of the line-of-sight velocity $\mbox{v}$ relative to the cluster mean velocity $\bar{\mbox{v}}$ as a function of the cluster-centric distance in Mpc (small black plus signs). Filled blue squares are the standard deviations of the line-of-sight velocity in bins of cluster-centric radii where the horizontal bars indicate the bin width 
and vertical bars the standard error of the standard deviation estimator following \citet{Ahn_2003}. The other 
vertical lines mark the core radius (dotted) and the virial radius $R_{\rm vir,200}$ (dashed). 
}
\label{fig:sigma_d}
\end{figure}

The radial velocity dispersion profile, that is, the line-of-sight velocity dispersion $\sigma_{\rm v}$ as a function of the cluster-centric distance, is an important tracer of the dynamical state of the cluster 
\citep[e.g.][]{Carlberg_1997, Hou_2009, Pimbblet_2014, Costa_2018}. 
Figure\,\ref{fig:sigma_d} shows the binned SCGS data of the radial velocity dispersion profile where 
the bin sizes were chosen in such a way that each bin contains 20 galaxies, with the only exception of 26 galaxies in the last interval. The velocity dispersion $\sigma_{\rm v}$ in each bin is related to the mean cluster velocity, 
and not to the average velocity within a bin. 
An early investigation of the velocity dispersion profile of the Perseus cluster, based on 119 measured redshifts, was presented by \citet{Struble_1979}. Some properties shown there are confirmed by our data\footnote{after re-scaling the cluster-centric distances by a factor 0.67 to correct for the difference in the adopted cluster distance}: the general trend of a decrease at approximately $0.1 \ldots 1$\,Mpc, a central dip, and local minima at $0.3 \ldots 0.4$\,Mpc and $0.8 \ldots 0.9$\,Mpc. We also confirm that the local minima occur at approximately the same radius as that at which local minima occur in the surface density and surface brightness profiles (Fig.\,\ref{fig:SD_profile}). This can also be seen in the distribution of the individual relative velocities of the cluster galaxies (black plus signs) in Fig.\,\ref{fig:sigma_d}.

\citet{Costa_2018} compared the composite velocity dispersion profiles of two samples of galaxy clusters (A\,426 not included), such with Gaussian velocity distributions (G) and such with non-Gaussian distributions (NG). They found significant differences: 
For the G sample, the velocity dispersion increases to $0.35\,R_{200}$ and monotonically decreases at larger radii.
For the NG sample, the velocity dispersion profile exhibits a central depression at $\sim 0.4\,R_{200}$ and increases from about $0.5\,R_{200}$ to $1.0\,R_{200}$, where $R_{200}$ is a proxy for the virial radius (see Sect.\,\ref{sect:virial_radius}, below). Such an increase was also found by \citet{Hou_2009} for NG galaxy groups. 
The trend seen in Fig.\,\ref{fig:sigma_d} is in better agreement with the NG composite. However, as demonstrated above, the observed velocity distribution is close to Gaussian, particularly in the central part.  
Deviations from the Gaussian are expected in the outskirts caused by the infall of galaxy groups. 
On the other hand, the increase in $\sigma_{\rm v}$ in the outer part at $\ga 1$\,Mpc may partly be caused by
background contamination. Of course, the radial velocity dispersion profile is strongly affected by the assumed cluster membership criterion. It is possible that the criterion shown in Fig.\,\ref{fig:membership} is over-simplifying, particularly in the outer part of the cluster region.

\subsection{Cluster redshift and line-of-sight velocity dispersion}\label{sect:redshift}

The top four rows of Table\,\ref{tab:virial_radius} list the numbers $N$ of galaxies, the mean observed heliocentric redshift $\bar{z}$, the mean heliocentric radial velocity $\bar{\rm v} = \bar{z}c$, and the velocity dispersion $\sigma_{\rm v}$ for all SCGS galaxies and for the sub-sample of E+S0 galaxies. 
The mean observed redshifts are in line with the `canonical' value $z= 0.0179$ from \citet{Struble_1999}. 
More recently, the Hitomi collaboration published a redshift measurement of $z = 0.01767 \pm 0.00003$ for the ICM in A\,426 and $z = 0.01728 \pm 0.00005$ for the stellar absorption lines in the spectrum of the BCG NGC\,1275
\citep{Hitomi_2018a, Hitomi_2018b}. For comparison, the mean value of the 53 NGC\,1275 redshift measurements listed in the NED is $z = 0.01752$ with a standard deviation of 0.00030.

\begin{table}[h]\centering
\caption{Mean redshift, line-of-sight velocity dispersion, cluster radius, and cluster mass from the spectroscopic cluster galaxy sample (SCGS).}
\begin{tabular}{lll}
\hline\hline
                                    & SCGS all               & SCGS E+S0     \\
\hline
$N$\hspace{2cm}                     & 286                    &   212        \\
$\bar{z}$                           & $0.0177\pm0.0002$      &  $0.0180\pm0.0003$   \\
$\bar{\rm v}$ (km/s)                & $5317$                 &  $5399$              \\
$\sigma_{\rm v}$ (km/s)             & $1163\pm49$            &  $1107\pm54$         \\
\hline
$\bar{z}_{\rm c}$                   & $0.0172\pm0.0002$      &  $0.0175\pm0.0002$   \\
$\bar{\rm v}_{\rm c}$ (km/s)        & $5160$                 &  $5242$              \\
$\sigma_{\rm v,c}$ (km/s)           & $1143\pm48$            &  $1088\pm53$         \\
\hline
$R_{\rm vir,200}$ (Mpc)             & $2.8\pm0.1$            &  $2.7\pm0.1$         \\
$R_{\rm vir,p}$ (Mpc)               & $2.4\pm0.5$            &  $2.1\pm0.4$       \\
$\mathcal{M}_{\rm vir,200} (\,10^{15} \mathcal{M}_\odot$)  & $2.6\pm0.3$ & $2.2\pm0.3$ \\
$\mathcal{M}_{\rm vir,p} (\,10^{15} \mathcal{M}_\odot$)    & $2.2\pm0.5$ & $1.8\pm0.4$ \\
$\mathcal{M}_{\rm pm}^{\ast} (\,10^{15} \mathcal{M}_\odot$)& $2.7\pm0.2$ & $2.2\pm0.2$ \\
\hline
\end{tabular}
\label{tab:virial_radius}
\tablefoot{$^\ast$ adopting $k_{\rm pm} = 32/\pi$ for isotropic orbits \citep{Heisler_1985}}
\end{table}

The next three rows list the redshifts $\bar{z}_{\rm c}$ and the mean velocities $\bar{\rm v}_{\rm c}$ corrected to the velocity centroid of the cosmic microwave background \citep{Struble_1999}
and the radial velocity dispersion in the frame of the cluster $\sigma_{\rm v,c} = \sigma_{\rm v}/(1+\bar{z}_{\rm c})$ \citep{Harrison_1974}. The uncertainty of the velocity dispersion is the standard deviation estimator \citep{Ahn_2003}.
The mean redshift with respect to the cosmic microwave background, $\bar{z}_{\rm c} = 0.0172$, 
corresponds to a cluster distance of 74\,Mpc, a distance modulus $DM = 34.34$\,mag and a scale of 0.36 kpc\,arcsec$^{-1}$. 
The angular size of the Abell radius is 100\,arcmin.
These values are used in the rest of this paper.

\subsection{Cluster mass and radius}\label{sect:virial_radius}

Under the assumption of virial equilibrium and spherical symmetry, the line-of-sight velocity dispersion can be related to the cluster virial mass $\mathcal{M}_{\rm vir}$ and virial radius $R_{\rm vir, \Delta}$
where the latter can be expressed in terms of the density contrast parameter $\Delta = \rho_{\rm vir}/\rho_{\rm crit}$
with $\rho_{\rm crit}$ as the critical density of the Universe \citep[e.g.][]{Lewis_2002}.
The radius $R_{\rm vir, 200}$ that encloses a mean density 200 times the critical density at $z$ is usually considered roughly equivalent to $R_{\rm vir}$ \citep[e.g.][]{Carlberg_1997}.
Another version of the virial radius estimator, $R_{\rm vir,p}$, is based on the projected distances between the galaxies \citep{Heisler_1985}. An alternative mass estimator is the projected mass $\mathcal{M}_{\rm pm}$ \citep{Bahcall_1981, Heisler_1985, Crook_2007}.

The estimated values of $R_{\rm vir, 200}$ and $R_{\rm vir,p}$, the virial mass based on $R_{\rm vir, 200}$ and $R_{\rm vir,p}$, and the projected mass $\mathcal{M}_{\rm pm}$ for the  Perseus cluster are listed in the lower part of Table\,\ref{tab:virial_radius}. We found $2.1 \ldots 2.8$\,Mpc for the virial radius.
For comparison, the catalogue of groups of galaxies in the 2MASS redshift survey lists A\,426  
with $\sigma_{\rm v} = 1061$\,km/s and $R_{\rm vir, p} = 2.54$\,Mpc, based on 117 group members \citep{Crook_2007}. 
Our result is also in good agreement with $R_{\rm vir} = 2.44$\,Mpc derived by \citet{Mathews_2006} from modelling  the hot gas distribution invoking hydrostatic equilibrium, but it is larger than  $1.79$\,Mpc from \citet{Simionescu_2011}.  
$R_{\rm vir, p}$ depends on the central concentration of the selected galaxies. As mentioned in Sect.\,\ref{sec:redshifts}, the selection effects in the spectroscopic sample are largely unknown, but there is very likely a selection bias towards the central cluster region.
Furthermore,  a bias towards brighter galaxies is expected to result in a smaller value of $R_{\rm vir, p}$ because of the known luminosity segregation in the Perseus cluster \citep[e.g.][Sect.\,\ref{sect:surf_density} below]{Kent_1983, Brunzendorf_1999}. 
We did indeed find a systematic increase in $R_{\rm p}$ with the mean W1 magnitude for the samples from Table\,\ref{tab:samples} with 2.25\,Mpc for MLS\_W1b and 2.95\,Mpc for MLS\_ur. A radius of 2.8\,Mpc corresponds to 
an angular radius of $130$\,arcmin and thus our survey field extends over approximately the full virial radius of the cluster, though the outer part ($> 100$\,arcmin) is not completely covered. 

All three methods result in a total mass estimator in the range $(1.8 \ldots 2.7)\,10^{15} \mathcal{M}_\odot$ (Table\,\ref{tab:virial_radius}, bottom). 
The projected mass estimator is in very good agreement with the virial mass estimator $\mathcal{M}_{\rm vir,200}$. 
(The uncertainty of $\mathcal{M}_{\rm pm}$ was derived by bootstrapping and is a lower limit only because it does not include the uncertainty of $k_{\rm pm}$.) 
The values of $\mathcal{M}_{\rm vir,200}$ and $\mathcal{M}_{\rm pm}$ derived for the early-type galaxies are identical with $\mathcal{M}_{\rm pm} = 2.2\,10^{15}\,\mathcal{M}_\odot$ reported by \citet{Crook_2007}. 
$\mathcal{M}_{\rm vir,p}$ is a factor of $1.2$ smaller, which is caused by the same selection effects that lead to a smaller value of $R_{\rm vir, p}$.
The mass estimation is based on the assumption that the radial velocity dispersion of the selected galaxies is a tracer of the cluster mass, which may not be the case if the velocity histogram of the galaxies differs significantly from a Gaussian distribution. Though the histograms in Fig.\,\ref{fig:v_distribution} indicate that the Gaussian fit is not perfect, we conclude from the $\chi^2$ test that we do not need to reject the assumption of a Gaussian distribution.

\subsection{Radial surface density profile}\label{sect:surf_density}

\begin{figure}[htbp]
\includegraphics[viewport= 40 30 600 800,width=6.4cm,angle=270]{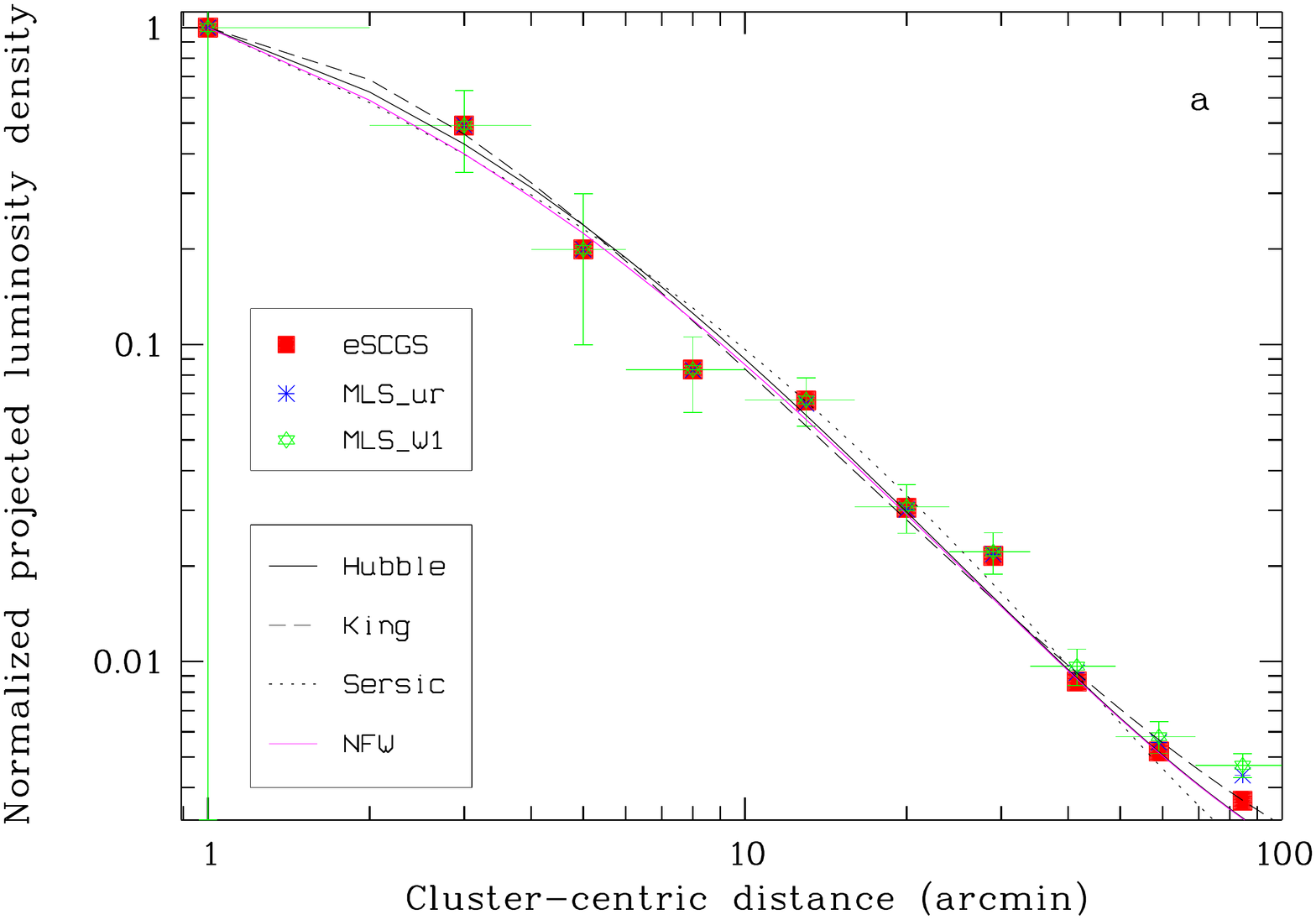}
\includegraphics[viewport= 40 30 600 800,width=6.4cm,angle=270]{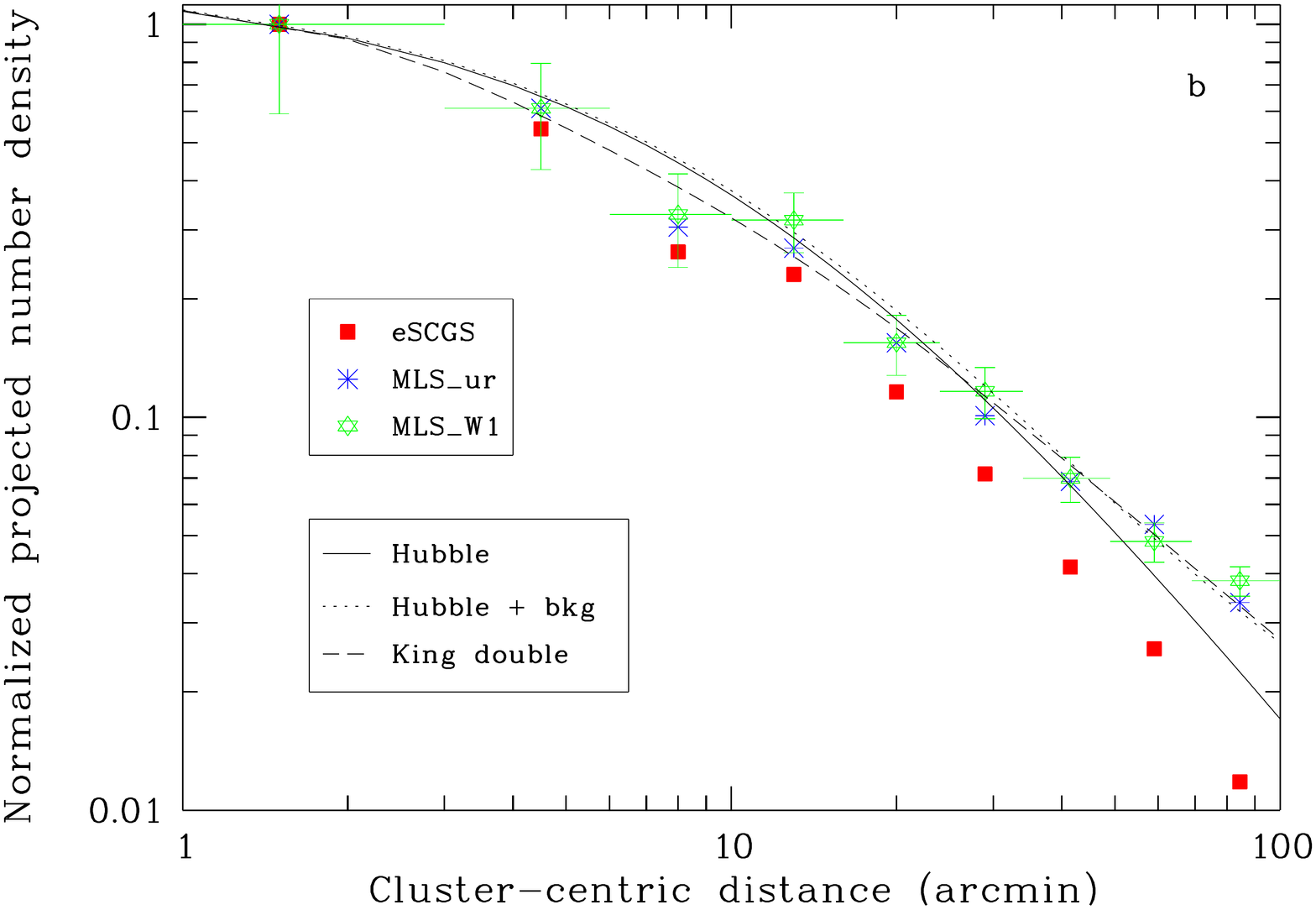}
\includegraphics[viewport= 40 30 600 800,width=6.4cm,angle=270]{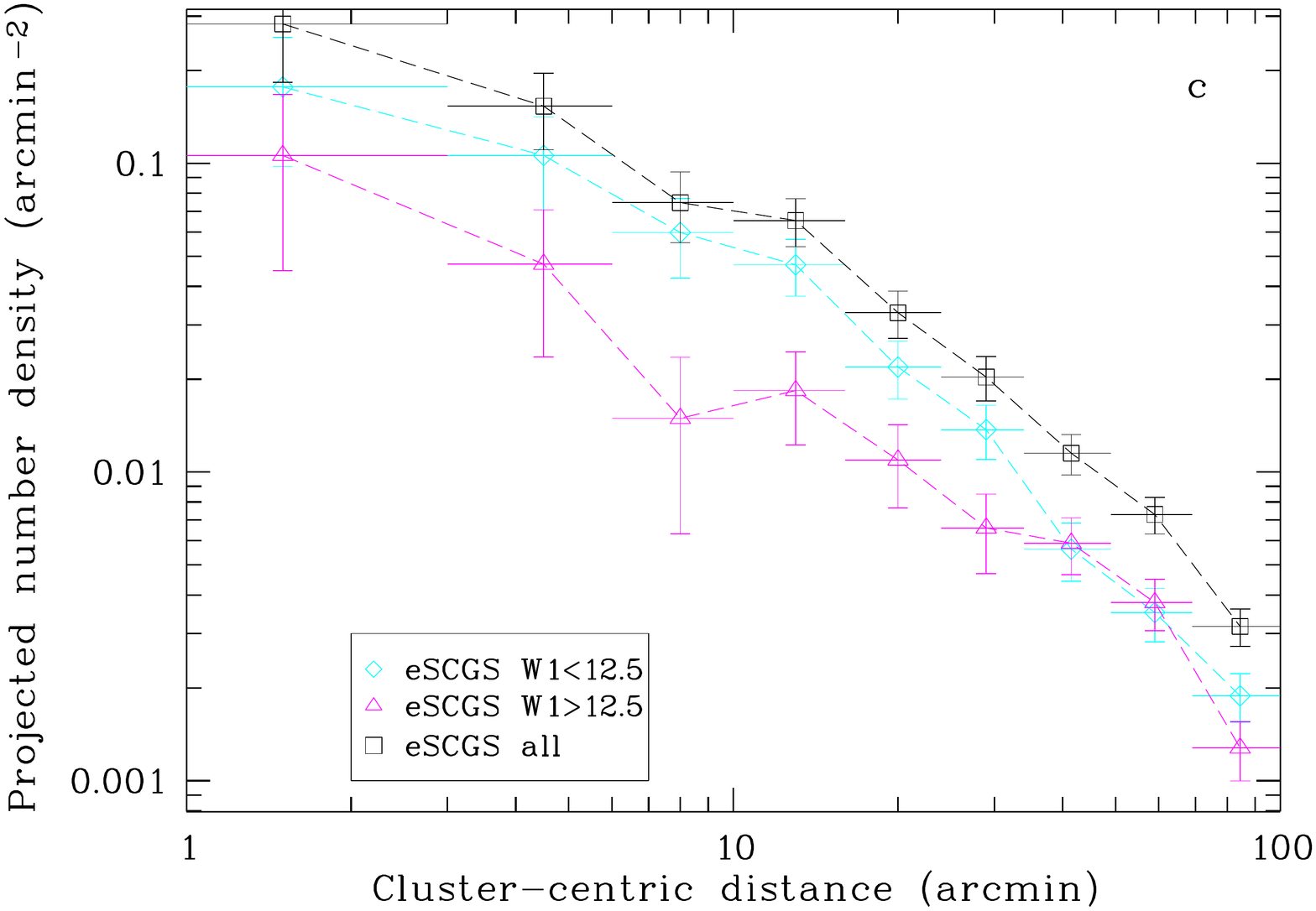}
\caption[hist_W1_lim_samples]{Radial projected profiles for the K-band surface brightness (a) and the number densities
(b and c). 
Bars indicate formal statistical errors (vertical) and binning intervals (horizontal).
The curves in panel a are for the model parameters from Table\,\ref{tab:rad_lprofile}.
In panel b, the Hubble model is for $(R_{\rm c}, \beta, \Sigma_{\rm b}) = (10\arcmin, 0.9, 0)$, 
the dotted curve includes a background fraction $f_{\rm b} = 0.15$. The dashed curve is a double-King profile
with $(R_{\rm c,1}, \beta_1, R_{\rm c,2}, \beta_2) = (3\arcmin, 1, 10\arcmin, 0.58)$ and $\Sigma_{\rm b} = 0$.
In panel c, the symbols are interconnected by dashed lines just to guide the eye. 
}
\label{fig:SD_profile}
\end{figure}

The projected galaxy density in rich galaxy clusters is usually described by King or Hubble profiles.
Alternatively to these `core profiles', theoretical profiles with `cusps' are in use: the  S\'ersic profile
and the NFW profile. Here we followed the approach by \citet{Adami_1998} and compared the observed radial profile 
with four different generalised theoretical models:
\begin{eqnarray}                                                    
&  \Sigma \propto & [1 + x]^{-2\beta} + \Sigma_{\rm b}                    \hspace{1.85cm} \mbox{(generalised Hubble)} \label{eq:Hubble} \\ 
&  \Sigma \propto & [1 + x^2]^{-\beta} + \Sigma_{\rm b}                   \hspace{1.81cm} \mbox{(generalised King)}   \label{eq:King} \\
&  \Sigma \propto & [R_{\rm s} f(x,c)/(1-x^2)]^{\,\beta} + \Sigma_{\rm b} \hspace{0.33cm} \mbox{(generalised NFW)}    \label{eq:NFW} \\
&  \Sigma \propto & \exp\{-b_n[x^{\, \beta}-1]\} + \Sigma_{\rm b}         \hspace{0.85cm} \mbox{(S\'ersic)},          \label{eq:Sersic} 
\end{eqnarray}
where $x = R/R_{\rm c}$ is the projected cluster-centric distance in units of a scale radius $R_{\rm c}$, which has  different meanings in the different models. For the Hubble and King model, respectively,  $R_{\rm c}$ corresponds to the core radius $R_0$. 
The projected NFW profile was taken from  \citet{Bartelmann_1996}; the function $f(x,c)$ is given there, with $c = R_{\rm vir}/R_{\rm c}$ being the concentration parameter and $R_{\rm vir}$ the virial radius. The term $b_n$ in the S\'ersic model, which is a generalised de Vaucouleurs model, is not a parameter but a function of the S\'ersic index $n = 1/\beta$ \citep[e.g.][]{Merritt_2006}. $\Sigma_{\rm b}$ is the projected background density.  

The models have three free parameters: $R_{\rm s}, \beta$, and $\Sigma_{\rm b}$. Two additional parameters are 
the cluster centre coordinates. Errors in the central position mainly affect the inner part of the derived cluster profile. 
The galaxy catalogue lists cluster-centric distances based on the median cluster centre from \citet{Ulmer_1992}. 
Here, we re-defined the cluster centre by the luminosity-weighted mean position of the catalogue galaxies 
in an iterative process. First, we identified a preliminary centre with the mean position in a 40 arcmin wide box centred on NGC\,1275. In a second step, we centred a box of the same size on this preliminary centre and re-computed the luminosity-weighted mean position. The distances to this new cluster centre at 
RA,\,DEC = $49\fdg932,\,41\fdg494$ lead to a steeper radial number density profile. Since a wrongly determined cluster centre tends to reduce or destroy a potential cusp \citep[e.g.][]{Adami_1998}, we believe that the new cluster-centric distances are more adequate in the context of the present section. 

Figure\,\ref{fig:SD_profile}a shows the projected radial K-band luminosity density profile overplotted by the best-fitting curves from Eqs.\,\ref{eq:Hubble}-\ref{eq:Sersic} for an adopted uniform background density equal to 25\% of the density in the largest $R$ bin (corresponding to a total background fraction $f_{\rm b} = 0.15$ for MLS\_W1).
The K-band luminosity $L_{\rm K}$ was computed from the absolute magnitude 
$M_{\rm K} = K - DM - k$ adopting
$M_{\rm K,\odot} = 3.27$ (Vega) \citep{Cox_2000}, where $K$ is the 2MASS K-band magnitude corrected for Galactic foreground extinction (Sect.\,\ref{sect:obs_photo}), $k = -6 \log (1+z)$ is the K-band $k$-correction \citep{Kochanek_2001}; a fixed distance modulus of $DM = 34.34$ is adopted for all galaxies. If no 2MASS photometry is available, $K$ was estimated from the mean relation $K = -1.2 + 1.08\cdot W1$ using the WISE W1 magnitude. 
The best-fit parameters (Table\,\ref{tab:rad_lprofile}) were found by minimising $\chi^2$ on 
a grid in the three-dimensional parameter space ($R_{\rm s}, \beta, \Sigma_{\rm b}$).   
According to the $\chi^2$ fitting test, the null hypothesis, that the observed radial profile is represented by the best fitting model, has not to be rejected on a 99\% level for a significance level $\alpha = 0.01$ and nine degrees of freedom. Adopting other values of $\Sigma_{\rm b}$ results in slightly different values of $R_{\rm s}$ and $\beta$ 
but similarly good fits. Generally, the core models reproduce the observed profiles better than the cusp models, 
in accordance with the results from the analysis of the number density profiles for a sample of rich clusters (A\,426 not included) by 
\citet{Adami_1998}. The $R_{\rm c}$ values (Table\,\ref{tab:rad_lprofile}) are three times smaller than those reported by \citet{Adami_1998}. This is most likely because the luminosity profile in Fig.\,\ref{fig:SD_profile}a is dominated by the most luminous galaxies in the cluster core region.

\begin{table}[h]
\caption{Best fit parameters for the projected radial surface brightness profile for sample MLS\_W1 with a
background fraction $f_b = 0.15$.
}
\begin{tabular}{lccr}
\hline\hline
Sample 
model              & $R_{\rm c}$ (\arcmin) & $\beta$ & $\chi^2$ \\
\hline
generalised Hubble &  2\farcm7$\pm  0.4$ & 0.98$\pm 0.18$  & 12.2 \\
generalised King   &  2\farcm0$\pm  0.5$ & 0.83$\pm 0.20$  &  8.6\\
generalised NFW    &  4\farcm3$\pm  1.0$ & 0.99$\pm 0.11$  & 13.8 \\
S\'ersic           & 27\farcm5$\pm 11.6$ & 3.37$\pm 0.90$  & 17.1 \\
\hline
\end{tabular}
\label{tab:rad_lprofile}
\end{table}

The projected radial number density profiles are shown in the middle panel of Fig.\,\ref{fig:SD_profile}. 
The observed data for the magnitude-limited samples are best reproduced by a Hubble profile with a 
background fraction $f_{\rm b} \approx 0.15$ and the profile parameters $\beta=0.9$ and $R_{\rm c} = 10\arcmin$, corresponding to $216$\,kpc. The average value of the Hubble core radius in the \citet{Adami_1998} sample is 
$\langle R_{\rm c} \rangle = 189 \pm 116$\,kpc. 
Compared with the spectroscopic sample, the observed profiles for the magnitude-limited samples flatten at $R \ga 20\arcmin$. Such a trend could be caused by background contamination. On the other hand, both a selection bias in the spectroscopic sample (Sect.\,\ref{sec:redshifts}) and the luminosity segregation do also contribute to this difference (Fig.\,\ref{fig:SD_profile}c). Hence, the magnitude-limited samples are expected to show a flatter profile. Alternatively, a perfect fit can be achieved by the superimposition of two King profiles, as is illustrated by the dashed curve in panel b. Such `double-beta' models have been invoked to describe the radial profile of the X-ray gas in clusters \citep{Mohr_1999, Holanda_2018}.

\subsection{Substructure}\label{sect:substructure}

\subsubsection{Projected local density}\label{sect:projected_density}

\begin{figure}[htbp]
\includegraphics[viewport= 0 0 519 374,width=9.0cm,angle=0]{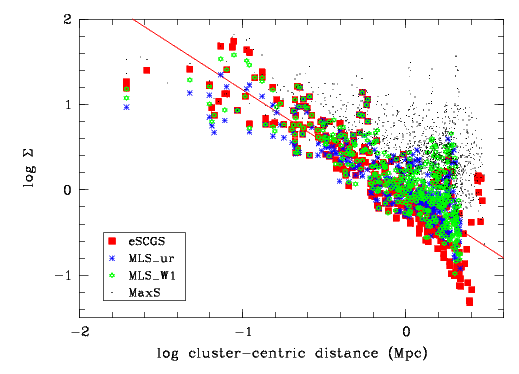}
\caption[ld_dc]{Local projected density parameter $\Sigma$ versus cluster-centric distance
for the galaxies in the samples eSCGS (red), MLS\_ur (green), MLS\_W1 (blue), and MaxS (black).
}
\label{fig:ld_dc}
\end{figure}

\begin{figure}[htbp]
\includegraphics[viewport= 0 0 550 530,width=9.2cm,angle=0]{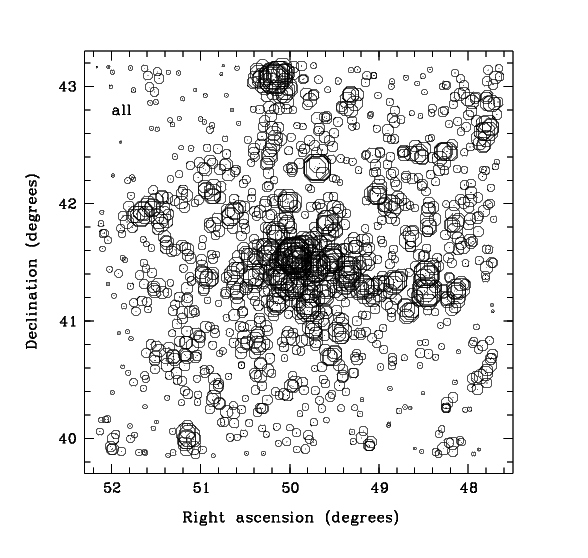}\\
\includegraphics[viewport= 0 0 550 530,width=9.2cm,angle=0]{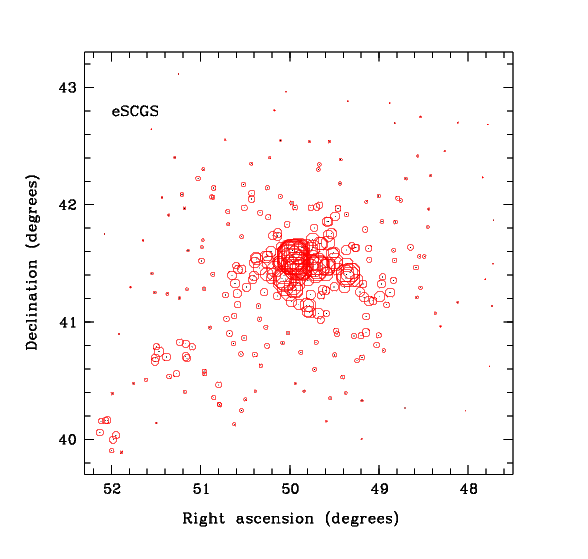}
\caption[projected_density_parameter]{
Bubble plot diagram for the local density parameter $\Sigma$ (Eq.\,\ref{eq:Sigma}) for the whole catalogue sample (top) and the eSCGS (bottom). 
}
\label{fig:local_density}
\end{figure}

We computed, for each galaxy $i$, the projected local density parameter 
\begin{equation}\label{eq:Sigma}
  \Sigma_i = k\,\frac{1}{2}\,\Bigg[\frac{4}{\pi d_{i,4}^2}+\frac{5}{\pi d_{i,5}^2}\Bigg]
\end{equation}
\citep{Baldry_2006, Rawle_2013}, where  $d_{i,4}$ and $d_{i,5}$ are the projected distances to the 4th and 5th nearest galaxy in arcmin and $k$ is an arbitrary constant. 
We set $k=100$, so that $\Sigma$ gives the number of galaxies in the local environment extrapolated to an area of 100 square arcmin if $d_i$ is given in arcmin. The density parameter was computed for each galaxy sample separately.

Figure\,\ref{fig:ld_dc} compares the projected local density parameter $\Sigma$ to the projected cluster-centric radius for the four samples. 
The diagonal line is the linear regression curve for the eSCGS galaxies. 
The generally good correspondence over about two orders of magnitude suggests that the Perseus cluster is mostly relaxed, although some over-density is indicated at cluster-centric distances 
$R \approx 0.5$, 1.7 and 2.8\,Mpc.
The bright magnitude-limited sample MLS\_W1b (not shown for clarity) shows a similar correlation, 
where the local densities tend to be smaller, of course. 
For the maximum sample (MaxS), on the other hand, the $\Sigma$ values tend to be larger and the correlations are weaker because of more and stronger density concentration at larger $R$.

Figure\,\ref{fig:local_density} shows bubble sky-plots where each galaxy is represented by a small black dot surrounded by a circle whose radius is set proportional to $\log (1+\Sigma$). 
The cluster centre is at RA,\,DE = $49\fdg9,\,41\fdg5$. A clustering of larger bubbles 
indicates a possible location of substructure in the projected distribution. 
Conspicuous structures in the whole catalogue sample are the core of the Perseus cluster, the well-known chain of bright galaxies in WSW direction from the centre, and the background cluster at $\mbox{RA, DEC}, z = 50\fdg1, 43\fdg1, 0.051$ (Sect.\,\ref{sect:background}). 
In addition to the chain of bright galaxies, the plot of the eSCGS galaxies also shows a nearly parallel structure south of the core. The same structure is seen in the two magnitude-limited samples.

\subsubsection{Three-dimensional (DS) test}

Substructures in the projected density distribution may be coincidental results of projection effects.  
The DS $\Delta$ test \citep{Dressler_1988} is one of the most sensitive three-dimensional tests \citep{Pinkney_1996} and has become a standard tool in cluster studies \citep[e.g.][]{Mulroy_2017, Golovich_2018}. 
For each galaxy $i$ the algorithm measures the dimensionless deviation
\begin{equation}\label{eq:delta}
  \delta_i = \sqrt{N_{\rm nn}+1}\,\sqrt{\frac{(\bar{\rm v}_{\rm loc}- \bar{\rm v})^2 +(\sigma_{\rm v,loc}-\sigma_{\rm v})^2}{\sigma_{\rm v}^2}}
\end{equation}
of local (index `loc') and global mean velocity $\bar{\rm v}$ and velocity dispersion $\sigma_{\rm v}$
where $N_{\rm nn}$ is the number of nearest neighbours (including the galaxy $i$ itself) in the plane of the sky. 
The result is usually presented in a bubble sky-plot.   
A clustering of larger bubbles indicates a most likely location of substructure.

\begin{figure}[htbp]
\includegraphics[viewport= 40 10 580 580,width=8.8cm,angle=270]{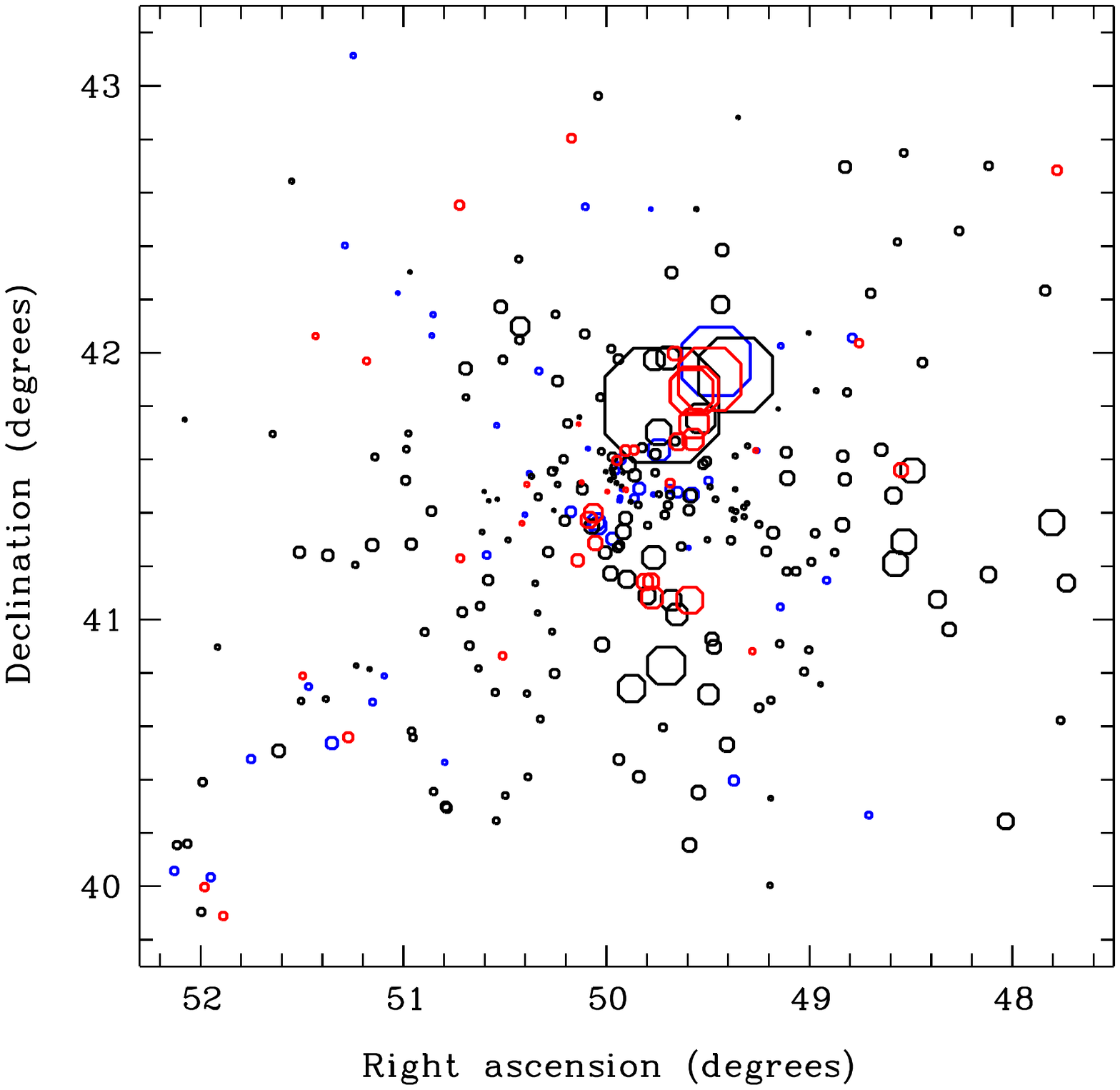}
\caption[Substructure_DS]{
Bubble plot diagram for the DS test. Galaxies are plotted by hexagons whose size is proportional to the test statistic $\delta$ on an exponential scale where the colours represent different intervals of relative velocities (see text).  
}
\label{fig:substructure_DS}
\end{figure}

Figure\,\ref{fig:substructure_DS} shows the bubble plot for the SCGS galaxies 
with $N_{\rm nn}=\sqrt{N}=17$ following \citet{Pinkney_1996}.
The three colours indicate three intervals of the relative velocity $\rm{v}_{\rm rel}$. 
Galaxies within $\pm  \sigma_{\rm v}$ from the mean cluster velocity are shown in black, galaxies in the blue or the red wing by blue or red symbols, respectively. With only one exception, all larger bubbles are either red or black. 
The largest deviations are seen about 20\arcmin\ NW of the cluster centre and about 20\arcmin\ south of the centre. 
Both two regions are seen also in Fig.\,\ref{fig:local_density} where the NW substructure is however only weakly indicated. The individual inspection does not reveal any other conspicuous properties of the galaxies in these regions.
The chain of bright galaxies seen in Fig.\,\ref{fig:local_density} does not appear as a substructure in the three-dimensional map. 
                                                                                                                                                                                                                                                                                                                                                                               The DS test is based on the null hypothesis of constant mean velocity and velocity dispersion as a function of position. 
The presence of substructure is quantified by the cumulative deviation $\Delta = \sum_i \delta_i$. 
We measured $\Delta = 382.0$ for the SCGS galaxies.
These results must be compared to the results from data files that are consistent with the null hypothesis of
no correlation between position and velocity. Following \citet{Dressler_1988} and \citet{Pinkney_1996}, we performed 
Monte Carlo simulations of the input data where the positions stayed fixed and the velocities were shuffled 
randomly with respect to the positions. 
From 2000 Monte Carlo runs we found $\Delta_{\rm MC}  = 313.6 \pm 32.1$, which means that the $\Delta$ value for the observed velocities from our spectroscopic cluster galaxy sample is $\approx 2 \sigma$ above the value expected under the null hypothesis. The simulated $\Delta$ was larger than the observed one in 52 runs ($2.6\%$). 
That is, the null hypothesis has to be rejected at a 2.6\% significance level, which is a marginal result.

%
\section{Properties of the cluster galaxy population}\label{sect:Cluster_galaxies}
%

\subsection{Morphological types}\label{sect:morph_seg}

\subsubsection{Morphological types and SDSS colour-magnitude diagram}\label{sect:cmd_morph}

\begin{figure*}[htbp]
\centering
\includegraphics[viewport= 0 0 540 540,width=5.6cm,angle=0]{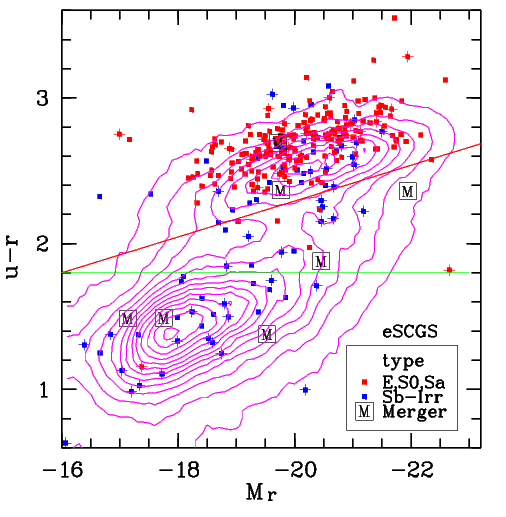}
\includegraphics[viewport= 0 0 540 540,width=5.6cm,angle=0]{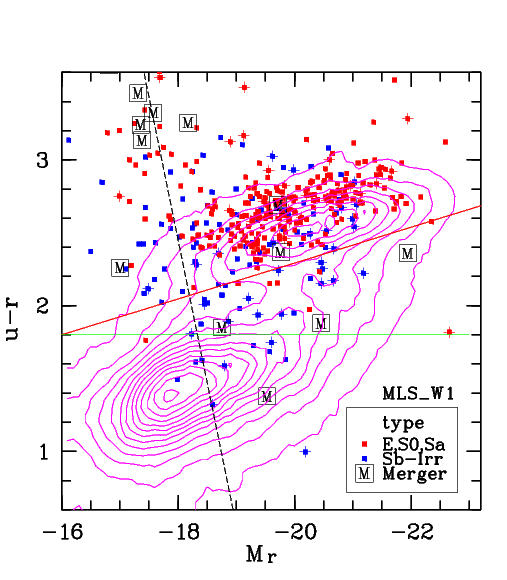}
\includegraphics[viewport= 0 0 540 540,width=5.6cm,angle=0]{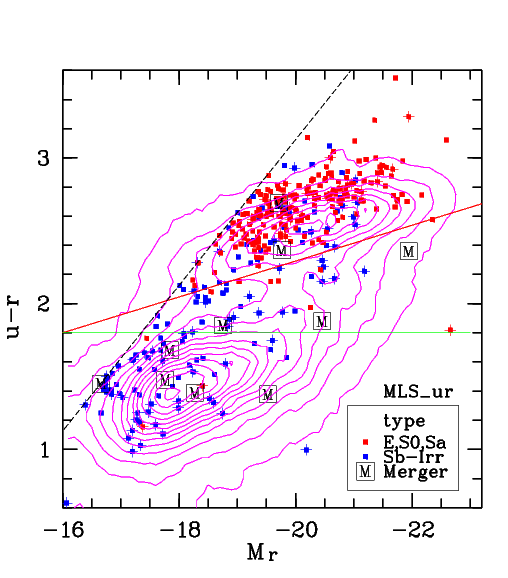}\\
\includegraphics[viewport= 0 0 540 540,width=5.6cm,angle=0]{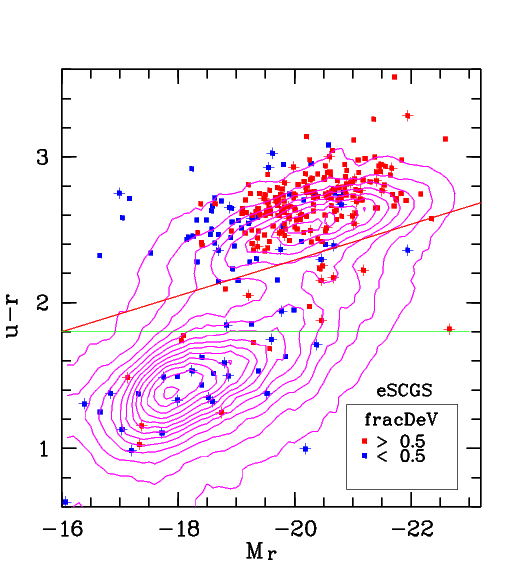}
\includegraphics[viewport= 0 0 540 540,width=5.6cm,angle=0]{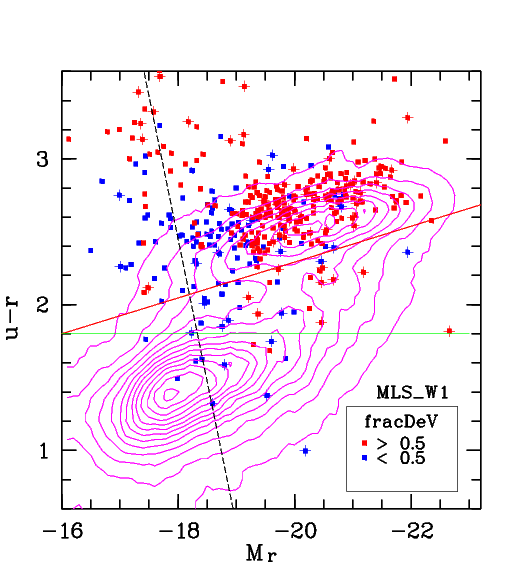}
\includegraphics[viewport= 0 0 540 540,width=5.6cm,angle=0]{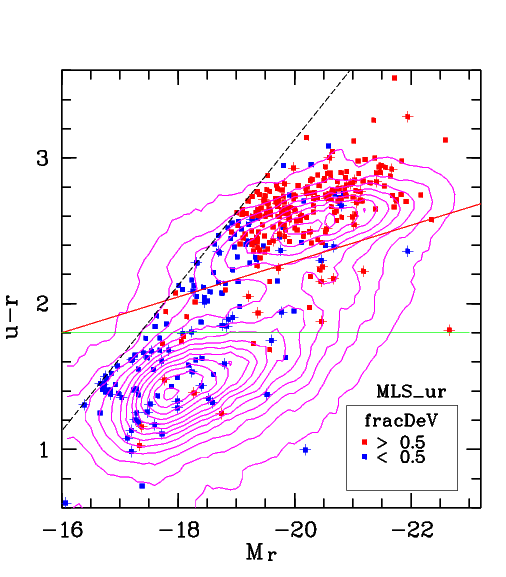}\\
\vspace{0.3cm}
\caption[CMD]{
Colour magnitude diagrams for the samples eSCGS, MLS\_ur, and MLS\_W1 (left to right).
Different colours indicate morphological types (top) or the {\tt fracDeV} type (bottom).
Morphological peculiarities are indicated by an overplottet plus sign, a framed `M' (top raw) indicates a merger system. 
The dashed lines mark the approximate border of the colour selection area for the magnitude-limited samples (see text). 
The magenta contours show the population density of nearby galaxies from SDSS.  
The solid lines mark the transition between red sequence and green valley (red) and green valley and blue cloud (green). 
}
\label{fig:CMD}
\end{figure*}

Figure\,\ref{fig:CMD} shows the colour-magnitude diagrams (CMD) for the three samples eSCGS, MLS\_ur and MLS\_W1, 
where $u$ and $r$ are the SDSS magnitudes corrected for galactic foreground extinction. 
The few faint galaxies with $M_{\rm r}<-16$ (about 0.5-2\%, dependent on the sample) were ignored. 
For comparison, the 19\,730 SDSS DR14 galaxies in the redshift range $z = 0.010 ... 0.026$ are overplotted as 
equally-spaced local point density contours with a grid size of $\Delta x, \Delta y = 0.1, 0.05$. 
In the top row, the morphological types from our visual classification are indicated by red squares for early types, blue squares for late types, 
and by `M' for mergers ({\tt cl1} = 9).
Disturbed systems (peculiarity flag {\tt pec} $\ge 1$) are indicated by asterisks. 
The second row shows the same diagrams, but the colour coding denotes the morphological classification into disc-dominated ({\tt fracDeV}  $<0.5$, blue) and bulge-dominated ({\tt fracDeV}  $>0.5$, red) systems. 
Because of the incompleteness of the SDSS data (Sect.\,\ref{sect:obs_photo})  not all sample members are plotted. 
For example, for the 313 eSCGS galaxies, the {\tt fracDeV} parameter is available for 300 and $u-r$ for 284 systems.
However, the proportion of late-type galaxies among the galaxies with available SDSS magnitudes (0.27) is very close to that for the whole sample (0.27). The same applies to the proportion of disc-dominated galaxies based on the {\tt fracDeV} parameter: 0.30 for the photometric sub-sample and 0.31 for the whole sample. 

For the cleanest of the three samples, the eSCGS, the following conclusions can be drawn from Fig.\,\ref{fig:CMD} (left column): Early-type galaxies are strongly concentrated in the red sequence.\footnote{There are two remarkable exceptions. One is the  cD galaxy \object{NGC\,1275} (\# 691) at $u-r = 1.82, M_r = -22.6$, which was classified here as a peculiar E galaxy. 
We found that its surface brightness profile fits the de Vaucouleurs $r^{1/4}$ law up to at least 60\,kpc in agreement with previous results \citep[e.g.][]{Prestwich_1997}. 
The other blue early-type galaxy is \object{J032311.4+402753} (\# 1020) at $u-r = 1.15, M_r = -17.3$, 
a blue compact galaxy that is surrounded by several smaller and fainter objects.} 
There is a systematic shift by approximately $0.2$\,mag of the cluster red sequence relative to the contour map towards redder colours. The cause is unknown to us, but it is most likely related to uncertainties of the SDSS photometry in the supplementary fields (see Sect.\,\ref{sect:obs_photo}).
Late-type galaxies are distributed over a wide range of colours, with about 50\% of them located in the blue cloud ($u-r < 2.2$). 
The repeated inspection of the SDSS images of all apparently red ($u-r>2.5$) late-type eSCGS galaxies revealed several plausible causes: edge-on discs with prominent dust lanes, 
incorrect photometry where the measurement was dominated by the bulge component or affected by foreground stars,
misclassified S0 galaxies (particularly edge-on systems and systems with a bar or ring),  
or true passive, red spirals. 
Bulge-dominated systems ({\tt fracDeV}  $>0.5$) are mainly found (93\%) in the red sequence, 
but the faint part of the red sequence at $M_r>-19$ is dominated by disc-dominated systems ({\tt fracDeV}  $<0.5$). 
The CMD for the bright sample MLS\_W1b is basically consistent with the bright part of the eSCGS CMD.

The MLS\_W1 CMD (central column)  shows conspicuous differences. 
First, the fainter ($M_r \ga -18$) blue galaxies are missing because of the 
selection bias introduced by the $W1$ magnitude cut in combination with the colour term in the $r-W1$ relation (Sect.\,\ref{sect:MLGS}). 
For the whole catalogue sample we find $r - W1 \approx  1.1 + 0.51\,(u-r)$. 
That is, at a given limiting magnitude $W1_{\rm lim}$, we selected galaxies in the red sequence ($u-r \approx 2.5$) down to a limiting r magnitude about $0.5$ mag fainter than for the blue cloud galaxies ($u-r \approx 1.5$). 
The corresponding $(u-r)$ versus $M_{\rm r,lim}$ relation is shown as a dashed black line. 
The other remarkable difference to the left-hand side of Fig.\,\ref{fig:CMD} is 
a cloud of faint ($r \ga 15$) and red ($u-r \ga 2.8$) galaxies. 
The re-inspection of the SDSS explorer pages supports the interpretation by photometric errors due to image crowding, the SDSS warning flag of unreliable photometry is set in many cases. 

The CMD of the MLS\_ur galaxies is characterised by the explicit colour selection, which excludes the faint red galaxies in the top left corner of the 
MLS\_W1 CMD and includes the galaxies from the blue cloud.

\begin{table}[h]
\caption{Statistical properties of the galaxies with $M_{\rm r} < -16$ in the samples from Table\,\ref{tab:samples} (see text).}
\begin{tabular}{lrccccc}
\hline\hline
Sample   
& $N$
& $r_{\rm l/e}$   
& $r_{\rm d/b}$
& $r_{\rm b/r}$ \\
\hline
MLS\_W1b    &  191 & 0.22 & 0.15 & 0.06  \\
SCGS        &  277 & 0.28 & 0.36 & 0.12  \\
eSCGS       &  297 & 0.36 & 0.44 & 0.19  \\
MLS\_ur     &  367 & 0.66 & 0.63 & 0.48  \\
MLS\_W1     &  403 & 0.39 & 0.48 & 0.12  \\
MaxS        & 1036 & 0.78 & 1.61 & 0.43  \\
\hline
\end{tabular}
\label{tab:samples_statistics}
\end{table}

Table\,\ref{tab:samples_statistics} lists statistical properties that characterise the mixture of morphological types in the samples from Table\,\ref{tab:samples}.
The second column gives the number $N$ of galaxies in the sample (after excluding the few faint galaxies with $M_{\rm r} >-16$). In the third column, the number ratio $r_{\rm l/e}=N_{\rm late}/N_{\rm early}$
of late-type to early-type galaxies is given based on the visual morphological classification.
Column four lists the ratio  $r_{\rm d/b}=N_{\rm fracDeV<0.5}/N_{\rm fracDeV>0.5}$ 
of disc-dominated to bulge-dominated galaxies based on the {\tt fracDeV}  parameter.
Finally, the ratio $r_{\rm b/r}=N_{\rm blue}/N_{\rm red}$ of the number of galaxies in the blue cloud ($u-r<2.2$) to the number of galaxies in the red sequence ($u-r>2.2$) is given in the last column. 
As in Table\,\ref{tab:samples}, the samples are sorted by the sample size $N$, which roughly correlates with the limiting magnitude. The proportion of blue, disc-dominated, late-type galaxies increases from top to bottom. An exception is MLS\_W1, which is biased against the fainter blue galaxies, as discussed above.  
The trend with the limiting magnitude can be explained by two effects. First, the blue cloud and the red sequence have distinct luminosity distributions (Fig.\,\ref{fig:CMD}). The different limiting magnitudes of the samples thus naturally lead to different portions. 
Secondly, a fainter magnitude limit is expected to be accompanied by a higher contamination by the background field, which has a higher percentage of late-type galaxies than the cluster.
The difference between the SCGS and eSCGS is easy to understand because all the galaxies in the eSCGS that are not SCGS members are H$\alpha$ emitters (Sect.\,\ref{sect:eSCGS}) and thus most likely blue star-forming galaxies. The percentage of H$\alpha$ emitters increases from 20\% in SCGS to 25\% in eSCGS.

\subsubsection{Morphology-density-position relation}\label{sect:radial_morph_seg}

\begin{figure*}[htbp]
\includegraphics[viewport= 0 10 560 540,width=6.4cm,angle=0]{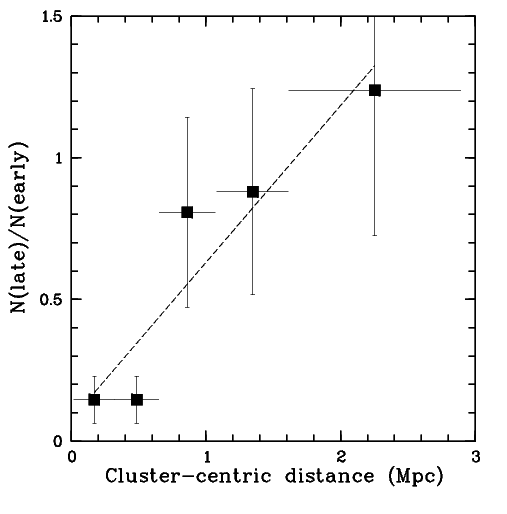}
\includegraphics[viewport= 0 10 560 540,width=6.4cm,angle=0]{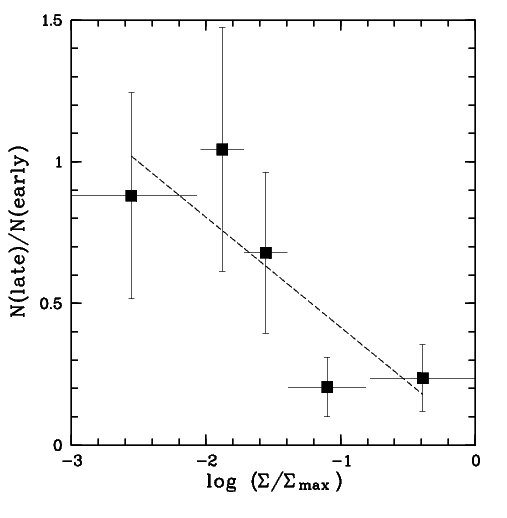}
\includegraphics[viewport= 0 10 560 540,width=6.4cm,angle=0]{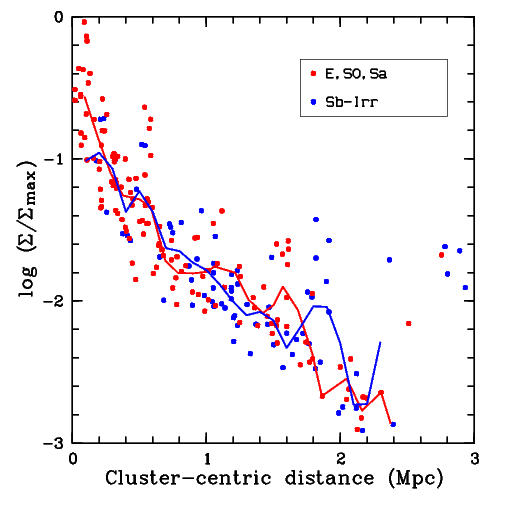}\\
\caption[Radial segregations]{Radial segregation in the eSCGS: number ratio of late-type to early-type galaxies. 
The left and middle panels show binned data where each bin contains roughly the same fraction of galaxies from the parent sample. Symbols are mean values, horizontal bars indicate the bin width, vertical bars the Poissonian errors per bin. The right panel displays the corresponding $\Sigma - R$ diagram where the polygons are the mean relations.
}
\label{fig:segregation_correlations_morph}
\end{figure*}

To investigate the relationship between morphology, density, and position,
we compared morphological parameters $X$ with positional properties $Y$, where 
$X$ is expressed either by the number ratio $r_{\rm l/e}$ of late types to early types 
or the number ratio $r_{\rm d/b}$ of disc-dominated to bulge-dominated galaxies based on the {\tt fracDeV} parameter.
$Y$ can be either the cluster-centric distance $R$, the normalised local projected density $\Sigma/\Sigma_{\rm max}$, 
or the normalised substructure parameter $\delta/\delta_{\rm max}$ (spectroscopic sample only). 

As a first step, Kendall's rank correlation coefficient $\tau$ was used to measure the ordinal association between $X$ and  the mean values $\bar{Y}$ in five $Y$ intervals with approximately the same number of galaxies per bin (Fig.\,\ref{fig:segregation_correlations_morph}). 
A general trend of a smaller ratio $r_{\rm l/e}$ in the central region and at higher $\Sigma$ is indicated for all samples, but the results are rather ambiguous. 
For the $r_{\rm l/e} - R$ relation, the null hypothesis of independence can be rejected at $\alpha = 0.05$
for MLS\_W1b, SCGS, and eSCGS but not for MLS\_ur, MLS\_W1, and MaxS. 
For the $r_{\rm l/e} - \Sigma$ relation, on the contrary, the null hypothesis can be rejected for MLS\_W1b, MLS\_ur, MLS\_W1, and MaxS but not for SCGS and eSCGS.
The discrepancy is strongest for MaxS, which might hint at substantial background structures at larger $R$ that  improve the correlation with $\Sigma$ but impair the correlation with $R$. 
The correlation diagrams for $X=r_{\rm d/b}$ show a larger scatter and the statistical test indicates a significant association only for one case, the $r_{\rm d/b} - R$ relation for MLS\_W1.   
The mean $\Sigma - R$ relations  of the late-type and early-type galaxies do not significantly differ from each other.
That is, the observed trend with $\Sigma$ is most likely a consequence of the trend with $R$ in combination with the 
projected cluster density profile (Fig.\,\ref{fig:segregation_correlations_morph}, right). 
The same applies to the comparison of disc-dominated and bulge-dominated galaxies.
There is no significant association with the three-dimensional local substructure parameter $\delta$. 

In a second approach, we subdivided the galaxy sample into two disjoint sub-samples S$_1$ and S$_2$ of comparable size as described in Table\,\ref{tab:samples_Z_test}.  
Pearson's Z test for the comparison of two proportions $X_1$ and $X_2$ was applied. 
The test statistic is
\begin{equation}
 \hat{z} = \frac{X_1 - X_2}{\sqrt{X_{\rm tot}(1-X_{\rm tot})}} \sqrt{\frac{N_1 N_2}{N_{\rm tot}}},
\end{equation}
where $X$ can be either the proportion $f_{\rm l}$ of late-type galaxies or the proportion $f_{\rm d}$ of disc galaxies, the indexes 1, 2 and `tot' refer to the sub-samples and the total sample (S$_1$ + S$_2$). The $\hat{z}$ values for the six $X,Y$ combinations and the six samples from Table\,\ref{tab:samples} are listed in Table\,\ref{tab:Z_test_morph}. We applied an upper-tailed test with the null hypothesis $H^0: X_1 \le X_2$ and  the alternative hypothesis $H^{A}: X_1 > X_2$. At a given error probability (significance level) $\alpha$, $H^0$ is rejected in favour of $H^{A}$ if $\hat{z} > z_\alpha$.

\begin{table}[h]
\caption{Subsamples S$_1$ and S$_2$. Here, $\langle \delta \rangle$ means the median of $\delta$.}
\begin{tabular}{ccc}
\hline\hline 
$Y$       &  S$_1$                                & S$_2$                                  \\
\hline
$R$       & $R = 1 \ldots 2$\,Mpc                 & $R \le 0.5$\,Mpc                       \\
$\Sigma$  & $\log \Sigma/\Sigma_{\rm max} < -1.3$ & $\log \Sigma/\Sigma_{\rm max} \ge -1.3$\\
$\delta$  & $\delta < \langle \delta \rangle$     & $\delta \ge \langle \delta \rangle$    \\
\hline
\end{tabular}
\label{tab:samples_Z_test}
\end{table}

If we set $\alpha = 0.05$ ($z_\alpha = 1.65)$, $H^0$ can be rejected for $Y=R$ or $\Sigma$ for all samples. 
Moreover, $H^0$ can be rejected at the low error probability $\alpha = 0.01$ ($z_\alpha = 2.33)$ in the cases of $f_{\rm l},R$, $f_{\rm l},\Sigma$, and $f_{\rm d},R$ for all samples and in the case of $f_{\rm d},\Sigma$  for SCGS, eSCGS, MLS\_W1, and MaxS. On the other hand, there is no reason to reject $H^{0}$ for the relations with $Y=\delta$.
The Z test clearly supports the finding that late-type galaxies prefer the outer parts of the cluster and lower local projected densities. On the other hand, there is no strong support for a continuous variation in the ratios $r_{\rm l/e}$ and $r_{\rm d/b}$ with $R$ or $\Sigma$. No association is indicated with the 3-dimensional substructure parameter $\delta$.

\begin{table}[h]
\caption{Z test statistic $\hat{z}$ for six $X,Y$  combinations.}
\begin{tabular}{lcccccc}
\hline\hline 
$X,Y$    &$f_{\rm l},R$&$f_{\rm l},\Sigma$&$f_{\rm l},\delta$&$f_{\rm d},R$&$f_{\rm d},\Sigma$&$f_{\rm d},\delta$ \\ 
\hline
MLS\_W1b & 4.28 & 4.00 &  -   & 2.54 & 2.32 &  -   \\
SCGS     & 4.61 & 3.66 & 0.69 & 2.69 & 2.56 & -1.04\\
eSCGS    & 5.11 & 4.57 &  -   & 2.86 & 3.02 &  -   \\
MLS\_ur  & 6.62 & 3.43 &  -   & 4.54 & 1.66 &  -   \\
MLS\_W1  & 6.15 & 4.24 &  -   & 3.75 & 2.90 &  -   \\
MaxS     & 8.41 & 5.07 &  -   & 5.15 & 3.88 &  -   \\
\hline
\end{tabular}
\label{tab:Z_test_morph}
\end{table}

\subsection{Galaxy luminosity function and total stellar mass}\label{sect:LF}

We focus on the K band luminosity function (LF) of the sub-sample MLS\_W1.
Infrared LFs are, compared to optical bands, much less affected by dust extinction (galactic foreground and intrinsic) and are less sensitive to $k$-corrections and the SF history in different galaxy types \citep[e.g.][]{Mobasher_1998, Andreon_2000}. 
A photometric incompleteness is introduced by the incompleteness of the 2MASS catalogue (Sect.\,\ref{sect:obs_photo}). 
For those 22 galaxies in MLS\_W1 (5\%) without 2MASS K-band magnitudes we simply used the WISE W1 magnitudes in combination with the mean relation $K = -1.2 + 1.08\,W1$ that was derived from our catalogue data. Absolute magnitudes $M_{\rm K}$ were computed as in Sect.\,\ref{sect:surf_density}.
The galaxies were binned in 0.5\,mag wide $M_{\rm K}$ intervals.
To correct for the background contamination we applied basically the same procedure as described in the context of Fig.\,\ref{fig:background_K} (Sect.\,\ref{sect:background_cont}). 
In each bin, the background fraction $f_{\rm b}$ was estimated and multiplied by the number of MLS\_W1 galaxies without known redshift. The thus estimated number of background systems was subtracted from the total number per bin. To correct for incompleteness in the galaxy selection, we applied correction factors derived from the comparison of the 2MASS galaxy counts (solid line in Fig.\,\ref{fig:background_K}) with the estimated number of background galaxies in MLS\_W1 in each bin.

\begin{figure}[htbp]
\centering
\includegraphics[viewport= 25 30 590 790,width=6.6cm,angle=270]{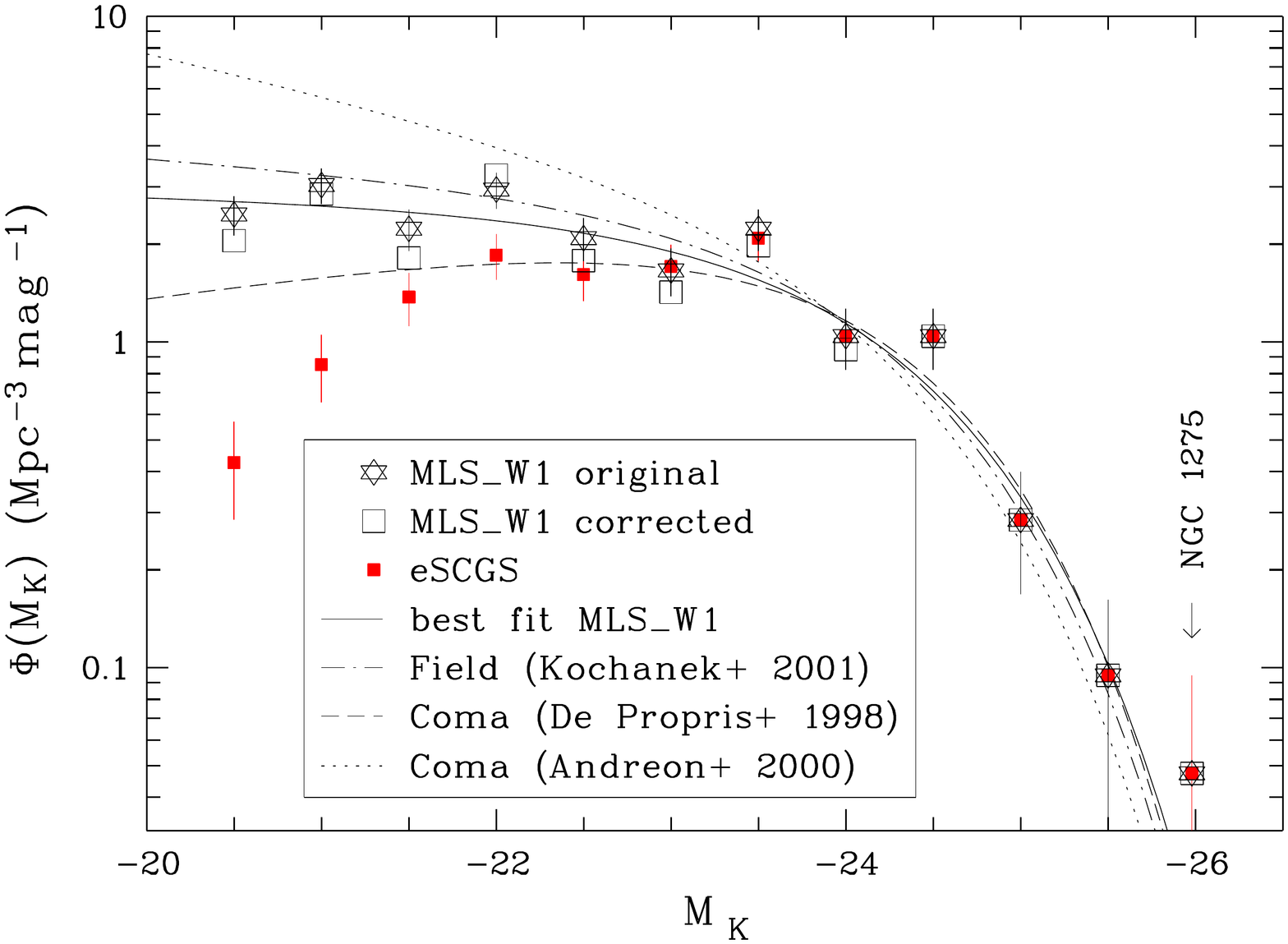}
\vspace{0.3cm}
\caption[CMD]{K-band LF for the galaxies from the magnitude-limited sample MLS\_W1 (stars with error bars) with the best-fit Schechter function (solid curve). The background-subtracted and incompleteness-corrected counts are shown as open squares. The LF from the galaxies with known redshift (eSCGS) is shown in red. For comparison, the LF from \citet{Kochanek_2001} (dashed-dotted) and the LFs for the Coma cluster from \citet{dePropris_1998} and \citet{Andreon_2000} (dashed and dotted) are overplotted, arbitrarily normalised at $M_{\rm K} = -24$.
}
\label{fig:LF}
\end{figure}

The resulting LF is shown in Fig.\,\ref{fig:LF}. 
For comparison, also the LF for the eSCGS is plotted (red symbols). The downturn of the latter towards fainter absolute magnitudes clearly reflects the incompleteness of the spectroscopic sample.  
The original (uncorrected) counts from the MLS\_W1 are indicated by star symbols, the corrected data (background and incompleteness) by open squares. These two opposite effects virtually cancel each other out.   
The corrected counts are on average slightly lower than the original ones, but this depends strongly on the adopted $N(K)$ relation for the background galaxies. 
Because we do not want to parametrise our lack of knowledge, we decided to fit a Schechter LF to the original counts only.
The logarithmic form of the Schechter function in terms of absolute magnitudes $M_{\rm K}$ is
\begin{equation}
 \log \Phi (M_{\rm K}) = C - \frac{\alpha+1}{2.5}(M_{\rm K}-M_{\rm K}^\ast) - \log{\rm e} \cdot 10^{-0.4(M_{\rm K}-M_{\rm K}^\ast)} 
\end{equation}
with $C = \log(0.4 \ln 10\cdot \Phi^\ast)$. The parameters for the best Schechter fit are listed in Table\,\ref{tab:LF_Schechter}. The galaxy NGC\,1275 is not fitted by the Schechter LF, as is generally known for BCGs \citep[e.g.][]{Lin_2004}. There is a good match between our best fit and the Schechter parameters $M_{\rm K}^\ast$ and $\alpha$ found by \citet{Kochanek_2001}  (dashed-dotted) for the general galaxy field from 2MASS data. 
The shape of the LF is also fairly similar to those found for the Coma cluster by
\citet{dePropris_1998} and \citet{Andreon_2000} based on H-band photometry.\footnote{We used the mean relation $H-K = 0.29\ (\pm0.07)$\,mag from our data to transform their Schechter magnitude $M_{\rm H}^\ast$ into the K band.}  
The Schechter functions for the parameters from \citet{dePropris_1998}, \citet{Andreon_2000}, and \citet{Kochanek_2001}  
corrected for the differences in the cosmological models, are also plotted, arbitrarily normalised at $M_{\rm K} = -24$. 
\citet{DePropris_2017} investigated the K-band LF for galaxies in 24 low-$z$ clusters and derived, independent of the environment, $K^\ast = 12.79\pm0.14$ for the full composite at $z=0.075$, which corresponds to   
$M_{\rm K}^\ast = -24.67\pm0.14$ if we adopt our $K-M_{\rm K}$ transformation.
These authors decided not to use 2MASS photometry and found by comparison with their photometry that 2MASS magnitudes are systematically $0.3$ mag fainter. Taking this difference into account, our Schechter absolute magnitude agrees with their composite LF within the given errors. On the other hand, our faint-end slope $\alpha$ is significantly flatter than their value of $\alpha=-1.41\pm0.10$. \citet{Kelvin_2014} argued that the total LF is best described by a double-Schechter form. If there is a second upturn at the faint end, the slope $\alpha$ of a single Schechter fit depends on the limiting magnitude of the available data. However, we feel that the main reason for the difference is the uncertainties in the incompleteness correction.

The integral of the Schechter function plus NGC\,1275 gives a total luminosity of $L_{\rm K, tot} = (12.4 \pm 1.6) 10^{12}\,L_\odot$. Galaxies fainter than $M_{\rm K} = -20$ contribute only $\la 2\%$,
the presumably missing galaxies at $M_{\rm K} \ga -22$ have no substantial effect and can be ignored.
We divided the sample into sub-samples according to morphological types where the
163 galaxies with ${\tt cl1} \le 3$ are considered to represent early types and the 97 galaxies with ${\tt cl1} \ge 5$ to represent late types. Compared to the entire sample, the Schechter magnitudes are fainter by approximately $0.4$\,mag for early types and by $1$\,mag for late types. The faint-end slope $\alpha$ is steeper for the late-type galaxies, but the Schechter fit is poor. These trends are in good qualitative agreement with the findings by \citet{DePropris_2017}. Early-type galaxies contribute 64\% to the total K-band luminosity, late-type galaxies 26\%, and the (mostly fainter) galaxies with the uncertain classification S/S0 contribute 10\%.

\begin{table}[htbp]
\caption{Parameters of the LF functions for MLS\_W1.}
\begin{tabular}{lccc}
\hline\hline     
                  &  all            &  early           &  late            \\                  
\hline
$N$               & 412             & 163              & 97              \\
$M_{\rm K}^\ast$  & $-24.25\pm0.15$ & $-24.10\pm0.15$  & $-23.50\pm0.20$  \\
$\alpha$          & $-1.03\pm0.15$  & $-0.65\pm0.15$   & $-1.05\pm0.20$   \\
$L_{\rm K, 12}$   & $12.4\pm1.6$    & $7.9\pm1.4$      & $3.2\pm1.1$    \\
$\mathcal{M_{\rm stars, 12}}$ & $9.9\pm1.3$    & $6.3\pm1.1$      & $2.6\pm0.9$    \\
\hline                
\end{tabular}\label{tab:LF_Schechter}
\tablefoot{$L_{\rm K, 12}$ and $\mathcal{M_{\rm stars, 12}}$ are the total K-band luminosity and the total mass in 
units of $10^{12}\,L_\odot$ and  $10^{12}\,\mathcal{M}_\odot$. 
}
\end{table}

Because the K-band luminosity is a good proxy for the stellar mass, we can estimate the total stellar mass in the Perseus cluster assuming a mean mass-to-luminosity ratio 
$\mathcal{M}/L$. From modelling of stellar populations with different stellar initial mass functions (IMF), \citet{Cole_2001} derived $\mathcal{M}/L = 1.32 \mathcal{M}_\odot/L_\odot$ for the Salpeter IMF and $0.73 \mathcal{M}_\odot/L_\odot$ for the Kennicutt IMF. For the assumption
$\mathcal{M}/L = 0.8 \mathcal{M}_\odot/L_\odot$ \citep{Graham_2013, Busch_2014} 
we estimate $\mathcal{M}_{\rm stars} = (9.9\pm1.3)\,10^{12} \mathcal{M}_\odot$ for MLS\_W1 where the uncertainty interval reflects only the uncertainty of $L_{\rm K}$, a larger additional uncertainty comes from the unknown IMF. A lower limit set by the spectroscopic sample for the whole survey field is $10.1\,10^{12} \mathcal{M}_\odot$.    
The stellar mass might be larger by a factor 1.3 if the 2MASS magnitudes are too faint by 0.3\,mag \citep{DePropris_2017}. Another correction factor of about $1.15\pm0.10$ would be necessary to account for the fraction of stars that were stripped from their host galaxies and feed the diffuse intra-cluster light \citep{Mihos_2016, Montes_2019}. Therewith, the total stellar mass is $(1.2\ldots 1.9)\,10^{13} \mathcal{M}_\odot$.

\subsection{SMBH mass function and total SMBH mass}\label{sect:SMBH_MF}

It is commonly assumed that any massive galaxy hosts a supermassive black hole (SMBH) and that 
the SMBH mass correlates tightly with the velocity
dispersion, the K-band luminosity, and the stellar mass of classical bulges and ellipticals 
\citep[Review:][]{Kormendy_2013}. 
SMBH masses  ${\mathcal M}_{\rm SMBH}$ can be estimated from the absolute magnitude of the bulges $M_{\rm K,b}$ adopting the relation
\begin{equation}\label{eq:SMBH_mass}
\log ({\mathcal M}_{\rm SMBH}/10^8{\mathcal M}_\odot) = 0.734 - 0.484(M_{\rm K,b} + 24.21)
\end{equation}
\citep{Kormendy_2013}, which is calibrated in the intervals 
$\log ({\mathcal M}_{\rm SMBH}/10^8{\mathcal M}_\odot) = -2 \ldots 1.3$ and $M_{\rm K,b} = -19 \ldots -27$.
The absolute K magnitude of the bulge is given by $ M_{\rm K,b} = M_{\rm K} - 2.5 \log (B/T)_{\rm K}$,
where the ratio  $B/T$ of the light from the bulge to the total light of the galaxy depends on the morphological type.  
Two-dimensional bulge-disc decompositions were performed for large galaxy samples from the SDSS legacy area
in several studies
\citep[e.g.][]{Simard_2011, Kelvin_2012, Meert_2015, Kim_2016} but unfortunately not for the special field of the Perseus cluster. The SDSS parameter {\tt fracDeV} does not directly represent the fraction of light from the bulge \citep{Shao_2015} but can be used as a quantitative measure of the type.   
\citet{Masters_2010} matched their sample of well-resolved spirals from the GalaxyZoo \citep{Lintott_2011} and the Millennium Galaxy Catalogue \citep[MGC,][]{Liske_2003} to construct a linear relation between {\tt fracDeV} from SDSS and $B/T$ as measured by the MGC. Here we used a combination of the 
$\langle B/T \rangle$--type relation from \citet{Graham_2013} and the $\langle B/T \rangle$--{\tt fracDeV} relation 
from \citet{Masters_2010}. For simplicity, $B/T=0$ was adopted for disc-dominated systems with ${\tt fracDeV} < 0.1$ and $B/T=1$ for E galaxies. In this way, we found that 231 galaxies (56\%) from the MLS\_W1 sample have
SMBH masses larger than $10^7 {\mathcal M}_\odot$.

\begin{figure}[htbp]
\centering
\includegraphics[viewport= 25 30 590 790,width=6.6cm,angle=270]{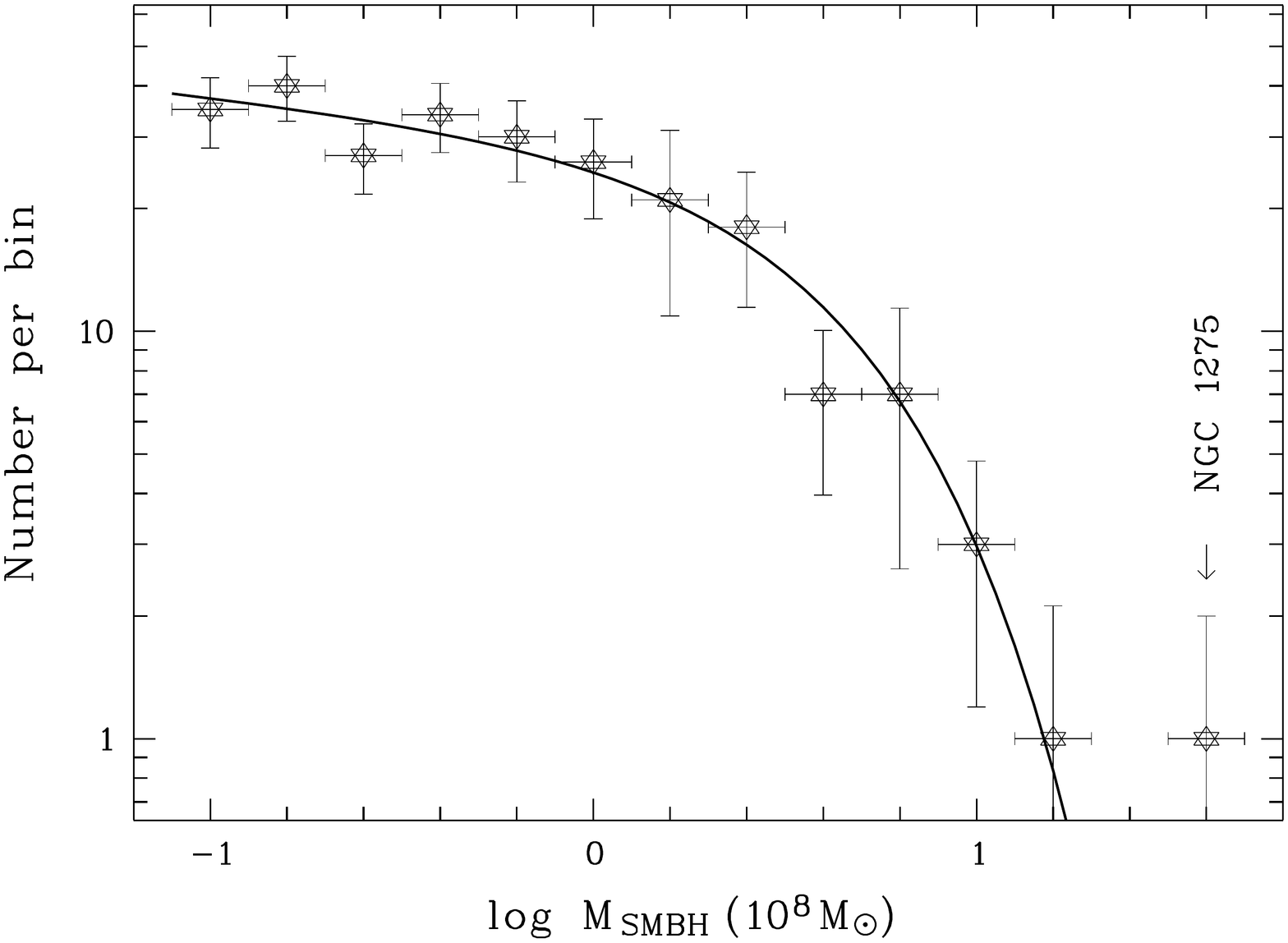}
\vspace{0.3cm}
\caption[CMD]{
SMBH mass function for the magnitude-limited sample MLS\_W1 fitted by a Schechter-type function.
Vertical bars represent counting errors, horizontal bars are the binning intervals. 
}
\label{fig:MF_SMBH}
\end{figure}

\begin{table}[htbp]
\caption{Parameters of the SMBH mass functions for MLS\_W1.}
\begin{tabular}{lccc}
\hline\hline     
                                                    &  all           &  early         &  late          \\                  
\hline
$N$                                                 &  250           &  142           &  59            \\
${\mathcal M}_{\rm SMBH,10}^\ast$                    & $4.8 \pm 1.0$  & $4.3 \pm 1.0$  & $1.5 \pm 0.5$  \\         
$\alpha$                                            & $-1.10\pm0.12$ & $-0.83\pm0.15$ & $-0.90\pm0.15$ \\
${\mathcal M}_{\rm SMBH,10}$                        & $3.3 \pm 0.8$  & $2.7 \pm 0.8$  & $0.4 \pm 0.1$  \\  
\hline                
\end{tabular}
\label{tab:SMBH_Schechter}
\tablefoot{
${\mathcal M}_{\rm SMBH,10}^\ast$ and ${\mathcal M}_{\rm SMBH,10}$ 
are the Schechter mass and the total mass of SMBHs in units of $10^{10}\,{\mathcal M}_\odot$.
}
\end{table}

The resulting SMBH mass function is displayed in Fig.\,\ref{fig:MF_SMBH} for the magnitude-limited sample MLS\_W1. 
The shape resembles the LF from Sect.\,\ref{sect:LF}, hence the mass distribution can be fitted by a Schechter function. 
The Schechter parameters and the total mass are listed in Table\,\ref{tab:SMBH_Schechter}, where $N$ is the number of galaxies involved in the Schechter fit.
We estimate a total SMBH mass of $3.3\,10^{10} {\mathcal M}_\odot$, corresponding to 0.3\% of the total stellar mass.
We note that the whole sample is not identical with the sum of the two sub-samples because the sub-samples do not include the (mostly fainter) galaxies of the uncertain type S/S0 that constitute $6$\% of the total SMBH mass. SMBHs of less than $10^7 {\mathcal M}_\odot$ contribute only $3$\%. The total SMBH mass in early-type galaxies is nearly 7 times the total SMBH mass in late-type galaxies.

AGN signify the growth of SMBHs by accretion of matter from the host galaxy, from a wet galaxy merger, or from the surroundings. As a back-of-the-envelope estimate we use the estimated total SMBH mass to compare the total energy released during the growth of the SMBHs $E_{\rm SMBH} = \epsilon c^2 {\mathcal M}_{\rm SMBH}/(1-\epsilon)$ with the internal kinetic energy of the ICM gas $E_{\rm gas} = 3{\mathcal M}_{\rm gas} k_{\rm B}T/2m_{\rm p}$ where $\epsilon$ is the radiative efficiency of the accretion process,
${\mathcal M}_{\rm gas}$ and $T$ are the mass and temperature of the hot ICM, $m_{\rm p}$ is the proton mass.
With ${\mathcal M}_{\rm SMBH} = 3.3\,10^{10} {\mathcal M}_\odot$ (Table\,\ref{tab:SMBH_Schechter}), 
${\mathcal M}_{\rm gas}/L_{\rm K} = 20 {\mathcal M}_\odot/L_\odot$ 
within $R \approx 2$\,Mpc \citep{Matsushita_2013}, $k_{\rm B}T = 5.5$\,keV \citep{Ebeling_1996}, 
and adopting the `canonical' value $\epsilon = 0.1$,
we find $E_{\rm SMBH} = 6\cdot 10^{56}\,\mbox{Ws} \approx 3\cdot E_{\rm gas}$.  
Provided that the energy output can efficiently couple onto the interstellar gas of the host galaxies or the ICM of the cluster, the AGN population must have played a non-negligible role in the cluster evolution, as had been suspected for a long time \citep[e.g.][]{Martini_2006, Hart_2009}.

%
\section{Active galaxies}\label{sect:Active}
%

\subsection{Star forming galaxies}\label{sect:SF}

\subsubsection{SFR and stellar mass from Chang et al. (2015)}

\begin{figure}[htbp]
\includegraphics[viewport= 20 30 740 570,width=8.6cm,angle=0]{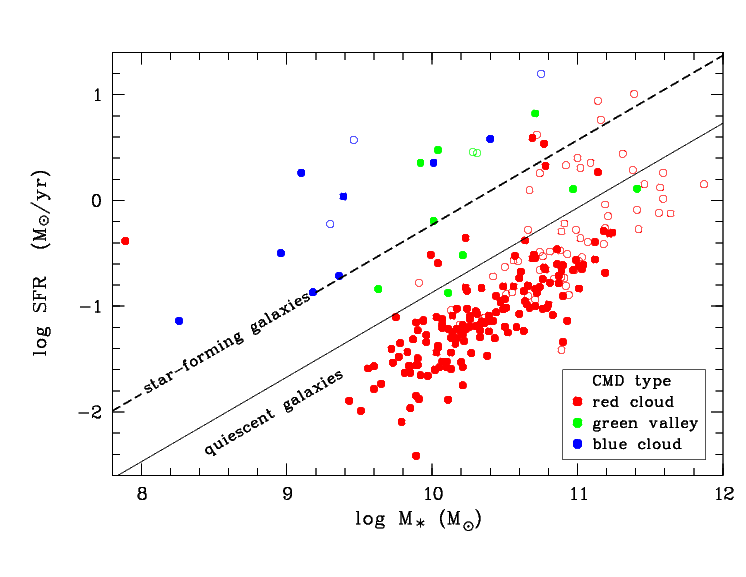}
\includegraphics[viewport= 20 20 740 570,width=8.6cm,angle=0]{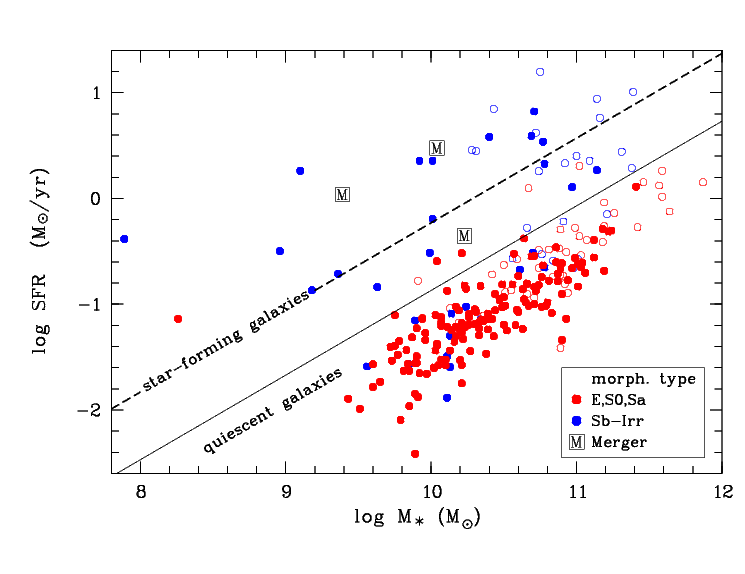}
\caption[SFR_M]{SFR versus stellar mass from \citet{Chang_2015} for 180 spectroscopic Perseus cluster members
(filled hexagons) and for 67 background galaxies (open hexagons). 
The colours indicate the position in the CMD (top) and the morphological type (bottom).
The dashed diagonal line is the median relation from \citet{Chang_2015}, the solid line marks the demarcation between quiescent galaxies (below) and the SF sequence (above).}
\label{fig:SFR-M}
\end{figure}

\citet{Chang_2015} combined SDSS and WISE photometry for the full SDSS spectroscopic galaxy
sample, creating spectral energy distributions (SED) that cover the wavelength range 
$0.4 - 22\ \mu$m for more than $8\,10^5$ galaxies. They employed SED modelling that consistently treats
stellar emission along with absorption and re-emission by interstellar dust and results in robust masses and star formation rates (SFR). The cross-correlation with our catalogue gives 247 matches, among them 
180 spectroscopic cluster members. 
Figure\,\ref{fig:SFR-M} shows the SFR versus stellar mass $\mathcal{M}_\ast$ for the cluster 
(filled symbols) and background galaxies (open symbols). The dashed diagonal line is the median 
$\log {\rm SFR} - \log \mathcal{M}_\ast$ from \citet{Chang_2015} with a downward scatter of 0.64 dex that is used here to define a demarcation (solid line) between star forming (SF) and quiescent galaxies. 
The two panels highlight the differences between the galaxies from the red sequence and the blue cloud (top) and between early and late morphological types from the visual inspection (bottom).

The top panel of Fig.\,\ref{fig:SFR-M} demonstrates that the cluster members in the quiescence region of the diagram are all allocated to the red sequence, with the only exception of two green valley galaxies close to the demarcation line. 
All blue cloud cluster galaxies and most galaxies from the green valley are found above the demarcation line.  
On the other hand, there are eight cluster galaxies in the SF area that are assigned to the red sequence.
This could be due to the lack of an intrinsic reddening correction. 
In fact, seven of these galaxies are spirals (see the panels below) with strong dust absorption and high inclination to the line of sight, among them is the FIR source IRAS\,03134+4008 (\# 397). 
A proper reddening correction would move most of these galaxies towards the blue cloud. 
SDSS gives the warning that the photometry may be unreliable for six out of these eight galaxies. 
The lower panel of Fig.\,\ref{fig:SFR-M} displays a strong correlation between morphological types and SF activity:
Early types prefer the quiescent region whereas most late type galaxies are located in the active region, with exceptions in both classes.

\subsubsection{WHAN diagram}\label{sect:SF_WHAN_diagram}

The optical spectra of SF regions are characterised by prominent emission lines. Emission-line classification schemes based on a small number of prominent lines provide a useful tool for an efficient differentiation between active galactic nuclei (AGN) and SF galaxies  \citep{Baldwin_1981}.
However, for many galaxies only a few of the strongest lines are recognised in SDSS spectra. 
To cope with the large population of these weak-line galaxies, \citet{CidFernandes_2011} proposed 
a robust and economic classification scheme, the WHAN diagram, which uses only the equivalent width (EW) of the H$\alpha$ line in combination with the line ratio
EW(\ion{N}{ii}) / EW(H$\alpha$).

\begin{figure}[htbp]
\includegraphics[viewport= 40 10 570 780,width=7.0cm,angle=270]{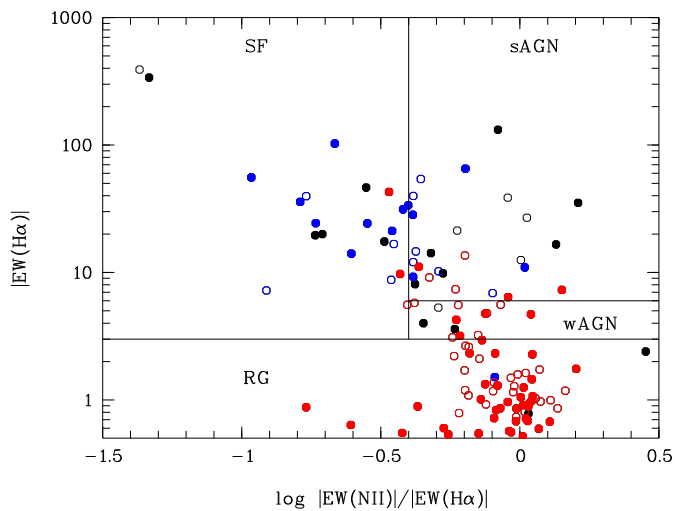}
\caption[WHAN diagram]{
WHAN diagram for 124 catalogued galaxies (filled symbols: 74 cluster members, open symbols: background systems). 
The four regions show the classification proposed by \citet{CidFernandes_2011}.
Blue and red symbols indicate the SF type based on the ${\rm SFR} - \mathcal{M}_\ast$ diagram  (Fig.\,\ref{fig:SFR-M}), galaxies not listed by \citet{Chang_2015} are in black.
}
\label{fig:WHAN}
\end{figure}

For the present study, most of the line data needed for the WHAN diagram were taken from the SDSS DR14 table {\tt galSpecLine}. For galaxies without SDSS spectroscopy but with spectroscopic observations performed in the framework of our programme, the line date were derived from the TLS or Calar Alto spectra by a manual de-blending technique: 
If H$\alpha$+[\ion{N}{ii}] emission is clearly indicated in the spectrum, the blended lines were modelled by the superimposition of four Gaussian components, namely the three narrow lines  H$\alpha$, [\ion{N}{ii}]\,$\lambda 6548$, [\ion{N}{ii}]\,$\lambda 6584$, and an additional broad  H$\alpha$ line component. 
Altogether 124 galaxies were found to have $|{\rm EW}({\rm H}\alpha)| > 0.5$\AA\ and 
log\,[EW(\ion{N}{ii}) / EW(H$\alpha$)] $> -1.5$, among them 74 cluster members.

Figure\,\ref{fig:WHAN} shows the WHAN diagram for 124 catalogue galaxies where 
the colour coding distinguishes star-forming (blue) from quiescent (red) galaxies based on 
the demarcation in Fig.\,\ref{fig:SFR-M} and the data from \citet{Chang_2015}.
Following \citet{CidFernandes_2011}, four different regions are indicated that contain  
pure star-forming galaxies (SF), strong AGN (sAGN), weak AGN (wAGN), and retired galaxies (RG). 
Typical representatives of sAGN and wAGN are Seyfert galaxies and LINERs, respectively. 
RGs are galaxies that have stopped SF and are ionised by evolved low-mass stars. 
Passive galaxies with $|{\rm EW}({\rm H}\alpha)| < 0.5$\AA\ and $|{\rm EW}({\rm \ion{N}{ii}})| < 0.5$\AA\ are not plotted.
As expected, quiescent galaxies with low SFR are strongly concentrated in the lower part of the WHAN diagram (RG and wAGN), but galaxies classified as star-forming in the ${\rm SFR} - \mathcal{M}_\ast$ diagram are distributed over the SF and sAGN area (blue symbols). Among the 15 cluster members classified as SF galaxies in Fig.\,\ref{fig:SFR-M}, nine belong to the SF region and four to the sAGN area in the WHAN diagram.

\subsubsection{WISE colour-colour diagram}

The WISE colour-colour diagram is a powerful tool for detecting SF activity in galaxy samples \citep[e.g.][]{Jarrett_2011, Stern_2012, Cluver_2017}. Galaxies with little hot dust emission occupy a narrow sequence at $W1 - W2 \approx -0.2 \ldots 0.5$\,mag, whereas the colour index $W2 - W3$ is a good indicator of the SF activity. 
Galaxies with little or absent SF populate the left-hand side of the sequence ($W2 - W3 < 2$\,mag) and SF galaxies  the right side ($W2 - W3 > 3.5$\,mag), the area in between is occupied by normal spiral discs with moderate SFR. Luminous infrared galaxies (LIRGs) and ultra-luminous infrared galaxies (ULIRGs) are typically found at the right-hand side. 
AGN with hot dust populate the colour space above the threshold $W1 - W2 = 0.8$\,mag. The reliability and completeness for such a SF-AGN separation are 95\% and 78\% for  $W2 < 15.05$ and drop steeply at fainter magnitudes \citep{Stern_2012, Assef_2013}. The constraint $W2 < 15.05$ is satisfied by 100\%, 96\% and 89\% of the galaxies in MLS\_W1, eSCGS, and MLS\_ur.

\begin{figure}[htbp]
\centering
\includegraphics[viewport= 0 30 730 510,width=9.0cm,angle=0]{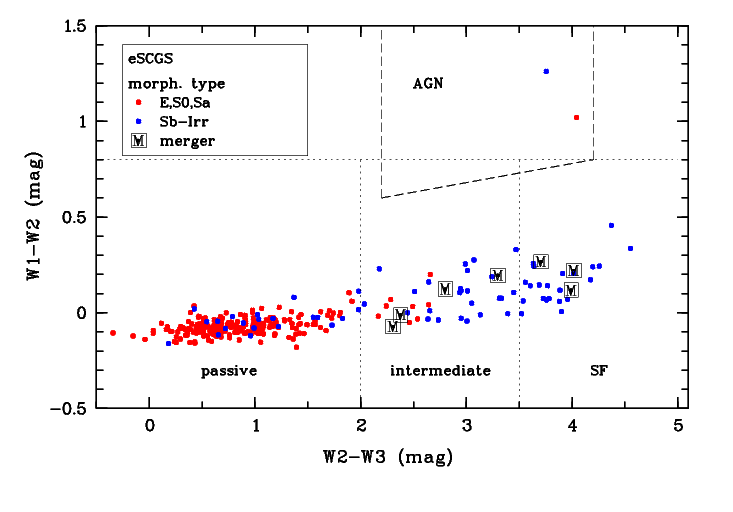}\\
\includegraphics[viewport= 0 30 730 510,width=9.0cm,angle=0]{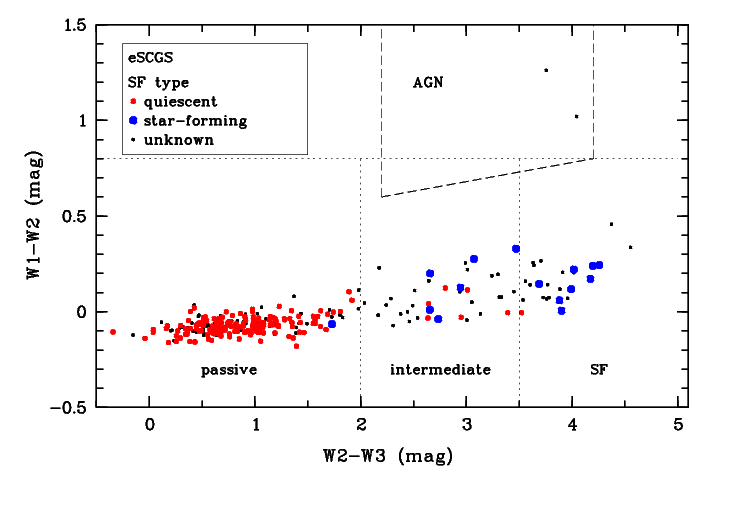}
\vspace{0.0cm}
\caption[CMD]{
WISE colour-colour diagram for the eSCGS galaxies. Different colours illustrate how galaxies separate by morphological type (top) and SF activity type (bottom)  as given in each panel. 
The dotted vertical demarcation lines are from \citet{Jarrett_2017}, the horizontal line marks the AGN threshold from \citet{Stern_2012}. The area indicated by the dashed lines is the AGN wedge from \citet{Jarrett_2011}.
}
\label{fig:WISE_ccd}
\end{figure}

Figure\,\ref{fig:WISE_ccd} shows the WISE colour-colour diagram for the eSCGS galaxies with $W2 < 15.05$. 
The corresponding diagrams for the samples MLS\_ur and MLS\_W1 look very similar. The vertical dotted lines are the demarcations between passive, early-type (left), intermediate (middle), and late-type SF galaxies (right) following \citet{Jarrett_2017}.  The horizontal dotted line is the AGN threshold from \citet{Stern_2012}, the dashed lines mark the AGN colour space as defined by \citet{Jarrett_2011}. The top panel shows a clear correlation between $W2 - W3$ and the morphological type. Early-types dominate at the left-hand side, late types at the right-hand side ($W2 - W3 \ga 2.5$) of the diagram. 
The bottom panel shows the distribution of the SF types according to the classification from Fig.\,\ref{fig:SFR-M}. 
There is a good correlation of the classification based on $W2 - W3$ with that one based on the data from \citet{Chang_2015}. 

It is well known that FIR emission is an efficient indicator of SF activity. A sample of 17 Perseus cluster galaxies identified with FIR sources from the IRAS PSC has been the subject of a previous study \citep{Meusinger_2000}.
One of these galaxies is out of the present survey field, the remaining 16 are distributed over the SF part of the WISE colour-colour diagram (10 galaxies), the intermediate part (4), and the AGN region (2).
The two galaxies with the largest colour index $W2-W3$ in Fig.\,\ref{fig:WISE_ccd}, J032505.4+403332 (\# 1146) and 
J032827.8+400917 (\# 1291),  are LIRGs from the IRAS sample. 
They do not belong to the \citet{Chang_2015} sample and are therefore not colour-coded as SF in the bottom panel.

\subsubsection{SFR and stellar mass from WISE data}\label{SFR_WISE}

\begin{figure}[htbp]
\includegraphics[viewport= 0 0 730 510,width=9.0cm,angle=0]{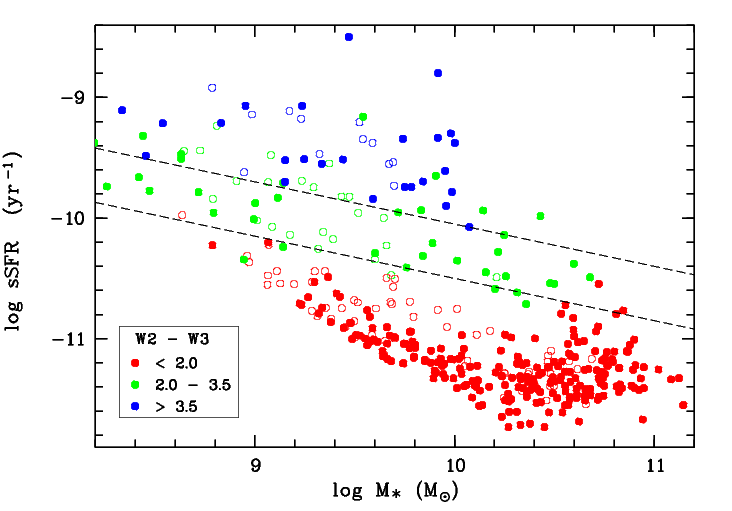}\\
\includegraphics[viewport= 0 0 730 510,width=9.0cm,angle=0]{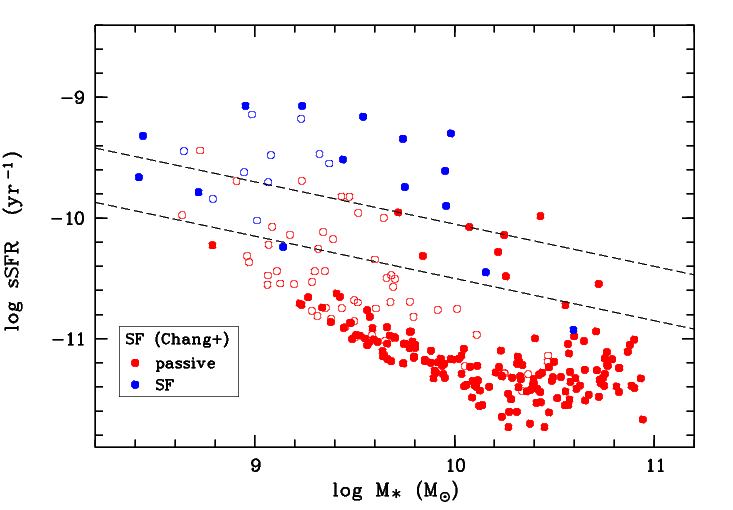}\\
\includegraphics[viewport= 0 0 730 510,width=9.0cm,angle=0]{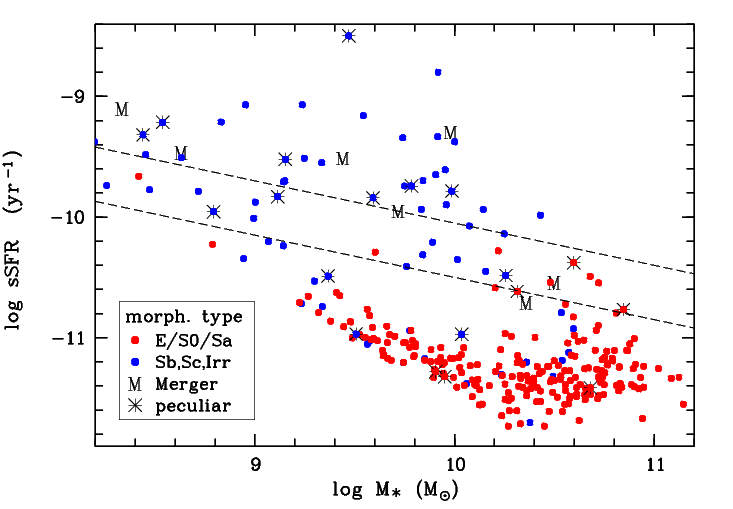}\\
\caption[SFR_M]{Specific star formation rate versus stellar mass from WISE data for the eSCGS sample (filled circles) and spectroscopically confirmed non-cluster members (open circles). 
The colours indicate the classification according to $W2-W3$ from Fig.\,\ref{fig:WISE_ccd} (top),
SF activity class from Fig.\,\ref{fig:SFR-M} (middle), and morphological properties, as indicated in each panel. 
The dashed diagonal lines roughly separate passive (bottom) from intermediate (middle) and SF (top) galaxies.
}
\label{fig:sSFR-M-WISE}
\end{figure}

Given the heterogeneous nature of our spectroscopic target selection, it is not clear whether the samples defined by the ${\rm SFR} - \mathcal{M}_\ast$ diagram (Fig.\,\ref{fig:SFR-M}) or the WHAN diagram (Fig.\,\ref{fig:WHAN}) are representative for an investigation of the spatial distribution of galaxy activity in  the cluster.
The WISE magnitude $W3$ can be taken as a proxy of the SFR and the combination of $W1$ with $W1-W2$ as a proxy of the stellar mass. \citet{Cluver_2017} found a good correlation between the 12\,$\mu$m luminosity, $\log L_{12 \mu{\rm m}}$ and $\log {\rm SFR}$ with a one-sigma scatter of 0.15 dex over a wide range in SFR and $\mathcal{M}_\ast$. 

We made use of the calibrated relations from 
\citet{Cluver_2017} and \citet{Jarrett_2017} to compute the SFR, the total stellar mass $\mathcal{M}_\ast$, and the specific SFR (sSFR = SFR/$\mathcal{M}_\ast$) for the non-AGN galaxies (i.e. $W1-W2 < 0.8$).  In Fig.\,\ref{fig:sSFR-M-WISE} we show the results for the spectroscopic sample in the sSFR - $\mathcal{M}_\ast$ plane in comparison with Figs.\,\ref{fig:SFR-M} and \ref{fig:WISE_ccd}. In the upper panel, the three ($W2-W3$)-selected classes are colour coded. There is a clear separation of SF from passive galaxies, which is indicated by the diagonal dashed lines. 
The results from the WISE data are also broadly consistent with the \citet{Chang_2015} data (middle panel) where a part of the scatter may result from the fact that the \citet{Chang_2015} data are based on distances computed from the SDSS redshifts, whereas we assumed that all cluster galaxies are at the same redshift.
In the bottom panel of Fig.\,\ref{fig:sSFR-M-WISE}, the colour coding is to distinguish between early-types, late-types, and mergers. In addition, galaxies with strong peculiarities (reliability flag $> 1$) are indicated by a star symbol.
In the SF region of the diagram, we find exclusively late-type galaxies and mergers. 
The intermediate region is dominated by late-type galaxies at masses 
$\mathcal{M}_\ast  \la 10^{10}\,\mathcal{M}_\odot$, but early-types dominate at higher masses. 
The only early-type galaxy in the low-mass part of the intermediate region is the blue compact dwarf galaxy J032311.4+402753 (\# 1020) at $\log (\mathcal{M}_\ast/\mathcal{M}_\odot) \approx 8.7$ that was mentioned already as an outlier in the SDSS colour-colour diagram (Sect.\,\ref{sect:cmd_morph}).

\begin{table}[h]
\caption{SF activity fractions for the samples from Table\,\ref{tab:samples} (see text).}
\begin{tabular}{lcccccc}
\hline\hline
Sample     
& $f_{\rm SF}^{(1)}$
& $f_{\rm inter}^{(1)}$
& $f_{\rm passive}^{(1)}$ 
& $f_{\rm SF}^{(2)}$
& $f_{\rm inter}^{(2)}$
& $f_{\rm passive}^{(2)}$ \\ 
\hline
MLS\_W1b     & 0.06 & 0.12  & 0.82  &  0.09 & 0.07 & 0.83   \\
SCGS         & 0.07 & 0.11  & 0.81  &  0.10 & 0.08 & 0.82   \\
eSCGS        & 0.08 & 0.13  & 0.78  &  0.12 & 0.08 & 0.79   \\
MLS\_ur      & 0.11 & 0.18  & 0.70  &  0.16 & 0.12 & 0.72   \\
MLS\_W1      & 0.10 & 0.18  & 0.72  &  0.16 & 0.11 & 0.73   \\
MaxS         & 0.13 & 0.34  & 0.52  &  0.24 & 0.18 & 0.58   \\
\hline
\end{tabular}
\label{tab:statistics_SF_1}
\end{table}

Table\,\ref{tab:statistics_SF_1} gives an overview of the proportions of galaxies in the three SF activity classes for the samples from Table\,\ref{tab:samples}.  
The columns 2-7 list the proportions $f_i^{(j)} = N_i/N_{\rm tot}$, where $i$ stands for one of the activity classes SF ($i=1$), intermediate ($i=2$) or passive ($i=3$) and $j$ marks whether this classification is based either on Fig.\,\ref{fig:WISE_ccd} ($j=1$) or Fig.\,\ref{fig:sSFR-M-WISE} ($j=2$). 
The proportion of passive galaxies is remarkably similar for both separation schemes ($j = 1,2$) but differs for the different samples. Whereas more than 80\% of the systems are classified as passive in the pure spectroscopic sample SCGS, the percentage is $77\% - 78\%$ in the eSCGS (after adding the galaxies from the H$\alpha$ survey) and drops to $70\% - 73\%$ in the two magnitude-limited samples MLS\_W1 and MLS\_ur. The higher percentage of 82\% in the bright magnitude-limited sample reflects again the selection bias discussed already in Sect.\,\ref{sect:cmd_morph}. The fraction of SF galaxies differs slightly between the two separation schemes with higher values for the separation based solely on the WISE data ($j=2$). Apart from MaxS, the SF fraction is largest for MLS\_W1 and MLS\_ur and smallest for MLS\_W1b, which is dominated by the bright early-type galaxies. 

To summarise the results for the three most representative samples of cluster galaxies, eSCGS, MLS\_W1, and MLS\_ur, it can be concluded that about $75\pm5$\% of the galaxies are passive and about $25\pm5$\% are actively star-forming where the latter can be subdivided into two approximately equal parts of moderately active (intermediate) or strongly active (SF) galaxies.
24 cluster members have SFR $> 1 \mathcal{M}_\odot\ \mbox{yr}^{-1}$ and host about 65\% of the total SF in the cluster. Highest SFRs ($\ga 10 \ \mathcal{M}_\odot\,\mbox{yr}^{-1}$) are found for the two IRAS sources \object{J032505.4+403332} (\#\,1146) and \object{J032827.8+400917} (\#\,1291) mentioned in the previous section. Both galaxies are morphologically distorted systems \citep{Meusinger_2000} at large projected cluster-centric distances of $R = 1.8$\,Mpc and 2.8\,Mpc. 

Table\,\ref{tab:statistics_SF_2} lists the integrated stellar mass, SFR and sSFR for all samples, 
where we assume again that all sample galaxies are at the distance of A\,426. 
The resulting mean SFR from the three main samples is $114\pm 6 \ \mathcal{M}_\odot$\,yr$^{-1}$. 
The total stellar mass in MLS\_W1 is about 25\% lower than the value estimated from the K-band LF (Table\,\ref{tab:LF_Schechter}). The bigger part of this discrepancy is most likely caused by differences in the calibration of the $M/L$ relations, whereas a smaller part comes from the exclusion of the galaxies fainter than $W1 = 14$ and the two bright galaxies NGC\,1275 and UGC\,2608 that were classified as AGN hosts.

\begin{table}[h]
\caption{Integrated stellar mass, SFR, and sSFR for the samples from Table\,\ref{tab:samples}.}
\begin{tabular}{lccc}
\hline\hline
Sample     
& $\mathcal{M}_\ast$
& SFR  
& sSFR  \\ 
& ($10^{12} \ \mathcal{M}_\odot$) 
& ($\mathcal{M}_\odot\ \mbox{yr}^{-1}$) 
& ($10^{-11} \mbox{yr}^{-1}$)   
\\ \hline
MLS\_W1b  & 6.7 &  92.2  &  1.4 \\
SCGS      & 7.2 & 113.7  &  1.6 \\
eSCGS     & 7.2 & 116.4  &  1.6 \\
MLS\_ur   & 6.9 & 100.2  &  1.4 \\
MLS\_W1   & 7.5 & 124.2  &  1.7 \\
MaxS      & 8.4 & 179.6  &  2.1 \\
\hline
\end{tabular}
\label{tab:statistics_SF_2}
\end{table}

\subsubsection{Final SF sample}\label{SF_sample}

To define a sample of cluster SF galaxies for the subsequent discussion, we employed a combination of the selection from the WHAN diagram with that from the WISE colour-colour diagram.  
The final sample  consists of 31 spectroscopically confirmed cluster members that are either classified as SF in 
Fig.\,\ref{fig:WHAN} (16) or are not classified in the WHAN diagram but fall into the SF region of the Fig.\,\ref{fig:WISE_ccd} (15).

\subsection{Active Galactic Nuclei}\label{sect:AGN}

\subsubsection{Identification of cluster AGN}

The 12th edition of the `Catalogue of Quasars and Active Nuclei' \citep{Veron-Cetty_2006} lists 
eight objects in our survey field, among them are three with $z < 0.03$: 
the Sy\,1.5 NGC\,1275 (\#691), the Sy\,2 UGC\,2608 (\#278), 
and the BL\,Lac V\,Zw\,331. 
The latter one is a background system ($z=0.0283, R = 65\farcm5$).
In this Section, we discuss the search for AGN in additional cluster galaxies via
the emission lines in optical spectra, non-thermal radio emission and
IR colours indicating a dusty torus.

\vspace{0.5cm}
\noindent
(a) Optical AGN  \label{sec:AGN_optical}

\begin{figure}[htbp]
\begin{center}
\includegraphics[viewport= 0 20 570 760,width=4.3cm,angle=0]{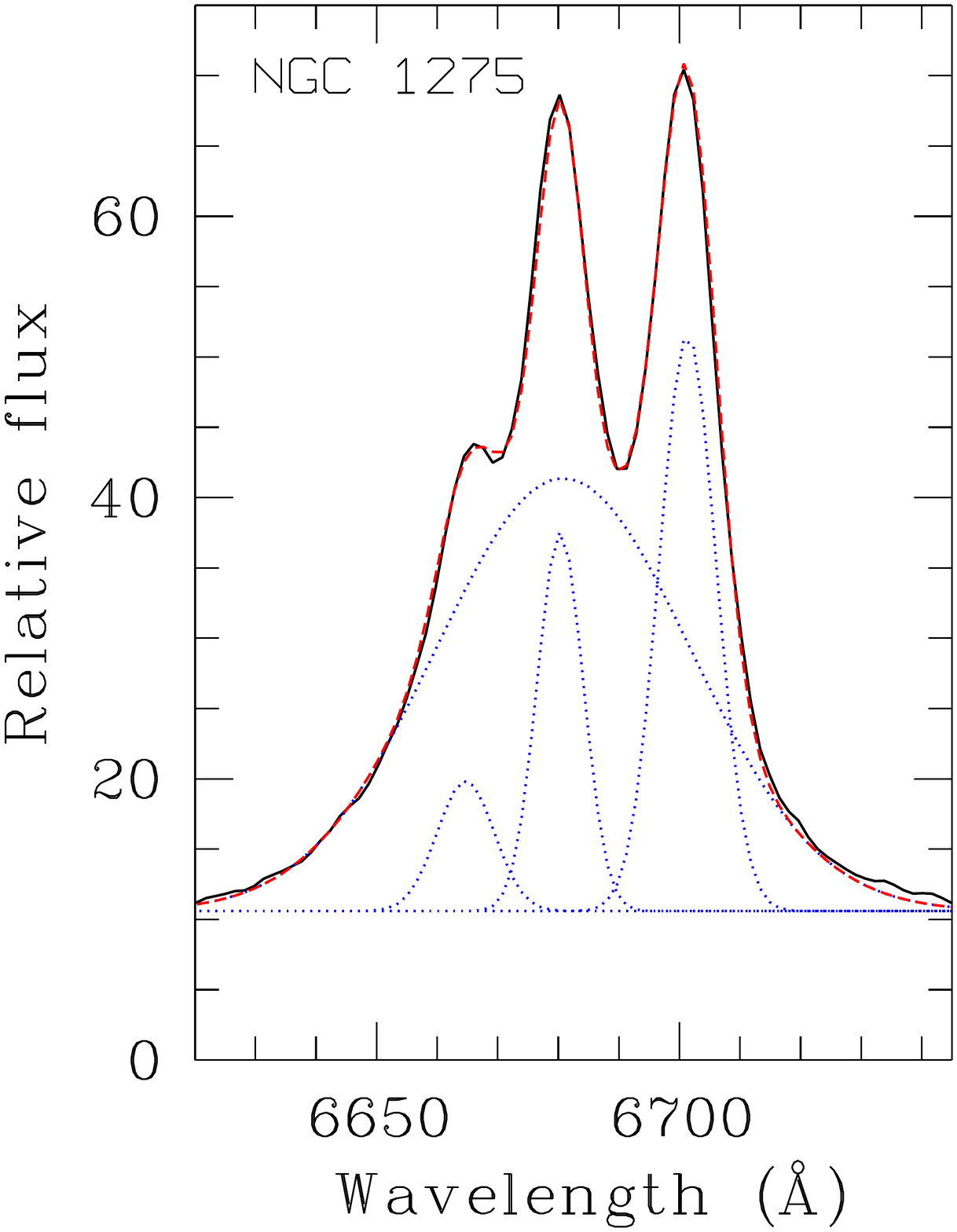}
\includegraphics[viewport= 0 20 570 760,width=4.3cm,angle=0]{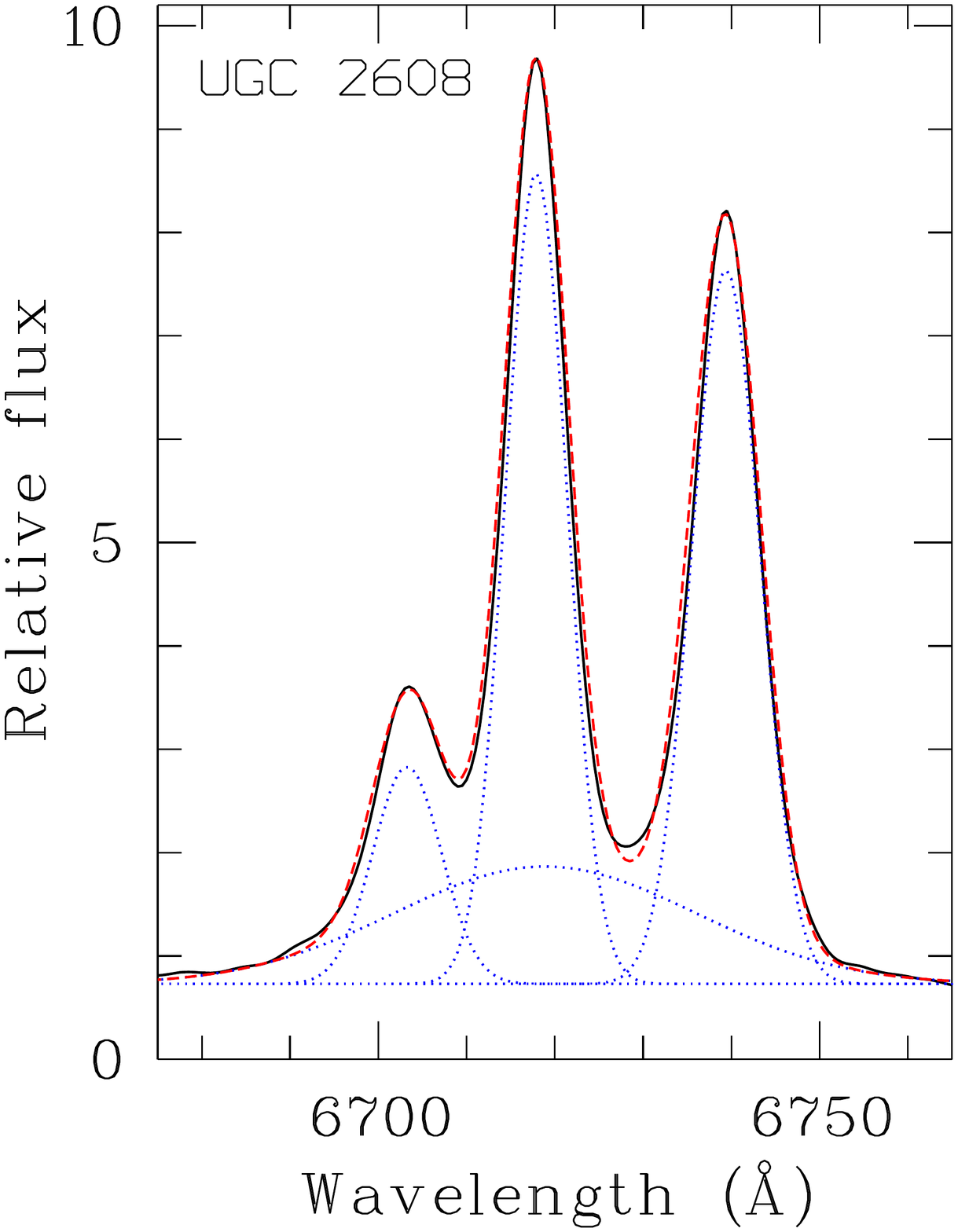}
\end{center}
\caption[Line deblending]{
Examples of Gaussian fits of the blended lines H$\alpha$+[\ion{N}{ii}]\,6548,6584 for NGC\,1275 (left) and UGC\,2608 (right). Three narrow and one broad Gaussian components are used (dotted blue curves). The slightly smoothed observed spectrum is shown in black, the sum of the four components is in red.
}
\label{fig:line_deblending}
\end{figure}

\begin{figure}[htbp]
\includegraphics[viewport= 40 10 570 780,width=7.0cm,angle=270]{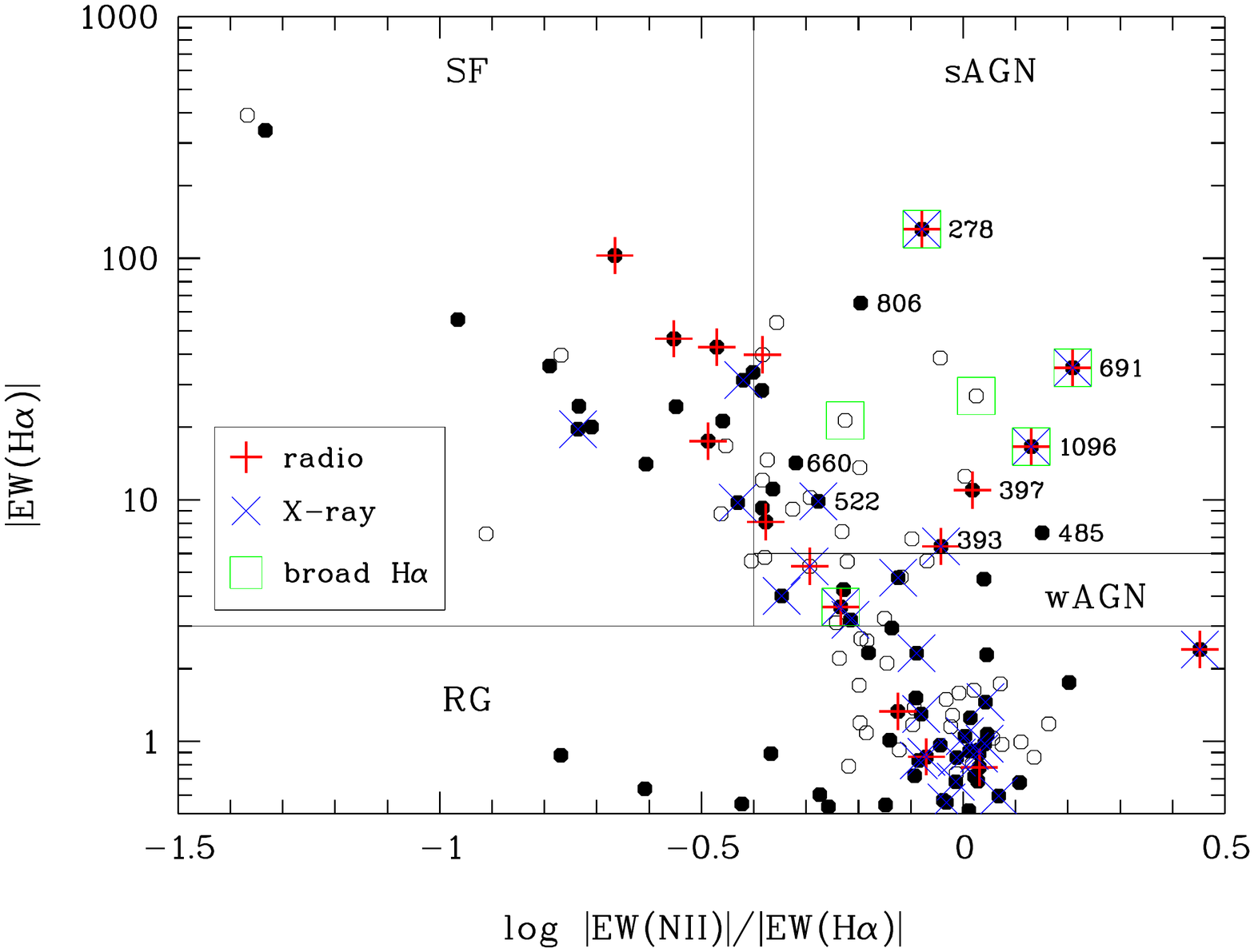}
\caption[WHAN diagram]{
Same as Fig.\,\ref{fig:WHAN}, but with additional symbols to indicate 
the presence of a broad H$\alpha$ line component (green open square),
an XMM X-ray source (blue cross), or a radio source from \citet{Miller_2001} (red plus sign). 
Some cluster members in the sAGN area are labelled with their catalogue numbers.
}
\label{fig:WHAN_Xray_radio}
\end{figure}

\noindent
Broad hydrogen Balmer line components signify type 1 AGNs. 
We identified a broad H$\alpha$ component not only in the two known cluster AGN
NGC\,1275 and UGC\,2608 (Fig.\,\ref{fig:line_deblending}),
but also in the edge-on disc galaxy UGC\,2715 (\#1096), which was not previously listed as an AGN.
Furthermore, a broad component was invoked to fit the  H$\alpha$ line profile in the spectrum of NGC\,1294 where the S/N is low however. Broad H$\alpha$ was also found in the two background systems UGC\,2724 (\#1152) and J032006.3+402159 (\#735). All six broad-line galaxies are classified as AGN in the WHAN diagram (Fig.\,\ref{fig:WHAN_Xray_radio}): NGC\,1275, UGC\,2608, UGC\,2715, and the two background sources as sAGN, whereas NGC\,1294 is a wAGN.

There are 13 cluster galaxies in the sAGN area of the WHAN diagram and six in the wAGN area. 
Several of these sources are close to the SF-AGN demarcation. 
An example is J031930.3+404430 (\# 660) with $\log\,{\rm EW}(\ion{N}{ii})/{\rm EW(H\alpha)} = -0.32$, a faint ($W2=16.01$), extended blue low-surface brightness galaxy of highly irregular morphology (Appendix\,\ref{sect:SF_individual}). Because there is no substantial optical core we ultimately classified this galaxy as SF. 
If we adopt the stronger AGN criterion 
log\,EW(\ion{N}{ii}) / EW(H$\alpha$) $>-0.25$, we would find seven sAGN and five wAGN in the cluster, all brighter than $W1= 14$. The total number of cluster members with spectra and $W1 < 14$ is 253.
Hence, we can conclude that about $5 - 8$\% of the brighter cluster galaxies with $W1< 14$ host an optical AGN (sAGN or wAGN) and about $3 - 5$\% of them host an sAGN.

The cluster sAGN occur preferably in late-type galaxies. 
The host is classified as spiral ({\tt cl1} = 6) for eight of them and as early-type  ({\tt cl1} $\le 4$) 
in four cases. One AGN galaxy (\#485) was classified as a merger ({\tt cl1=9}).
On the other hand, cluster wAGN seem to prefer early-type galaxies: four were classified as {\tt cl1} $\le 4$ and two as  {\tt cl1} = 6.  Morphological peculiarities are common among AGN hosts (Sect.\,\ref{sect:pec-SF-AGN}).

\vspace{0.5cm}
\noindent
(b) Radio galaxies \label{sec:AGN_radio}

\begin{figure}[htbp]
\includegraphics[viewport= 35 35 600 800,width=6.8cm,angle=270]{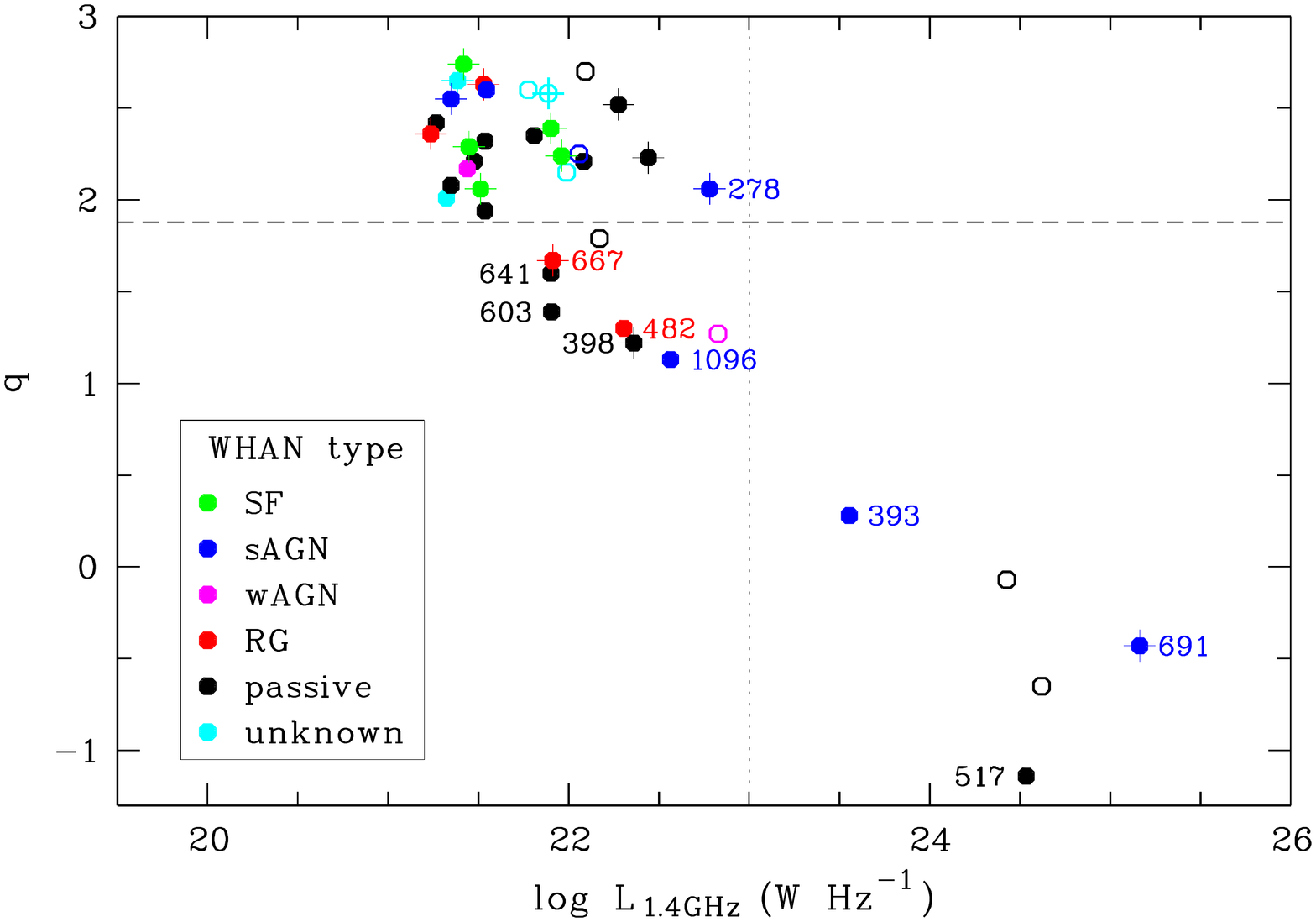}
\includegraphics[viewport= 35 35 600 800,width=6.8cm,angle=270]{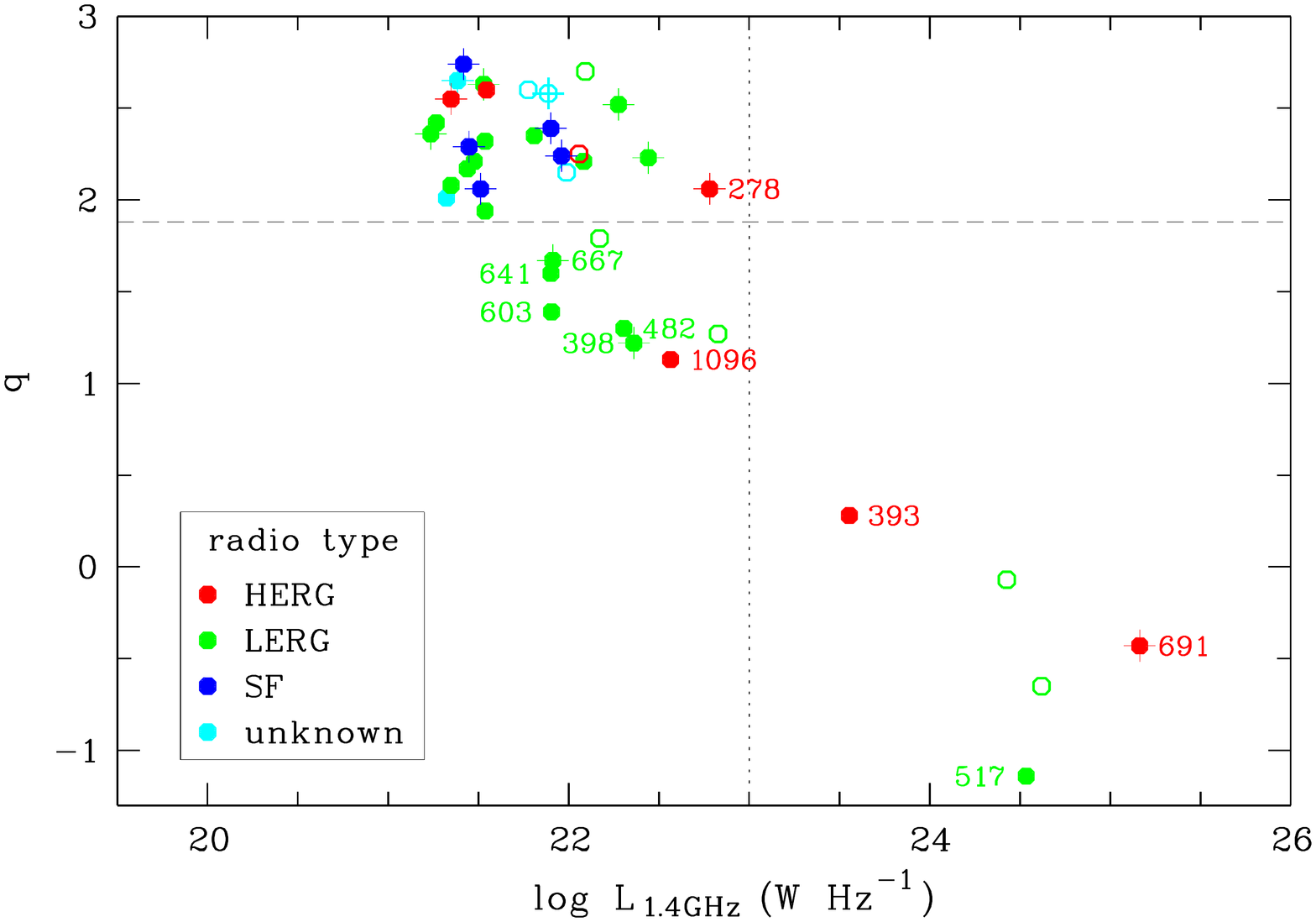}
\caption[L_q_diagram]{Radio-FIR slope $q$ versus 1.4 GHz radio luminosity for the NVSS sources from \citet{Miller_2001}. 
The colours represent the WHAN type (top) and radio galaxy type (bottom), cluster members are indicated by filled, background galaxies by open circles. Over-plotted plus signs indicate morphological peculiarities.
The dotted vertical line represents a (statistical) luminosity threshold for AGN, 
the dashed horizontal line separates SF galaxies (top) from AGN (bottom).  
Cluster members outside the `clump' around $q = 2.3, \log L_{\rm 1.4 GHz} = 22$ are labelled with their catalogue numbers.  
}
\label{fig:L_q_diagram}
\end{figure}

\noindent
The identification of radio sources is based on observations with the Westerbork Telescope at 610 MHz
\citep{Gisler_1979}, the NRAO VLA Sky Survey \citep[NVSS;][]{Condon_1998} at 1.4 GHz, 
and the GMRT Sky Survey (TGSS) project at 150\,MHz \citep{Intema_2017}.
\citet{Miller_2001} used the NVSS to compile a catalogue of 467 radio galaxies in the fields of 18 nearby clusters in an area of about 3\,Mpc from the cluster core. 
The NVSS guarantees nearly uniform-sensitivity coverage over wide fields.
With 52 sources, A\,426 is the cluster with the largest number of radio galaxies in this catalogue. 
43 of these radio sources were identified with optical galaxies at the cluster redshift and 9 with background galaxies. 
The estimated probability of an identification with an optical galaxy by chance superposition is only 0.5\%.
We cross-matched the list from \citet{Miller_2001} with our catalogue
and identified 40 galaxies. All identified radio galaxies have spectroscopic redshifts, 31 are classified as cluster members and 9 as background systems. None of these NVSS sources was classified by \citet{Miller_2001} as extended.

The radio flux of SF galaxies is well known to strongly correlate with the FIR flux. 
The FIR-to-radio flux density ratio 
$ q = \log (S_{\rm FIR}/S_{\rm 1.4\,GHz})$ is $2.3 \pm 0.2$ for SF galaxies \citep{Helou_1985},
whereas AGN have an excess of radio emission and thus $q \la 2$. 
Hence, $q$ can be used, in conjunction with the absolute radio luminosity, to distinguish AGN from SF galaxies. 
Following \citet{Sadler_2002}, AGN dominate the radio luminosity distribution for  
$L_{\rm 1.4\,GHz} > 10^{23}$\,W\,Hz$^{-1}$.\footnote{The monochromatic 1.4\,GHz luminosity is defined as 
$L_{\rm 1.4\,GHz} = 4\pi\,d_{\rm L}^2\,(1+z)^{-1+\alpha}\,S_{\rm 1.4\,GHz}$, 
where $d_{\rm L}$ is the luminosity distance, 
$S_{\rm 1.4\,GHz}$ the radio flux density at 1.4\,GHz, 
and $\alpha$ is the radio spectral index where we assumed $S_\nu \propto \nu^{-\alpha}$ and $\alpha = 0.8$ for AGN.}
However, such a luminosity threshold differentiates in a statistical sense only, 
AGN can have a lower radio luminosity \citep[e.g.][]{Park_2017}.

\citet{Miller_2001} presented $q$ values utilising the IRAS catalogues or the IPAC XSCANPI package.  
Figure\,\ref{fig:L_q_diagram} shows the estimated $q$ parameter versus $L_{\rm 1.4\,GHz}$. 
The colour indicates the optical spectral type from the WHAN diagram (top) or the radio galaxy type (bottom). The radio type classification delineates the radio sources into those dominated by SF activity and those dominated by an AGN \citep{Ching_2017} where the spectrum of the latter can be characterised either by strong or high-excitation emission lines (high-excitation emission line radio galaxy, HERG) or by weak or low-excitation lines (low-excitation emission line radio galaxy, LERG). The difference between HERG and LERG is thought to be driven by the accretion mechanism, corresponding to radiative-mode AGN and jet-mode AGN (see Sect.\,\ref{sec:AGN_sample} below). Here, we simply identified HERGs with radio galaxies of the WHAN type sAGN and LERGs with wAGN, RGs and passive galaxies.

About 70\% of the cluster radio galaxies are concentrated around $\log\,L_{\rm 1.4\,GHz} \approx 22 \pm 1$ and $q > 1.8$ 
(Fig.\,\ref{fig:L_q_diagram}) with a mean value $q = 2.32 \pm 0.22$ . Two remarkable properties of this sub-sample are the exceptionally high percentage of morphological peculiarities (55\%) and the dominance of late-type galaxies (70\%).
All radio galaxies classified as SF in the WHAN diagram are found here, but there are also other WHAN types, 
including UGC\,2608 and another two HERGs. 
Adopting $q < 1.9$ as a criterion, nine radio sources from \citet{Miller_2001} are classified as AGN.

To ensure a secure detection of the radio AGN, we rechecked the NVSS images in combination with an inspection of the image cutouts from the Alternative Data Release (ADR) of the TIFR GMRT Sky Survey (TGSS) project at 150\,MHz \citep{Intema_2017}. With respect to its relatively high resolution and low noise, TGSS is considered a metre-wavelength equivalent to the centimetre-wavelength NVSS. Four of the nine selected radio AGN (PGC\,12254, NGC\,1270, NGC\,1272, PGC\,12405) were found not to show a significant (S/N\,$>5$) radio signal in either survey and were thus rejected from the AGN sample (yet plotted in Fig.\,\ref{fig:L_q_diagram}).

Radio sources with $L_{\rm 1.4\,GHz} > 10^{23}$\,W\,Hz$^{-1}$ are rare and have much lower $q$ values. 
For $q  \la 2$ the dominant morphology is early-type. 
The three cluster members in the bottom right corner of Fig.\,\ref{fig:L_q_diagram} are the bright radio source Per\,A 
(NGC\,1275, \# 691) and the two well-known head-tail galaxies IC\,310 (\# 393) and NGC\,1265 (\# 517).
The latter is an optically passive galaxy, the other two are classified as sAGN in the WHAN diagram and are thus HERGs. 
The region $q<1.9$ and $L_{\rm 1.4\,GHz} < 10^{23}$\,W\,Hz$^{-1}$ is most likely dominated by weak radio-mode AGN. 
An interesting example is the peculiar early-type galaxy IC\,311  (\# 398), which is discussed in more detail in Sect.\,\ref{sect:pec-SF-AGN}.

\vspace{0.5cm}   
\noindent   
(c) AGN in the WISE colour-colour diagram \label{sec:AGN_IR}

\begin{figure}[htbp]
\centering
\includegraphics[viewport= 0 10 730 505,width=9.2cm,angle=0]{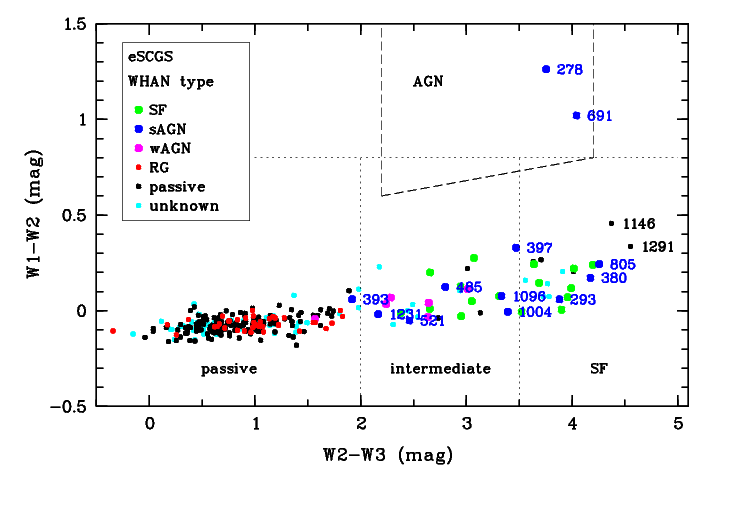}\\
\includegraphics[viewport= 0 10 730 505,width=9.2cm,angle=0]{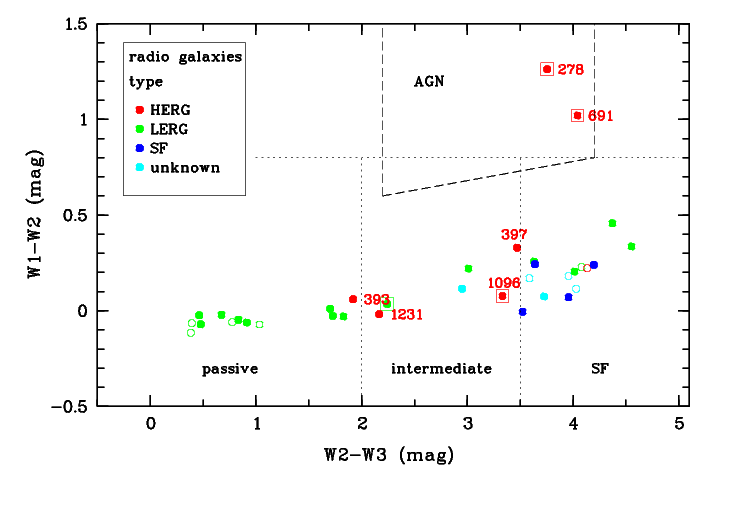}
\vspace{0.0cm}
\caption[CMD]{
WISE colour-colour diagram for the cluster galaxies (top) and the radio galaxies (bottom). The colours indicate the classification based on the WHAN diagram (top) and the radio galaxy type (bottom). In the bottom panel, filled circles represent cluster members, open circles are background galaxies, and framed symbols signify the presence of broad H$\alpha$ emission.
}
\label{fig:WISE_ccd_WHAN}
\end{figure}

\noindent
Figure\,\ref{fig:WISE_ccd_WHAN} shows the WISE colour-colour diagram in the same style as Fig.\,\ref{fig:WISE_ccd}. 
In the top panel, the established cluster galaxies with $W2 < 15.05$ are plotted, colour-coded by the WHAN-based activity type. For the two magnitude-limited samples MLS\_W1 and MLS\_ur, the diagrams look very similar. 

For QSOs, which are luminous AGN, an efficient selection threshold is given by the colour criterion $W1 - W2 \ga 0.8$ \citep{Assef_2010, Stern_2012}. The WISE QSO region is of course only sparsely populated in Fig.\,\ref{fig:WISE_ccd_WHAN}. The two galaxies with very warm, accretion-dominated colours $W1 - W2 > 0.8$\,mag are NGC\,1275 (\# 691) and UGC\,2608 (\# 278), which display both broad H$\alpha$ emission, radio emission with low $q$ values and hard X-ray counterparts. 
The spectra of low-luminosity AGN are diluted by stellar radiation from the host galaxy.
That is, the WISE colours depend on the AGN-to-host ratio where  $W1 - W2$ becomes smaller with a decreasing ratio
for early-type host galaxies. If the flux from the AGN is less than half of the host flux in the WISE bands, the integrated colour index would hardly exceed 
$W1 - W2 \approx 0.5$ \citep{Stern_2012}.

The sAGN outside the WISE QSO region and all wAGN have diverse $W2-W3$ colours and are distributed across the `main sequence' in the WISE diagram, suggesting a combination of SF, AGN, and emission from the older stellar populations.   
Galaxies with optical spectra that indicate dominant SF activity populate the intermediate and SF region of the WISE main sequence, whereas both passive galaxies and RGs are strongly concentrated on the left-hand side, that is, in the passive part.

The bottom panel of Fig.\,\ref{fig:WISE_ccd_WHAN} shows the WISE diagram for the radio sources, colour-coded by the radio types introduced above. Again, the SF-dominated radio galaxies populate the SF part of the diagram whereas 
HERGs are found in both the QSO and the intermediate part. LERGs are scattered across the whole main sequence with the majority in the passive part, indicating the dominant emission from the old stellar population.

\vspace{0.5cm}   
\noindent  
(d) X-ray AGN \label{sec:AGN_Xray}

\noindent
The detection of an X-ray point source (XPS) in the centre of a galaxy is another efficient technique to identify AGN. 
\citet{Santra_2007} analysed XPSs coincident with member galaxies in A\,426 on a very deep Chandra image of the cluster core region. Their result is consistent with the nuclei of all galaxies with $M_{\rm B} < -18$ being active at low radiative efficiency. The `Third XMM-Newton serendipitous source catalogue' \citep[3XMM-DR8,][]{Rosen_2016} contains 531\,454 unique X-ray sources but covers only a few percent of the sky. 
With a search radius of 5\arcsec, we identified XMM X-ray counterparts for 42 confirmed cluster galaxies.
Approximately $70$\%  of them are located at $R \le 30$\arcmin, compared to $40$\% for all spectroscopic cluster members. 
This indicates a strong bias in the XMM data. 
Because of the inhomogeneous field coverage by the X-ray source catalogues we decided to ignore X-ray selected AGN 
for the subsequent statistical analysis.

\begin{table*}[htbp]
\caption{AGN galaxies in A\,426.}
\begin{tabular}{rcllccrrcrcccc}
\hline\hline 
 Nr. &  
 Name & 
 Other name &
 Sel &
 WT  &
 bl  &
 $q$ \ \ &
 $\log L_{1.4}$& 
 RT  &
 $W1-W2$&
 Xray   &
 cl1    &
 pec \\
 (1) &
 (2) &
$\qquad$ (3) &
 (4) &
 (5) &
 (6) &
 (7) &
 (8) \ \  &
 (9) &
(10)$\quad$ &
(11) &
(12) &
(13) \\
\hline 
 278 & J031501.4+420209 & UGC 2608  & o/ir   & 2 & 1 &$ 2.06$& 22.78 & HERG &$ 1.26\quad $ & 1 & 6 & 9 \\  
 293 & J031520.6+413645 & PGC 12097 & o      & 2 &   &$     $&       &      &$ 0.06\quad $ &   & 6 & 0 \\  
 344 & J031601.0+420428 & UGC 2618  & o      & 3 &   &$     $&       &      &$ 0.11\quad $ & 1 & 6 & 9 \\  
 380 & J031634.5+410251 & PGC 12166 & o      & 2 &   &$     $&       &      &$ 0.17\quad $ &   & 6 & 9 \\  
 393 & J031643.0+411930 & IC  310   & o/r    & 2 &   &$ 0.28$& 23.55 & HERG &$ 0.06\quad $ & 1 & 3 & 0 \\  
 397 & J031645.9+401948 &           & o      & 2 &   &$ 2.60$& 21.54 & HERG &$ 0.33\quad $ &   & 6 & 0 \\  
 398 & J031646.7+400013 & IC  311   & r      & 5 &   &$ 1.22$& 22.36 & LERG &$ 0.01\quad $ &   & 1 & 3 \\  
 444 & J031727.2+412419 & NGC 1260  & o      & 3 &   &$     $&       &      &$ 0.07\quad $ & 1 & 3 & 4 \\  
 485 & J031755.2+405536 & UGC 2642  & o      & 2 &   &$     $&       &      &$ 0.12\quad $ &   & 9 & 6 \\  
 517 & J031815.8+415128 & NGC 1265  & r      & 5 &   &$-1.14$& 24.53 & LERG &$-0.02\quad $ & 1 & 2 & 0 \\  
 521 & J031817.7+414031 & PGC 12290 & o      & 2 &   &$     $&       &      &$-0.05\quad $ & 1 & 4 & 0 \\  
 533 & J031822.5+412435 & PGC 12295 & o      & 3 &   &$     $&       &      &$-0.04\quad $ & 1 & 3 & 0 \\  
 656 & J031927.4+413807 & UGC 2665  & o      & 3 &   &$     $&       &      &$-0.03\quad $ &   & 6 & 3 \\  
 691 & J031948.7+413042 & NGC 1275  & o/r/ir & 2 & 1 &$-0.43$& 25.16 & HERG &$ 1.02\quad $ & 1 & 1 & 3 \\  
 711 & J031954.4+420053 &           & o      & 3 &   &$     $&       &      &$ 0.04\quad $ &   & 4 & 1 \\  
 805 & J032041.4+424815 & PGC 12535 & o      & 2 &   &$     $&       &      &$ 0.24\quad $ &   & 6 & 1 \\  
 924 & J032140.0+412138 & NGC 1294  & o      & 3 & 1 &$ 2.17$& 21.44 & LERG &$ 0.04\quad $ & 1 & 3 & 0 \\  
1004 & J032253.0+411348 &           & o      & 2 &   &$     $&       &      &$-0.00\quad $ &   & 6 & 0 \\  
1096 & J032422.8+404719 & UGC 2715  & o/r    & 2 & 1 &$ 1.13$& 22.56 & HERG &$ 0.08\quad $ & 1 & 6 & 0 \\  
1231 & J032634.7+414143 &           & o      & 2 &   &$ 2.55$& 21.35 & HERG &$-0.02\quad $ &   & 2 & 3 \\  
\hline
\end{tabular}
\tablefoot{
(1) catalogue number, 
(2) catalogue name, 
(3) other name from NED, 
(4) AGN selection technique: o = optical, r = radio, ir = infrared, 
(5) spectral type from WHAN diagram: 1 = SF, 2 = sAGN, 3 = wAGN, 4 = RG, 
(6) broad emission line flag, 
(7) log FIR-to-radio flux density ratio,
(8) monochromatic 1.4\,GHz luminosity (W\,Hz$^{-1}$), 
(9) radio galaxy type,
(10) WISE colour index,
(11) XMM X-ray detection flag,
(12) morphological type, 
(13) morphological peculiarity class.
}
\label{tab:AGN_all}
\end{table*}

\subsubsection{Final AGN sample}\label{sec:AGN_sample}

Altogether, we classified 20 systems as AGN galaxies among the catalogued galaxies with $W1 < 14$ that are not spectroscopically established as background systems. A summary of relevant data is presented in Table\,\ref{tab:AGN_all}. 

Most (18) AGN were selected from the WHAN diagram where 12 systems were classified as sAGN and six as wAGN. 
Seven optical AGN are also detected at 1.4\,GHz, but only three of them have $q < 1.9$ and would thus be classified as radio-detected AGN. 
Nevertheless, we classify all sAGN with radio counterparts as HERGs and all wAGN with radio-counterparts as LERGs. Another two radio sources have $q < 1.9$ and are related to optical spectra indicating RG or passive galaxies. These radio-detected AGN are also classified as LERGs. 
The WISE colour-colour diagram establishes the sAGN NGC\,1275 (\# 691) and UGC\,2608 (\# 278) as luminous AGN but does not contribute to the discovery of additional AGN.

The AGN fraction of the cluster galaxies depends on which galaxy sample is considered. 
All 20 selected AGN galaxies belong to both the SCGS, eSCGS, MLS\_W1, and MLS\_ur, corresponding to an AGN fraction of $f_{\rm AGN} = 0.06\pm0.01$.
However, because all of them are brighter than $W1 \approx 12.5$, we can also take the bright sub-sample MLS\_W1b as the basis, which results in $f_{\rm AGN} = 0.10$ for $W1 \le 12.5$. 
This is most likely a lower limit only. 
First, some AGN may be wrongly excluded by the criterion $q < 1.9$, either because $q$ may be larger than 1.9 for some AGN or $q$ was over-estimated. 
Reasons for the latter can be uncertainties in the FIR flux or a co-existent starburst. 
In fact, there are three sAGN with $q > 2$ in Fig.\,\ref{fig:L_q_diagram}, among them UGC\,2608 (\# 278), that show broad H$\alpha$ emission (Fig.\,\ref{fig:line_deblending}). 
Secondly, radio AGN below the NVSS flux limit are not included. 
Using the European VLBI network, \citet{Park_2017} detected radio emission compact on the parsec scale in five early-type galaxies in the central $10\arcmin$ of the Perseus cluster and concluded that the radio sources are linked to jets from accreting SMBHs in low-luminosity AGN rather than to SF. 
Their sample includes NGC\,1270, which was excluded here as a result of the re-inspection of the NVSS image (Sect.\,\ref{sec:AGN_radio}), the galaxy VZw\,339 that is not in the list of radio sources by \citet{Miller_2001} 
and the two galaxies NGC\,1277 and NGC\,1278 that are NVSS sources in our catalogue but were not classified as AGN because of their high $q$ values.

Table\,\ref{tab:AGN_morph} lists the number $N$ of galaxies, the number ratio $r_{\rm l/e}$  of late types to early types, and the mean cluster-centric distance $R$ for the samples of AGN and SF galaxies. 
The SF sample was defined in Sect.\,\ref{SF_sample}.
The entries can be compared with Tables\,\ref{tab:samples} and \ref{tab:samples_statistics} for the parent galaxy samples.
The distribution of the morphological types of the AGN hosts differs from that of the parent sample.
Whereas the number ratio of late types to early types is $r_{\rm l/e} = 0.22$ for the MLS\_W1b galaxies (Table\,\ref{tab:samples_statistics}), the ratio is four times larger for the AGN sample. 
There are, however, strong differences between the different AGN types: most optically selected AGN have late-type hosts
($r_{\rm l/e} = 1.12$), with the highest ratio for sAGN ($r_{\rm l/e} = 1.75$). 
For radio AGN, on the other hand, the ratio is smaller, $r_{\rm l/e} = 0.60$.

\begin{table}[htbp]
\caption{Properties of the AGN host galaxy samples and the SF sample.}
\begin{tabular}{lrrc}
\hline\hline 
Sample                  
&$N$
&$r_{\rm l/e}$
& $\langle R\rangle$\,($\arcmin$) \\
\hline
all optical AGN        & 18 & 1.12 &  $42\pm7$ \\  
- sAGN                 & 12 & 1.75 &  $49\pm6$ \\  
- wAGN                 &  6 & 0.50 &  $26\pm6$ \\ 
\hline
all radio AGN          &  8 & 0.60 &  $55\pm11$ \\ 
- HERGs                &  6 & 1.00 &  $54\pm12$ \\  
- LERGs                &  2 & 0.00 &  $61\pm35$ \\  
\hline
all AGN                & 20 & 0.90 &  $44\pm6$ \\  
- rad. mode$^\ast$     & 11 & 2.33 &  $54\pm6$ \\
- jet mode             &  8 & 0.33 &  $35\pm10$ \\
\hline     
SF                     & 31 &  -   &  $71\pm6$ \\ 
\hline
\end{tabular}
\tablefoot{$^\ast$ without the BCG NGC\,1275}
\label{tab:AGN_morph}
\end{table}

The AGN population in the present-day Universe can be divided into two disjoint categories: the radiative-mode AGN and the jet-mode AGN \citep{Heckman_2014}. 
In the radiative mode, the energetic output is powered by the gravitational energy of the accreted gas in a radiatively efficient disc and is mainly released in form of the photons from the `big blue bump' in the UV-to-optical part of the electromagnetic spectrum. 
A radiatively efficient, optically thick and geometrically thin accretion disc is thought to exist around the SMBH if the accretion rate approaches the Eddington limit. At lower accretion rates of less than a few percent of the Eddington rate, a radiatively inefficient accretion flow from an optically thin, geometrically thick disc is formed. In the jet mode, the produced radiation flux is much lower and the primary energetic output is in the form of collimated jets, the AGN is observed as a radio galaxy.

We applied the classification scheme from \cite{Heckman_2014} to delineate the AGN from Table\,\ref{tab:AGN_all} 
into  jet-mode AGN and radiative-mode AGN. 
All LERGs and wAGN were considered to represent the jet mode, whereas sAGN were identified with radiative-mode AGN. 
The most luminous AGN in A426, hosted by the BCG NGC\,1275, is classified here as radiative-mode type. 
BCGs, the most luminous elliptical galaxies in the present-day Universe and strong radio sources, 
are usually situated close to the bottom of the potential of relaxed, cool-core clusters, like A\,426. 
The cooling flow onto the central galaxy provides a unique situation for feeding the SMBH and, therefore,
we excluded NGC\,1275 from the statistics in Table\,\ref{tab:AGN_morph} and in Sect.\,\ref{sect:spatial_AGN_SF} below.
Radiative-mode AGN and jet-mode AGN differ from each other both in the distribution of morphological types and in their 
spatial distribution.

\subsection{Spatial (projected) distribution of AGN and SF galaxies}\label{sect:spatial_AGN_SF}

The different mean cluster-centric distances $\langle R\rangle$ for jet-mode AGN, radiative-mode AGN, and SF galaxies (Table\,\ref{tab:AGN_morph}) indicate different radial distributions. 
For comparison, the mean cluster-centric distance of all eSCGS galaxies is $\langle R\rangle = 45\arcmin$ (Table\,\ref{tab:samples}). Jet-mode AGN have a smaller $\langle R\rangle = 35\arcmin$.
The mean cluster-centric distance of the radiative-mode AGN, $\langle R\rangle = 54\arcmin$, is 1.5 times larger. The SF sample is even less concentrated towards the cluster centre with
$\langle R\rangle = 71\arcmin$.

\begin{figure}[htbp]
\includegraphics[viewport = 0 0 535 509,width=8.7cm,angle=0]{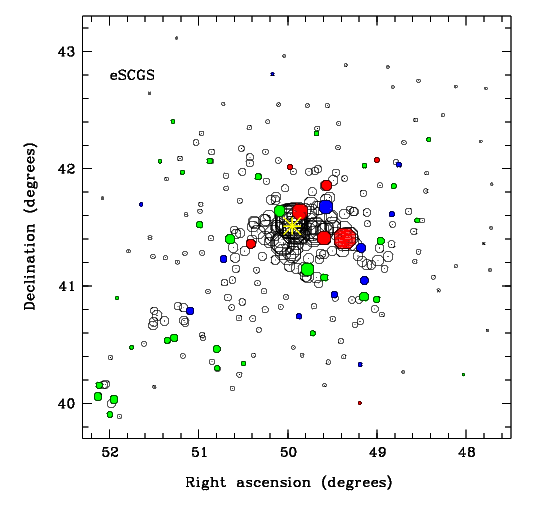}
\caption[local density with AGN]{
Local density bubble plot diagram for the eSCGS as Fig.\,\ref{fig:local_density}, 
but with colour coding for SF galaxies (green), radiative-mode AGN (blue), jet-mode AGN (red). 
The BCG AGN galaxy NGC\,1275 is indicated by the yellow asterisk.
}
\label{fig:local_density_eSCGS_AGN_SF}
\end{figure}

Figure\,\ref{fig:local_density_eSCGS_AGN_SF} shows the distributions of the two AGN types and of the SF sample over the field for the eSCGS sample. The plots for the magnitude-limited samples look similar.  
Jet-mode AGN prefer smaller projected cluster-centric distances and higher projected densities than radiative mode AGN and SF galaxies. Both radiative-mode AGN and SF galaxies are distributed smoothly over the field without
a significant concentration towards the cluster centre. 
We note a concentration of AGN  along the chain of bright galaxies in the east-west direction.
There is a concentration of SF galaxies in the bottom left corner that might be related to the filament from SE to NW through A\,426 (Fig.\,\ref{fig:supercluster}). 

\begin{figure}[htbp]
\includegraphics[viewport = 20 40 580 790,width=6.7cm,angle=270]{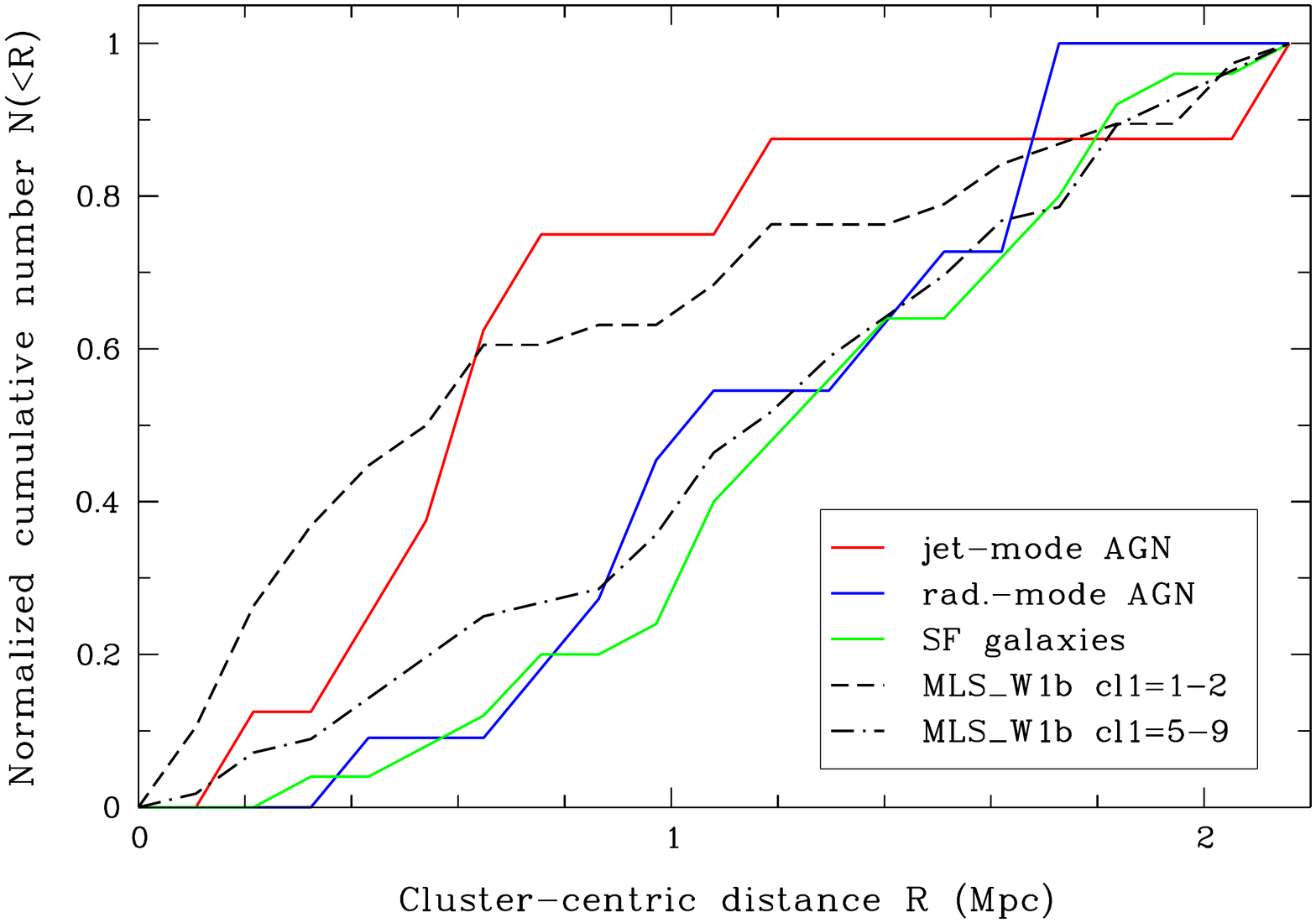} 
\caption[R distribution_AGN]{Cumulative distribution of the cluster-centric distances $R$ for 
cluster galaxies classified as jet-mode AGN (red), radiative-mode AGN (blue), or SF (green).
For comparison, the distributions of the bright early-type and late-type galaxies from MLS\_W1b are plotted (black).
}
\label{fig:R_distribution_AGN}
\end{figure}

Given that the different activity types occur in different morphological types, different projected distributions are 
qualitatively expected because of the radial morphological segregation (Sect.\,\ref{sect:radial_morph_seg}).
Figure\,\ref{fig:R_distribution_AGN} shows the cumulative distributions of $R$ for the two AGN samples and the SF sample. For comparison, the distributions of the bright E and S0 galaxies (${\tt cl1}=1-2$) and the late-type  (${\tt cl1}=5-9$) galaxies are plotted. We applied the Kolmogorov-Smirnov two-sample test to check the null hypothesis $H^0$ that both samples come from a common distribution against the alternative hypothesis that the two samples do not come from the same distribution. $H^0$ has to be rejected for the comparison of jet-mode AGN and SF galaxies ($\alpha = 0.05$), 
but not for the comparison of radiative-mode versus jet-mode AGN or SF galaxies. 
The $R$ distributions of the radiative-mode AGN and the SF galaxies are very similar to that of the 
late-type galaxies. On the other hand, jet-mode AGN are radially distributed like E galaxies.

\begin{figure}[htbp]
\includegraphics[viewport = 20 40 580 790,width=6.7cm,angle=270]{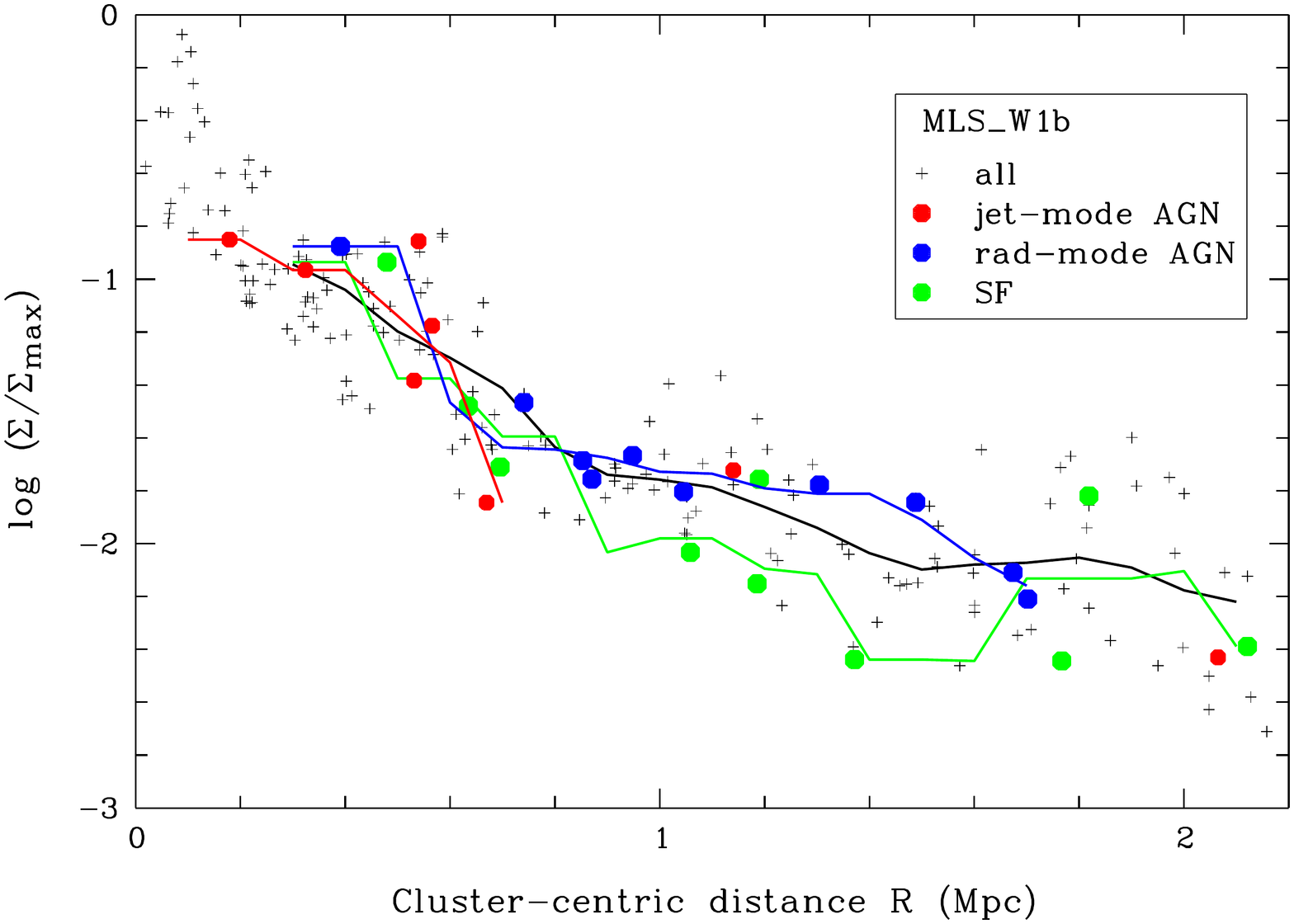} 
\caption[Sigma-R_AGN]{Normalised local density parameter $\Sigma$ as a function of the cluster-centric distance for the 
two AGN types and the SF sample. The MLS\_W1b sample is plotted for comparison.
Symbols indicate individual galaxies, polygons the mean relations for the sub-samples.
}
\label{fig:Sigma_R_AGN}
\end{figure}

Figure\,\ref{fig:Sigma_R_AGN} shows the $\Sigma-R$ relations for the two AGN samples, the SF sample, and, for comparison, all MLS\_W1b galaxies. At fixed $R$, the mean $\Sigma$ values do not differ greatly. 
This means that the trend with $\Sigma$ can be completely explained by the radial segregation. At $R \approx 0.8 - 1.6$\,Mpc, the SF galaxies seem to prefer an environment of slightly lower $\Sigma$ than the radiative-mode AGN. However, given the small sample size, it is not clear whether this is a real trend.  

Analogous to Sect.\,\ref{sect:radial_morph_seg}, we looked for correlations between the ratios $r_{\rm rAGN/jAGN} = N_{\rm rAGN}/N_{\rm jAGN}$ and $r_{\rm sf/passive} = N_{\rm sf}/N_{\rm passive}$ on the one hand
with the cluster-centric distance $R$, the local projected density $\Sigma$, and the 3-dimensional substructure parameter $\delta$ on the other hand.
The null hypothesis $H^0$ of independence between $X$ and $Y$ was tested against the alternative hypothesis 
$H^A$ of a negative or positive association between $X$ and $Y$, 
where $X = r_{\rm rAGN/jAGN}$ or $r_{\rm sf/passive}$ and $Y = R, \Sigma$, or $\delta$. 
Because of the small sample size, Fisher's exact (one-sided) test was used to analyse 2x2 contingency tables. 
For $Y = R$, we compared the ratio $X$ from the central region at $R \le 0.6$\,Mpc 
with that from the outer region  $R > 0.6$\,Mpc. 
For $Y = \Sigma$ or $\delta$, the sample was subdivided into three nearly equal-sized sub-samples of 
low, intermediate, or high $Y$ to compare the ratio $X$ from the low-$Y$ with that from the high-$Y$ sub-sample.  
Table\,\ref{tab:Fisher_exact} lists $H^A$ and the resulting $P$ values for the six $X-Y$ combinations. 
The smaller the value of $P$, the greater the evidence of rejecting $H^0$ in favour of $H^A$. 
We concluded that  $H^0$ can be rejected for $Y = R$ and $\Sigma$ but not for the substructure parameter $\delta$.

\begin{table}[h] 
\caption{Results from Fisher's exact test of independence between $X$ ans $Y$ for the eSCGS galaxies.}
\begin{tabular}{llcc}
\hline\hline 
$X$                  &   $Y$        & $H^A$                    & $P$  \\
\hline
$r_{\rm rAGN/jAGN}$  & $R$          & $X$ larger for large $Y$ & 0.018  \\
$r_{\rm rAGN/jAGN}$  & $\Sigma$     & $X$ larger for small $Y$ & 0.084  \\
$r_{\rm rAGN/jAGN}$  & $\delta$     & $X$ larger for large $Y$ & 0.500  \\
\hline
$r_{\rm sf/passive}$ & $R$          & $X$ larger for large $Y$ & 0.021  \\
$r_{\rm sf/passive}$ & $\Sigma$     & $X$ larger for small $Y$ & 0.003  \\
$r_{\rm sf/passive}$ & $\delta$     & $X$ larger for large $Y$ & 0.645  \\
\hline
\end{tabular}
\label{tab:Fisher_exact}
\end{table}

%
%
\section{Distorted galaxies and mergers}\label{sect:peculiar}
%

\subsection{Peculiarity fraction and (projected) distribution}\label{sect:pec-statistics}

In the following, the ratio of the number of peculiar galaxies and mergers to the number of all galaxies in the sample is referred to as peculiarity fraction $f_{\rm p}$.
Here, peculiar galaxies are either systems classified as morphological type merger ({\tt cl1} = 9) or as
any other morphological type but with a peculiarity flag that indicates strong lopsidedness, a minor or major merger or a collisional ring ({\tt pec} = $2 \ldots 6$ or 8). We did not include the somewhat uncertain cases of weak lopsidedness and M51 type where the classification might be affected by projection effects. In addition, a few galaxies without morphological type estimation ({\tt cl1} = 0) or with low reliability of peculiarity (flag\_pec $< 2$) were excluded from the parent sample. Both in the spectroscopic sample and the magnitude-limited samples, peculiar galaxies account for 7-9\%. We observe an anti-correlation of $f_{\rm p}$ with the sample-averaged magnitude, 
which might indicate that the differences are caused by the different proportions of fainter galaxies with a lower probability for detecting faint peculiar structures.  

\begin{figure}[htbp]
\includegraphics[viewport= 0 40 580 790,width=6.7cm,angle=270]{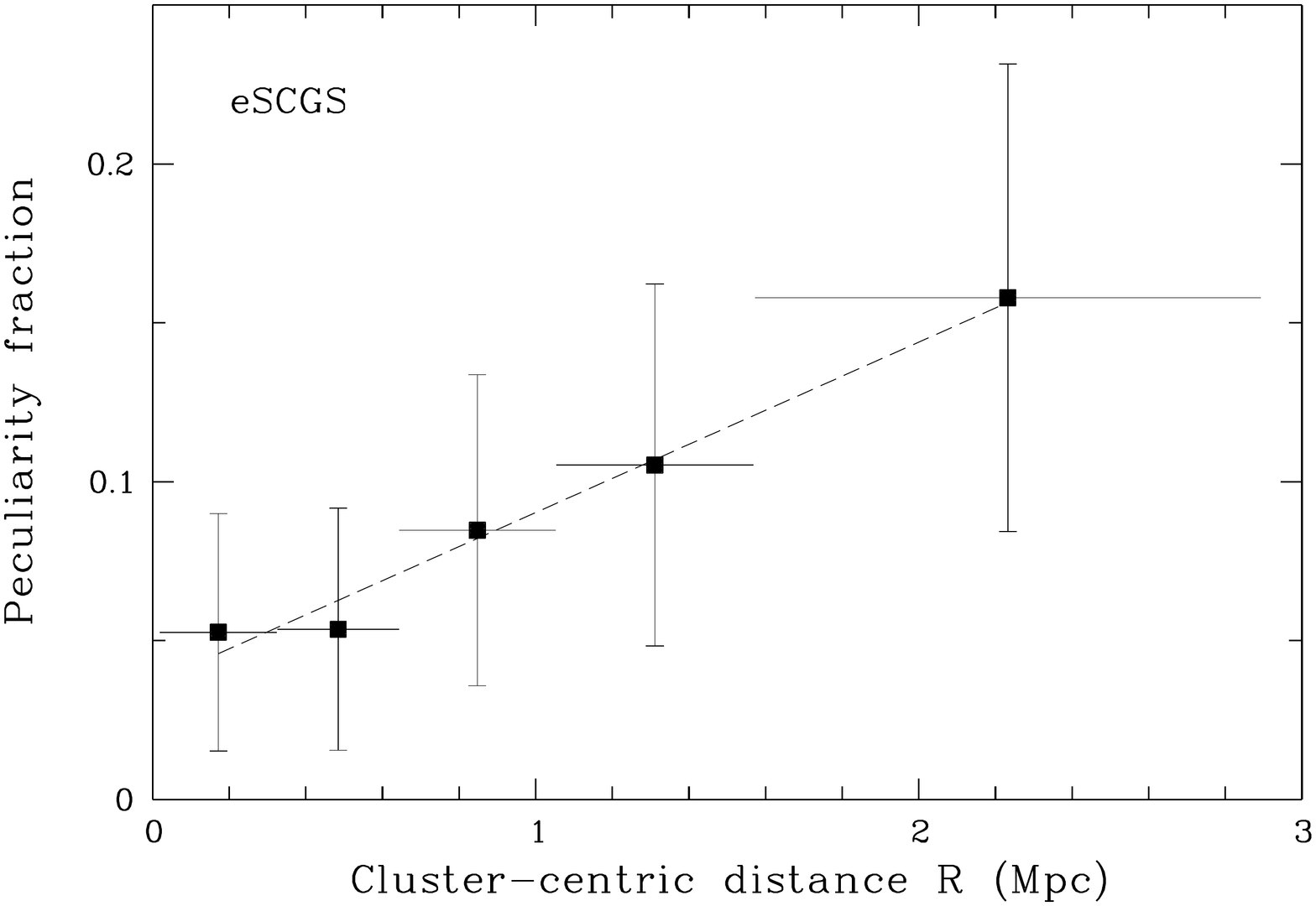}\\
\includegraphics[viewport= 0 40 580 790,width=6.7cm,angle=270]{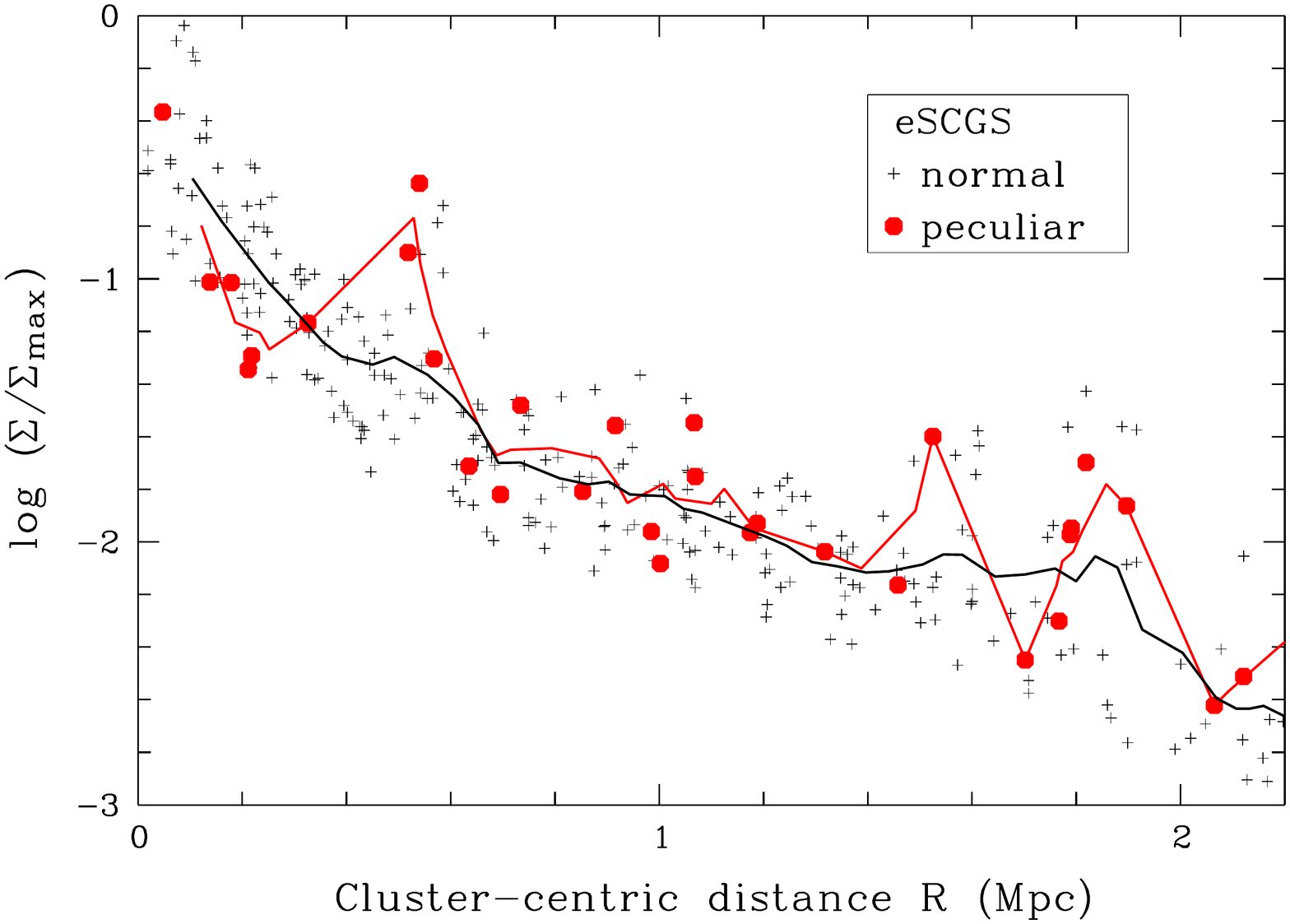} 
\caption[Radial segregations]{Top: peculiarity fraction of the eSCGS galaxies in five $R$ bins. 
Vertical bars are counting errors, horizontal bars mark the bin widths.
Each bin contains approximately the same proportion of galaxies from the parent sample.
Bottom: $\Sigma - R$ diagram for normal (black) and peculiar (red) galaxies. The polygons signify the mean relations.
}
\label{fig:segregation_correlations_morph_pec}
\end{figure}

As in Sect.\,\ref{sect:radial_morph_seg}, we looked for trends of $f_{\rm p}$ with $R, \Sigma$ and $\delta$. Kendall's  rank correlation test indicates a significant correlation with $R$ but only for the eSCGS 
(Fig.\,\ref{fig:segregation_correlations_morph_pec},\,top);
there is no significant trend with $\Sigma$ (Fig.\,\ref{fig:segregation_correlations_morph_pec},\,bottom) or $\delta$ for any sample. We compared the peculiarity fraction $f_{\rm p,1}$ in the sub-sample S$_1$ from the inner ($R<0.5$\,Mpc) region with $f_{\rm p,2}$ in the sub-sample S$_2$ from the outer ($R\ge0.5$\,Mpc) region (Table\,\ref{tab:samples_stat_pec}).  
and found $f_{\rm p,1} < f_{\rm p,2}$ for all samples. The Z test (Sect.\,\ref{sect:radial_morph_seg})
indicates that the null hypothesis $H^0: f_{\rm p,1} \ge f_{\rm p,2}$ can be rejected in favour of the alternative hypothesis $H^{\rm A}: f_{\rm p,1} < f_{\rm p,2}$ for eSCGS and MLS\_ur at $\alpha = 0.05 \ (z_\alpha = 1.65)$ and for MLS\_W1 at $\alpha = 0.07 \  (z_\alpha = 1.48)$. Among the galaxies classified as mergers ({\tt cl1} = 9), only one system (\# 613 = PGC 12358 at $z = 0.011$) has a projected cluster-centric distance less than the core radius.

\begin{table}[h]
\caption{Peculiarity fractions $f_{\rm p}$ and test statistic $\hat{z}$ for the comparison of the sub-samples $f_{\rm p,1}$ and $f_{\rm p,2}$. 
}
\begin{tabular}{lcccc}
\hline\hline
Sample   
& $f_{\rm p, tot}$ 
& $f_{\rm p,1}$
& $f_{\rm p,2}$ 
& $\hat{z}$ \\
\hline
MLS\_W1b     & 0.082 & 0.043 & 0.104 & 1.47 \\
SCGS         & 0.083 & 0.034 & 0.107 & 2.06 \\
eSCGS        & 0.086 & 0.033 & 0.110 & 2.22 \\
MLS\_ur      & 0.081 & 0.036 & 0.095 & 1.71 \\
MLS\_W1      & 0.068 & 0.033 & 0.079 & 1.54 \\
MaxS         & 0.060 & 0.043 & 0.062 & 0.81 \\
\hline
SF           & 0.419 & 0.000 & 0.448 & - \\ 
AGN rad. mode& 0.182 & 0.000 & 0.200 & - \\
AGN jet mode & 0.250 & 0.167 & 0.333 & - \\
\hline
\end{tabular}
\label{tab:samples_stat_pec}
\end{table}

At first glance, it may come as a surprise that the peculiarity fraction is lower in the cluster core region. 
Such a trend can be expected however as a consequence of the morphological segregation in combination with the fact that (dynamically cool) discs are fragile and more susceptible than (dynamically hot) ellipsoidal systems. 
In fact, the peculiarity fraction $f_{\rm p,e}$ of the sub-samples of late-type galaxies is about $5\pm3$ times larger than $f_{\rm p,l}$ from the sub-samples of early-type galaxies, dependent on the parent sample and the definition of peculiarity. A hypothetical total peculiarity fraction can be modelled by  
$h_{\rm p}(R) = f_{\rm e}(R)\cdot f_{\rm p,e} + f_{\rm l}(R)\cdot f_{\rm p,l}$,
where the proportions $f_{\rm e}$ and $f_{\rm l}$ of early-type galaxies and late-type galaxies 
are assumed to depend on $R$, but the type-dependent peculiarity fractions $f_{\rm p,e}$ and $f_{\rm p,l}$ are not.
We computed $h_{\rm p,1}$ and $h_{\rm p,2}$ for the sub-samples S$_1$ and S$_2$ as above, estimating $f_{\rm e}$ and $f_{\rm l}$ from the corresponding sub-sample S$_1$ or S$_2$, 
but $f_{\rm p,e}$ and $f_{\rm p,l}$ from the total sample. 
The general result $h_{\rm p,1} < h_{\rm p,2}$ supports the assumption that the higher peculiarity fraction in the outer region is caused by the morphological segregation. However, 
the ratio $h_{\rm p,2}/h_{\rm p,1} = 1.8 \ldots 2.1$ is smaller than the observed ratio $f_{\rm p,2}/f_{\rm p,1} \approx 2.4 \ldots 3.4$. This difference hints at additional effects. 
It is known from numerical simulations that galaxy interactions in a cluster core differ from those in the field or in the outer parts of a cluster  \citep{Mihos_2004}. Slow encounters in the outer parts produce long-living phenomena, while faster encounters in the core result in short-lived perturbations. In addition, the lifetime of the tidal debris is shorter in the core region where the tidal field of the cluster quickly strips the loosely bound material. As a consequence, the interaction fraction is underestimated in the cluster core if derived from the presence of tidal tracers.

\subsection{Distorted SF and AGN galaxies}\label{sect:pec-SF-AGN}

Morphological peculiarities of 17 IRAS galaxies in the Perseus cluster have been described by \citet{Meusinger_2000}. In the Appendix\,\ref{sect:individual} of the present paper, we present additional 18 peculiar cluster galaxies.
 
In Figs.\,\ref{fig:SFR-M} and \ref{fig:sSFR-M-WISE}, mergers and peculiar morphology that is indicative of gravitational distortions are clearly more frequent in the upper regions of the diagrams. The sample from Fig.\,\ref{fig:SFR-M} contains three mergers (\# 379, 485, 887), all classified as SF galaxies with moderate sSFR between 0.09 and 0.85 \,Gyr$^{-1}$. 
In Fig.\,\ref{fig:sSFR-M-WISE}, seven mergers ({\tt cl1} = 9) are plotted where five are in the SF region and two in the intermediate region. The SF sample from Sect.\,\ref{SF_sample} (Tables\,\ref{tab:AGN_morph} and \ref{tab:samples_stat_pec}) includes eight IRAS galaxies.
Four SF galaxies were classified as mergers. With only one exception (the blue compact dwarf J032311.4+402753 mentioned already in Sects.\,\ref{sect:morph_seg} and \ref{SFR_WISE}), all SF galaxies were found to show late-type morphology.
The peculiarity fraction of the SF sample, $f_{\rm p} = 0.45$, is five times larger than in the parent galaxy samples (Table\,\ref{tab:samples_stat_pec}). All peculiar SF galaxies are at $R > 0.5$\,Mpc. We did not perform the Z test because of the small sample size. 

Morphological peculiarities were registered for a remarkably high percentage of the AGN hosts. 
We applied the Z test for the comparison of the peculiarity fractions in the extended spectroscopic sample eSCGS, $f_{\rm p}$, with that from the sub-sample of AGN galaxies, $f_{\rm p, AGN}$.
The null hypothesis $H^0: f_{\rm p} \ge f_{\rm p, AGN}$ has to be rejected in favour of the alternative hypothesis $H^0: f_{\rm p} < f_{\rm p, AGN}$ at an error probability of less than 1\% ($\alpha \le 0.01$).

There is a clear difference between the morphological compositions of the samples of jet-mode and radiative-mode AGN (Table\,\ref{tab:AGN_morph}). 
The hosts of radiative-mode AGN are dominated by late-type galaxies ($r_{\rm l/e} = 2.33$), whereas jet-mode AGN galaxies are dominated by early types  ($r_{\rm l/e} = 0.20$).
The peculiarity fraction is at least twice as high as that of the parent samples not only for the sample of radiative-mode AGN galaxies but also for the hosts of the jet-mode AGN.  
Likewise, \citet{Ehlert_2015} concluded that there is some evidence that X-ray AGN in clusters may be preferably hosted in galaxies with disturbed morphology. 
Larger data samples, more robust and more detailed classification schemes will be required to further investigate which type of perturbations may be responsible to which extent for the triggering of AGN activity in galaxy clusters.
As for the parent samples, we found that the peculiarity fraction of AGN hosts is larger in the outer cluster region ($R>0.5$\,Mpc) compared to the inner region  (Table\,\ref{tab:samples_stat_pec}). We did not preform the Z test because of the small sample sizes. 

Compared to the SF sample, the peculiarity fraction of AGN hosts is smaller and strong morphological features indicative of major mergers are less frequent and/or fainter.
The merger fraction of radiative-mode AGN $f_{\rm m} = 0.09$ is at least three times larger than that of the galaxy samples (Table\,\ref{tab:samples_statistics}), but this result is not robust because of the small sample size.  
The only AGN host galaxy with the morphological type {\tt cl1} = 9 (i.e. merger) is UGC\,2642 (\# 485), see below. 
For the BCG NGC\,1275, it is not clear whether its rich phenomenology is related to a recent (minor) merger \citep[e.g.][Sect.\,\ref{sect:Introduction}]{Conselice_2001}. 
NGC\,1260 (\# 444) is close to J031732.0+412442 (\# 454) in projection and it might thus be speculated that it is a merger in an early stage, but the evidence is sparse.

%
%
\section{Summary and conclusion}\label{sect:summary}
%

We compiled a new catalogue of galaxies in the 10 square degrees field of the Perseus cluster centred on the BCG NGC\,1275. 
Altogether 1294 galaxies were selected primarily from images taken with the CCD camera in the Schmidt focus of the Tautenburg 2\,m telescope and verified on SDSS images. Morphological types were obtained for nearly 90\% of the galaxies by the eyeball inspection of the Schmidt images in the B band and in the narrow H$\alpha$ band, in combination with the SDSS colour images and additional images taken at the Calar Alto observatory for selected galaxies. The morphological classification was supported by the surface brightness profile analysis for 143 galaxies and the SDSS profile parameter {\tt fracDeV}. The catalogue incorporates extinction-corrected magnitudes from  SDSS, 2MASS, and WISE, 1.4\,GHz flux densities from NVSS, X-ray flux densities from XMM-Newton, and spectroscopic data from various sources.  
Spectroscopic redshifts are available for 384 galaxies, including 41 redshifts from observations carried out in the framework of the present study. Redshift distances are therewith known for about 60\% of the galaxies with $W1 < 14$ and for about 85\% of the galaxies with $W1<12.5$.  

We constructed different galaxy samples (Sect.\,\ref{sect:galaxy_samples}).
The spectroscopic cluster galaxy sample (SCGS) consists of 286 galaxies with spectroscopic redshifts 
compatible with cluster membership. An extended spectroscopic cluster galaxy sample (eSCGS) of 313 Perseus cluster galaxies includes additional cluster members identified from the H$\alpha$ images. Because of the  heterogeneous nature of the spectroscopic data, we also constructed magnitude-limited samples where the limits were set either in the WISE W1 band (MLS\_W1) or in the SDSS u and r bands (MLS\_ur). The following conclusions can be drawn from the analysis of these samples.

The sample-averaged heliocentric redshift from the SCGS is $\bar{z} = 0.0177\pm0.0039$ (Sect.\,\ref{sect:redshift}), in good agreement with the ICM redshift $z_{\rm ICM} = 0.01767\pm0.00003$ derived by the \citet{Hitomi_2018a}.
The line-of-sight velocity distribution can be considered as Gaussian, though the outer part of the radial velocity dispersion profile shows a certain similarity with that of non-Gaussian clusters (Sect.\,\ref{sect:velocity_distribution}).
 
The radial density profiles of the magnitude-limited samples were found to be in a generally good agreement with the generalised Hubble, King, NFW, and de Vaucouleurs profiles where the core profiles (Hubble, King) provide a slightly better fit (Sect.\,\ref{sect:surf_density}) .  
We found weak indications for local substructures, both from the line-of-sight velocity map, the projected local density map and the three-dimensional DS test (Sect.\,\ref{sect:substructure}).

The virial mass of the Perseus cluster is $(2.4\pm0.5)\,10^{15}\,\mathcal{M}_\odot$, in line with the projected mass estimator of $2.7\,10^{15}\,\mathcal{M}_\odot$. The virial radius is estimated to $2.6\pm0.4$\,Mpc, which is roughly equal to the maximum extension of our survey field (Sect.\,\ref{sect:virial_radius}). 
The K-band luminosity function is well represented by a Schechter LF with $M_{\rm K}^\ast = −24.25 \pm 0.15$ and $\alpha = -1.03 \pm 0.15$. Early-type and late-type galaxies have different Schechter parameters that are in line with \citet{DePropris_2017}. 
The total K-band luminosity and stellar mass are estimated to $L_{\rm K} = (12.4\pm1.6)\,10^{12}\,L_\odot$ and $\mathcal{M}_\ast \approx (1 - 2)\,10^{13} \ \mathcal{M}_\odot$, which corresponds to $\la 1$\% of the virial mass (Sect.\,\ref{sect:LF}). 
The total mass in SMBHs is $\mathcal{M}_{\rm SMBH} \approx 3\,10^{10}\,\mathcal{M}_\odot$ (Sect.\,\ref{sect:SMBH_MF}).

The morphological mixture varies between the samples owing to different selection effects. The number ratio of late-type to early-type galaxies is $r_{\rm l/e} = 0.64$ for MLS\_ur and 0.39 for MLS\_W1. There is a general trend towards a lower $r_{\rm l/e}$ in the central region and at higher surface density, but no significant correlation was found with the three-dimensional local substructure parameter.  The number ratio of disc-dominated to bulge-dominated systems is $0.5\pm0.1$ (Sect.\,\ref{sect:morph_seg}).

Based on WISE and SDSS colours, about $75 \pm 5\%$ of the galaxies were classified as passive and $25 \pm 5\%$ of them as actively star-forming where the latter is divided in approximately equal parts into moderately active (intermediate) and strongly active (SF). With the only exception of a blue compact galaxy, all SF galaxies exhibit late-type morphology. The total SFR of the Perseus cluster is estimated to approximately $115\,\mathcal{M}_\odot$\,yr$^{-1}$ (Sect.\,\ref{sect:SF}).

We identified 20  (mostly low-luminosity) AGN from optical, infrared and radio data. The AGN sample consists of two approximately equal-sized distinct sub-samples of jet-mode and radiative-mode AGN (Sect.\,\ref{sect:AGN}). 
Jet-mode AGN clearly prefer early-type hosts (75\%) and the central cluster region with its high local (projected) density. 
Radiative-mode AGN are preferably found in late-type galaxies (70\%) outside the cluster core, similar to the SF galaxies (Sect.\,\ref{sect:spatial_AGN_SF}). 

About 8\% of the cluster galaxies were found to show morphological peculiarities indicative of some kind of substantial distortion.
About 80\% of the peculiar galaxies populate the outer cluster region at $R \ge 0.5$\,Mpc
where the peculiarity fraction $f_{\rm p}$ is about three times larger than at $R < 0.5$\,Mpc.  
This result can be explained by the rarity of fragile discs in the cluster core in combination with expected differences in the observability of tidal structures between the dense core and the outer cluster (Sect.\,\ref{sect:pec-statistics}).  
More data, in particular higher resolution images, and detailed modelling of the individual galaxies are necessary to assign the observed peculiar structures to the various types of interactions expected in the cluster environment.
The peculiarity fraction of the SF sample is about five times higher than that of the parent galaxy samples. All peculiar SF galaxies are found at $R \ge 0.5$\,Mpc (Sect.\,\ref{sect:pec-SF-AGN}). 
For the AGN sample $f_{\rm p}$ is also larger than that of the parent samples but smaller than that of the SF sample (Sect.\,\ref{sect:pec-SF-AGN}).
A few galaxies with conspicuous morphological peculiarities were presented and discussed individually (Appendix\,\ref{sect:individual}). In several cases, the peculiar morphology seems to indicate galaxy mergers. 
A remarkable example is the peculiar elliptical galaxy IC\,311, which seems to host a jet-mode AGN and is probably a down-scaled version of NGC\,5128 (Cen\,A) and one of the nearest known GPS radio sources. We also presented J032853.7+400211 as a so far unnoticed candidate for the rare class of Hoag-type ring galaxies.

\begin{acknowledgements}
The anonymous referee is thanked for helpful comments and suggestions.
This research is partly based on observations carried out with the 2.2 m telescope and the 3.5 https://www.aanda.org/for-authors/language-editing/8-frequent-changesm telescope of the 
Spanish-German Astronomical Centre on Calar Alto, Spain, during several campaigns. We would like to thank the Calar Alto staff for their kind support. 
This work is partly based also on observations carried out under project number D31-02 with the IRAM 30\,m telescope. IRAM is supported by INSU/CNRS (France), MPG (Germany) and IGN (Spain). The authors are grateful to the IRAM staff at Pico Veleta, Spain, for performing the observations in Director's time. 
\\

This research has made use of data products from the Sloan
Digital Sky Survey (SDSS). Funding for the SDSS and SDSS-II has been provided by
the Alfred P. Sloan Foundation, the Participating Institutions
(see below), the National Science Foundation, the National
Aeronautics and Space Administration, the U.S. Department
of Energy, the Japanese Monbukagakusho, the Max Planck
Society, and the Higher Education Funding Council for
England. The SDSS Web site is http://www.sdss.org/.
The SDSS is managed by the Astrophysical Research
Consortium (ARC) for the Participating Institutions.
The Participating Institutions are: the American
Museum of Natural History, Astrophysical Institute
Potsdam, University of Basel, University of Cambridge
(Cambridge University), Case Western Reserve University,
the University of Chicago, the Fermi National
Accelerator Laboratory (Fermilab), the Institute
for Advanced Study, the Japan Participation Group,
the Johns Hopkins University, the Joint Institute
for Nuclear Astrophysics, the Kavli Institute for
Particle Astrophysics and Cosmology, the Korean
Scientist Group, the Los Alamos National Laboratory,
the Max-Planck-Institute for Astronomy (MPIA),
the Max-Planck-Institute for Astrophysics (MPA),
the New Mexico State University, the Ohio State
University, the University of Pittsburgh, University
of Portsmouth, Princeton University, the United
States Naval Observatory, and the University of
Washington. \\

This publication has made extensive use of the NASA/IPAC Infrared Science Archive (IRSA) and the NASA/IPAC Extragalactic Database (NED), operated by the Jet Propulsion Laboratories/California Institute of Technology, under contract with the National Aeronautic and Space Administration. This publication has also made use of the VizieR catalogue access tool, CDS, Strasbourg, France, 
In particular, we acknowledge the use of data products from the Wide-field Infrared Survey Explorer, 
which is a joint project of the University of California, Los Angeles, 
and the Jet Propulsion Laboratory/California Institute of Technology, 
funded by the National Aeronautics and Space Administration, and from the Two Micron All Sky Survey, which is a 
joint project of the University of Massachusetts and the Infrared Processing and
Analysis Center/California Institute of Technology, funded by the National Aeronautics 
and Space Administration and the National Science Foundation. 

\end{acknowledgements}

\bibliographystyle{aa} 
\bibliography{literature} 

\begin{thebibliography}{210}
\expandafter\ifx\csname natexlab\endcsname\relax\def\natexlab#1{#1}\fi

\bibitem[{{Abadi} {et~al.}(1999){Abadi}, {Moore}, \& {Bower}}]{Abadi_1999}
{Abadi}, M.~G., {Moore}, B., \& {Bower}, R.~G. 1999, \mnras, 308, 947

\bibitem[{{Abazajian} {et~al.}(2009){Abazajian}, {Adelman-McCarthy},
  {Ag{\"u}eros}, {Allam}, {Allende Prieto}, {An}, {Anderson}, {Anderson},
  {Annis}, {Bahcall}, \& et~al.}]{Abazajian_2009}
{Abazajian}, K.~N., {Adelman-McCarthy}, J.~K., {Ag{\"u}eros}, M.~A., {et~al.}
  2009, \apjs, 182, 543

\bibitem[{{Abolfathi} {et~al.}(2018){Abolfathi}, {Aguado}, {Aguilar}, {Allende
  Prieto}, {Almeida}, {Ananna}, {Anders}, {Anderson}, {Andrews}, {Anguiano},
  {Arag{\'o}n-Salamanca}, {Argudo-Fern{\'a}ndez}, {Armengaud}, {Ata},
  {Aubourg}, {Avila-Reese}, {Badenes}, {Bailey}, {Balland}, {Barger},
  {Barrera-Ballesteros}, {Bartosz}, {Bastien}, {Bates}, {Baumgarten},
  {Bautista}, {Beaton}, {Beers}, {Belfiore}, {Bender}, {Bernardi}, {Bershady},
  {Beutler}, {Bird}, {Bizyaev}, {Blanc}, {Blanton}, {Blomqvist}, {Bolton},
  {Boquien}, {Borissova}, {Bovy}, {Bradna Diaz}, {Brandt}, {Brinkmann},
  {Brownstein}, {Bundy}, {Burgasser}, {Burtin}, {Busca}, {Ca{\~n}as},
  {Cano-D{\'\i}az}, {Cappellari}, {Carrera}, {Casey}, {Cervantes Sodi}, {Chen},
  {Cherinka}, {Chiappini}, {Choi}, {Chojnowski}, {Chuang}, {Chung}, {Clerc},
  {Cohen}, {Comerford}, {Comparat}, {Correa do Nascimento}, {da Costa},
  {Cousinou}, {Covey}, {Crane}, {Cruz-Gonzalez}, {Cunha}, {da Silva Ilha},
  {Damke}, {Darling}, {Davidson}, {Dawson}, {de Icaza Lizaola}, {de la
  Macorra}, {de la Torre}, {De Lee}, {de Sainte Agathe}, {Deconto Machado},
  {Dell'Agli}, {Delubac}, {Diamond-Stanic}, {Donor}, {Downes}, {Drory}, {du Mas
  des Bourboux}, {Duckworth}, {Dwelly}, {Dyer}, {Ebelke}, {Davis Eigenbrot},
  {Eisenstein}, {Elsworth}, {Emsellem}, {Eracleous}, {Erfanianfar},
  {Escoffier}, {Fan}, {Fern{\'a}ndez Alvar}, {Fernandez-Trincado}, {Fernand o
  Cirolini}, {Feuillet}, {Finoguenov}, {Fleming}, {Font-Ribera}, {Freischlad},
  {Frinchaboy}, {Fu}, {G{\'o}mez Maqueo Chew}, {Galbany}, {Garc{\'\i}a
  P{\'e}rez}, {Garcia-Dias}, {Garc{\'\i}a-Hern{\'a}ndez}, {Garma Oehmichen},
  {Gaulme}, {Gelfand }, {Gil-Mar{\'\i}n}, {Gillespie}, {Goddard}, {Gonz{\'a}lez
  Hern{\'a}ndez}, {Gonzalez-Perez}, {Grabowski}, {Green}, {Grier}, {Gueguen},
  {Guo}, {Guy}, {Hagen}, {Hall}, {Harding}, {Hasselquist}, {Hawley}, {Hayes},
  {Hearty}, {Hekker}, {Hernand ez}, {Hernandez Toledo}, {Hogg},
  {Holley-Bockelmann}, {Holtzman}, {Hou}, {Hsieh}, {Hunt}, {Hutchinson},
  {Hwang}, {Jimenez Angel}, {Johnson}, {Jones}, {J{\"o}nsson}, {Jullo}, {Khan},
  {Kinemuchi}, {Kirkby}, {Kirkpatrick}, {Kitaura}, {Knapp}, {Kneib},
  {Kollmeier}, {Lacerna}, {Lane}, {Lang}, {Law}, {Le Goff}, {Lee}, {Li}, {Li},
  {Lian}, {Liang}, {Lima}, {Lin}, {Long}, {Lucatello}, {Lundgren}, {Mackereth},
  {MacLeod}, {Mahadevan}, {Maia}, {Majewski}, {Manchado}, {Maraston},
  {Mariappan}, {Marques-Chaves}, {Masseron}, {Masters}, {McDermid}, {McGreer},
  {Melendez}, {Meneses-Goytia}, {Merloni}, {Merrifield}, {Meszaros}, {Meza},
  {Minchev}, {Minniti}, {Mueller}, {Muller-Sanchez}, {Muna}, {Mu{\~n}oz},
  {Myers}, {Nair}, {Nand ra}, {Ness}, {Newman}, {Nichol}, {Nidever},
  {Nitschelm}, {Noterdaeme}, {O'Connell}, {Oelkers}, {Oravetz}, {Oravetz},
  {Ort{\'\i}z}, {Osorio}, {Pace}, {Padilla}, {Palanque-Delabrouille},
  {Palicio}, {Pan}, {Pan}, {Parikh}, {P{\^a}ris}, {Park}, {Peirani},
  {Pellejero-Ibanez}, {Penny}, {Percival}, {Perez-Fournon}, {Petitjean},
  {Pieri}, {Pinsonneault}, {Pisani}, {Prada}, {Prakash}, {Queiroz}, {Raddick},
  {Raichoor}, {Barboza Rembold}, {Richstein}, {Riffel}, {Riffel}, {Rix},
  {Robin}, {Rodr{\'\i}guez Torres}, {Rom{\'a}n-Z{\'u}{\~n}iga}, {Ross},
  {Rossi}, {Ruan}, {Ruggeri}, {Ruiz}, {Salvato}, {S{\'a}nchez}, {S{\'a}nchez},
  {Sanchez Almeida}, {S{\'a}nchez-Gallego}, {Santana Rojas}, {Santiago},
  {Schiavon}, {Schimoia}, {Schlafly}, {Schlegel}, {Schneider}, {Schuster},
  {Schwope}, {Seo}, {Serenelli}, {Shen}, {Shen}, {Shetrone}, {Shull}, {Silva
  Aguirre}, {Simon}, {Skrutskie}, {Slosar}, {Smethurst}, {Smith}, {Sobeck},
  {Somers}, {Souter}, {Souto}, {Spindler}, {Stark}, {Stassun}, {Steinmetz},
  {Stello}, {Storchi-Bergmann}, {Streblyanska}, {Stringfellow}, {Su{\'a}rez},
  {Sun}, {Szigeti}, {Taghizadeh-Popp}, {Talbot}, {Tang}, {Tao}, {Tayar},
  {Tembe}, {Teske}, {Thakar}, {Thomas}, {Tissera}, {Tojeiro}, {Tremonti},
  {Troup}, {Urry}, {Valenzuela}, {van den Bosch}, {Vargas-Gonz{\'a}lez},
  {Vargas-Maga{\~n}a}, {Vazquez}, {Villanova}, {Vogt}, {Wake}, {Wang},
  {Weaver}, {Weijmans}, {Weinberg}, {Westfall}, {Whelan}, {Wilcots}, {Wild},
  {Williams}, {Wilson}, {Wood-Vasey}, {Wylezalek}, {Xiao}, {Yan}, {Yang},
  {Ybarra}, {Y{\`e}che}, {Zakamska}, {Zamora}, {Zarrouk}, {Zasowski}, {Zhang},
  {Zhao}, {Zhao}, {Zheng}, {Zheng}, {Zhou}, {Zhu}, {Zinn}, \&
  {Zou}}]{Abolfathi_2018}
{Abolfathi}, B., {Aguado}, D.~S., {Aguilar}, G., {et~al.} 2018, \apjs, 235, 42

\bibitem[{{Adami} {et~al.}(1998){Adami}, {Mazure}, {Katgert}, \&
  {Biviano}}]{Adami_1998}
{Adami}, C., {Mazure}, A., {Katgert}, P., \& {Biviano}, A. 1998, \aap, 336, 63

\bibitem[{{Adelman-McCarthy} {et~al.}(2008){Adelman-McCarthy}, {Ag{\"u}eros},
  {Allam}, {Allende Prieto}, {Anderson}, {Anderson}, {Annis}, {Bahcall},
  {Bailer-Jones}, {Baldry}, {Barentine}, {Bassett}, {Becker}, {Beers}, {Bell},
  {Berlind}, {Bernardi}, {Blanton}, {Bochanski}, {Boroski}, {Brinchmann},
  {Brinkmann}, {Brunner}, {Budav{\'a}ri}, {Carliles}, {Carr}, {Castander},
  {Cinabro}, {Cool}, {Covey}, {Csabai}, {Cunha}, {Davenport}, {Dilday}, {Doi},
  {Eisenstein}, {Evans}, {Fan}, {Finkbeiner}, {Friedman}, {Frieman},
  {Fukugita}, {G{\"a}nsicke}, {Gates}, {Gillespie}, {Glazebrook}, {Gray},
  {Grebel}, {Gunn}, {Gurbani}, {Hall}, {Harding}, {Harvanek}, {Hawley},
  {Hayes}, {Heckman}, {Hendry}, {Hindsley}, {Hirata}, {Hogan}, {Hogg}, {Hyde},
  {Ichikawa}, {Ivezi{\'c}}, {Jester}, {Johnson}, {Jorgensen}, {Juri{\'c}},
  {Kent}, {Kessler}, {Kleinman}, {Knapp}, {Kron}, {Krzesinski}, {Kuropatkin},
  {Lamb}, {Lampeitl}, {Lebedeva}, {Lee}, {French Leger}, {L{\'e}pine}, {Lima},
  {Lin}, {Long}, {Loomis}, {Loveday}, {Lupton}, {Malanushenko}, {Malanushenko},
  {Mandelbaum}, {Margon}, {Marriner}, {Mart{\'{\i}}nez-Delgado}, {Matsubara},
  {McGehee}, {McKay}, {Meiksin}, {Morrison}, {Munn}, {Nakajima}, {Neilsen},
  {Newberg}, {Nichol}, {Nicinski}, {Nieto-Santisteban}, {Nitta}, {Okamura},
  {Owen}, {Oyaizu}, {Padmanabhan}, {Pan}, {Park}, {Peoples}, {Pier}, {Pope},
  {Purger}, {Raddick}, {Re Fiorentin}, {Richards}, {Richmond}, {Riess}, {Rix},
  {Rockosi}, {Sako}, {Schlegel}, {Schneider}, {Schreiber}, {Schwope}, {Seljak},
  {Sesar}, {Sheldon}, {Shimasaku}, {Sivarani}, {Allyn Smith}, {Snedden},
  {Steinmetz}, {Strauss}, {SubbaRao}, {Suto}, {Szalay}, {Szapudi}, {Szkody},
  {Tegmark}, {Thakar}, {Tremonti}, {Tucker}, {Uomoto}, {Vanden Berk},
  {Vandenberg}, {Vidrih}, {Vogeley}, {Voges}, {Vogt}, {Wadadekar}, {Weinberg},
  {West}, {White}, {Wilhite}, {Yanny}, {Yocum}, {York}, {Zehavi}, \&
  {Zucker}}]{Adelman_2008}
{Adelman-McCarthy}, J.~K., {Ag{\"u}eros}, M.~A., {Allam}, S.~S., {et~al.} 2008,
  \apjs, 175, 297

\bibitem[{{Adelman-McCarthy} {et~al.}(2007){Adelman-McCarthy}, {Ag{\"u}eros},
  {Allam}, {Anderson}, {Anderson}, {Annis}, {Bahcall}, {Bailer-Jones},
  {Baldry}, {Barentine}, {Beers}, {Belokurov}, {Berlind}, {Bernardi},
  {Blanton}, {Bochanski}, {Boroski}, {Bramich}, {Brewington}, {Brinchmann},
  {Brinkmann}, {Brunner}, {Budav{\'a}ri}, {Carey}, {Carliles}, {Carr},
  {Castander}, {Connolly}, {Cool}, {Cunha}, {Csabai}, {Dalcanton}, {Doi},
  {Eisenstein}, {Evans}, {Evans}, {Fan}, {Finkbeiner}, {Friedman}, {Frieman},
  {Fukugita}, {Gillespie}, {Gilmore}, {Glazebrook}, {Gray}, {Grebel}, {Gunn},
  {de Haas}, {Hall}, {Harvanek}, {Hawley}, {Hayes}, {Heckman}, {Hendry},
  {Hennessy}, {Hindsley}, {Hirata}, {Hogan}, {Hogg}, {Holtzman}, {Ichikawa},
  {Ichikawa}, {Ivezi{\'c}}, {Jester}, {Johnston}, {Jorgensen}, {Juri{\'c}},
  {Kauffmann}, {Kent}, {Kleinman}, {Knapp}, {Kniazev}, {Kron}, {Krzesinski},
  {Kuropatkin}, {Lamb}, {Lampeitl}, {Lee}, {Leger}, {Lima}, {Lin}, {Long},
  {Loveday}, {Lupton}, {Mandelbaum}, {Margon}, {Mart{\'{\i}}nez-Delgado},
  {Matsubara}, {McGehee}, {McKay}, {Meiksin}, {Munn}, {Nakajima}, {Nash},
  {Neilsen}, {Newberg}, {Nichol}, {Nieto-Santisteban}, {Nitta}, {Oyaizu},
  {Okamura}, {Ostriker}, {Padmanabhan}, {Park}, {Peoples}, {Pier}, {Pope},
  {Pourbaix}, {Quinn}, {Raddick}, {Re Fiorentin}, {Richards}, {Richmond},
  {Rix}, {Rockosi}, {Schlegel}, {Schneider}, {Scranton}, {Seljak}, {Sheldon},
  {Shimasaku}, {Silvestri}, {Smith}, {Smol{\v c}i{\'c}}, {Snedden}, {Stebbins},
  {Stoughton}, {Strauss}, {SubbaRao}, {Suto}, {Szalay}, {Szapudi}, {Szkody},
  {Tegmark}, {Thakar}, {Tremonti}, {Tucker}, {Uomoto}, {Vanden Berk},
  {Vandenberg}, {Vidrih}, {Vogeley}, {Voges}, {Vogt}, {Weinberg}, {West},
  {White}, {Wilhite}, {Yanny}, {Yocum}, {York}, {Zehavi}, {Zibetti}, \&
  {Zucker}}]{Adelman_2007}
{Adelman-McCarthy}, J.~K., {Ag{\"u}eros}, M.~A., {Allam}, S.~S., {et~al.} 2007,
  \apjs, 172, 634

\bibitem[{Ahn \& Fessler(2003)}]{Ahn_2003}
Ahn, S. \& Fessler, J.~A. 2003, Standard Errors of Mean, Variance, and Standard
  Deviation Estimators

\bibitem[{{Alam} {et~al.}(2015){Alam}, {Albareti}, {Allende Prieto}, {Anders},
  {Anderson}, {Anderton}, {Andrews}, {Armengaud}, {Aubourg}, {Bailey}, \&
  et~al.}]{Alam_2015}
{Alam}, S., {Albareti}, F.~D., {Allende Prieto}, C., {et~al.} 2015, \apjs, 219,
  12

\bibitem[{{Andreon}(1994)}]{Andreon_1994}
{Andreon}, S. 1994, \aap, 284, 801

\bibitem[{{Andreon} \& {Pell{\'o}}(2000)}]{Andreon_2000}
{Andreon}, S. \& {Pell{\'o}}, R. 2000, \aap, 353, 479

\bibitem[{{Assef} {et~al.}(2010){Assef}, {Kochanek}, {Brodwin}, {Cool},
  {Forman}, {Gonzalez}, {Hickox}, {Jones}, {Le Floc'h}, {Moustakas}, {Murray},
  \& {Stern}}]{Assef_2010}
{Assef}, R.~J., {Kochanek}, C.~S., {Brodwin}, M., {et~al.} 2010, \apj, 713, 970

\bibitem[{{Assef} {et~al.}(2013){Assef}, {Stern}, {Kochanek}, {Blain},
  {Brodwin}, {Brown}, {Donoso}, {Eisenhardt}, {Jannuzi}, {Jarrett}, {Stanford},
  {Tsai}, {Wu}, \& {Yan}}]{Assef_2013}
{Assef}, R.~J., {Stern}, D., {Kochanek}, C.~S., {et~al.} 2013, \apj, 772, 26

\bibitem[{{Bahcall} \& {Tremaine}(1981)}]{Bahcall_1981}
{Bahcall}, J.~N. \& {Tremaine}, S. 1981, \apj, 244, 805

\bibitem[{{Baldry} {et~al.}(2006){Baldry}, {Balogh}, {Bower}, {Glazebrook},
  {Nichol}, {Bamford}, \& {Budavari}}]{Baldry_2006}
{Baldry}, I.~K., {Balogh}, M.~L., {Bower}, R.~G., {et~al.} 2006, \mnras, 373,
  469

\bibitem[{{Baldry} {et~al.}(2004){Baldry}, {Glazebrook}, {Brinkmann},
  {Ivezi{\'c}}, {Lupton}, {Nichol}, \& {Szalay}}]{Baldry_2004}
{Baldry}, I.~K., {Glazebrook}, K., {Brinkmann}, J., {et~al.} 2004, \apj, 600,
  681

\bibitem[{{Baldwin} {et~al.}(1981){Baldwin}, {Phillips}, \&
  {Terlevich}}]{Baldwin_1981}
{Baldwin}, J.~A., {Phillips}, M.~M., \& {Terlevich}, R. 1981, \pasp, 93, 5

\bibitem[{{Balogh} {et~al.}(2004){Balogh}, {Baldry}, {Nichol}, {Miller},
  {Bower}, \& {Glazebrook}}]{Balogh_2004}
{Balogh}, M.~L., {Baldry}, I.~K., {Nichol}, R., {et~al.} 2004, \apjl, 615, L101

\bibitem[{{Bartelmann}(1996)}]{Bartelmann_1996}
{Bartelmann}, M. 1996, \aap, 313, 697

\bibitem[{{Barton} {et~al.}(2000){Barton}, {Geller}, \& {Kenyon}}]{Barton_2000}
{Barton}, E.~J., {Geller}, M.~J., \& {Kenyon}, S.~J. 2000, \apj, 530, 660

\bibitem[{{Batuski} \& {Burns}(1985)}]{Batuski_1985}
{Batuski}, D.~J. \& {Burns}, J.~O. 1985, \apj, 299, 5

\bibitem[{{Behroozi} {et~al.}(2014){Behroozi}, {Wechsler}, {Lu}, {Hahn},
  {Busha}, {Klypin}, \& {Primack}}]{Behroozi_2014}
{Behroozi}, P.~S., {Wechsler}, R.~H., {Lu}, Y., {et~al.} 2014, \apj, 787, 156

\bibitem[{{Bertin} \& {Arnouts}(1996)}]{Bertin_1996}
{Bertin}, E. \& {Arnouts}, S. 1996, \aaps, 117, 393

\bibitem[{{Bialas} {et~al.}(2015){Bialas}, {Lisker}, {Olczak}, {Spurzem}, \&
  {Kotulla}}]{Bialas_2015}
{Bialas}, D., {Lisker}, T., {Olczak}, C., {Spurzem}, R., \& {Kotulla}, R. 2015,
  \aap, 576, A103

\bibitem[{{Biviano} {et~al.}(2011){Biviano}, {Fadda}, {Durret}, {Edwards}, \&
  {Marleau}}]{Biviano_2011}
{Biviano}, A., {Fadda}, D., {Durret}, F., {Edwards}, L.~O.~V., \& {Marleau}, F.
  2011, \aap, 532, A77

\bibitem[{{Blanton} {et~al.}(2003){Blanton}, {Hogg}, {Bahcall}, {Baldry},
  {Brinkmann}, {Csabai}, {Eisenstein}, {Fukugita}, {Gunn}, {Ivezi{\'c}},
  {Lamb}, {Lupton}, {Loveday}, {Munn}, {Nichol}, {Okamura}, {Schlegel},
  {Shimasaku}, {Strauss}, {Vogeley}, \& {Weinberg}}]{Blanton_2003}
{Blanton}, M.~R., {Hogg}, D.~W., {Bahcall}, N.~A., {et~al.} 2003, \apj, 594,
  186

\bibitem[{{Boehringer} {et~al.}(1993){Boehringer}, {Voges}, {Fabian}, {Edge},
  \& {Neumann}}]{Boehringer_1993}
{Boehringer}, H., {Voges}, W., {Fabian}, A.~C., {Edge}, A.~C., \& {Neumann},
  D.~M. 1993, \mnras, 264, L25

\bibitem[{{Boroson}(1990)}]{Boroson_1990}
{Boroson}, T.~A. 1990, \apj, 360, 465

\bibitem[{{Boselli} \& {Gavazzi}(2006)}]{Boselli_2006}
{Boselli}, A. \& {Gavazzi}, G. 2006, \pasp, 118, 517

\bibitem[{{Braine} {et~al.}(1995){Braine}, {Kruegel}, {Sievers}, \&
  {Wielebinski}}]{Braine_1995}
{Braine}, J., {Kruegel}, E., {Sievers}, A., \& {Wielebinski}, R. 1995, \aap,
  295, L55

\bibitem[{{Branduardi-Raymont} {et~al.}(1981){Branduardi-Raymont}, {Fabricant},
  {Feigelson}, {Gorenstein}, {Grindlay}, {Soltan}, \&
  {Zamorani}}]{Branduardi_1981}
{Branduardi-Raymont}, G., {Fabricant}, D., {Feigelson}, E., {et~al.} 1981,
  \apj, 248, 55

\bibitem[{{Brentjens}(2011)}]{Brentjens_2011}
{Brentjens}, M.~A. 2011, \aap, 526, A9

\bibitem[{{Brosch}(1985)}]{Brosch_1985}
{Brosch}, N. 1985, \aap, 153, 199

\bibitem[{{Brown} {et~al.}(2014){Brown}, {Moustakas}, {Smith}, {da Cunha},
  {Jarrett}, {Imanishi}, {Armus}, {Brandl}, \& {Peek}}]{Brown_2014}
{Brown}, M.~J.~I., {Moustakas}, J., {Smith}, J.-D.~T., {et~al.} 2014, \apjs,
  212, 18

\bibitem[{{Brunzendorf} \& {Meusinger}(1999)}]{Brunzendorf_1999}
{Brunzendorf}, J. \& {Meusinger}, H. 1999, \aaps, 139, 141

\bibitem[{{Bucknell} {et~al.}(1979){Bucknell}, {Godwin}, \&
  {Peach}}]{Bucknell_1979}
{Bucknell}, M.~J., {Godwin}, J.~G., \& {Peach}, J.~V. 1979, \mnras, 188, 579

\bibitem[{{Busch} {et~al.}(2014){Busch}, {Zuther}, {Valencia-S.}, {Moser},
  {Fischer}, {Eckart}, {Scharw{\"a}chter}, {Gadotti}, \&
  {Wisotzki}}]{Busch_2014}
{Busch}, G., {Zuther}, J., {Valencia-S.}, M., {et~al.} 2014, \aap, 561, A140

\bibitem[{{Buta} \& {Combes}(1996)}]{Buta_1996}
{Buta}, R. \& {Combes}, F. 1996, \fcp, 17, 95

\bibitem[{{Carlberg} {et~al.}(1997){Carlberg}, {Yee}, {Ellingson}, {Morris},
  {Abraham}, {Gravel}, {Pritchet}, {Smecker-Hane}, {Hartwick}, {Hesser},
  {Hutchings}, \& {Oke}}]{Carlberg_1997}
{Carlberg}, R.~G., {Yee}, H.~K.~C., {Ellingson}, E., {et~al.} 1997, \apjl, 476,
  L7

\bibitem[{{Chang} {et~al.}(2015){Chang}, {van der Wel}, {da Cunha}, \&
  {Rix}}]{Chang_2015}
{Chang}, Y.-Y., {van der Wel}, A., {da Cunha}, E., \& {Rix}, H.-W. 2015, \apjs,
  219, 8

\bibitem[{{Chen} {et~al.}(2017){Chen}, {Ho}, {Mandelbaum}, {Bahcall},
  {Brownstein}, {Freeman}, {Genovese}, {Schneider}, \& {Wasserman}}]{Chen_2017}
{Chen}, Y.-C., {Ho}, S., {Mandelbaum}, R., {et~al.} 2017, \mnras, 466, 1880

\bibitem[{{Ching} {et~al.}(2017){Ching}, {Croom}, {Sadler}, {Robotham},
  {Brough}, {Baldry}, {Bland-Hawthorn}, {Colless}, {Driver}, {Holwerda},
  {Hopkins}, {Jarvis}, {Johnston}, {Kelvin}, {Liske}, {Loveday}, {Norberg},
  {Pracy}, {Steele}, {Thomas}, \& {Wang}}]{Ching_2017}
{Ching}, J.~H.~Y., {Croom}, S.~M., {Sadler}, E.~M., {et~al.} 2017, \mnras, 469,
  4584

\bibitem[{{Churazov} {et~al.}(2003){Churazov}, {Forman}, {Jones}, \&
  {B{\"o}hringer}}]{Churazov_2003}
{Churazov}, E., {Forman}, W., {Jones}, C., \& {B{\"o}hringer}, H. 2003, \apj,
  590, 225

\bibitem[{{Cid Fernandes} {et~al.}(2011){Cid Fernandes}, {Stasi{\'n}ska},
  {Mateus}, \& {Vale Asari}}]{CidFernandes_2011}
{Cid Fernandes}, R., {Stasi{\'n}ska}, G., {Mateus}, A., \& {Vale Asari}, N.
  2011, \mnras, 413, 1687

\bibitem[{{Cluver} {et~al.}(2017){Cluver}, {Jarrett}, {Dale}, {Smith},
  {August}, \& {Brown}}]{Cluver_2017}
{Cluver}, M.~E., {Jarrett}, T.~H., {Dale}, D.~A., {et~al.} 2017, \apj, 850, 68

\bibitem[{{Cole} {et~al.}(2001){Cole}, {Norberg}, {Baugh}, {Frenk},
  {Bland-Hawthorn}, {Bridges}, {Cannon}, {Colless}, {Collins}, {Couch},
  {Cross}, {Dalton}, {De Propris}, {Driver}, {Efstathiou}, {Ellis},
  {Glazebrook}, {Jackson}, {Lahav}, {Lewis}, {Lumsden}, {Maddox}, {Madgwick},
  {Peacock}, {Peterson}, {Sutherland}, \& {Taylor}}]{Cole_2001}
{Cole}, S., {Norberg}, P., {Baugh}, C.~M., {et~al.} 2001, \mnras, 326, 255

\bibitem[{{Condon} {et~al.}(1998){Condon}, {Cotton}, {Greisen}, {Yin},
  {Perley}, {Taylor}, \& {Broderick}}]{Condon_1998}
{Condon}, J.~J., {Cotton}, W.~D., {Greisen}, E.~W., {et~al.} 1998, \aj, 115,
  1693

\bibitem[{{Conselice} {et~al.}(2001){Conselice}, {Gallagher}, \&
  {Wyse}}]{Conselice_2001}
{Conselice}, C.~J., {Gallagher}, III, J.~S., \& {Wyse}, R.~F.~G. 2001, \aj,
  122, 2281

\bibitem[{{Conselice} {et~al.}(2002){Conselice}, {Gallagher}, \&
  {Wyse}}]{Conselice_2002}
{Conselice}, C.~J., {Gallagher}, III, J.~S., \& {Wyse}, R.~F.~G. 2002, \aj,
  123, 2246

\bibitem[{{Conselice} {et~al.}(2003){Conselice}, {Gallagher}, \&
  {Wyse}}]{Conselice_2003}
{Conselice}, C.~J., {Gallagher}, III, J.~S., \& {Wyse}, R.~F.~G. 2003, \aj,
  125, 66

\bibitem[{{Costa} {et~al.}(2018){Costa}, {Ribeiro}, \& {de
  Carvalho}}]{Costa_2018}
{Costa}, A.~P., {Ribeiro}, A.~L.~B., \& {de Carvalho}, R.~R. 2018, \mnras, 473,
  L31

\bibitem[{{Cox}(2000)}]{Cox_2000}
{Cox}, A.~N. 2000, {Allen's astrophysical quantities}

\bibitem[{{Crone Odekon} {et~al.}(2018){Crone Odekon}, {Hallenbeck}, {Haynes},
  {Koopmann}, {Phi}, \& {Wolfe}}]{Odekon_2018}
{Crone Odekon}, M., {Hallenbeck}, G., {Haynes}, M.~P., {et~al.} 2018, \apj,
  852, 142

\bibitem[{{Crook} {et~al.}(2007){Crook}, {Huchra}, {Martimbeau}, {Masters},
  {Jarrett}, \& {Macri}}]{Crook_2007}
{Crook}, A.~C., {Huchra}, J.~P., {Martimbeau}, N., {et~al.} 2007, \apj, 655,
  790

\bibitem[{{Davies} {et~al.}(2019){Davies}, {Nersesian}, {Baes}, {Bianchi},
  {Casasola}, {Cassar{\`a}}, {Clark}, {De Looze}, {De Vis}, \&
  {Evans}}]{Davies_2019}
{Davies}, J.~I., {Nersesian}, A., {Baes}, M., {et~al.} 2019, \aap, 626, A63

\bibitem[{{de Bruyn} \& {Brentjens}(2005)}]{deBruyn_2005}
{de Bruyn}, A.~G. \& {Brentjens}, M.~A. 2005, \aap, 441, 931

\bibitem[{{De Propris}(2017)}]{DePropris_2017}
{De Propris}, R. 2017, \mnras, 465, 4035

\bibitem[{{De Propris} \& {Pritchet}(1998)}]{dePropris_1998}
{De Propris}, R. \& {Pritchet}, C.~J. 1998, \aj, 116, 1118

\bibitem[{{de Rijcke} {et~al.}(2009){de Rijcke}, {Penny}, {Conselice},
  {Valcke}, \& {Held}}]{Rijcke_2009}
{de Rijcke}, S., {Penny}, S.~J., {Conselice}, C.~J., {Valcke}, S., \& {Held},
  E.~V. 2009, \mnras, 393, 798

\bibitem[{{Di Matteo} {et~al.}(2007){Di Matteo}, {Combes}, {Melchior}, \&
  {Semelin}}]{DiMatteo_2007}
{Di Matteo}, P., {Combes}, F., {Melchior}, A.-L., \& {Semelin}, B. 2007, \aap,
  468, 61

\bibitem[{{Diaferio} \& {Geller}(1997)}]{Diaferio_1997}
{Diaferio}, A. \& {Geller}, M.~J. 1997, \apj, 481, 633

\bibitem[{{Dressler}(1980)}]{Dressler_1980}
{Dressler}, A. 1980, \apj, 236, 351

\bibitem[{{Dressler} \& {Shectman}(1988)}]{Dressler_1988}
{Dressler}, A. \& {Shectman}, S.~A. 1988, \aj, 95, 985

\bibitem[{{Duc} \& {Renaud}(2013)}]{Duc_2013}
{Duc}, P.-A. \& {Renaud}, F. 2013, in Lecture Notes in Physics, Berlin Springer
  Verlag, Vol. 861, Lecture Notes in Physics, Berlin Springer Verlag, ed.
  J.~{Souchay}, S.~{Mathis}, \& T.~{Tokieda}, 327

\bibitem[{{Eales} {et~al.}(2018){Eales}, {Baes}, {Bourne}, {Bremer}, {Brown},
  {Clark}, {Clements}, {de Vis}, {Driver}, \& {Dunne}}]{Eales_2018}
{Eales}, S.~A., {Baes}, M., {Bourne}, N., {et~al.} 2018, \mnras, 481, 1183

\bibitem[{{Ebeling} {et~al.}(1996){Ebeling}, {Voges}, {Bohringer}, {Edge},
  {Huchra}, \& {Briel}}]{Ebeling_1996}
{Ebeling}, H., {Voges}, W., {Bohringer}, H., {et~al.} 1996, \mnras, 281, 799

\bibitem[{{Edge} {et~al.}(1990){Edge}, {Stewart}, {Fabian}, \&
  {Arnaud}}]{Edge_1990}
{Edge}, A.~C., {Stewart}, G.~C., {Fabian}, A.~C., \& {Arnaud}, K.~A. 1990,
  \mnras, 245, 559

\bibitem[{{Ehlert} {et~al.}(2015){Ehlert}, {Allen}, {Brandt}, {Canning}, {Luo},
  {Mantz}, {Morris}, {von der Linden}, \& {Xue}}]{Ehlert_2015}
{Ehlert}, S., {Allen}, S.~W., {Brandt}, W.~N., {et~al.} 2015, \mnras, 446, 2709

\bibitem[{{Fabian}(1994)}]{Fabian_1994}
{Fabian}, A.~C. 1994, \araa, 32, 277

\bibitem[{{Fabian}(2012)}]{Fabian_2012}
{Fabian}, A.~C. 2012, \araa, 50, 455

\bibitem[{{Fabian} {et~al.}(2008){Fabian}, {Johnstone}, {Sanders}, {Conselice},
  {Crawford}, {Gallagher}, \& {Zweibel}}]{Fabian_2008}
{Fabian}, A.~C., {Johnstone}, R.~M., {Sanders}, J.~S., {et~al.} 2008, \nat,
  454, 968

\bibitem[{{Fabian} {et~al.}(2011{\natexlab{a}}){Fabian}, {Sanders}, {Allen},
  {Canning}, {Churazov}, {Crawford}, {Forman}, {Gabany}, {Hlavacek-Larrondo},
  \& {Johnstone}}]{Fabian_2011a}
{Fabian}, A.~C., {Sanders}, J.~S., {Allen}, S.~W., {et~al.} 2011{\natexlab{a}},
  \mnras, 418, 2154

\bibitem[{{Fabian} {et~al.}(2011{\natexlab{b}}){Fabian}, {Sanders}, {Williams},
  {Lazarian}, {Ferland}, \& {Johnstone}}]{Fabian_2011b}
{Fabian}, A.~C., {Sanders}, J.~S., {Williams}, R.~J.~R., {et~al.}
  2011{\natexlab{b}}, \mnras, 417, 172

\bibitem[{{Fadda} {et~al.}(2008){Fadda}, {Biviano}, {Marleau},
  {Storrie-Lombardi}, \& {Durret}}]{Fadda_2008}
{Fadda}, D., {Biviano}, A., {Marleau}, F.~R., {Storrie-Lombardi}, L.~J., \&
  {Durret}, F. 2008, \apjl, 672, L9

\bibitem[{{Finkelman} \& {Brosch}(2011)}]{Finkelman_2011}
{Finkelman}, I. \& {Brosch}, N. 2011, \mnras, 413, 2621

\bibitem[{{Finkelman} {et~al.}(2011){Finkelman}, {Moiseev}, {Brosch}, \&
  {Katkov}}]{Finkelman_et_al_2011}
{Finkelman}, I., {Moiseev}, A., {Brosch}, N., \& {Katkov}, I. 2011, \mnras,
  418, 1834

\bibitem[{{Frith} {et~al.}(2003){Frith}, {Busswell}, {Fong}, {Metcalfe}, \&
  {Shanks}}]{Frith_2003}
{Frith}, W.~J., {Busswell}, G.~S., {Fong}, R., {Metcalfe}, N., \& {Shanks}, T.
  2003, \mnras, 345, 1049

\bibitem[{{Froebrich} \& {Meusinger}(2000)}]{Froebrich_2000}
{Froebrich}, D. \& {Meusinger}, H. 2000, \aaps, 145, 229

\bibitem[{{Gendron-Marsolais} {et~al.}(2018){Gendron-Marsolais},
  {Hlavacek-Larrondo}, {Martin}, {Drissen}, {McDonald}, {Fabian}, {Edge},
  {Hamer}, {McNamara}, \& {Morrison}}]{Gendron_Marsolais_2018}
{Gendron-Marsolais}, M., {Hlavacek-Larrondo}, J., {Martin}, T.~B., {et~al.}
  2018, \mnras, 479, L28

\bibitem[{{Gendron-Marsolais} {et~al.}(2017){Gendron-Marsolais},
  {Hlavacek-Larrondo}, {van Weeren}, {Clarke}, {Fabian}, {Intema}, {Taylor},
  {Blundell}, \& {Sanders}}]{Gendron_Marsolais_2017}
{Gendron-Marsolais}, M., {Hlavacek-Larrondo}, J., {van Weeren}, R.~J., {et~al.}
  2017, \mnras, 469, 3872

\bibitem[{{Gisler} \& {Miley}(1979)}]{Gisler_1979}
{Gisler}, G.~R. \& {Miley}, G.~K. 1979, \aap, 76, 109

\bibitem[{{Golovich} {et~al.}(2018){Golovich}, {Dawson}, {Wittman}, {van
  Weeren}, {Andrade-Santos}, {Jee}, {Benson}, {de Gasperin}, {Venturi},
  {Bonafede}, {Sobral}, {Ogrean}, {Lemaux}, {Brada{\v c}}, {Br{\"u}ggen}, \&
  {Peter}}]{Golovich_2018}
{Golovich}, N., {Dawson}, W.~A., {Wittman}, D.~M., {et~al.} 2018, arXiv
  e-prints [\eprint[arXiv]{1806.10619}]

\bibitem[{{Graham} \& {Scott}(2013)}]{Graham_2013}
{Graham}, A.~W. \& {Scott}, N. 2013, \apj, 764, 151

\bibitem[{{Gunn} \& {Gott}(1972)}]{Gunn_1972}
{Gunn}, J.~E. \& {Gott}, J.~Richard, I. 1972, \apj, 176, 1

\bibitem[{{Hamer} {et~al.}(2016){Hamer}, {Edge}, {Swinbank}, {Wilman},
  {Combes}, {Salom{\'e}}, {Fabian}, {Crawford}, {Russell}, {Hlavacek-Larrondo},
  {McNamara}, \& {Bremer}}]{Hamer_2016}
{Hamer}, S.~L., {Edge}, A.~C., {Swinbank}, A.~M., {et~al.} 2016, \mnras, 460,
  1758

\bibitem[{{Harrison} {et~al.}(2018){Harrison}, {Costa}, {Tadhunter},
  {Fl{\"u}tsch}, {Kakkad}, {Perna}, \& {Vietri}}]{Harrison_2018}
{Harrison}, C.~M., {Costa}, T., {Tadhunter}, C.~N., {et~al.} 2018, Nature
  Astronomy, 2, 198

\bibitem[{{Harrison}(1974)}]{Harrison_1974}
{Harrison}, E.~R. 1974, \apjl, 191, L51

\bibitem[{{Hart} {et~al.}(2009){Hart}, {Stocke}, \& {Hallman}}]{Hart_2009}
{Hart}, Q.~N., {Stocke}, J.~T., \& {Hallman}, E.~J. 2009, \apj, 705, 854

\bibitem[{{Hatch} {et~al.}(2006){Hatch}, {Crawford}, {Johnstone}, \&
  {Fabian}}]{Hatch_2006}
{Hatch}, N.~A., {Crawford}, C.~S., {Johnstone}, R.~M., \& {Fabian}, A.~C. 2006,
  \mnras, 367, 433

\bibitem[{{Heckman} \& {Best}(2014)}]{Heckman_2014}
{Heckman}, T.~M. \& {Best}, P.~N. 2014, \araa, 52, 589

\bibitem[{{Heisler} {et~al.}(1985){Heisler}, {Tremaine}, \&
  {Bahcall}}]{Heisler_1985}
{Heisler}, J., {Tremaine}, S., \& {Bahcall}, J.~N. 1985, \apj, 298, 8

\bibitem[{{Helou} {et~al.}(1985){Helou}, {Soifer}, \&
  {Rowan-Robinson}}]{Helou_1985}
{Helou}, G., {Soifer}, B.~T., \& {Rowan-Robinson}, M. 1985, \apjl, 298, L7

\bibitem[{{Hitomi Collaboration} {et~al.}(2018{\natexlab{a}}){Hitomi
  Collaboration}, {Aharonian}, {Akamatsu}, {Akimoto}, {Allen}, {Angelini},
  {Audard}, {Awaki}, {Axelsson}, \& {Bamba}}]{Hitomi_2018a}
{Hitomi Collaboration}, {Aharonian}, F., {Akamatsu}, H., {et~al.}
  2018{\natexlab{a}}, \pasj, 70, 9

\bibitem[{{Hitomi Collaboration} {et~al.}(2018{\natexlab{b}}){Hitomi
  Collaboration}, {Aharonian}, {Akamatsu}, {Akimoto}, {Allen}, {Angelini},
  {Audard}, {Awaki}, {Axelsson}, {Bamba}, {Bautz}, {Blandford}, {Brenneman},
  {Brown}, {Bulbul}, {Cackett}, {Chernyakova}, {Chiao}, {Coppi}, {Costantini},
  {de Plaa}, {de Vries}, {den Herder}, {Done}, {Dotani}, {Ebisawa}, {Eckart},
  {Enoto}, {Ezoe}, {Fabian}, {Ferrigno}, {Foster}, {Fujimoto}, {Fukazawa},
  {Furuzawa}, {Galeazzi}, {Gallo}, {Gandhi}, {Giustini}, {Goldwurm}, {Gu},
  {Guainazzi}, {Haba}, {Hagino}, {Hamaguchi}, {Harrus}, {Hatsukade}, {Hayashi},
  {Hayashi}, {Hayashida}, {Hiraga}, {Hornschemeier}, {Hoshino}, {Hughes},
  {Ichinohe}, {Iizuka}, {Inoue}, {Inoue}, {Ishida}, {Ishikawa}, {Ishisaki},
  {Iwai}, {Kaastra}, {Kallman}, {Kamae}, {Kataoka}, {Katsuda}, {Kawai},
  {Kelley}, {Kilbourne}, {Kitaguchi}, {Kitamoto}, {Kitayama}, {Kohmura},
  {Kokubun}, {Koyama}, {Koyama}, {Kretschmar}, {Krimm}, {Kubota}, {Kunieda},
  {Laurent}, {Lee}, {Leutenegger}, {Limousin}, {Loewenstein}, {Long}, {Lumb},
  {Madejski}, {Maeda}, {Maier}, {Makishima}, {Markevitch}, {Matsumoto},
  {Matsushita}, {McCammon}, {McNamara}, {Mehdipour}, {Miller}, {Miller},
  {Mineshige}, {Mitsuda}, {Mitsuishi}, {Miyazawa}, {Mizuno}, {Mori}, {Mori},
  {Mukai}, {Murakami}, {Mushotzky}, {Nakagawa}, {Nakajima}, {Nakamori},
  {Nakashima}, {Nakazawa}, {Nobukawa}, {Nobukawa}, {Noda}, {Odaka}, {Ohashi},
  {Ohno}, {Okajima}, {Ota}, {Ozaki}, {Paerels}, {Paltani}, {Petre}, {Pinto},
  {Porter}, {Pottschmidt}, {Reynolds}, {Safi-Harb}, {Saito}, {Sakai}, {Sasaki},
  {Sato}, {Sato}, {Sato}, {Sawada}, {Schartel}, {Serlemitsos}, {Seta},
  {Shidatsu}, {Simionescu}, {Smith}, {Soong}, {Stawarz}, {Sugawara}, {Sugita},
  {Szymkowiak}, {Tajima}, {Takahashi}, {Takahashi}, {Takeda}, {Takei},
  {Tamagawa}, {Tamura}, {Tanaka}, {Tanaka}, {Tanaka}, {Tashiro}, {Tawara},
  {Terada}, {Terashima}, {Tombesi}, {Tomida}, {Tsuboi}, {Tsujimoto}, {Tsunemi},
  {Tsuru}, {Uchida}, {Uchiyama}, {Uchiyama}, {Ueda}, {Ueda}, {Uno}, {Urry},
  {Ursino}, {Watanabe}, {Werner}, {Wilkins}, {Williams}, {Yamada}, {Yamaguchi},
  {Yamaoka}, {Yamasaki}, {Yamauchi}, {Yamauchi}, {Yaqoob}, {Yatsu}, {Yonetoku},
  {Zhuravleva}, {Zoghbi}, \& {Kawamuro}}]{Hitomi_2018b}
{Hitomi Collaboration}, {Aharonian}, F., {Akamatsu}, H., {et~al.}
  2018{\natexlab{b}}, \pasj, 70, 13

\bibitem[{{Hoag}(1950)}]{Hoag_1950}
{Hoag}, A.~A. 1950, \aj, 55, 170

\bibitem[{{Holanda}(2018)}]{Holanda_2018}
{Holanda}, R.~F.~L. 2018, Astroparticle Physics, 99, 1

\bibitem[{{Holincheck} {et~al.}(2016){Holincheck}, {Wallin}, {Borne},
  {Fortson}, {Lintott}, {Smith}, {Bamford}, {Keel}, \&
  {Parrish}}]{Holincheck_2016}
{Holincheck}, A.~J., {Wallin}, J.~F., {Borne}, K., {et~al.} 2016, \mnras, 459,
  720

\bibitem[{{Hopkins} {et~al.}(2008){Hopkins}, {Cox}, {Kere{\v s}}, \&
  {Hernquist}}]{Hopkins_2008}
{Hopkins}, P.~F., {Cox}, T.~J., {Kere{\v s}}, D., \& {Hernquist}, L. 2008,
  \apjs, 175, 390

\bibitem[{{Horne}(1986)}]{Horne_1986}
{Horne}, K. 1986, \pasp, 98, 609

\bibitem[{{Hou} {et~al.}(2009){Hou}, {Parker}, {Harris}, \&
  {Wilman}}]{Hou_2009}
{Hou}, A., {Parker}, L.~C., {Harris}, W.~E., \& {Wilman}, D.~J. 2009, \apj,
  702, 1199

\bibitem[{{Hoyle} {et~al.}(2005){Hoyle}, {Rojas}, {Vogeley}, \&
  {Brinkmann}}]{Hoyle_2005}
{Hoyle}, F., {Rojas}, R.~R., {Vogeley}, M.~S., \& {Brinkmann}, J. 2005, \apj,
  620, 618

\bibitem[{{Huchra} {et~al.}(1999){Huchra}, {Vogeley}, \&
  {Geller}}]{Huchra_1999}
{Huchra}, J.~P., {Vogeley}, M.~S., \& {Geller}, M.~J. 1999, \apjs, 121, 287

\bibitem[{{Hulsen} {et~al.}(2008){Hulsen}, {de Vlieg}, \&
  {Alkema}}]{Hulsen_2008}
{Hulsen}, T., {de Vlieg}, J., \& {Alkema}, W. 2008, BMC Genomics, 9, 488

\bibitem[{{Ichinohe} {et~al.}(2019){Ichinohe}, {Simionescu}, {Werner},
  {Fabian}, \& {Takahashi}}]{Ichinohe_2019}
{Ichinohe}, Y., {Simionescu}, A., {Werner}, N., {Fabian}, A.~C., \&
  {Takahashi}, T. 2019, \mnras, 483, 1744

\bibitem[{{Intema} {et~al.}(2017){Intema}, {Jagannathan}, {Mooley}, \&
  {Frail}}]{Intema_2017}
{Intema}, H.~T., {Jagannathan}, P., {Mooley}, K.~P., \& {Frail}, D.~A. 2017,
  \aap, 598, A78

\bibitem[{{Jackson} {et~al.}(2007){Jackson}, {Battye}, {Browne}, {Joshi},
  {Muxlow}, \& {Wilkinson}}]{Jackson_2007}
{Jackson}, N., {Battye}, R.~A., {Browne}, I.~W.~A., {et~al.} 2007, \mnras, 376,
  371

\bibitem[{{Jarrett} {et~al.}(2000){Jarrett}, {Chester}, {Cutri}, {Schneider},
  {Skrutskie}, \& {Huchra}}]{Jarrett_2000}
{Jarrett}, T.~H., {Chester}, T., {Cutri}, R., {et~al.} 2000, \aj, 119, 2498

\bibitem[{{Jarrett} {et~al.}(2017){Jarrett}, {Cluver}, {Magoulas}, {Bilicki},
  {Alpaslan}, {Bland-Hawthorn}, {Brough}, {Brown}, {Croom}, {Driver},
  {Holwerda}, {Hopkins}, {Loveday}, {Norberg}, {Peacock}, {Popescu}, {Sadler},
  {Taylor}, {Tuffs}, \& {Wang}}]{Jarrett_2017}
{Jarrett}, T.~H., {Cluver}, M.~E., {Magoulas}, C., {et~al.} 2017, \apj, 836,
  182

\bibitem[{{Jarrett} {et~al.}(2011){Jarrett}, {Cohen}, {Masci}, {Wright},
  {Stern}, {Benford}, {Blain}, {Carey}, {Cutri}, {Eisenhardt}, {Lonsdale},
  {Mainzer}, {Marsh}, {Padgett}, {Petty}, {Ressler}, {Skrutskie}, {Stanford},
  {Surace}, {Tsai}, {Wheelock}, \& {Yan}}]{Jarrett_2011}
{Jarrett}, T.~H., {Cohen}, M., {Masci}, F., {et~al.} 2011, \apj, 735, 112

\bibitem[{{Kaviraj}(2014)}]{Kaviraj_2014}
{Kaviraj}, S. 2014, \mnras, 440, 2944

\bibitem[{{Kelvin} {et~al.}(2012){Kelvin}, {Driver}, {Robotham}, {Hill},
  {Alpaslan}, {Baldry}, {Bamford}, {Bland-Hawthorn}, {Brough}, {Graham},
  {H{\"a}ussler}, {Hopkins}, {Liske}, {Loveday}, {Norberg}, {Phillipps},
  {Popescu}, {Prescott}, {Taylor}, \& {Tuffs}}]{Kelvin_2012}
{Kelvin}, L.~S., {Driver}, S.~P., {Robotham}, A.~S.~G., {et~al.} 2012, \mnras,
  421, 1007

\bibitem[{{Kelvin} {et~al.}(2014){Kelvin}, {Driver}, {Robotham}, {Taylor},
  {Graham}, {Alpaslan}, {Baldry}, {Bamford}, {Bauer}, {Bland-Hawthorn},
  {Brown}, {Colless}, {Conselice}, {Holwerda}, {Hopkins}, {Lara-L{\'o}pez},
  {Liske}, {L{\'o}pez-S{\'a}nchez}, {Loveday}, {Norberg}, {Phillipps},
  {Popescu}, {Prescott}, {Sansom}, \& {Tuffs}}]{Kelvin_2014}
{Kelvin}, L.~S., {Driver}, S.~P., {Robotham}, A.~S.~G., {et~al.} 2014, \mnras,
  444, 1647

\bibitem[{{Kent} \& {Sargent}(1983)}]{Kent_1983}
{Kent}, S.~M. \& {Sargent}, W.~L.~W. 1983, \aj, 88, 697

\bibitem[{{Kim} {et~al.}(2018){Kim}, {Kim}, {Choi}, {Lee}, {de Grijs}, {Lee},
  \& {Hwang}}]{Kim_2018}
{Kim}, E., {Kim}, S.~S., {Choi}, Y.-Y., {et~al.} 2018, \mnras, 479, 562

\bibitem[{{Kim} {et~al.}(2016){Kim}, {Oh}, {Jeong}, {Arag{\'o}n-Salamanca},
  {Smith}, \& {Yi}}]{Kim_2016}
{Kim}, K., {Oh}, S., {Jeong}, H., {et~al.} 2016, \apjs, 225, 6

\bibitem[{{Kochanek} {et~al.}(2001){Kochanek}, {Pahre}, {Falco}, {Huchra},
  {Mader}, {Jarrett}, {Chester}, {Cutri}, \& {Schneider}}]{Kochanek_2001}
{Kochanek}, C.~S., {Pahre}, M.~A., {Falco}, E.~E., {et~al.} 2001, \apj, 560,
  566

\bibitem[{{Kormendy} \& {Ho}(2013)}]{Kormendy_2013}
{Kormendy}, J. \& {Ho}, L.~C. 2013, \araa, 51, 511

\bibitem[{{Koyama} {et~al.}(2011){Koyama}, {Kodama}, {Hayashi}, {Tadaki},
  {Nakata}, {Tanaka}, {Shimasaku}, \& {Okamura}}]{Koyama_2011}
{Koyama}, Y., {Kodama}, T., {Hayashi}, M., {et~al.} 2011, in Galaxy Formation,
  P113

\bibitem[{{Larson} \& {Tinsley}(1978)}]{Larson_1978}
{Larson}, R.~B. \& {Tinsley}, B.~M. 1978, \apj, 219, 46

\bibitem[{{Larson} {et~al.}(1980){Larson}, {Tinsley}, \&
  {Caldwell}}]{Larson_1980}
{Larson}, R.~B., {Tinsley}, B.~M., \& {Caldwell}, C.~N. 1980, \apj, 237, 692

\bibitem[{{Lee} {et~al.}(2017){Lee}, {Chung}, {Tonnesen}, {Kenney}, {Wong},
  {Vollmer}, {Petitpas}, {Crowl}, \& {van Gorkom}}]{Lee_2017}
{Lee}, B., {Chung}, A., {Tonnesen}, S., {et~al.} 2017, \mnras, 466, 1382

\bibitem[{{Lewis} {et~al.}(2002){Lewis}, {Balogh}, {De Propris}, {Couch},
  {Bower}, {Offer}, {Bland-Hawthorn}, {Baldry}, {Baugh}, {Bridges}, {Cannon},
  {Cole}, {Colless}, {Collins}, {Cross}, {Dalton}, {Driver}, {Efstathiou},
  {Ellis}, {Frenk}, {Glazebrook}, {Hawkins}, {Jackson}, {Lahav}, {Lumsden},
  {Maddox}, {Madgwick}, {Norberg}, {Peacock}, {Percival}, {Peterson},
  {Sutherland}, \& {Taylor}}]{Lewis_2002}
{Lewis}, I., {Balogh}, M., {De Propris}, R., {et~al.} 2002, \mnras, 334, 673

\bibitem[{{Liao} \& {Gu}(2020)}]{Liao_2020}
{Liao}, M. \& {Gu}, M. 2020, \mnras, 491, 92

\bibitem[{{Lin} {et~al.}(2004){Lin}, {Mohr}, \& {Stanford}}]{Lin_2004}
{Lin}, Y.-T., {Mohr}, J.~J., \& {Stanford}, S.~A. 2004, \apj, 610, 745

\bibitem[{{Lintott} {et~al.}(2011){Lintott}, {Schawinski}, {Bamford}, {Slosar},
  {Land}, {Thomas}, {Edmondson}, {Masters}, {Nichol}, {Raddick}, {Szalay},
  {Andreescu}, {Murray}, \& {Vandenberg}}]{Lintott_2011}
{Lintott}, C., {Schawinski}, K., {Bamford}, S., {et~al.} 2011, \mnras, 410, 166

\bibitem[{{Liske} {et~al.}(2003){Liske}, {Lemon}, {Driver}, {Cross}, \&
  {Couch}}]{Liske_2003}
{Liske}, J., {Lemon}, D.~J., {Driver}, S.~P., {Cross}, N.~J.~G., \& {Couch},
  W.~J. 2003, \mnras, 344, 307

\bibitem[{{Liu} {et~al.}(2019){Liu}, {Hao}, {Wang}, \& {Yang}}]{Liu_2019}
{Liu}, C., {Hao}, L., {Wang}, H., \& {Yang}, X. 2019, \apj, 878, 69

\bibitem[{{Lynds}(1970)}]{Lynds_1970}
{Lynds}, R. 1970, \apj, 159, L151

\bibitem[{{Man} {et~al.}(2019){Man}, {Peng}, {Kong}, {Guo}, {Zhang}, \&
  {Dou}}]{Man_2019}
{Man}, Z.-y., {Peng}, Y.-j., {Kong}, X., {et~al.} 2019, \mnras, 1665

\bibitem[{{Markevitch} \& {Vikhlinin}(2007)}]{Markevitch_2007}
{Markevitch}, M. \& {Vikhlinin}, A. 2007, \physrep, 443, 1

\bibitem[{{Martin} {et~al.}(2018){Martin}, {Kaviraj}, {Devriendt}, {Dubois}, \&
  {Pichon}}]{Martin_2018}
{Martin}, G., {Kaviraj}, S., {Devriendt}, J.~E.~G., {Dubois}, Y., \& {Pichon},
  C. 2018, \mnras, 480, 2266

\bibitem[{{Martini} {et~al.}(2006){Martini}, {Kelson}, {Kim}, {Mulchaey}, \&
  {Athey}}]{Martini_2006}
{Martini}, P., {Kelson}, D.~D., {Kim}, E., {Mulchaey}, J.~S., \& {Athey}, A.~A.
  2006, \apj, 644, 116

\bibitem[{{Masters} {et~al.}(2010){Masters}, {Nichol}, {Bamford}, {Mosleh},
  {Lintott}, {Andreescu}, {Edmondson}, {Keel}, {Murray}, {Raddick},
  {Schawinski}, {Slosar}, {Szalay}, {Thomas}, \& {Vandenberg}}]{Masters_2010}
{Masters}, K.~L., {Nichol}, R., {Bamford}, S., {et~al.} 2010, \mnras, 404, 792

\bibitem[{{Mathews} {et~al.}(2006){Mathews}, {Faltenbacher}, \&
  {Brighenti}}]{Mathews_2006}
{Mathews}, W.~G., {Faltenbacher}, A., \& {Brighenti}, F. 2006, \apj, 638, 659

\bibitem[{{Matsushita} {et~al.}(2013){Matsushita}, {Sakuma}, {Sasaki}, {Sato},
  \& {Simionescu}}]{Matsushita_2013}
{Matsushita}, K., {Sakuma}, E., {Sasaki}, T., {Sato}, K., \& {Simionescu}, A.
  2013, \apj, 764, 147

\bibitem[{{Meert} {et~al.}(2015){Meert}, {Vikram}, \& {Bernardi}}]{Meert_2015}
{Meert}, A., {Vikram}, V., \& {Bernardi}, M. 2015, \mnras, 446, 3943

\bibitem[{{Melnick} \& {Sargent}(1977)}]{Melnick_1977}
{Melnick}, J. \& {Sargent}, W.~L.~W. 1977, \apj, 215, 401

\bibitem[{{Merritt} {et~al.}(2006){Merritt}, {Graham}, {Moore}, {Diemand}, \&
  {Terzi{\'c}}}]{Merritt_2006}
{Merritt}, D., {Graham}, A.~W., {Moore}, B., {Diemand}, J., \& {Terzi{\'c}}, B.
  2006, \aj, 132, 2685

\bibitem[{{Meusinger} {et~al.}(2000){Meusinger}, {Brunzendorf}, \&
  {Krieg}}]{Meusinger_2000}
{Meusinger}, H., {Brunzendorf}, J., \& {Krieg}, R. 2000, \aap, 363, 933

\bibitem[{{Mihos}(2004)}]{Mihos_2004}
{Mihos}, J.~C. 2004, in Clusters of Galaxies: Probes of Cosmological Structure
  and Galaxy Evolution, ed. J.~S. {Mulchaey}, A.~{Dressler}, \& A.~{Oemler},
  277

\bibitem[{{Mihos}(2016)}]{Mihos_2016}
{Mihos}, J.~C. 2016, in IAU Symposium, Vol. 317, The General Assembly of Galaxy
  Halos: Structure, Origin and Evolution, ed. A.~{Bragaglia}, M.~{Arnaboldi},
  M.~{Rejkuba}, \& D.~{Romano}, 27--34

\bibitem[{{Mihos} \& {Hernquist}(1996)}]{Mihos_1996}
{Mihos}, J.~C. \& {Hernquist}, L. 1996, \apj, 464, 641

\bibitem[{{Miller} \& {Owen}(2001)}]{Miller_2001}
{Miller}, N.~A. \& {Owen}, F.~N. 2001, \apjs, 134, 355

\bibitem[{{Minkowski}(1957)}]{Minkowski_1957}
{Minkowski}, R. 1957, in IAU Symposium, Vol.~4, Radio astronomy, ed. H.~C. {van
  de Hulst}, 107

\bibitem[{{Mishra} {et~al.}(2019){Mishra}, {Wadadekar}, \&
  {Barway}}]{Mishra_2019}
{Mishra}, P.~K., {Wadadekar}, Y., \& {Barway}, S. 2019, \mnras, 1546

\bibitem[{{Mobasher} \& {Trentham}(1998)}]{Mobasher_1998}
{Mobasher}, B. \& {Trentham}, N. 1998, \mnras, 293, 315

\bibitem[{{Mohr} {et~al.}(1999){Mohr}, {Mathiesen}, \& {Evrard}}]{Mohr_1999}
{Mohr}, J.~J., {Mathiesen}, B., \& {Evrard}, A.~E. 1999, \apj, 517, 627

\bibitem[{{Monet} {et~al.}(2003){Monet}, {Levine}, {Canzian}, {Ables}, {Bird},
  {Dahn}, {Guetter}, {Harris}, {Henden}, \& {Leggett}}]{Monet_2003}
{Monet}, D.~G., {Levine}, S.~E., {Canzian}, B., {et~al.} 2003, \aj, 125, 984

\bibitem[{{Montes} \& {Trujillo}(2019)}]{Montes_2019}
{Montes}, M. \& {Trujillo}, I. 2019, Monthly Notices of the Royal Astronomical
  Society, 482, 2838

\bibitem[{{Moore} {et~al.}(1996){Moore}, {Katz}, {Lake}, {Dressler}, \&
  {Oemler}}]{Moore_1996}
{Moore}, B., {Katz}, N., {Lake}, G., {Dressler}, A., \& {Oemler}, A. 1996,
  \nat, 379, 613

\bibitem[{{Moorman} {et~al.}(2016){Moorman}, {Moreno}, {White}, {Vogeley},
  {Hoyle}, {Giovanelli}, \& {Haynes}}]{Moorman_2016}
{Moorman}, C.~M., {Moreno}, J., {White}, A., {et~al.} 2016, \apj, 831, 118

\bibitem[{{Moss} \& {Whittle}(2000)}]{Moss_2000}
{Moss}, C. \& {Whittle}, M. 2000, \mnras, 317, 667

\bibitem[{{Moss} \& {Whittle}(2006)}]{Moss_2006}
{Moss}, C. \& {Whittle}, M. 2006, VizieR Online Data Catalog, 735

\bibitem[{{Mulroy} {et~al.}(2017){Mulroy}, {McGee}, {Gillman}, {Smith},
  {Haines}, {D{\'e}mocl{\`e}s}, {Okabe}, \& {Egami}}]{Mulroy_2017}
{Mulroy}, S.~L., {McGee}, S.~L., {Gillman}, S., {et~al.} 2017, \mnras, 472,
  3246

\bibitem[{{O'Dea}(1998)}]{O'Dea_1998}
{O'Dea}, C.~P. 1998, \pasp, 110, 493

\bibitem[{{Oosterloo} \& {van Gorkom}(2005)}]{Oosterloo_2005}
{Oosterloo}, T. \& {van Gorkom}, J. 2005, \aap, 437, L19

\bibitem[{{Osterbrock} \& {Martel}(1992)}]{Osterbrock_1992}
{Osterbrock}, D.~E. \& {Martel}, A. 1992, \pasp, 104, 76

\bibitem[{{Padilla} \& {Strauss}(2008)}]{Padilla_2008}
{Padilla}, N.~D. \& {Strauss}, M.~A. 2008, \mnras, 388, 1321

\bibitem[{{Park} {et~al.}(2008){Park}, {Gott}, \& {Choi}}]{Park_2008}
{Park}, C., {Gott}, J.~Richard, I., \& {Choi}, Y.-Y. 2008, \apj, 674, 784

\bibitem[{{Park} {et~al.}(2017){Park}, {Yang}, {Oonk}, \& {Paragi}}]{Park_2017}
{Park}, S., {Yang}, J., {Oonk}, J.~B.~R., \& {Paragi}, Z. 2017, \mnras, 465,
  3943

\bibitem[{{Peng} {et~al.}(2015){Peng}, {Maiolino}, \& {Cochrane}}]{Peng_2015}
{Peng}, Y., {Maiolino}, R., \& {Cochrane}, R. 2015, \nat, 521, 192

\bibitem[{{Penny} {et~al.}(2014){Penny}, {Forbes}, {Pimbblet}, \&
  {Floyd}}]{Penny_2014}
{Penny}, S.~J., {Forbes}, D.~A., {Pimbblet}, K.~A., \& {Floyd}, D. J.~E. 2014,
  \mnras, 443, 3381

\bibitem[{{Pfister} {et~al.}(2013){Pfister}, {Schwarz}, {Janczyk}, {Dale}, \&
  {Freeman}}]{Pfister_2013}
{Pfister}, R., {Schwarz}, K.~A., {Janczyk}, M., {Dale}, R., \& {Freeman}, J.~B.
  2013, Frontiers in Psychology, 4, 1

\bibitem[{{Pimbblet} {et~al.}(2014){Pimbblet}, {Penny}, \&
  {Davies}}]{Pimbblet_2014}
{Pimbblet}, K.~A., {Penny}, S.~J., \& {Davies}, R.~L. 2014, \mnras, 438, 3049

\bibitem[{{Pinkney} {et~al.}(1996){Pinkney}, {Roettiger}, {Burns}, \&
  {Bird}}]{Pinkney_1996}
{Pinkney}, J., {Roettiger}, K., {Burns}, J.~O., \& {Bird}, C.~M. 1996, \apjs,
  104, 1

\bibitem[{{Polletta} {et~al.}(2007){Polletta}, {Tajer}, {Maraschi},
  {Trinchieri}, {Lonsdale}, {Chiappetti}, {Andreon}, {Pierre}, {Le F{\`e}vre},
  {Zamorani}, {Maccagni}, {Garcet}, {Surdej}, {Franceschini}, {Alloin},
  {Shupe}, {Surace}, {Fang}, {Rowan-Robinson}, {Smith}, \&
  {Tresse}}]{Polletta_2007}
{Polletta}, M., {Tajer}, M., {Maraschi}, L., {et~al.} 2007, \apj, 663, 81

\bibitem[{{Postman} \& {Geller}(1984)}]{Postman_1984}
{Postman}, M. \& {Geller}, M.~J. 1984, \apj, 281, 95

\bibitem[{{Poulain} {et~al.}(1992){Poulain}, {Nieto}, \&
  {Davoust}}]{Poulain_1992}
{Poulain}, P., {Nieto}, J.~L., \& {Davoust}, E. 1992, \aaps, 95, 129

\bibitem[{{Prestwich} {et~al.}(1997){Prestwich}, {Joy}, {Luginbuhl},
  {Sulkanen}, \& {Newberry}}]{Prestwich_1997}
{Prestwich}, A.~H., {Joy}, M., {Luginbuhl}, C.~B., {Sulkanen}, M., \&
  {Newberry}, M. 1997, \apj, 477, 144

\bibitem[{{Rawle} {et~al.}(2013){Rawle}, {Lucey}, {Smith}, \&
  {Head}}]{Rawle_2013}
{Rawle}, T.~D., {Lucey}, J.~R., {Smith}, R.~J., \& {Head}, J.~T.~C.~G. 2013,
  \mnras, 433, 2667

\bibitem[{{Ricciardelli} {et~al.}(2017){Ricciardelli}, {Cava}, {Varela}, \&
  {Tamone}}]{Ricciardelli_2017}
{Ricciardelli}, E., {Cava}, A., {Varela}, J., \& {Tamone}, A. 2017, \apjl, 846,
  L4

\bibitem[{{Rodr{\'\i}guez} \& {Padilla}(2013)}]{Rodriguez_2013}
{Rodr{\'\i}guez}, S. \& {Padilla}, N.~D. 2013, \mnras, 434, 2153

\bibitem[{{Roediger} {et~al.}(2006){Roediger}, {Br{\"u}ggen}, \&
  {Hoeft}}]{Roediger_2006}
{Roediger}, E., {Br{\"u}ggen}, M., \& {Hoeft}, M. 2006, \mnras, 371, 609

\bibitem[{{Rojas} {et~al.}(2004){Rojas}, {Vogeley}, {Hoyle}, \&
  {Brinkmann}}]{Rojas_2004}
{Rojas}, R.~R., {Vogeley}, M.~S., {Hoyle}, F., \& {Brinkmann}, J. 2004, \apj,
  617, 50

\bibitem[{{Rosen} {et~al.}(2016){Rosen}, {Webb}, {Watson}, {Ballet}, {Barret},
  {Braito}, {Carrera}, {Ceballos}, {Coriat}, {Della Ceca}, {Denkinson},
  {Esquej}, {Farrell}, {Freyberg}, {Gris{\'e}}, {Guillout}, {Heil},
  {Koliopanos}, {Law-Green}, {Lamer}, {Lin}, {Martino}, {Michel}, {Motch},
  {Nebot Gomez-Moran}, {Page}, {Page}, {Page}, {Pakull}, {Pye}, {Read},
  {Rodriguez}, {Sakano}, {Saxton}, {Schwope}, {Scott}, {Sturm}, {Traulsen},
  {Yershov}, \& {Zolotukhin}}]{Rosen_2016}
{Rosen}, S.~R., {Webb}, N.~A., {Watson}, M.~G., {et~al.} 2016, \aap, 590, A1

\bibitem[{{Ryle} \& {Windram}(1968)}]{Ryle_1968}
{Ryle}, M. \& {Windram}, M.~D. 1968, \mnras, 138, 1

\bibitem[{{Sadler} {et~al.}(2002){Sadler}, {Jackson}, {Cannon}, {McIntyre},
  {Murphy}, {Bland-Hawthorn}, {Bridges}, {Cole}, {Colless}, {Collins}, {Couch},
  {Dalton}, {De Propris}, {Driver}, {Efstathiou}, {Ellis}, {Frenk},
  {Glazebrook}, {Lahav}, {Lewis}, {Lumsden}, {Maddox}, {Madgwick}, {Norberg},
  {Peacock}, {Peterson}, {Sutherland}, \& {Taylor}}]{Sadler_2002}
{Sadler}, E.~M., {Jackson}, C.~A., {Cannon}, R.~D., {et~al.} 2002, \mnras, 329,
  227

\bibitem[{{Sakai} {et~al.}(2012){Sakai}, {Kennicutt}, \& {Moss}}]{Sakai_2012}
{Sakai}, S., {Kennicutt}, Jr., R.~C., \& {Moss}, C. 2012, \apjs, 199, 36

\bibitem[{{Santra} {et~al.}(2007){Santra}, {Sanders}, \&
  {Fabian}}]{Santra_2007}
{Santra}, S., {Sanders}, J.~S., \& {Fabian}, A.~C. 2007, \mnras, 382, 895

\bibitem[{{Scharw{\"a}chter} {et~al.}(2013){Scharw{\"a}chter}, {McGregor},
  {Dopita}, \& {Beck}}]{Scharwachter_2013}
{Scharw{\"a}chter}, J., {McGregor}, P.~J., {Dopita}, M.~A., \& {Beck}, T.~L.
  2013, \mnras, 429, 2315

\bibitem[{{Schlafly} \& {Finkbeiner}(2011)}]{Schlafly_2011}
{Schlafly}, E.~F. \& {Finkbeiner}, D.~P. 2011, \apj, 737, 103

\bibitem[{{Schweizer} {et~al.}(1987){Schweizer}, {Ford}, {Jedrzejewski}, \&
  {Giovanelli}}]{Schweizer_1987}
{Schweizer}, F., {Ford}, W.~Kent, J., {Jedrzejewski}, R., \& {Giovanelli}, R.
  1987, \apj, 320, 454

\bibitem[{{Shao} {et~al.}(2015){Shao}, {Disseau}, {Yang}, {Hammer}, {Puech},
  {Rodrigues}, {Liang}, \& {Deng}}]{Shao_2015}
{Shao}, X., {Disseau}, K., {Yang}, Y.~B., {et~al.} 2015, \aap, 579, A57

\bibitem[{{Shimwell} {et~al.}(2019){Shimwell}, {Tasse}, {Hardcastle}, {Mechev},
  {Williams}, {Best}, {R{\"o}ttgering}, {Callingham}, {Dijkema}, {de Gasperin},
  {Hoang}, {Hugo}, {Mirmont}, {Oonk}, {Prandoni}, {Rafferty}, {Sabater},
  {Smirnov}, {van Weeren}, {White}, {Atemkeng}, {Bester}, {Bonnassieux},
  {Br{\"u}ggen}, {Brunetti}, {Chy{\.z}y}, {Cochrane}, {Conway}, {Croston},
  {Danezi}, {Duncan}, {Haverkorn}, {Heald}, {Iacobelli}, {Intema}, {Jackson},
  {Jamrozy}, {Jarvis}, {Lakhoo}, {Mevius}, {Miley}, {Morabito}, {Morganti},
  {Nisbet}, {Orr{\'u}}, {Perkins}, {Pizzo}, {Schrijvers}, {Smith}, {Vermeulen},
  {Wise}, {Alegre}, {Bacon}, {van Bemmel}, {Beswick}, {Bonafede}, {Botteon},
  {Bourke}, {Brienza}, {Calistro Rivera}, {Cassano}, {Clarke}, {Conselice},
  {Dettmar}, {Drabent}, {Dumba}, {Emig}, {En{\ss}lin}, {Ferrari}, {Garrett},
  {G{\'e}nova-Santos}, {Goyal}, {G{\"u}rkan}, {Hale}, {Harwood}, {Heesen},
  {Hoeft}, {Horellou}, {Jackson}, {Kokotanekov}, {Kondapally},
  {Kunert-Bajraszewska}, {Mahatma}, {Mahony}, {Mandal}, {McKean}, {Merloni},
  {Mingo}, {Miskolczi}, {Mooney}, {Nikiel-Wroczy{\'n}ski}, {O'Sullivan},
  {Quinn}, {Reich}, {Roskowi{\'n}ski}, {Rowlinson}, {Savini}, {Saxena},
  {Schwarz}, {Shulevski}, {Sridhar}, {Stacey}, {Urquhart}, {van der Wiel},
  {Varenius}, {Webster}, \& {Wilber}}]{Shimwell_2019}
{Shimwell}, T.~W., {Tasse}, C., {Hardcastle}, M.~J., {et~al.} 2019, \aap, 622,
  A1

\bibitem[{{Simard} {et~al.}(2011){Simard}, {Mendel}, {Patton}, {Ellison}, \&
  {McConnachie}}]{Simard_2011}
{Simard}, L., {Mendel}, J.~T., {Patton}, D.~R., {Ellison}, S.~L., \&
  {McConnachie}, A.~W. 2011, \apjs, 196, 11

\bibitem[{{Simionescu} {et~al.}(2011){Simionescu}, {Allen}, {Mantz}, {Werner},
  {Takei}, {Morris}, {Fabian}, {Sanders}, {Nulsen}, {George}, \&
  {Taylor}}]{Simionescu_2011}
{Simionescu}, A., {Allen}, S.~W., {Mantz}, A., {et~al.} 2011, Science, 331,
  1576

\bibitem[{{Simionescu} {et~al.}(2012){Simionescu}, {Werner}, {Urban}, {Allen},
  {Fabian}, {Sanders}, {Mantz}, {Nulsen}, \& {Takei}}]{Simionescu_2012}
{Simionescu}, A., {Werner}, N., {Urban}, O., {et~al.} 2012, \apj, 757, 182

\bibitem[{{Skrutskie} {et~al.}(2006){Skrutskie}, {Cutri}, {Stiening},
  {Weinberg}, {Schneider}, {Carpenter}, {Beichman}, {Capps}, {Chester}, \&
  {Elias}}]{Skrutzkie_2006}
{Skrutskie}, M.~F., {Cutri}, R.~M., {Stiening}, R., {et~al.} 2006, \aj, 131,
  1163

\bibitem[{{Soboleva} {et~al.}(1983){Soboleva}, {Temirova}, {Timofeeva}, \&
  {Aliakberov}}]{Soboleva_1983}
{Soboleva}, N.~S., {Temirova}, A.~V., {Timofeeva}, G.~M., \& {Aliakberov},
  K.~D. 1983, Soviet Astronomy Letters, 9, 305

\bibitem[{{Stern} {et~al.}(2012){Stern}, {Assef}, {Benford}, {Blain}, {Cutri},
  {Dey}, {Eisenhardt}, {Griffith}, {Jarrett}, {Lake}, {Masci}, {Petty},
  {Stanford}, {Tsai}, {Wright}, {Yan}, {Harrison}, \& {Madsen}}]{Stern_2012}
{Stern}, D., {Assef}, R.~J., {Benford}, D.~J., {et~al.} 2012, \apj, 753, 30

\bibitem[{{Strateva} {et~al.}(2001){Strateva}, {Ivezi{\'c}}, {Knapp},
  {Narayanan}, {Strauss}, {Gunn}, {Lupton}, {Schlegel}, {Bahcall}, \&
  {Brinkmann}}]{Strateva_2001}
{Strateva}, I., {Ivezi{\'c}}, {\v{Z}}., {Knapp}, G.~R., {et~al.} 2001, \aj,
  122, 1861

\bibitem[{{Struble}(1979)}]{Struble_1979}
{Struble}, M.~F. 1979, \aj, 84, 27

\bibitem[{{Struble} \& {Rood}(1999)}]{Struble_1999}
{Struble}, M.~F. \& {Rood}, H.~J. 1999, \apjs, 125, 35

\bibitem[{{Toomre} \& {Toomre}(1972)}]{Toomre_1972}
{Toomre}, A. \& {Toomre}, J. 1972, \apj, 178, 623

\bibitem[{{Tremblay} {et~al.}(2015){Tremblay}, {O'Dea}, {Baum}, {Mittal},
  {McDonald}, {Combes}, {Li}, {McNamara}, {Bremer}, {Clarke}, {Donahue},
  {Edge}, {Fabian}, {Hamer}, {Hogan}, {Oonk}, {Quillen}, {Sanders},
  {Salom{\'e}}, \& {Voit}}]{Tremblay_2015}
{Tremblay}, G.~R., {O'Dea}, C.~P., {Baum}, S.~A., {et~al.} 2015, \mnras, 451,
  3768

\bibitem[{{Ulmer} {et~al.}(1992){Ulmer}, {Wirth}, \& {Kowalski}}]{Ulmer_1992}
{Ulmer}, M.~P., {Wirth}, G.~D., \& {Kowalski}, M.~P. 1992, \apj, 397, 430

\bibitem[{{van den Bergh}(1976)}]{vandenBergh_1976}
{van den Bergh}, S. 1976, \apj, 206, 883

\bibitem[{{Verdugo} {et~al.}(2008){Verdugo}, {Ziegler}, \&
  {Gerken}}]{Verdugo_2008}
{Verdugo}, M., {Ziegler}, B.~L., \& {Gerken}, B. 2008, \aap, 486, 9

\bibitem[{{V{\'e}ron-Cetty} \& {V{\'e}ron}(2006)}]{Veron-Cetty_2006}
{V{\'e}ron-Cetty}, M.-P. \& {V{\'e}ron}, P. 2006, \aap, 455, 773

\bibitem[{{Vollmer} {et~al.}(2010){Vollmer}, {Gassmann}, {Derri{\`e}re},
  {Boch}, {Louys}, {Bonnarel}, {Dubois}, {Genova}, \&
  {Ochsenbein}}]{Vollmer_2010}
{Vollmer}, B., {Gassmann}, B., {Derri{\`e}re}, S., {et~al.} 2010, \aap, 511,
  A53

\bibitem[{{Vulcani} {et~al.}(2019){Vulcani}, {Poggianti}, {Moretti},
  {Gullieuszik}, {Fritz}, {Franchetto}, {Fasano}, {Bettoni}, \&
  {Jaff{\'e}}}]{Vulcani_2019}
{Vulcani}, B., {Poggianti}, B.~M., {Moretti}, A., {et~al.} 2019, \mnras, 487,
  2278

\bibitem[{{Wakamatsu}(1990)}]{Wakamatsu_1990}
{Wakamatsu}, K.-I. 1990, \apj, 348, 448

\bibitem[{{Whitbourn} \& {Shanks}(2014)}]{Whitbourn_2014}
{Whitbourn}, J.~R. \& {Shanks}, T. 2014, \mnras, 437, 2146

\bibitem[{{Wilkinson} {et~al.}(2017){Wilkinson}, {Pimbblet}, \&
  {Stott}}]{Wilkinson_2017}
{Wilkinson}, C.~L., {Pimbblet}, K.~A., \& {Stott}, J.~P. 2017, \mnras, 472,
  1447

\bibitem[{{Willett} {et~al.}(2013){Willett}, {Lintott}, {Bamford}, {Masters},
  {Simmons}, {Casteels}, {Edmondson}, {Fortson}, {Kaviraj}, {Keel}, {Melvin},
  {Nichol}, {Raddick}, {Schawinski}, {Simpson}, {Skibba}, {Smith}, \&
  {Thomas}}]{Willett_2013}
{Willett}, K.~W., {Lintott}, C.~J., {Bamford}, S.~P., {et~al.} 2013, \mnras,
  435, 2835

\bibitem[{{Wittmann} {et~al.}(2019){Wittmann}, {Kotulla}, {Lisker}, {Grebel},
  {Conselice}, {Janz}, \& {Penny}}]{Wittmann_2019}
{Wittmann}, C., {Kotulla}, R., {Lisker}, T., {et~al.} 2019, \apjs, 245, 10

\bibitem[{{Wittmann} {et~al.}(2017){Wittmann}, {Lisker}, {Ambachew Tilahun},
  {Grebel}, {Conselice}, {Penny}, {Janz}, {Gallagher}, {Kotulla}, \&
  {McCormac}}]{Wittmann_2017}
{Wittmann}, C., {Lisker}, T., {Ambachew Tilahun}, L., {et~al.} 2017, \mnras,
  470, 1512

\bibitem[{{Wright} {et~al.}(2010){Wright}, {Eisenhardt}, {Mainzer}, {Ressler},
  {Cutri}, {Jarrett}, {Kirkpatrick}, {Padgett}, {McMillan}, {Skrutskie},
  {Stanford}, {Cohen}, {Walker}, {Mather}, {Leisawitz}, {Gautier}, {McLean},
  {Benford}, {Lonsdale}, {Blain}, {Mendez}, {Irace}, {Duval}, {Liu}, {Royer},
  {Heinrichsen}, {Howard}, {Shannon}, {Kendall}, {Walsh}, {Larsen}, {Cardon},
  {Schick}, {Schwalm}, {Abid}, {Fabinsky}, {Naes}, \& {Tsai}}]{Wright_2010}
{Wright}, E.~L., {Eisenhardt}, P.~R.~M., {Mainzer}, A.~K., {et~al.} 2010, \aj,
  140, 1868

\bibitem[{{York} {et~al.}(2000){York}, {Adelman}, {Anderson}, {Anderson},
  {Annis}, {Bahcall}, {Bakken}, {Barkhouser}, {Bastian}, {Berman}, {Boroski},
  {Bracker}, {Briegel}, {Briggs}, {Brinkmann}, {Brunner}, {Burles}, {Carey},
  {Carr}, {Castander}, {Chen}, {Colestock}, {Connolly}, {Crocker}, {Csabai},
  {Czarapata}, {Davis}, {Doi}, {Dombeck}, {Eisenstein}, {Ellman}, {Elms},
  {Evans}, {Fan}, {Federwitz}, {Fiscelli}, {Friedman}, {Frieman}, {Fukugita},
  {Gillespie}, {Gunn}, {Gurbani}, {de Haas}, {Haldeman}, {Harris}, {Hayes},
  {Heckman}, {Hennessy}, {Hindsley}, {Holm}, {Holmgren}, {Huang}, {Hull},
  {Husby}, {Ichikawa}, {Ichikawa}, {Ivezi{\'c}}, {Kent}, {Kim}, {Kinney},
  {Klaene}, {Kleinman}, {Kleinman}, {Knapp}, {Korienek}, {Kron}, {Kunszt},
  {Lamb}, {Lee}, {Leger}, {Limmongkol}, {Lindenmeyer}, {Long}, {Loomis},
  {Loveday}, {Lucinio}, {Lupton}, {MacKinnon}, {Mannery}, {Mantsch}, {Margon},
  {McGehee}, {McKay}, {Meiksin}, {Merelli}, {Monet}, {Munn}, {Narayanan},
  {Nash}, {Neilsen}, {Neswold}, {Newberg}, {Nichol}, {Nicinski}, {Nonino},
  {Okada}, {Okamura}, {Ostriker}, {Owen}, {Pauls}, {Peoples}, {Peterson},
  {Petravick}, {Pier}, {Pope}, {Pordes}, {Prosapio}, {Rechenmacher}, {Quinn},
  {Richards}, {Richmond}, {Rivetta}, {Rockosi}, {Ruthmansdorfer}, {Sandford},
  {Schlegel}, {Schneider}, {Sekiguchi}, {Sergey}, {Shimasaku}, {Siegmund},
  {Smee}, {Smith}, {Snedden}, {Stone}, {Stoughton}, {Strauss}, {Stubbs},
  {SubbaRao}, {Szalay}, {Szapudi}, {Szokoly}, {Thakar}, {Tremonti}, {Tucker},
  {Uomoto}, {Vanden Berk}, {Vogeley}, {Waddell}, {Wang}, {Watanabe},
  {Weinberg}, {Yanny}, {Yasuda}, \& {SDSS Collaboration}}]{York_2000}
{York}, D.~G., {Adelman}, J., {Anderson}, Jr., J.~E., {et~al.} 2000, \aj, 120,
  1579

\bibitem[{{Yu} {et~al.}(2015){Yu}, {Lim}, {Ohyama}, {Chan}, \&
  {Broadhurst}}]{Yu_2015}
{Yu}, A. P.~Y., {Lim}, J., {Ohyama}, Y., {Chan}, J. C.~C., \& {Broadhurst}, T.
  2015, \apj, 814, 101

\bibitem[{{Zinger} {et~al.}(2018){Zinger}, {Dekel}, {Kravtsov}, \&
  {Nagai}}]{Zinger_2018}
{Zinger}, E., {Dekel}, A., {Kravtsov}, A.~V., \& {Nagai}, D. 2018, \mnras, 475,
  3654

\end{thebibliography}


\begin{appendix}

\onecolumn
%
\section{Catalogue description}\label{sect:catalogue}
%

The  catalogue contains 41 columns of 1294 galaxies. The content in each column is described in Table\,\ref{tab:catalogue}.

\begin{table*}[hbpt]
\caption{Catalogue columns.}
\begin{tabular}{lll}
\hline\hline
01 &  Number              & Running catalogue number                                                                        \\ 
02 &  ID                  & Object designation hmmss.s+ddmmss (J2000.0)                                                     \\ 
03 &  RA                  & Right ascension in decimal degrees (J2000)                                                      \\ 
04 &  DE                  & Declination in decimal degrees (J2000)                                                          \\ 
05 &  {\tt otherName}     & Cross ID                                                                                        \\ 
06 &  {\tt BM99}          & Catalogue number in the \citet{Brunzendorf_1999} catalogue                                      \\ 
07 &  $R$                 & Cluster-centric distance in arcmin                                                              \\ 
08 &  $z$                 & Spectroscopic redshift                                                                          \\ 
09 &  {\tt source\_z}     & Source of spectroscopic redshift                                                                 \\ 
   &                      & (1,2: TLS-CA survey, 3: SDSS, 4: other sources from NED, 5: \citet{Sakai_2012})                 \\ 
10 &  {\tt Halpha}        & H$\alpha$ detection flag                                                                        \\ 
   &                      & (0-3 from TLS survey where 0:no,1:weak,...,3:strong; 4 from \citet{Sakai_2012})                 \\ 
11 &  {\tt cl1}           & Numerical descriptor of morphological class                                                     \\ 
   &                      & (0: unclassified, 1: E, 2: E/S0, 3: S0, 4: S/S0, 5: Sa, 6: Sb/Sc, 7: S/Irr, 8: Irr, 9: merger)  \\ 
12 &  {\tt flag\_cl1}     & Reliability flag for {\tt cl1} (1: low, 2: moderate, 3: high)                                   \\ 
13 &  {\tt cl2}           & Description of morphological type                                                               \\ 
14 &  {\tt pec}           & Numerical descriptor of the peculiarity type                                                    \\ 
   &                      & (1: weakly lopsided, 2: strongly lopsided, 3: minor merger,  4: early major merger,             \\ 
   &                      & 5: intermediate major merger, 6: late major merger, 7: M51 type, 8: collisional ring, 9: others) \\ 
15 &  {\tt flag\_pec}     & Reliability flag for {\tt pec} (1: low, 2: moderate, 3: high)                                   \\ 
16 &  {\tt fracDeV}       & Coefficient of the de Vaucouleurs term in the surface brightness profile fit                    \\ 
   &                      & (clipped between zero and one)                                                                  \\ 
17 &  {\tt source\_fD}    & Source of {\tt fracDeV}  (1: SDSS, 2: TLS-CA)                                                   \\ 
18 &  {\tt umag}          & SDSS model magnitude in the u band                                                              \\ 
19 &  {\tt gmag}          & SDSS model magnitude in the g band                                                              \\ 
20 &  {\tt rmag}          & SDSS model magnitude in the r band                                                              \\ 
21 &  {\tt imag}          & SDSS model magnitude in the i band                                                              \\ 
22 &  {\tt zmag}          & SDSS model magnitude in the z band                                                              \\ 
23 &  {\tt Jmag}          & 2MASS XSC magnitude in the J band                                                               \\ 
24 &  {\tt Hmag}          & 2MASS XSC magnitude in the H band                                                               \\ 
25 &  {\tt Kmag}          & 2MASS XSC magnitude in the K band                                                               \\ 
26 &  {\tt W1mag}         & WISE magnitude in the W1 band                                                                   \\ 
27 &  {\tt W2mag}         & WISE magnitude in the W2 band                                                                   \\ 
28 &  {\tt W3mag}         & WISE magnitude in the W3 band                                                                   \\ 
29 &  {\tt W4mag}         & WISE magnitude in the W4 band                                                                   \\ 
30 &  {\tt EBV)}          & Galactic foreground reddening $E(B-V)$ from \citet{Schlafly_2011}                               \\ 
31 &  {\tt clustMemb}     & Spectroscopic cluster membership flag (1: established member, 0: established non-member)       \\ 
32 &  {\tt MLS\_W1\_flag} & Membership flag for the MLS\_W1 sample (0:non-member, 1: member)                                \\ 
33 &  {\tt MLS\_ur\_flag} & Membership flag for the MLS\_ur sample (0:non-member, 1: member)                                \\ 
34 &  {\tt absMag\_r}     & Absolute SDSS r band magnitude                                                                  \\ 
35 &  {\tt WHAN\_type}    & Spectral type from the WHAN diagram                                                             \\ 
   &                      & (1: SF, 2: sAGN, 3: wAGN, 4: retired)                                                           \\ 
36 &  {\tt log\_SFR}      & Decadic logarithm of star formation rate ($\mathcal{M}_\odot$ yr$^{-1}$)                        \\ 
37 &  {\tt log\_Mstar}    & Decadic logarithm of stellar mass  ($\mathcal{M}_\odot$)                                        \\ 
38 &  {\tt bl\_flag}      & Broad emission line flag (1: broad lines indicated)                                             \\ 
39 &  {\tt log\_XMMflux}  & Decadic logarithm of mean XMM flux in 0.2-12 keV  (mW\,m$^{-2}$)                                \\ 
40 &  {\tt log\_NVSSflux} & Decadic logarithm of integrated 1.4 GHz flux density (mJy)                                      \\ 
41 &  $q$                 & FIR-to-radio flux density ratio from \citet{Miller_2001}                                        \\ 
\hline
\end{tabular}
\label{tab:catalogue}
\end{table*}

\newpage

\twocolumn

%
\section{Individual peculiar galaxies}\label{sect:individual}
%

\subsection{Individual peculiar AGN host galaxies}

The AGN galaxy sample from Table\,\ref{tab:AGN_all} includes 12 galaxies with a peculiarity parameter ${\tt pec} > 0$. The half of the sub-sample of peculiar AGN galaxies belongs to the IRAS sample that was discussed previously \citep{Meusinger_2000}, including the hosts of the two optically brightest cluster AGN NGC\,1275 (\#691) and UGC\,2608 (\#278). In addition, remarks on the morphology of UGC\,2618, PGC\,12166, J031645.9+401948, NGC\,1260 and PGC\,12535 can be found there. Here we present the six remaining peculiar systems from Table\,\ref{tab:AGN_all}.\\

\begin{figure}[htbp]
\fbox{\includegraphics[viewport= 0 0 281 281,width=4.18cm,angle=0]{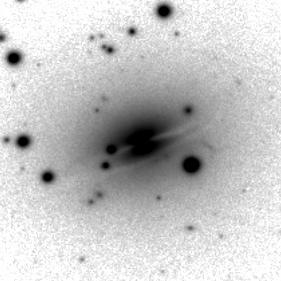}}
\includegraphics[viewport= 0 15 540 555,width=4.45cm,angle=0]{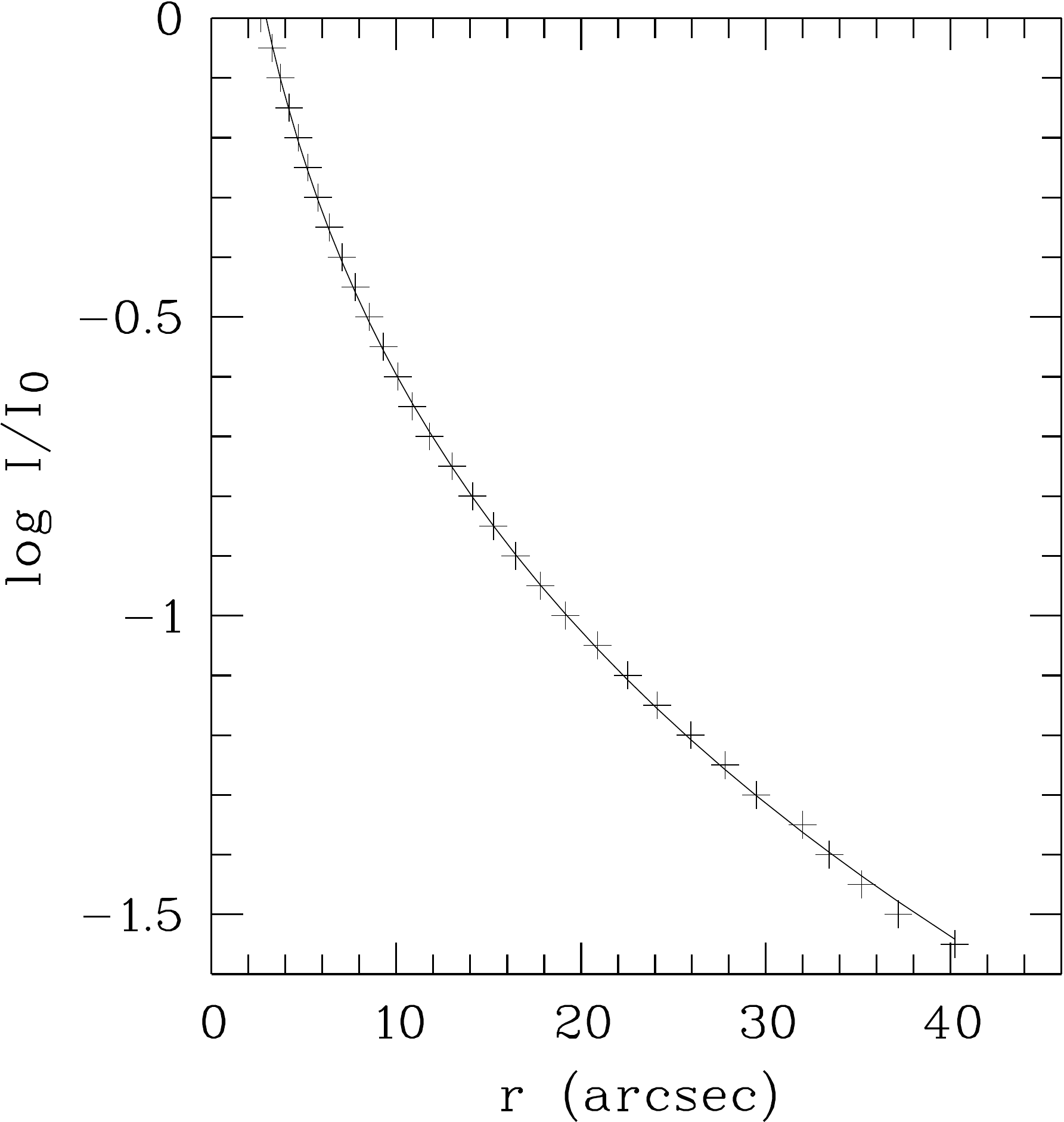}
\vspace{0.2cm}
\caption{Surface brightness distribution of J031646.7+400013 (\#398). 
Left: composite BRgri image (side length $2$\arcmin). 
Right: Radial surface brightness profile from the cleaned optical image (crosses) compared 
with the best-matching de Vaucouleurs profile (solid curve).
}
\label{fig:App_398_1}
\end{figure}

\begin{figure}[htbp]
\fbox{\includegraphics[viewport= 0 0 555 555,width=4.05cm,angle=0]{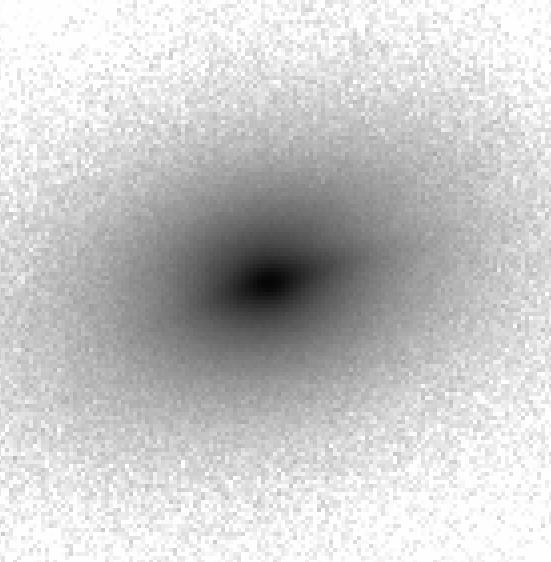}}
\includegraphics[viewport= 0 12 420 440,width=4.35cm,angle=0]{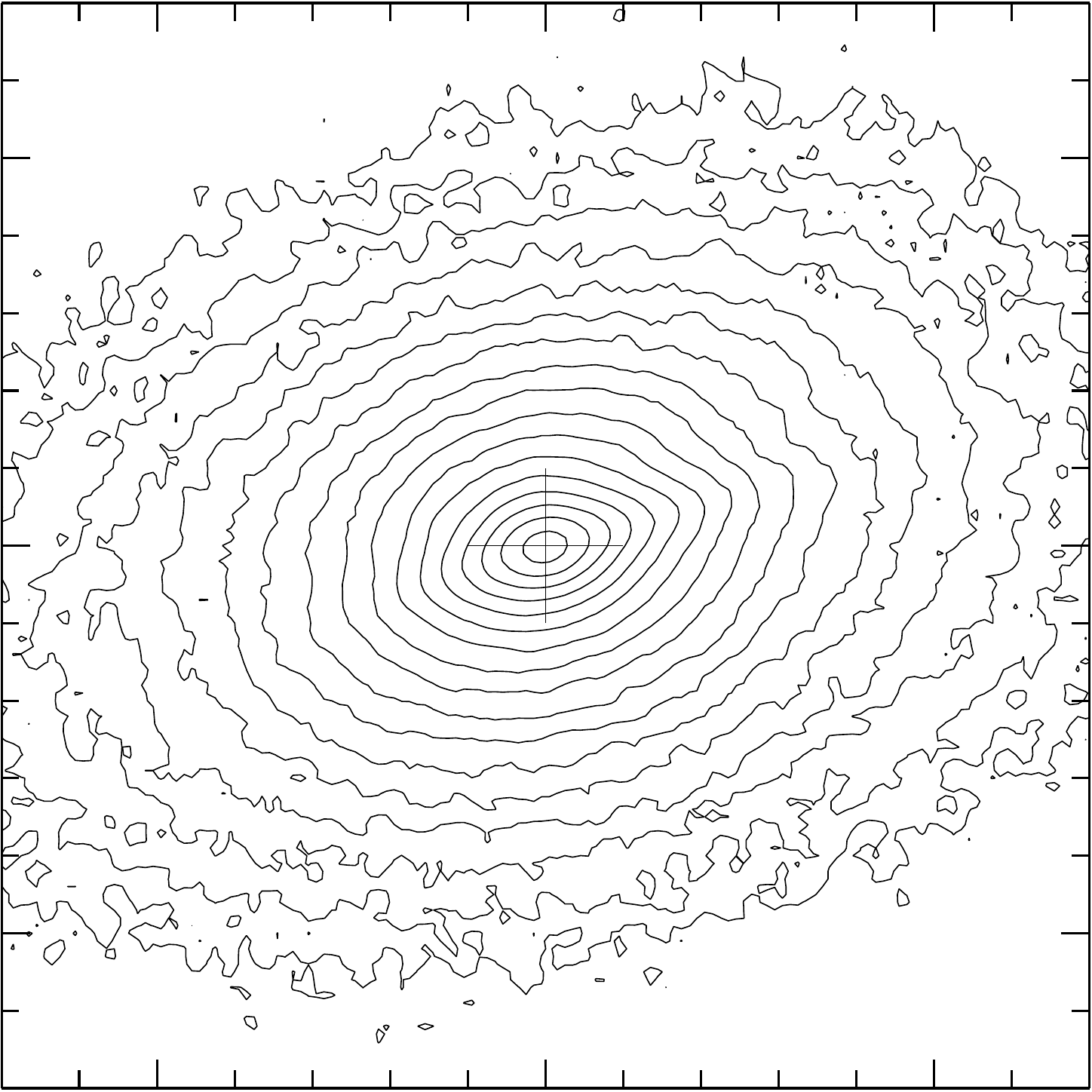}
\vspace{0.2cm}
\caption{K band surface brightness distribution of J031646.7+400013. 
Left: direct image (side length $14$\arcsec). 
Right: contour plot (logarithmic scale). N up, E left.
}
\label{fig:App_398_2}
\end{figure}

\noindent
(1) \ \object{J031646.7+400013} (\#398, IC\,311)\\
This galaxy is classified as a LERG.
In the optical, the eye-catching morphological feature is an unusual absorption structure in an otherwise
nearly featureless system of approximately elliptical shape (Fig.\,\ref{fig:App_398_1} left). 
An extended ($\sim 20$\,kpc), strong dust lane hides the centre of the galaxy. 
A second, slightly weaker dust lane runs roughly parallel.
IC\,311 was sometimes classified as spiral. 
We do not find any evidence of an extended stellar disc in the optical image. 
The global distribution of the stellar light is ellipsoidal, slightly lopsided with some very faint fuzzy structures.
The radial optical surface brightness profile (after subtraction of foreground stars) is nearly perfectly matched by de Vaucouleurs'  $r^{1/4}$ law (Fig.\,\ref{fig:App_398_1} right).
Images of IC\,311  in the J, H and K bands at sub-arcsec resolution were obtained with the OmegaCass camera at the 3.5\,m telescope at Calar Alto  by one of us (D. A.). 
The K-band image  (Fig.\,\ref{fig:App_398_2}) reveals a compact NIR core. In addition, a weak, linearly extended (diameter $\la 2$\,kpc) emission structure is seen that is positionally coincident with the bottom side of the upper dust lane. 
A plausible interpretation of this peculiar morphology is a ring-like or disc-like component containing lots of cold interstellar matter caused by a previous capture of a gas-rich system by a more massive E galaxy, reminiscent of the prototypical nearby AGN galaxy NGC\,5128 (Cen\,A).

\begin{figure}[htbp]
\includegraphics[viewport= -10 0 510 500,width=4.45cm,angle=0]{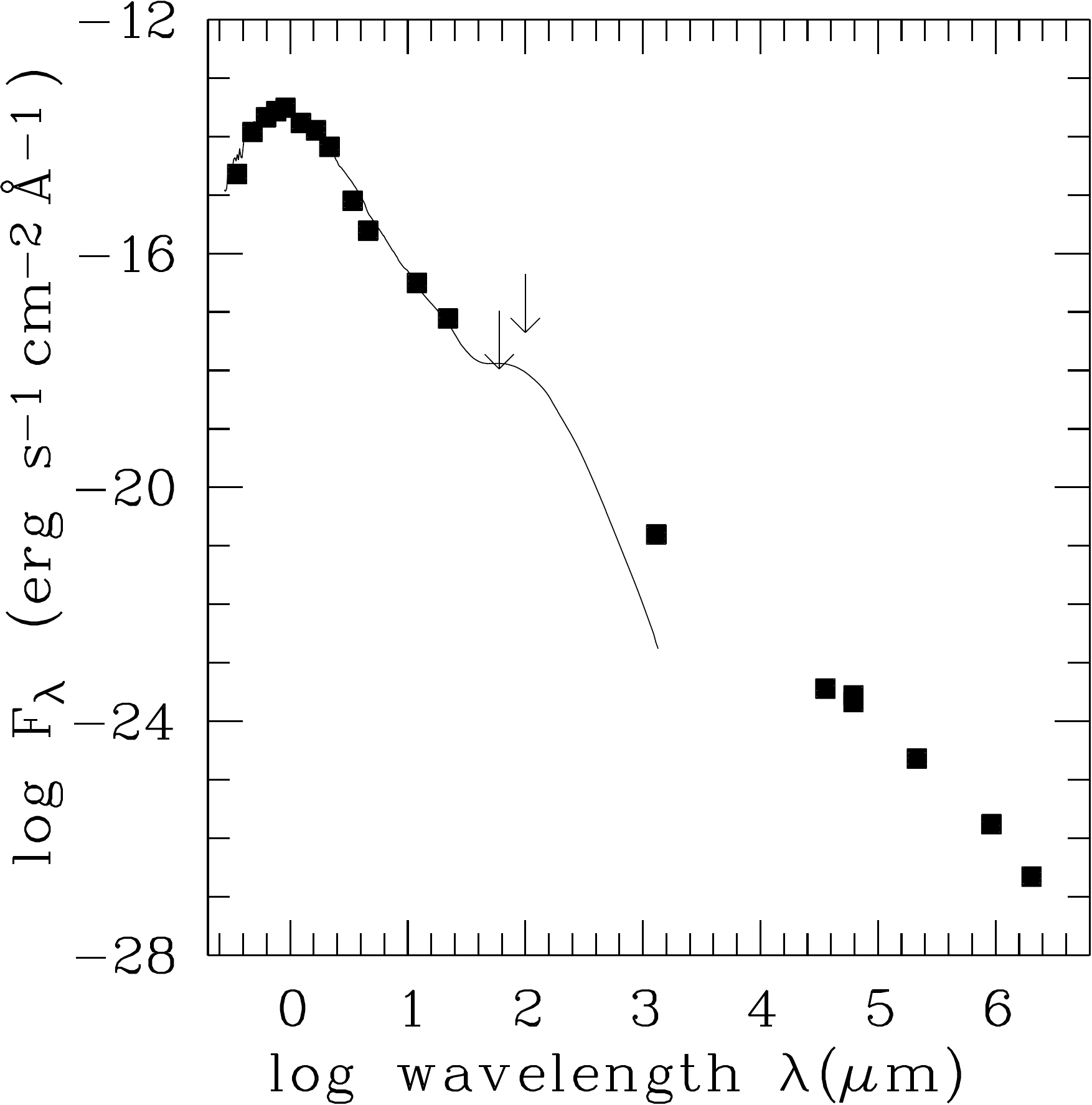}
\includegraphics[viewport= 0 0 500 510,width=4.27cm,angle=0]{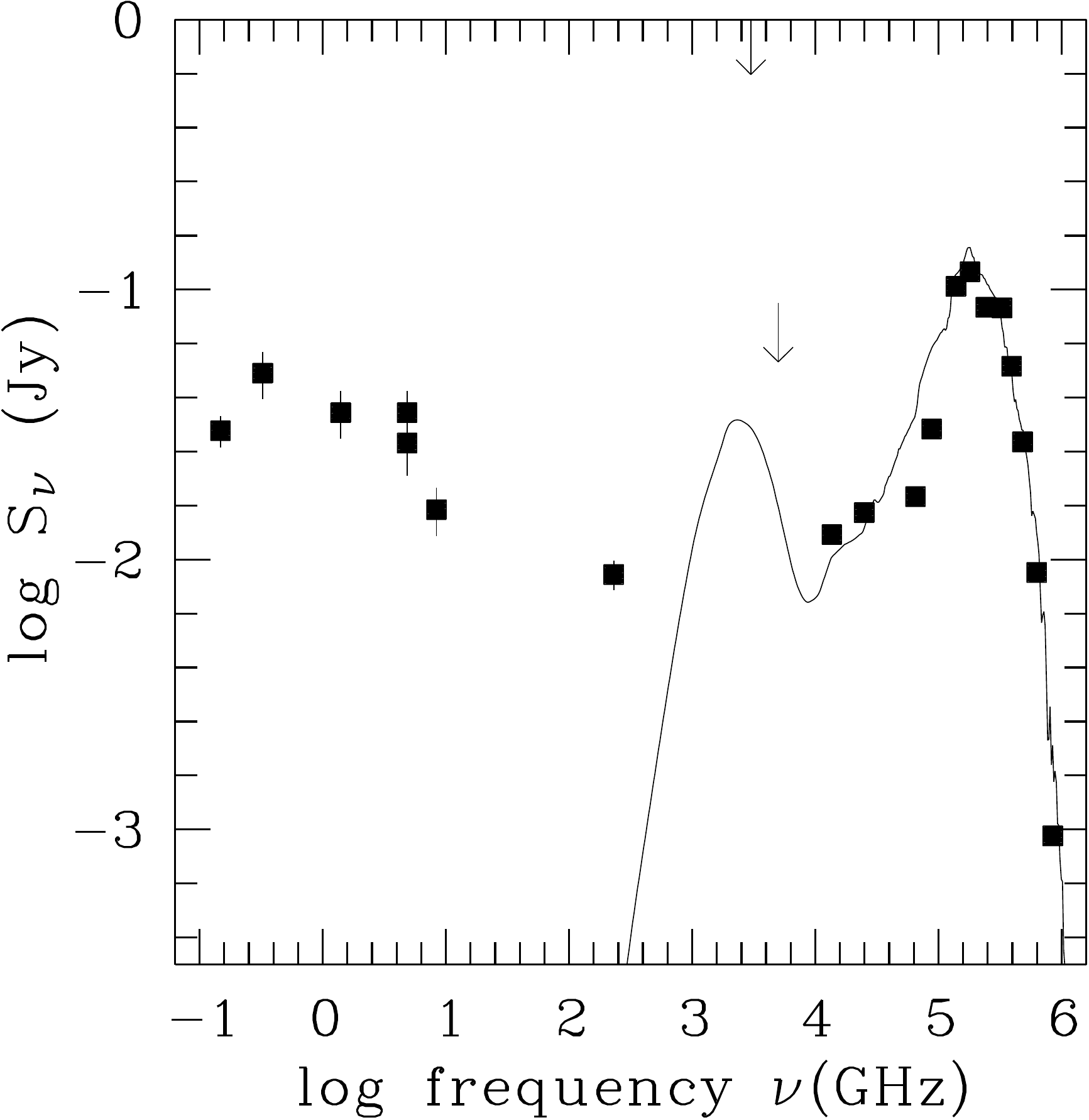}
\vspace{0.2cm}
\caption{Observed SED of J031646.7+400013 in two different styles (filled symbols, downward arrows for upper limits) and fitted galaxy template (solid curve). 
}
\label{fig:App_398_3}
\end{figure}

Figure\,\ref{fig:App_398_3} shows the SED based on broad-band magnitudes from SDSS, 2MASS and WISE. In addition, estimated upper limits of IRAS fluxes at 60\,$\mu$m and 100\,$\mu$m from IPAC XSCANPI \citep[taken from][]{Miller_2001} are plotted. 
For most of the measured fluxes the error bars are smaller than the symbol sizes, but it can be assumed that the real errors are larger because of the low Galactic latitude. We used the SWIRE library of galaxy template spectra from
\citet{Polletta_2007}\footnote{http://www.sedfitting.org/Data.html} 
for a spectroscopic classification. 
Instead of correcting the observed fluxes, the uncorrected data were plotted in the observer frame and the template was redshifted and reddened using the Galactic reddening law with $E(B-V) = 0.175$. 
A 13 Gyr old single stellar population template (Ell 13) provides the best fit in the optical and NIR but is too faint in the MIR. On the other hand, the 2\,Gyr old E (Ell 2) provides the best match from NIR and MIR but has too much emission at shorter wavelengths. As a compromise, we plotted a composite model where the younger (2 Gyr) stellar population contributes 5\% to the K-band flux and is assumed to be reddened with $E(B-V) = 1.0$ (assuming Galactic dust). A more detailed modelling is beyond the scope of the present paper. 

IC\,311 was observed in January 2003 with the bolometer of the 30\,m IRAM telescope on Pico Veleta, Spain.
We derived a continuum flux of $8.7\pm1.1$\,mJy at 1.3\,mm, about two orders of magnitude higher than predicted by the stellar population model. 
NGC\,5128 scaled to the distance of the Perseus cluster would have an only 2.5 times higher flux. 
Thus, IC\,311 seems in fact to be a down-scaled version of NGC\,5128. 
Fitting the SED at 1.3\,mm would obviously require to assume a substantial cold dust component of about $20$\,K \citep[e.g.][]{Braine_1995}.

The centre of IC\,311 is positionally coincident with an unresolved NVSS radio source. 
The SPECFIND V2.0 catalogue \citep{Vollmer_2010} lists six measurements at four different radio frequencies 
(325 MHz, 1.4 GHz, 4.85 GHz, and 8.4 GHz). Three different data are available at 4.85 GHz where 
one data point was excluded by the SPECFIND procedure as the flux does not match the spectrum. 
The best fit of the remaining five data points results in a radio spectral index $\alpha = 0.27$ (for $S_\nu \propto \nu^{-\alpha}$) signifying 
a flat radio spectrum typical of the compact cores of radio AGN. 
The degree of polarisation is low \citep[7\% at 8.4\,GHz,][]{Jackson_2007}.
The clear detection of a point-like source in the radio image from the TGSSADR Image Cutout Service \citep{Intema_2017} provides an additional data point at 148\,MHz. We estimated a flux density of $(30\pm 4)$\,mJy, but we note that the true value might be about 20\% lower because the source is located in a region affected by imaging artefacts. The resulting radio spectrum $S_\nu(\nu)$ (Fig.\,\ref{fig:App_398_3} right) seems to have a peak between one GHz and a few 
hundreds MHz, thus resembling gigahertz peaked-spectrum (GPS) and compact steep-spectrum (CSS) radio sources \citep{O'Dea_1998}. 
Flux densities from the ongoing LOFAR HBA 120-168\,MHz wide-area survey \citep{Shimwell_2019} will be helpful to establish the shape of the spectrum. 
The two main competing models for the GPS and CSS sources are they are young and evolving radio sources or they are old sources confined by interaction with dense gas in their environment \citep{O'Dea_1998}. 
Remarkably, IC\,311 shares several other properties of GPS/CSS AGN: compact radio morphology, low polarisation, low X-ray luminosity, and peculiar morphology of the host galaxy. If confirmed, IC\,311 would be one of nearest members of this rare object type. The largest sample of GPS and CSS radio sources \citep{Liao_2020} contains only two sources at comparable redshift, both at $z=0.025$.
\\

\noindent
(2) \ \object{J031755.2+405536} (\#485, UGC\,2642) 

\begin{figure}[htbp]
\centering
\fbox{\includegraphics[viewport= 0 0 370 350,width=4.1cm,angle=0]{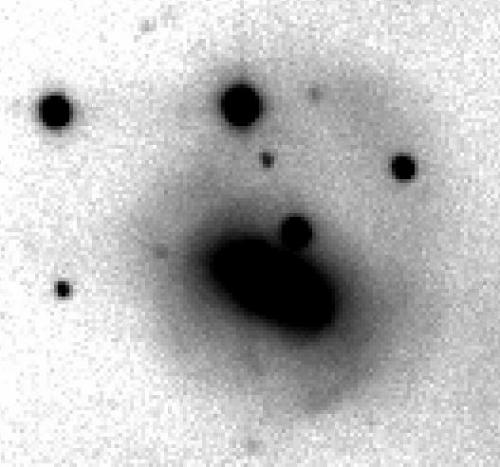}}
\fbox{\includegraphics[viewport= 0 0 260 247,width=4.1cm,angle=0]{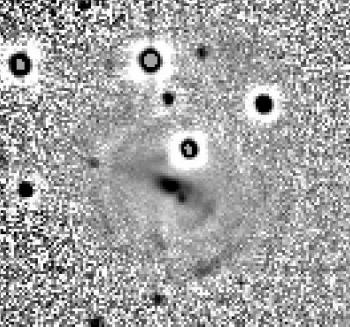}}
\caption[CMD]{
J031755.2+405536 (\#485).
Left: composite Rgri image (side length $\sim 1$\arcmin).
Right: unsharp masked R band image.
N up, E left.} 
\label{fig:App_485}
\end{figure}

\noindent
In addition to its pronounced bar, this galaxy shows an extended lopsided morphology and a system of faint arc-like structures (Fig.\,\ref{fig:App_485}). Both properties most indicate a previous accretion of a dynamically cold stellar system. 
There is perhaps a second, fainter nucleus about $2.5 \arcsec$ ($\sim 0.9$\,kpc) SW of the brighter core. The system is classified as passive based on the SFR data from \citet{Chang_2015} (Fig\,\ref{fig:SFR-M}) and as intermediate based on the WISE colours (Fig.\,\ref{fig:WISE_ccd}). 
However, the optical spectrum positions the core of this galaxy into the sAGN area of the WHAN diagram (Fig.\,\ref{fig:WHAN}).\\

\noindent
(3) \ \object{J031927.4+413807}  (\#656, UGC\,2665)

\begin{figure}[htbp]
\centering
\fbox{\includegraphics[viewport= 0 0 399 399,width=4.1cm,angle=0]{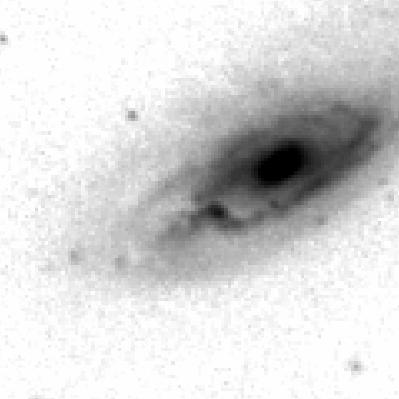}}
\fbox{\includegraphics[viewport= 0 0 426 426,width=4.1cm,angle=0]{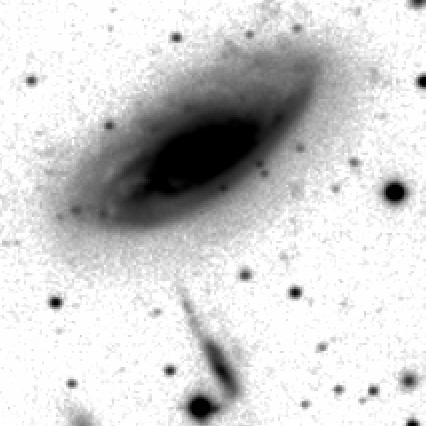}}
\vspace{0.1cm}
\caption[CMD]{
J031927.4+413807 (\#656).
Left: SDSS gr composite of the distorted part of the disc (side length $0\farcm8$). 
Right: Rgri composite image of a larger field  (side length $1\farcm4$) showing the galaxy
together with a smaller neighbour. 
N up, E left.}
\label{fig:App_656}
\end{figure}

\noindent
This is a giant disc galaxy with an irregular, asymmetric spiral pattern starting from a blue inner ring at a galactocentric distance of about $3$\,kpc. The eastern part of the disc, outside the ring, shows some irregular structure (Fig.\,\ref{fig:App_656}, left), which might have been caused by the interaction with the smaller galaxy that is seen to the south of J031927.4+413807 (Fig.\,\ref{fig:App_656}, right). The latter shows an extended low-surface brightness potentially tidal tail that points directly towards the distorted region. 
J031927.4+413807 has a star-like nucleus and is spectroscopically classified as a wAGN. It is detected as a H$\alpha$ source, its SDSS magnitudes are consistent with a green valley galaxy. 
The projected cluster-centric distance amounts to about $180$\,kpc, less than the core radius.

\vspace{0.5cm}
\noindent
(4) \ \object{J031934.2+413450} (\#667, PGC\,12405)

\begin{figure}[htbp]
\centering
\fbox{\includegraphics[viewport= 0 0 350 360,width=3.6cm,angle=0]{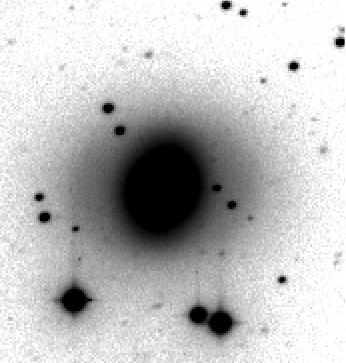}}
\includegraphics[viewport= 0 25 520 450,width=4.8cm,angle=0]{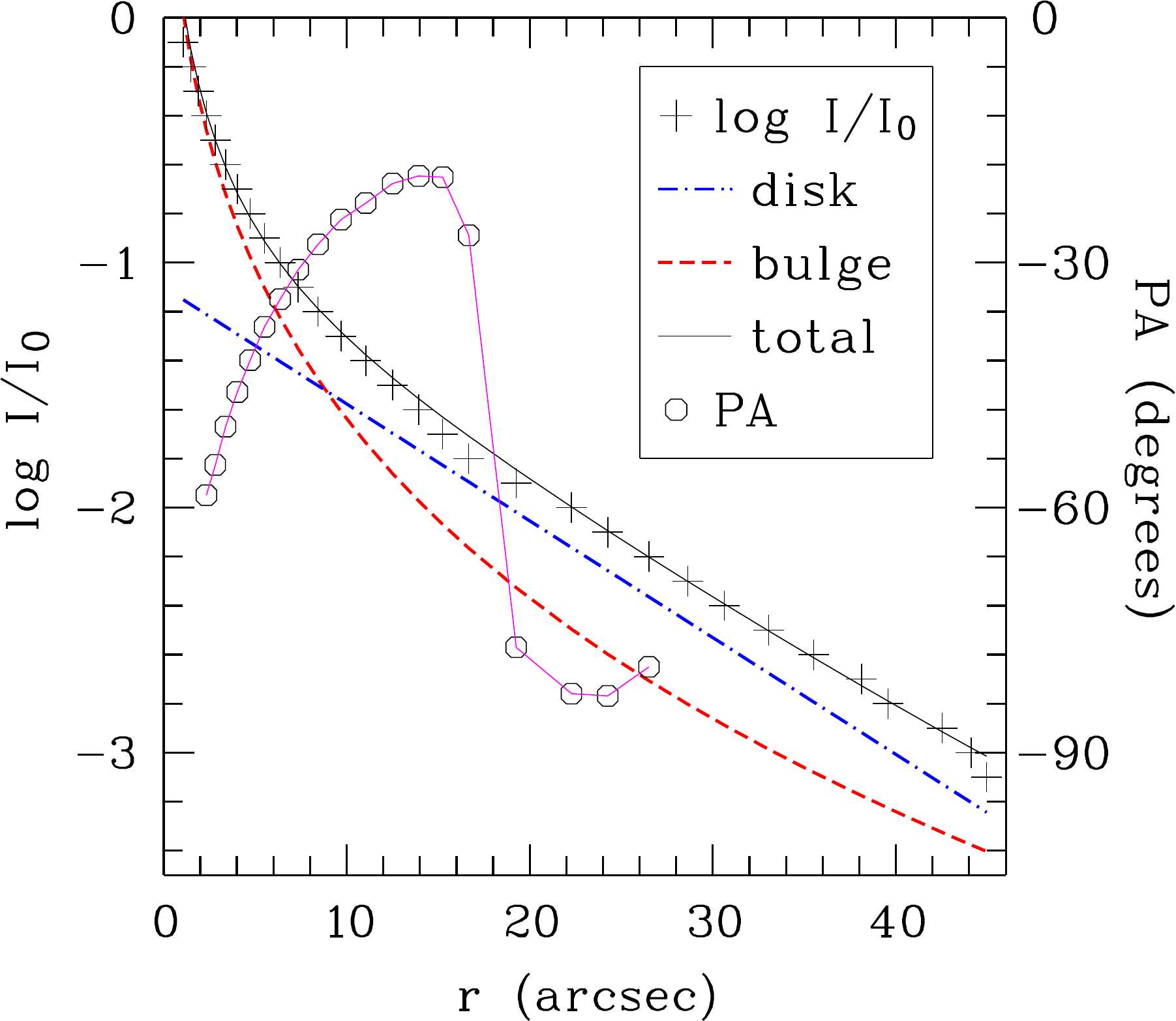}
\vspace{0.2cm}
\caption[CMD]{
J031934.2+413450 (\#667).
Left: R band image at high contrast.
Right: surface brightness and position angle PA as function of distance $r$ from the centre along the major axis after subtraction of the foreground stars. 
N up, E left.}
\label{fig:App_667}
\end{figure}

\noindent
This galaxy was identified with a radio source \citep{Miller_2001} but was excluded from our AGN list after the inspection of NVSS and TGSS images. On the other hand, it was classified as a X-ray AGN by \citet{Santra_2007} and thus it is included here in the AGN subsection.
At first sight, this galaxy looks like an elliptical with a strong isophote twist (Fig.\,\ref{fig:App_667}, left), similar to the Perseus cluster galaxy presented by \citet[][their Fig.\,12]{Conselice_2002}. 
However, the right panel of Fig.\,\ref{fig:App_667} shows that the outer part of the radial surface brightness profile follows an exponential law. As a whole, the radial surface brightness profile is well fitted by the combination of a bright, elongated ellipsoidal bulge with a slightly inclined, extended disc of low surface brightness. 
The major axis of the projected disc is tilted at an angle of about $60\degr$ to the major axis of the bulge.
Non-axialsymmetric bulges are known to efficiently feed gas towards the centre and to trigger nuclear activity 
\citep{Kim_2018}.

\vspace{0.5cm}
\noindent
(5) \ \object{J031954.4+420053} (\#711)

\noindent
Because viewed nearly edge-on, the morphological classification of this galaxy remains uncertain (S/S0). A dust lane is seen crossing the centre, but there is no indication of spiral structure. In the high-contrast image, a small low-surface brightness feature is indicated at the northern edge of the disc. 
The spectral classification as wAGN is based on the WHAN diagram.

\begin{figure}[htbp]
\centering
\fbox{\includegraphics[viewport= 0 0 141 141,width=4.1cm,angle=0]{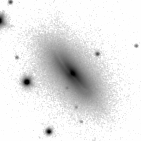}}
\fbox{\includegraphics[viewport= 0 0 141 141,width=4.1cm,angle=0]{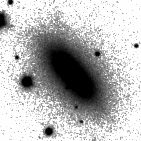}}
\vspace{0.2cm}
\caption[CMD]{
J031954.4+420053 (\#711).
Left: SDSS gri image at low contrast (side length $0\farcm9$).
Right: the same at high contrast.
N up, E left.}
\label{fig:App_281}
\end{figure}

\vspace{0.5cm}
\noindent
(6) \ \object{J032634.7+414143} (\#1231)

\begin{figure}[htbp]
\centering
\fbox{\includegraphics[viewport= 0 0 620 620,width=4.1cm,angle=0]{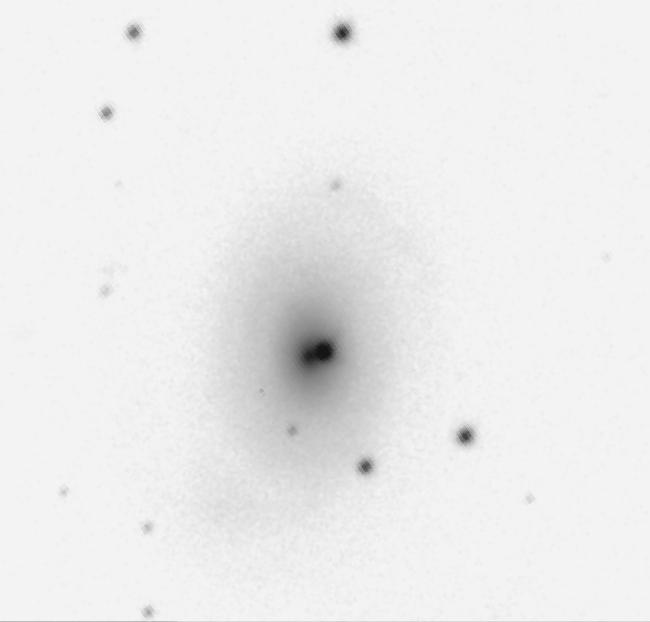}}
\fbox{\includegraphics[viewport= 0 0 350 350,width=4.1cm,angle=0]{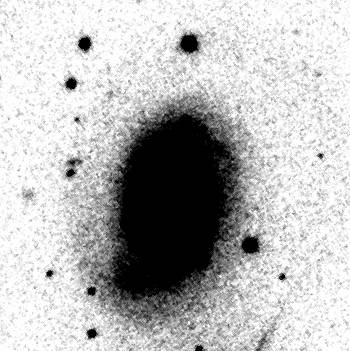}}
\vspace{0.2cm}
\caption[CMD]{
J032634.7+414143 (\#1231).
Left: SDSS gri image at low contrast (side length $1$\arcmin).
Right: the same at high contrast.
N up, E left.}
\label{fig:App_1231}
\end{figure}

\noindent
This HERG is morphologically classified as a peculiar S0 or E. The peculiarities are diffuse extended components at either side of the galaxy (Fig.\ref{fig:App_1231}). At high contrast (right panel), a one-armed spiral structure of low surface brightness becomes visible that surrounds the main body and might indicate a stellar stream from the tidal relic from an infalling smaller system. The star-like object close to the centre seen in the left panel is very likely a foreground star.

\subsection{Individual peculiar SF galaxies}\label{sect:SF_individual}

\begin{figure}[htbp]
\fbox{\includegraphics[viewport= 0 0 625 624,width=4.2cm,angle=0]{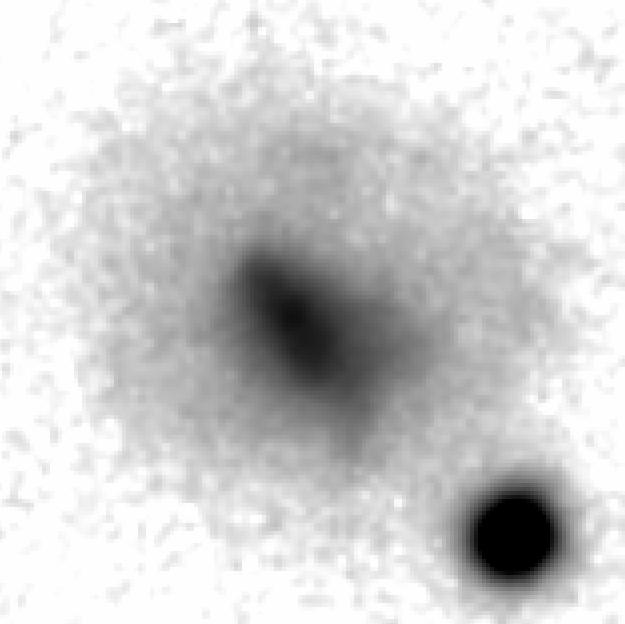}}
\fbox{\includegraphics[viewport= 0 0 565 565,width=4.2cm,angle=0]{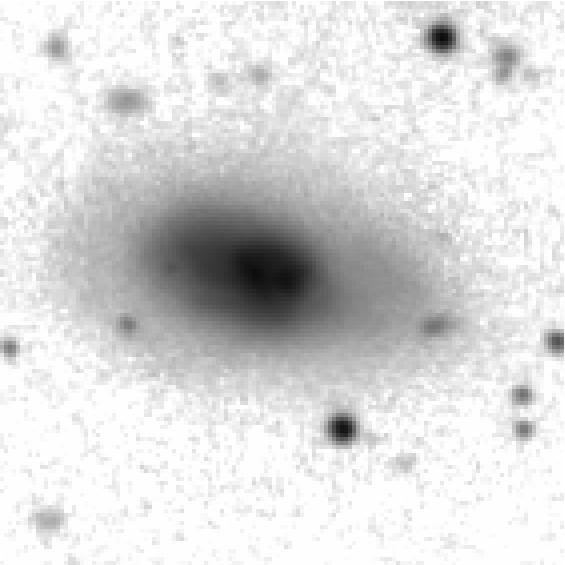}}\\
\fbox{\includegraphics[viewport= 0 0 450 450,width=4.2cm,angle=0]{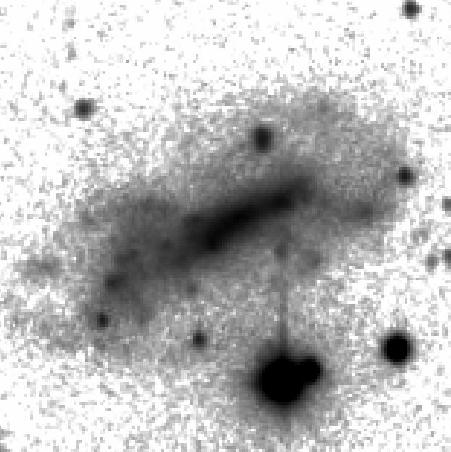}}
\fbox{\includegraphics[viewport= 0 0 348 348,width=4.2cm,angle=0]{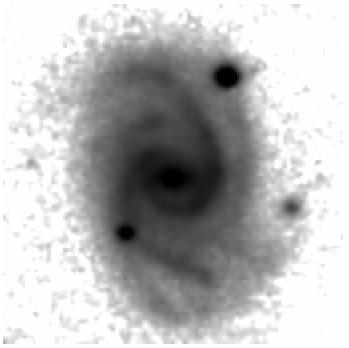}}\\
\fbox{\includegraphics[viewport= 0 0 361 361,width=4.2cm,angle=0]{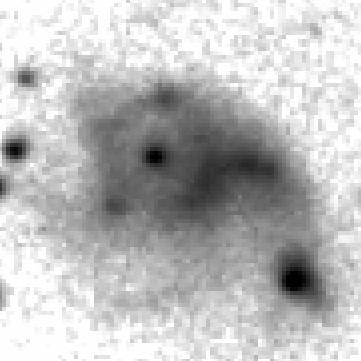}}
\fbox{\includegraphics[viewport= 0 0 438 438,width=4.2cm,angle=0]{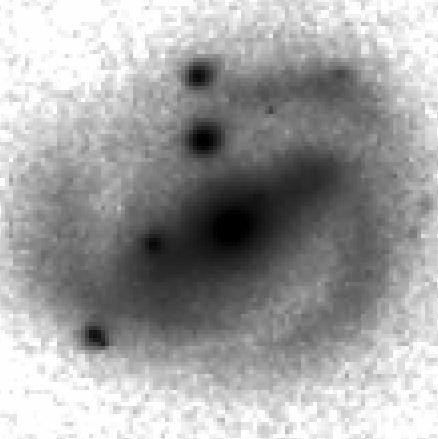}}\\
\vspace{0.2cm}
\caption{Six distorted SF galaxies in the Perseus cluster (logarithmic intensity scale; N up, E left). 
Top (left to right):  J031411.7+413336 ($0\farcm5$), J031633.9+420133 ($0\farcm9$). 
Middle: J031930.3+404430 ($1\farcm0$), J032235.4+412357 ($0\farcm6$).
Bottom: J032700.4+402838 ($0\farcm6$), J032739.9+405349 ($0\farcm7$). 
The number in parentheses is the edge length of the field in arcmin.
} 
\label{fig:SF-pec}
\end{figure}

Figure\,\ref{fig:SF-pec} presents six morphologically peculiar galaxies that do not belong to the IRAS sample but have MIR colours characteristic of active star formation. A short description of the peculiarities is given below.\\

\noindent
(1) \ \object{J031411.7+413336} (\#234)\\
The bright lenticular central component is embedded in an extended low-surface brightness halo that shows 
some brighter lopsided structures in the SW and an arc-like feature in the NW corner. 
These structures probably indicate the late stage of a major merger.\\

\noindent
(2) \ \object{J031633.9+420133} (\#379)\\
This is probably another late major merger.
The inner structure is non-centric and complex with two brightness peaks at a separation of about $4\arcsec$,  
the outer part appears slightly lopsided. With its strong H$\alpha$ emission and substantial Balmer absorption (EW(H$\delta$) = -4\,\AA) this galaxy resembles the type of E+A (post-starburst) galaxies. The SDSS colours place it into the blue cloud. \\

\noindent
(3) \ \object{J031930.3+404430} (\#660, UGC\,2664)\\
This is one of the faintest spectroscopic cluster members ($z = 0.0178$) in our catalogue. 
In the WHAN diagram, this blue galaxy lies in the sAGN part but close to the SF border. Because there is no substantial core and SF extends over the whole galaxy, the classification as a SF galaxy appears to be more applicable. 
It is however not included in the SF sample from Sect.\,\ref{SF_sample} because of its faint magnitude $W2 = 16.01$. 
The morphology is very unusual: a bar-like component with several knots and irregular structures at either end, perhaps forming two ring-like components. \\

\noindent
(4) \ \object{J032235.4+412357} (\#986)\\
The spiral structure of this two-armed spiral galaxy is clearly distorted, perhaps by a minor merger with a small galaxy seen between the core and the northern spiral arm. 
In addition, there seems to be a fainter double of the spiral structure, shifted in western direction by a few arcseconds.  
Because of the relatively small projected cluster-centric distance of $730$\,kpc, one might speculate that this structure is related to ram pressure stripping.\\

\noindent
(5) \ \object{J032700.4+402838} (\#1244)\\
The compact core of this galaxy is surrounded by a concentric ring-like structure with several knots. The SW part of the ring is brighter and seems to be related to a second component to the SW of the core and ring. This very peculiar system is probably a collisional ring galaxy.

\noindent
(6) \ \object{J032739.9+405349} (\#1269)\\
This is an unusual three-armed barred spiral galaxy. Two arms begin at either side of the bar, while the third arm is probably a tidally disrupted infalling galaxy.\\

\subsection{Other conspicuous galaxies}

\begin{figure}[htbp]
\fbox{\includegraphics[viewport= 0 0 504 504,width=4.2cm,angle=0]{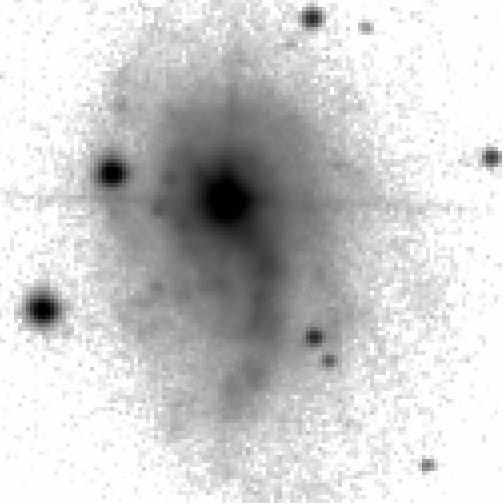}}
\fbox{\includegraphics[viewport= 0 2 380 380,width=4.2cm,angle=0]{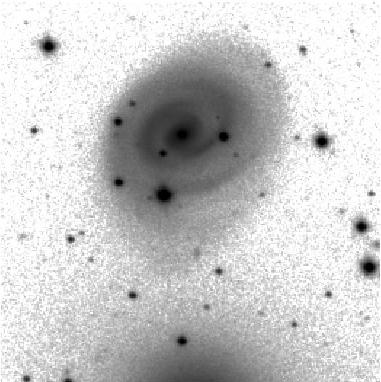}}\\
\fbox{\includegraphics[viewport= 0 0 497 497,width=4.2cm,angle=0]{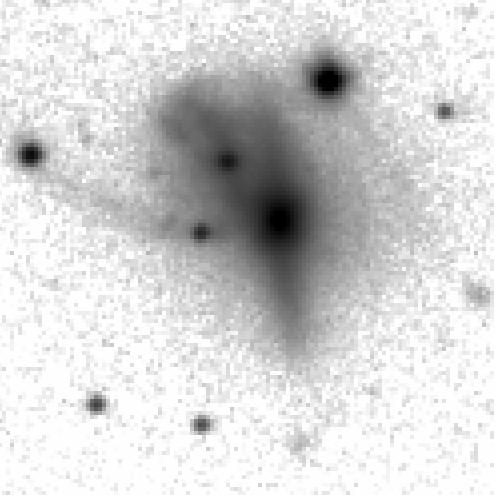}}
\fbox{\includegraphics[viewport= 0 0 474 474,width=4.2cm,angle=0]{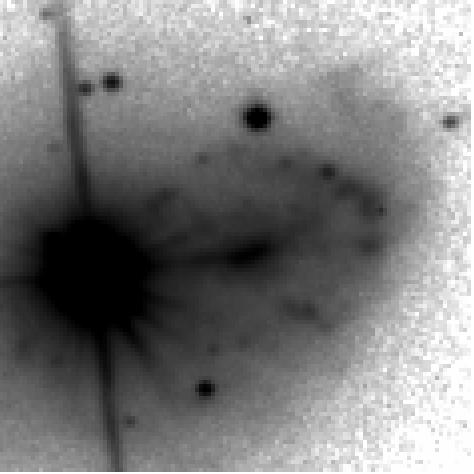}}\\
\fbox{\includegraphics[viewport= 0 0 182 182,width=4.2cm,angle=0]{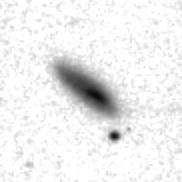}}
\fbox{\includegraphics[viewport= 0 0 520 520,width=4.2cm,angle=0]{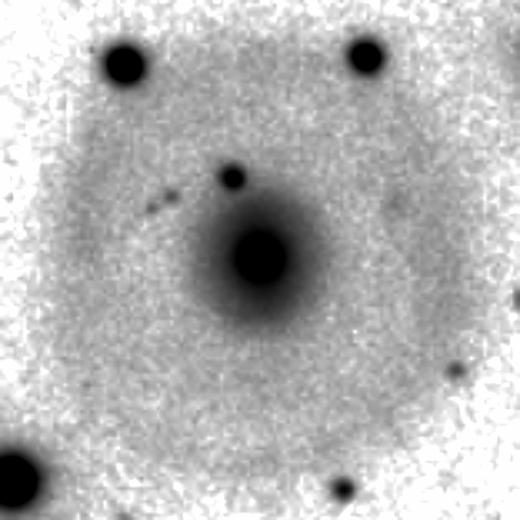}}\\
\vspace{0.2cm}
\caption{Another six peculiar Perseus cluster galaxies (logarithmic intensity scale; N up, E left). 
Top (left to right):  J031821.6+410422 ($0\farcm8$), J031845.2+412919 ($1\farcm8$).
Middle: J031904.7+412807 ($0\farcm8$), J032524.7+403213 ($0\farcm8$).
Bottom: J032810.7+400948 ($0\farcm6$), J032853.7+400211 ($1\farcm0$).
The number in parentheses is the edge length of the field in arcmin.
} 
\label{fig:Passive-pec}
\end{figure}

Figure\,\ref{fig:Passive-pec} presents six morphologically peculiar systems that were not classified as SF galaxies or AGN hosts. A short description of the peculiarities is given below.\\

\noindent
(1) \ \object{J031821.6+410422}  (\#531, BM\,183)\\
The spiral structure of this galaxy is clearly distorted, though a detailed evaluation of the morphology is difficult due to the bright foreground star superimposed with the central region of the galaxy. In the WHAN diagram, the system lies in the SF region. The WISE colours classify it as an intermediate system.\\

\noindent
(2) \ \object{J031845.2+412919}  (\#203, NGC\,1268)\\
The outer arms of this two-armed spiral galaxy are asymmetric, perhaps caused by a tidal interaction with the giant E galaxy NGC\,1267 to the south of NGC\,1268. Both systems belong to the chain of bright galaxies in A\,426. In the WISE colour-colour diagram, NGC\,1277 lies in the intermediate region.\\

\noindent
(3) \ \object{J031904.7+412807}  (\#613, PGC\,12358)\\
This system was originally classified as a merger. However, the complex morphology turned out to be caused by  
the chance superposition of a Perseus cluster galaxy at $z = 0.0134$ with a background system at $z = 0.0519$.
The cluster galaxy is an edge-on disc without any obvious peculiarities. The background system is perhaps a merger.\\

\noindent
(4) \ \object{J032524.7+403213}  (\#1163, PGC\,12797)\\
Although the analysis is hampered by a bright foreground star, the image clearly reveals a ring with a substantial number of SF regions embedded in an irregular structure. The system lies in the SF region of the WHAN diagram and is positionally coincident with a radio source with $q = 2.24$. \\

\noindent
(5) \ \object{J032810.7+400948}  (\#1285, BM\,622)\\
This small galaxy is strongly lopsided. With $W2-W3 = 3.3$ it belongs to the intermediate class but close to the SF regime. \\

\noindent
(6) \ \object{J032853.7+400211}  (UGC\,2751)\\
This galaxy is just outside the area covered by the present catalogue but was listed as a Perseus cluster galaxy by \citet{Brunzendorf_1999}.  
We decided to add it to this selection of peculiar galaxies because it may be another candidate of the very rare class of Hoag-type ring galaxies. The morphology resembles Hoag's object \citep{Hoag_1950} in that it appears to show a nearly complete ring surrounding a roundish core. The outer diameter of the ring is approximately $20$\,kpc if the galaxy ($z = 0.0241$) is at the distance of A\,426. The spectrum of the core is dominated by an old, red stellar population, the ring appears slightly bluer. Possible scenarios for the ring formation in Hoag-type galaxies include resonances in the disc, collision with a small gas-rich galaxy, or the accretion of cold gas \citep{Brosch_1985, Schweizer_1987, Buta_1996, Finkelman_2011, Finkelman_et_al_2011}. 
In contrast to Hoag's object, but similar to the Hoag-type galaxy NGC\,6028 \citep{Wakamatsu_1990},
the core of UGC\,2751 appears to be slightly elongated, reminiscent of a faint bar embedded in a lens and there is luminous material between the core and the ring with a nearly exponential surface brightness profile (scale length $\approx 1.6$\,kpc).

\end{appendix}

\listofobjects

\end{document}